\documentclass[aip,jcp,reprint,superscriptaddress]{revtex4-2}

\usepackage{amsmath}
\usepackage{amssymb}
\usepackage{bm}
\usepackage{booktabs}
\usepackage{graphicx}
\usepackage{listings}
\usepackage[version=4]{mhchem} 
\usepackage{hyperref}
\usepackage[capitalize]{cleveref}
\crefname{appendix}{Appendix}{Appendices}
\Crefname{appendix}{Appendix}{Appendices}

\newcommand{\Exc}{E_\text{xc}}
\newcommand{\exc}{e_\text{xc}}
\newcommand{\vxc}{v_\text{xc}}
\newcommand{\fxc}{f_\text{xc}}
\newcommand{\kxc}{k_\text{xc}}
\newcommand{\lxc}{l_\text{xc}}
\newcommand{\rr}{\mathbf{r}}
\newcommand{\xck}{\textsc{libxckernel}}

\begin{document}

\title{Automatic generation of exchange--correlation response kernels}

\author{Susi Lehtola}
\email{susilehtola@gmail.com}
\affiliation{Department of Chemistry, University of Helsinki,
  P.\,O.\,Box 55, FIN-00014 Helsinki, Finland}

\date{\today}

\begin{abstract}
  Computing density-functional response properties requires
  contracting derivatives of the exchange--correlation (xc) energy
  with perturbed densities on a grid. While the xc functional's
  derivatives have long been available to high orders from libraries
  such as Libxc, the surrounding contraction layer---the chain rule
  that combines them with the grid-based basis-function and density
  data into the total xc contribution---has instead been hand-derived
  and hand-coded in every program, for every functional family, spin
  case, and property. We present \xck{}, a library that automatically
  generates this layer by symbolic differentiation. The xc
  functional's ingredients are expressed as sesquilinear forms in the
  density matrix, so the xc energy can be differentiated to any order
  while the functional derivatives remain opaque. The basis functions
  and the orbital coefficients may both be complex. The chain rule
  then produces any xc matrix element: Fock matrices, orbital
  Hessians, and response terms of any order. Terms sharing a pattern
  are collapsed for efficiency. Extensible plugins emit the
  expressions as NumPy Einstein sums, compiled C kernels with
  C++/Fortran interfaces, or host-program specific source
  code. \xck{} is free and open-source software under the
  BSD-3-Clause license, enabling rapid implementation of
  functionality missing from many electronic structure programs. We
  demonstrate it by extending Psi4 with meta-generalized gradient
  approximation (mGGA) response kernels for orbital stability
  analysis and time-dependent density-functional theory (TD-DFT), GGA
  and mGGA analytic nuclear Hessians, and exact quadrature grid
  response, and by extending GPAW with mGGA and triplet TD-DFT
  kernels and gradient-corrected kernels for periodic dielectric
  response.
\end{abstract}

\maketitle

\section{Introduction}
\label{sec:intro}

Molecular response properties, such as forces on the nuclei,
vibrational frequencies, infrared and Raman intensities, electronic excitation
energies, (hyper)polarizabilities, and nuclear magnetic shielding
tensors, can be
computed as derivatives of the total energy with respect to external
perturbations.\cite{Casida1995__155, Olsen1985_JCP_3235,
  Helgaker2012_CR_543, Norman2018_CR_7208}
In semi-local or (range-separated) hybrid density-functional
theory\cite{Hohenberg1964_PR_864, Kohn1965_PR_1133}
(DFT), this total energy is given by
\begin{equation}
  \label{eq:etot}
  E = E_\text{T} + E_\text{V} + E_\text{J} + E_\text{K} + \Exc
  + E_\text{nuc},
\end{equation}
where $E_\text{T}$ is the kinetic energy, $E_\text{V}$ is the
interaction energy with the external potential, $E_\text{J}$ is the
Coulomb energy, $E_\text{K}$ is the exact-exchange energy (for hybrid
functionals), $\Exc$ is the exchange--correlation (xc) energy, and
$E_\text{nuc}$ is the internuclear repulsion energy, which does not depend on the electronic
structure.

We will now discuss the evaluation of the energy terms. We accomplish
this by discretizing the electronic problem: the unknown Kohn--Sham
orbitals are expanded as a linear combination of known basis
functions $\{\chi_\alpha\}$,
\begin{equation}
  \label{eq:basis}
  \psi_i(\rr) = \sum_\alpha C_{\alpha i} \chi_\alpha(\rr).
\end{equation}
The occupied orbitals then define the density matrix
\begin{equation}
  \label{eq:dm}
  P_{\alpha\beta} = \sum_i o_i C^*_{\alpha i} C_{\beta i},
\end{equation}
where $o_i$ are the orbitals' occupation numbers. The density matrix
in turn determines the electron density everywhere in space as
\begin{equation}
  \label{eq:nr}
  n(\rr) = \sum_{\alpha\beta} P_{\alpha\beta}\, \chi^*_\alpha(\rr)
  \chi_\beta(\rr).
\end{equation}
\Cref{eq:basis,eq:dm,eq:nr} are valid for any type of basis set
(which can be real or complex): for example, atom-centered functions
as in the linear combination of atomic orbitals (LCAO) approach,
finite elements, and plane waves. The orbital coefficients may also
be either real or complex. The expressions in this work are
written in the general complex form; the conjugations can
be omitted for real quantities.
For notational simplicity, we will employ the
spin-restricted formalism, in which the spin-up and spin-down
orbitals coincide; the spin-unrestricted case is analogous.

The occupation numbers enter only through \cref{eq:dm}, and are
otherwise unconstrained. No special handling is required for the
use of fractional occupations, such as in occupation smearing commonly
used in the solid state, nor for the larger occupations
that arise when non-Abelian spatial symmetry is imposed and a single
spatial orbital stands for a degenerate set: an atomic $p$, $d$, or $f$
shell holds up to 6, 10, or 14 electrons,\cite{Lehtola2019_IJQC_25945}
while non-$\sigma$ orbitals ($\pi$, $\delta$, $\varphi$, \ldots) of a
linear molecule hold up to 4.\cite{Lehtola2019_IJQC_25944}

The kinetic and external-potential energies
\begin{equation}
  \label{eq:onel}
  E_\text{T} = \sum_{\alpha\beta} P_{\alpha\beta} T_{\alpha\beta},
  \qquad
  E_\text{V} = \sum_{\alpha\beta} P_{\alpha\beta} V_{\alpha\beta}
\end{equation}
are given in terms of $T_{\alpha\beta}$ and
$V_{\alpha\beta}$, the matrix elements of the kinetic-energy
operator $-\tfrac12 \nabla^2$ and of the external (nuclear attraction)
potential, respectively. The Coulomb and exact-exchange energies
are given by
\begin{align}
  E_\text{J} ={}& \frac{1}{2} \sum_{\alpha\beta\gamma\delta}
  P_{\alpha\beta} P_{\gamma\delta}\, (\alpha\beta|\gamma\delta),
  \label{eq:coul}\\
  E_\text{K} ={}& -\frac{1}{4} \sum_{\alpha\beta\gamma\delta}
  P_{\alpha\beta} P_{\gamma\delta}\, (\alpha\delta|\gamma\beta)_\text{x} .
  \label{eq:exx}
\end{align}
The two-electron integrals
appearing in \cref{eq:coul,eq:exx} are defined by
\begin{equation}
  \label{eq:eri}
\begin{split}
  (\alpha\beta|\gamma\delta)_u = \iint &
  \chi^*_\alpha(\rr) \chi_\beta(\rr)\, u(|\rr - \rr'|) \\
  & \times \chi^*_\gamma(\rr') \chi_\delta(\rr')
  \,\mathrm{d}^3 r \,\mathrm{d}^3 r',
\end{split}
\end{equation}
where $u(r)$ is the interaction kernel: the Coulomb integrals
$(\alpha\beta|\gamma\delta)$ carry the kernel $u(r) = 1/r$, while the
exchange integrals $(\alpha\delta|\gamma\beta)_\text{x}$ carry an
exchange kernel $u_\text{x}(r)$, whose form is part of the definition
of the functional. The present formalism is agnostic to this choice,
and covers semi-local functionals
($u_\text{x}(r) = 0$), global hybrids\cite{Becke1993_JCP_1372,
Becke1993_JCP_5648} such as B3LYP\cite{Stephens1994_JPC_11623}
($u_\text{x}(r) = c_\text{x}/r$, where $c_\text{x}$ is the fraction of
exact exchange), and range-separated
hybrids\cite{Savin1995_IJQC_327, Leininger1997_CPL_151,
Iikura2001_JCP_3540} such as CAM-B3LYP,\cite{Yanai2004_CPL_51}
HSE,\cite{Heyd2003_JCP_8207, Heyd2006_JCP_219906} and
$\omega$B97M-V.\cite{Mardirossian2016_JCP_214110} The
exact-exchange term is not considered further in this work.

Having specified the total energy, we can now discuss the
computation of its derivatives. The one- and two-electron energies
of \cref{eq:onel,eq:coul,eq:exx} are linear and quadratic in the density
matrix $\mathbf{P}$, respectively; their derivatives with respect
to $\mathbf{P}$ are trivial. The topic of this work is the $\mathbf{P}$
dependence of the xc energy
\begin{equation}
  \label{eq:Exc}
  \Exc[n] = \int \exc\big(n(\rr), \gamma(\rr), \nabla^2 n(\rr), \tau(\rr)\big)
  \, \mathrm{d}^3 r ,
\end{equation}
whose ingredients are again simple functions of the density matrix: the
electron density $n$ of \cref{eq:nr} and its Laplacian $\nabla^2
n$; the reduced gradient
\begin{equation}
  \label{eq:gamma}
  \gamma(\rr) = \nabla n(\rr) \cdot \nabla n(\rr),
\end{equation}
and the kinetic energy density
\begin{equation}
  \label{eq:tau}
  \tau(\rr) = \frac{1}{2} \sum_{\alpha\beta} P_{\alpha\beta}\,
  \nabla\chi^*_\alpha(\rr) \cdot \nabla\chi_\beta(\rr).
\end{equation}
The electron density is
always real, and so are all the other ingredients of the xc
functional, such as the reduced gradient of \cref{eq:gamma} and the
kinetic energy density of \cref{eq:tau}. When the basis functions
are real but the orbital coefficients are complex, the density
matrix of \cref{eq:dm} is Hermitian: its real part is symmetric,
and its imaginary part is antisymmetric. In this case, only the real part of the density matrix
contributes to the density of \cref{eq:nr}, the reduced gradient of
\cref{eq:gamma}, and the kinetic energy density of \cref{eq:tau},
since the basis-function term is symmetric when the basis
functions are real.

In contrast to \cref{eq:onel,eq:coul,eq:exx}, \cref{eq:Exc} is
nonlinear in the density matrix, which complicates the evaluation
of its derivatives. Furthermore, the reduced gradient of
\cref{eq:gamma} is quadratic in the density, which further
complicates the derivative expressions.

The nonlinearity of $\exc$ also means that the integral of
\cref{eq:Exc} cannot be evaluated in closed form. Regardless of the
implementation, the xc energy is therefore practically always
evaluated by numerical
quadrature,\cite{Becke1988_JCP_2547, Murray1993_MP_997,
Gill1993_CPL_506, Treutler1995_JCP_346, Stratmann1996_CPL_213,
Towler1996_CPC_181, Boerrigter1988_IJQC_87, Velde1991_PRB_7888,
Velde1992_JCP_84, Franchini2013_JCC_1819, Ihm1979_JPCSSP_4409,
Payne1992_RMP_1045}
\begin{equation}
  \label{eq:quad}
  \Exc \approx \sum_g w_g\, \exc\big(n(\rr_g), \gamma(\rr_g),
  \nabla^2 n(\rr_g), \tau(\rr_g)\big),
\end{equation}
where the $\rr_g$ are the quadrature nodes and $w_g$ the corresponding
weights. Specifically, molecular and periodic LCAO codes typically
build their grids from atom-centered radial and angular quadratures
using fuzzy Voronoi partitioning,\cite{Becke1988_JCP_2547,
  Murray1993_MP_997, Gill1993_CPL_506, Treutler1995_JCP_346,
  Stratmann1996_CPL_213, Towler1996_CPC_181} while true Voronoi
partitionings are also sometimes
employed.\cite{Boerrigter1988_IJQC_87, Velde1991_PRB_7888,
  Velde1992_JCP_84, Franchini2013_JCC_1819} Plane-wave codes evaluate
the semilocal density functional on a uniform real-space
grid.\cite{Ihm1979_JPCSSP_4409, Payne1992_RMP_1045} All relevant DFT
implementations fit the formalism of \cref{eq:quad}, which thus only
presumes that the density and other DFT ingredients are available
pointwise.

With the energy expression in hand, we can turn to the evaluation of
molecular properties, whose starting point is a self-consistent field
(SCF) calculation. The usual approach in molecular electronic
structure calculations within the LCAO approach is to determine the
orbitals from the Roothaan--Hall equations\cite{Roothaan1951_RMP_69,
  Hall1951_PRSAMPES_541}
\begin{equation}
  \label{eq:rh}
  \mathbf{F}(\mathbf{P})\, \mathbf{C} =
  \mathbf{S} \mathbf{C} \boldsymbol{\epsilon},
\end{equation}
where $F_{\alpha\beta}(\mathbf{P})$ is the Fock matrix, $\boldsymbol{\epsilon}$ is the
diagonal matrix of the orbital energies, and the overlap matrix is
\begin{equation}
  \label{eq:S}
  S_{\alpha\beta} = \int \chi^*_\alpha(\rr) \chi_\beta(\rr)
  \,\mathrm{d}^3 r .
\end{equation}
As $\mathbf{F}$ depends on the density matrix, and thereby on the
orbitals, \cref{eq:rh} must be solved
self-consistently; we refer to our recent
works\cite{Lehtola2020_M_1218, Lehtola2025_JPCA_5651} for
discussion of modern solution approaches.

At this stage, we
encounter our first xc kernel: the SCF step requires the xc
contribution to the Kohn--Sham--Fock matrix, which is given by the
first density-matrix derivative of the xc energy, $F^{\rm xc}_{\alpha
  \beta} = \partial E_{\rm xc} / \partial P_{\alpha \beta}$.
In a compact basis, in which the density and Fock matrices can be
stored in memory, the matrix elements of the potential are obtained
by direct differentiation of \cref{eq:quad}:
\begin{equation}
  \label{eq:vxcmat}
\begin{split}
  F^\text{xc}_{\alpha\beta} = \sum_g w_g \bigg[ &
  \frac{\partial \exc}{\partial n}\, \chi^*_\alpha \chi_\beta
  + 2\, \frac{\partial \exc}{\partial \gamma}\,
  \nabla n \cdot \nabla (\chi^*_\alpha \chi_\beta) \\
  & + \frac{\partial \exc}{\partial (\nabla^2 n)}\,
  \nabla^2 (\chi^*_\alpha \chi_\beta) \\
  & + \frac{1}{2}\, \frac{\partial \exc}{\partial \tau}\,
  \nabla \chi^*_\alpha \cdot \nabla \chi_\beta \bigg]_{\rr_g},
\end{split}
\end{equation}
where all factors are evaluated at the grid point $\rr_g$. The working
equations of \cref{eq:vxcmat} were derived by
\citet{Baerends1973_CP_41} for the local density approximation
(LDA), by \citet{Kobayashi1991_PRA_5810} and
\citet{Pople1992_CPL_557} for generalized-gradient
approximation (GGA) functionals, and by \citet{Neumann1996_MP_1} for
meta-GGA (mGGA) functionals; see also the discussion in
Ref.~\onlinecite{Lehtola2020_M_1218} on this topic.

Once the ground-state wave function has been found via an SCF
calculation, ground-state response properties can be obtained by
perturbing the solution. As examples, forces are computed as first
derivatives of the energy with respect to the nuclear
geometry,\cite{Pople1992_CPL_557} while static dipole
polarizabilities are second derivatives of the energy with respect to
a uniform external electric field.\cite{Helgaker2012_CR_543} Many
response properties require even higher derivatives.

A major issue is that the application of the chain rule to compute
the derivatives of \cref{eq:quad} leads to a rapid growth in the
number of terms and in the number of derivatives of the density
functional. For
example, a spin-polarized $\tau$-mGGA functional like
r$^2$SCAN\cite{Furness2020_JPCL_8208, Furness2020_JPCL_9248} has 7 input ingredients
($n_\uparrow$, $n_\downarrow$, $\gamma_{\uparrow \uparrow}$,
$\gamma_{\uparrow \downarrow}$, $\gamma_{\downarrow \downarrow}$,
$\tau_\uparrow$, $\tau_\downarrow$) that lead to 7, 28, 84, and 210
symmetry-unique derivatives of the first, second, third, and fourth
orders, respectively.

The evaluation of these pointwise functional derivatives is
already a solved problem: Libxc\cite{Lehtola2018_S_1} provides
efficient symbolically generated expressions for all of these
derivatives (4th derivatives were added in
version 6).\cite{LibxcWeb} The library is now used by over 50
electronic structure programs.\cite{LibxcWeb} Libxc started out with hand-written implementations of all
functionals and all of their
derivatives,\cite{Marques2012_CPC_2272} but later switched to symbolic
generation of derivatives,\cite{Lehtola2018_S_1} as generating
hundreds of derivative expressions for hundreds of functionals by
hand is a doomed endeavor.

In this work, the motivation that originally drove us to adopt
automatic code generation in Libxc motivates automating the
consumption of the functional derivatives: the expressions needed to
compute response properties are complex and verbose, and error-free
implementations are best obtained by generating them. As already
alluded to above, the derivatives of $\Exc$ with respect to an
arbitrary perturbation reduce to contractions of the pointwise
functional derivatives $\partial^n \exc / \partial n^a\, \partial
\gamma^b\, \partial \tau^c \cdots$ with perturbed densities and
basis-function values on the quadrature grid; these expressions are
straightforward to derive automatically. As all
of the resulting basic mathematical expressions are reusable between
various basis set infrastructures, programming languages and
paradigms, a single extensible open-source\cite{Lehtola2022_WIRCMS_1610}
code generator can be used
to implement any kind of response kernel in any electronic structure
program, eliminating the need for duplicate work that has held back
the progress of the field.\cite{Lehtola2023_JCP_180901}

We thus present \xck{}, a reusable
library\cite{Lehtola2023_JCP_180901} that generates the full
contraction layer
required to compute arbitrary xc kernels. The key organizing
principle is a strict division of labor. An external functional
library (e.g., Libxc\cite{Lehtola2018_S_1} or
XCFun\cite{Ekstroem2010_JCTC_1971}) computes the pointwise
derivative tower
$\partial^n \exc / \partial\{n, \gamma, \nabla^2 n, \tau\}^n$, which
\xck{} treats as opaque per-point input arrays. \xck{} takes in
these derivatives as well as the wave-function and basis-function
data, and combines them together into optimal expressions for the
xc kernels.
These kernel expressions are obtained by mechanical derivation with
the chain and Leibniz rules as an Einstein sum expression over
basis-function values, basis-function derivatives, quadrature
weights, named functional derivatives, and (perturbed) density
matrices.
The library can thus be
viewed as a \emph{symbolic differentiation} engine for xc kernels
that complements Libxc: Libxc supplies the pointwise functional
derivatives, while \xck{} supplies the basis-set contractions that
turn them into matrix elements. In contrast to automatic
differentiation, which propagates numerical derivatives at run time,
\xck{} differentiates the expressions symbolically, once, so that
they can be deduplicated, collapsed, and optimized before any code
is generated.
The final expressions are emitted as executable code via
customizable code writer plugins.

The idea of generating the contraction layer is not entirely without
precedent. After we had finished our implementation and examined codes
that might want to subsume our library, we ran across a small Python
generator in the source tree of DIRAC\cite{Saue2020_JCP_204104}
(\texttt{src/openrsp/generate\_xc\_response.py}), which emits Fortran
code contracting XCFun partial derivatives with perturbed-variable
products through the fifth functional derivative for closed-shell
spin-restricted LDA and GGA calculations, and which powers the
response module of DIRAC. According to DIRAC's git version history, the
script was written by Radovan Bast in less than a month in late
2009; its contents have not changed since. The
tool was never extended to generate
spin-polarized or mGGA response blocks, and they remain
unavailable in DIRAC to this day. The generator is also unable to differentiate the basis functions
or the quadrature grids and weights, and it emits its expressions
term by term, without deduplication or collapse, which---as
quantified in \cref{sec:collapsecount}---becomes unwieldy for
mGGAs. Moreover, the generator does not emit code to perform
contractions, which is still performed using a hand-written
implementation. Also, the generator does not appear to ever have been packaged
for use by other programs.

As to the formal description of the approach, the 2010 paper on the
application of automatic differentiation by
\citet{Ekstroem2010_JCTC_1971} (where Bast was third author) focused
mainly on the evaluation of the pointwise derivatives of the
xc functional itself using the XCFun library described in that
work.\cite{Ekstroem2010_JCTC_1971} In spite of this,
\citet{Ekstroem2010_JCTC_1971} described the problem studied in this
work exactly: since the expansion of the xc potential in the density
variables does not vanish for second- and higher-order corrections,
repeated application of the chain rule leads to a large number of
derivatives of both the functional itself with respect to its
variables, and of these variables with respect to the perturbing field
strengths. Perturbations that modify the overlap of the basis
functions, such as geometric displacements, or magnetic fields when
London orbitals are used, add further derivatives of the basis
functions themselves. Moreover,
\citet{Ekstroem2010_JCTC_1971} note that the xc contributions to the
linear response function had by then been presented in the
literature many times over, citing nine earlier derivations in
varying notations,\cite{Bauernschmitt1996_CPL_454, Jamorski1996_JCP_5134,
Petersilka1996_PRL_1212, Stratmann1998_JCP_8218,
Tozer1998_JCP_10180, Hirata1999_CPL_375, Goerling1999_JCP_2785,
Gisbergen1999_CPC_119, Rinkevicius2003_JCP_34} while their own appendix
adds another, giving the recursion relations for the corresponding
LDA and GGA expressions through quartic response. They anticipated automation,
observing that higher orders and the spin-density
variables could be achieved with code generation.

A third route has also been taken. In the analytic cubic and
quartic force fields of \citet{Ringholm2014_JCP_34103}, the
contractions of the functional derivatives with the perturbed
generalized densities are not programmed at all, but obtained by
feeding the perturbed densities into
XCFun\cite{Ekstroem2010_JCTC_1971} as the coefficients of a
generalized density Taylor series, which the library contracts
internally against the Taylor expansion of the functional. The
automatic differentiation is thereby repeated at every grid point of
every evaluation, and the resulting expressions never exist in a
form that could be inspected, deduplicated, or collapsed.

The layout of the present work is as follows.
\Cref{sec:theory} presents the theory of the xc derivatives and
reviews the related literature on geometric
derivatives (\cref{sec:geomderiv}), stability analysis
(\cref{sec:stability}), time-dependent DFT (\cref{sec:tddft}),
magnetic properties (\cref{sec:magnetic}), higher orders of
response (\cref{sec:higherorders}), the real-space
Kohn--Sham potential (\cref{sec:realspace}), and further
properties and ingredient extensions (\cref{sec:extensions}).
\Cref{sec:implementation} describes the implementation: the
ingredient primitives, the response contractions of arbitrary
order, spin resolution, and the pattern collapse and fusion
(\cref{sec:collapse}), followed by the monomial representation, the
emitter backends, and the kernel catalog (\cref{sec:library}).
\Cref{sec:compdet} lists the computational details of the
demonstration calculations, for the Gaussian-basis programs
(\cref{sec:compgauss}) and for GPAW\cite{Mortensen2024_JCP_92503}
(\cref{sec:compgpaw}).
\Cref{sec:results} presents the results. We first quantify the term
explosion and the effect of the collapse and fusion
(\cref{sec:collapsecount}), and compare the approach to the two
existing alternatives: the generator in
DIRAC\cite{Saue2020_JCP_204104} (\cref{sec:diraccomparison}) and
runtime automatic differentiation
(\cref{sec:xcfuncomparison}). We then demonstrate the new
functionality implemented with the generated kernels in
Psi4\cite{Smith2020_JCP_184108} (\cref{sec:psi4demo})---the new
features themselves (\cref{sec:psi4new}), excitation energies and
stability analysis (\cref{sec:psi4exc}), and geometric derivatives
(\cref{sec:geomresults})---and in GPAW (\cref{sec:gpaw}), comprising
stress tensors (\cref{sec:gpawstress}), excitation energies
(\cref{sec:gpawexc}), and the dielectric response
(\cref{sec:gpawdiel}).
\Cref{sec:conclusions} closes with a summary and discussion,
including new ingredients and future directions
(\cref{sec:newingredients}).
\Cref{app:tmgga} illustrates the verbosity of the hand-derived
expressions that this work supersedes.

\section{Theory}
\label{sec:theory}

We begin by recalling how derivatives of the total energy are
obtained in practice. The energy of \cref{eq:etot} depends on a
perturbation $x$ both explicitly and through the density matrix, so
that the first derivative is given by
\begin{equation}
  \label{eq:dEdx}
\begin{split}
  \frac{\mathrm{d}E}{\mathrm{d}x} &= \frac{\partial E}{\partial x}
  + \sum_{\alpha\beta} \frac{\partial E}{\partial P_{\alpha\beta}}
  \frac{\partial P_{\alpha\beta}}{\partial x} \\
  &= \frac{\partial E}{\partial x}
  + \sum_{\alpha\beta} F_{\alpha\beta}
  \frac{\partial P_{\alpha\beta}}{\partial x},
\end{split}
\end{equation}
where the second step identifies the Fock matrix $F_{\alpha\beta} =
\partial E / \partial P_{\alpha\beta}$ of \cref{eq:rh},
and $\partial \mathbf{P} / \partial x$ is the response of the density
matrix to the perturbation. Evaluation of \cref{eq:dEdx} at the
converged density matrix then yields the property associated with
the perturbation $x$.

First derivatives are simple to evaluate. The first term of
\cref{eq:dEdx} is the
explicit derivative of the energy at a fixed density matrix, which
is assembled from derivative integrals. In the second term, the response
$\partial \mathbf{P} / \partial x$ decomposes into a free part,
which describes the variational freedom of the wave function, and
a part forced by the orthonormality constraint. The free part is a
rotation between the occupied and unoccupied orbitals, and its
contribution vanishes, as the converged Fock matrix is diagonal in
the molecular-orbital basis by \cref{eq:rh}. The forced part arises when the basis
functions depend on the perturbation---as atom-centered basis
functions depend on the nuclear positions, and plane waves depend
on the shape and size of the unit cell---so that the orbital
orthonormality changes. This part is fixed algebraically by the
perturbed overlap matrix, and its contribution reduces to a trace of the
energy-weighted density matrix against the perturbed overlap
matrix (see \cref{sec:geomderiv} for an example). First
derivatives are thereby assembled entirely from unperturbed
quantities.

Higher derivatives, in contrast, also require the response of the
wave function. The partial derivatives in \cref{eq:dEdx} are
themselves functions of the perturbation and of the density
matrix, and the second differentiation acts on them by the
chain rule before any evaluation at a fixed density matrix.
Differentiating \cref{eq:dEdx} once more thus yields
\begin{equation}
  \label{eq:d2E}
\begin{split}
  \frac{\mathrm{d}^2 E}{\mathrm{d}x\, \mathrm{d}y} ={}&
  \frac{\partial^2 E}{\partial x\, \partial y}
  + \sum_{\alpha\beta}
  \frac{\partial^2 E}{\partial x\, \partial P_{\alpha\beta}}
  \frac{\partial P_{\alpha\beta}}{\partial y} \\
  &+ \sum_{\alpha\beta}
  \frac{\partial^2 E}{\partial y\, \partial P_{\alpha\beta}}
  \frac{\partial P_{\alpha\beta}}{\partial x} \\
  &+ \sum_{\alpha\beta\gamma\delta}
  \frac{\partial^2 E}{\partial P_{\alpha\beta}\, \partial P_{\gamma\delta}}
  \frac{\partial P_{\alpha\beta}}{\partial x}
  \frac{\partial P_{\gamma\delta}}{\partial y} \\
  &+ \sum_{\alpha\beta} \frac{\partial E}{\partial P_{\alpha\beta}}
  \frac{\partial^2 P_{\alpha\beta}}{\partial x\, \partial y},
\end{split}
\end{equation}
which involves the first-order responses $\partial \mathbf{P} /
\partial x$ and $\partial \mathbf{P} / \partial y$. These
responses are obtained by solving the coupled-perturbed Kohn--Sham
(CPKS) equations, whose Hartree--Fock counterpart dates back to
Peng\cite{Peng1941_PRSAMPES_499} and
Dalgarno,\cite{Dalgarno1959_PRSAMPES_282} and whose explicit form
will be presented in \cref{sec:magnetic}. The perturbed density matrices are
contracted in \cref{eq:d2E} with mixed
derivatives $\partial^2 E / \partial x\, \partial P_{\alpha\beta}$ and
second density-matrix derivatives $\partial^2 E / \partial
P_{\alpha\beta}\, \partial P_{\gamma\delta}$.

The last term of \cref{eq:d2E}
again contracts the energy gradient $\partial E / \partial
P_{\alpha\beta} = F_{\alpha\beta}$ with a density-matrix response,
and it is handled like the analogous term of \cref{eq:dEdx}: the
free part of the second-order response is eliminated by the
stationarity argument, and the orthonormality-forced part is again
fixed by the derivatives of the overlap matrix. The second-order
response $\partial^2 \mathbf{P} / \partial x\, \partial y$
therefore never needs to be solved for second-order derivatives.

The pattern continues at higher orders, but with an important
structural change. Differentiating
\cref{eq:d2E} once more yields, among terms analogous to those
already encountered, the contraction
\begin{equation}
  \label{eq:d3E}
  \sum_{\alpha\beta\gamma\delta\mu\nu}
  \frac{\partial^3 E}{\partial P_{\alpha\beta}\,
    \partial P_{\gamma\delta}\, \partial P_{\mu\nu}}
  \frac{\partial P_{\alpha\beta}}{\partial x}
  \frac{\partial P_{\gamma\delta}}{\partial y}
  \frac{\partial P_{\mu\nu}}{\partial z}
\end{equation}
of the third density-matrix derivative with three first-order
responses. Because all of the other terms of \cref{eq:etot} are at
most quadratic in the density matrix, the third density-matrix
derivative receives contributions from the xc energy \emph{alone}.
The other terms of \cref{eq:etot} still contribute to the third
total derivative, but only through their first and second
density-matrix derivatives, whose contractions are analogous to
those at lower orders.
From the third order onwards, the new response kernels are thus
purely an xc problem.

Third derivatives govern, for example, first
hyperpolarizabilities and two-photon absorption cross
sections\cite{Gisbergen1998_JCP_10644, Salek2002_JCP_9630,
Norman2018_CR_7208} and
excited-state gradients.\cite{Furche2002_JCP_7433} By the
$(2n+1)$ rule, the responses through order $n$ determine the
energy derivatives through order
$2n+1$.\cite{Hylleraas1930_ZP_209, Helgaker2012_CR_543} Thus, third derivatives are
still determined by the first-order responses, while going to
fourth derivatives, which govern second
hyperpolarizabilities\cite{Gisbergen1998_JCP_10644,
Jansik2005_JCP_54107, Norman2018_CR_7208}
and excited-state Hessians,\cite{Liu2011_JCP_184111} for example,
requires the second-order density-matrix response, which then also
suffices for the fifth derivatives. The third- and higher-order
response implementations are precisely those that few codes
support at the moment.

The compact appearance of \crefrange{eq:dEdx}{eq:d3E} is
deceptive. The equations are written in terms of partial
derivatives of the total energy with respect to the density matrix
and the perturbations, and the complications hide inside these
derivatives of the xc energy. Every density-matrix derivative acts
through
the ingredients of the functional by the chain rule. As discussed
in \cref{sec:intro}, the number of pointwise functional
derivatives grows rapidly with the derivative order. Moreover,
because the reduced gradient $\gamma$ of \cref{eq:gamma} is
quadratic in the density matrix, the product rule generates
further terms at every order: each new derivative acts both on the
functional-derivative factors, raising their order, and on the
density-matrix-dependent ingredient factors left behind by the
previous derivatives. The resulting explosion in the number of
terms is thus not visible in \crefrange{eq:dEdx}{eq:d3E}; it
unfolds once the ingredient definitions are inserted.

The nature
of the perturbation also matters: if the perturbation affects
neither the basis functions nor the quadrature grid, the explicit
xc derivatives $\partial \Exc / \partial x$ and their higher
mixed analogues vanish, as the xc energy then depends on the
perturbation only through the density matrix. If the perturbation
does affect the basis functions, further
contributions arise in which the derivatives act on the basis
functions within the ingredients.
Perturbations that affect the quadrature grid add even more
terms, as the quadrature weights are nonlinear functions of the
nuclear positions, whose derivatives are nonzero at every
order. Building the
derivative stack of \crefrange{eq:dEdx}{eq:d3E} is therefore only
the start of the job: the biggest task lies in going through all
of the various terms that contribute to a given derivative of the
xc energy with respect to the density matrix, the perturbations,
or both. Both of these steps---forming the derivative stack, and
enumerating and assembling the terms it generates---are what
\xck{} automates, as we will discuss in \cref{sec:response}.

In the following, we adopt Libxc's naming convention for the
pointwise derivatives of the xc energy. As was already specified
in \cref{eq:Exc}, $\exc$ is the xc energy density. Its first
derivative (the xc potential) is denoted $\vxc$, the
second (the xc kernel) $\fxc$, and the third and fourth
derivatives $\kxc$ and $\lxc$, respectively. All $n$:th
derivatives are included on an equal footing: for example, $\vxc$
comprises $\partial \exc / \partial n$, $\partial \exc /
\partial \gamma$, $\partial \exc / \partial (\nabla^2 n)$, and
$\partial \exc / \partial \tau$, as well as their individual
spin components, and the higher derivatives are organized
analogously. The higher derivatives are moreover deduplicated
using the symmetry of the mixed derivatives, for example,
$\partial^2 \exc / \partial X\, \partial Y = \partial^2 \exc /
\partial Y\, \partial X$, where $X$ and $Y$ denote any two
ingredients.

We continue the discussion by grounding our work on further
literature to orient the reader on the ecosystem where \xck{}
fits. Specifically, we wish to illustrate the diverse problems
that can be approached with \xck{}. The various applications are
discussed in the chronological order of the original literature.

\subsection{Geometric derivatives}
\label{sec:geomderiv}

Among the first DFT properties to appear were the analytic geometric
derivatives. Their wave-function-theory roots are older still:
analytic forces---first derivatives of the energy with respect to the
nuclear positions---were described for Hartree--Fock theory by Pulay
already in 1969.\cite{Pulay1969_MP_197} Analytic DFT forces were
formulated and first evaluated in the LCAO $X_\alpha$ method by
\citet{Satoko1981_CPL_111} in 1981; the first
implementation applied broadly to molecules, including
transition-metal complexes, was that of
\citet{Versluis1988_JCP_322} in 1988, with the Gaussian-basis
counterpart following in 1989.\cite{Fournier1989_JCP_6371}

For a
variational SCF solution, the force expression takes the form
\begin{equation}
  \label{eq:force}
  \frac{\mathrm{d}E}{\mathrm{d}\mathbf{R}_I} =
  \frac{\partial E}{\partial \mathbf{R}_I}
  - \sum_{\alpha\beta} W_{\alpha\beta}
  \frac{\partial S_{\alpha\beta}}{\partial \mathbf{R}_I},
\end{equation}
where $\mathbf{R}_I$ denotes the coordinates of the $I$-th
nucleus.
\Cref{eq:force} is \cref{eq:dEdx} specialized to nuclear
displacements. The first term differentiates \cref{eq:etot} at a
fixed density matrix; for $\Exc$, this includes the derivatives of
the basis functions,\cite{Versluis1988_JCP_322, Pople1992_CPL_557}
as well as of the quadrature nodes and weights. The quadrature
contributions were recognized early on.
\citet{Fournier1990_JCP_5422} included them in his derivative
theory for fitted local-spin-density energies, pointing out that
the quadrature weights must be smooth functions of the nuclear
coordinates for the derivatives to exist.
\citet{Delley1991_JCP_7245} likewise formulated the quadrature
contributions to the gradient, but omitted them in
practice, noting that the omission leaves a residual of the order
of $10^{-3}$ a.u.\ in the gradient at the energy minimum.

The second term in \cref{eq:force} is the orthonormality-forced part: demanding
that the orbitals stay orthonormal when the basis functions move
and using \cref{eq:rh} leads to \cref{eq:force}. This term is
known as the Pulay term, and it contains the overlap matrix of \cref{eq:S} and the
energy-weighted density matrix
\begin{equation}
  \label{eq:W}
  W_{\alpha\beta} = \sum_i o_i \epsilon_i C^*_{\alpha i} C_{\beta i}.
\end{equation}
The Pulay term vanishes if the perturbation does not
change the basis functions, as would be the case for a uniform
external electric field, for example.

The xc contribution to the first term of \cref{eq:force} consists of
two parts,
\begin{equation}
  \label{eq:excforce}
  \frac{\partial \Exc}{\partial \mathbf{R}_I} = \sum_g \left[
    \frac{\partial w_g}{\partial \mathbf{R}_I}\, \exc(\rr_g)
    + w_g\, \frac{\partial \exc(\rr_g)}{\partial \mathbf{R}_I} \right],
\end{equation}
namely the gradient of the quadrature weight and the gradient of the
xc energy density. The chain rule expands the latter as
\begin{equation}
  \label{eq:excchain}
  \frac{\partial \exc(\rr_g)}{\partial \mathbf{R}_I} =
  \sum_{k \in \{n, \gamma, \nabla^2 n, \tau\}}
  \frac{\partial \exc}{\partial k}\bigg|_{\rr_g}
  \frac{\partial k(\rr_g)}{\partial \mathbf{R}_I}
\end{equation}
in terms of the components of $\vxc$ and the geometric derivatives
of the ingredients. At a fixed density matrix, the ingredient
derivatives consist of basis-function gradients, e.g.,
\begin{equation}
  \label{eq:dndX}
  \frac{\partial n(\rr)}{\partial \mathbf{R}_I} =
  \sum_{\alpha\beta} P_{\alpha\beta}
  \left( \frac{\partial \chi^*_\alpha(\rr)}{\partial \mathbf{R}_I}
  \chi_\beta(\rr)
  + \chi^*_\alpha(\rr) \frac{\partial \chi_\beta(\rr)}{\partial \mathbf{R}_I}
  \right).
\end{equation}
Geometric perturbations thus add
geometry-differentiated basis functions and quadrature-weight
derivatives as further operands of the contractions.

Harmonic vibrational frequencies are computed analogously from the
second geometric derivatives, which contract the singlet $\fxc$ kernel
with the nuclear-perturbed densities of the CPKS equations. The perturbed Hartree--Fock theory for
nuclear displacements was formulated by \citet{Gerratt1968_JCP_1719}
in 1968, with the general spin-orbital formulation and the first
efficient implementation given by \citet{Pople1979_IJQC_225} in 1979.
The Kohn--Sham case followed a decade later:
\citet{Fournier1990_JCP_5422} derived the second and third
derivatives of the local-spin-density energy in 1990,
\citet{Komornicki1993_JCP_1398} gave a comprehensive treatment of
Kohn--Sham gradients and Hessians in January 1993, and
\citet{Johnson1993_CPL_133} reported the first complete
implementation for gradient-corrected functionals in December of the
same year, described in detail in 1994.\cite{Johnson1994_JCP_7429}
\citet{Johnson1993_CPL_133} also showed
that the derivatives of the quadrature weights are important for
precise results, revisiting the conclusion of
\citet{Delley1991_JCP_7245} (we examine this in
\cref{sec:results}).

Anharmonic force fields require the cubic force constants
$\mathrm{d}^3 E / \mathrm{d}\mathbf{R}_I\,
\mathrm{d}\mathbf{R}_J\, \mathrm{d}\mathbf{R}_K$ and their
quartic counterparts, which combine the nuclear-perturbed densities
of the CPKS equations with the $\kxc$ and $\lxc$ kernels and the
geometric collocation operands; the widely used second-order
perturbative treatment instead obtains the semi-diagonal quartic
constants by numerical differentiation of analytic
Hessians.\cite{Barone2004_JCP_14108}
\Citet{Ringholm2014_JCP_34103} reported the first analytic cubic and
quartic force constants at the Kohn--Sham level, using an open-ended
formulation of the energy derivatives in the atomic-orbital basis, to
which we return in \cref{sec:higherorders}.

\subsection{Stability analysis}
\label{sec:stability}

When integer occupation numbers $o_i$ are employed, the orbitals
separate cleanly into occupied and virtual ones. The question of
whether a converged SCF solution is a true minimum is then decided by
the second derivative of the energy with respect to orbital
rotations. The
stability conditions were formulated by
\citet{Thouless1960_NP_225} in 1960, recast in the language of
quantum chemistry by \citet{Cizek1967_JCP_3976} in 1967, and given a
practical classification into internal and external instabilities by
\citet{Seeger1977_JCP_3045} in 1977. The Kohn--Sham analogue was published by Bauernschmitt and
Ahlrichs in 1996.\cite{Bauernschmitt1996_JCP_9047} The
orbital-rotation Hessian is built from the matrices
\begin{align}
  A_{ia,jb} ={}& \delta_{ij}\delta_{ab}(\epsilon_a - \epsilon_i)
  + (ai|jb) - (ab|ji)_\text{x} + K^\text{xc}_{ia,jb},
  \label{eq:Amat}\\
  B_{ia,jb} ={}& (ai|bj) - (aj|bi)_\text{x} + K^\text{xc}_{ia,bj},
  \label{eq:Bmat}
\end{align}
which are given here in the spin-orbital basis. Here, $i, j$ ($a,
b$) label occupied (virtual) orbitals, and the two-electron
integrals of \cref{eq:eri} are understood in the molecular-orbital
basis.

The SCF solution is a local minimum if the stability matrices are
positive definite. For a closed-shell reference, these are the
singlet and triplet forms of $\mathbf{A} + \mathbf{B}$, in which
the two spin channels are rotated in phase and out of phase,
respectively; the triplet form controls the external
(spin-symmetry-breaking) instabilities. The xc contribution consists
of the kernel matrix elements
\begin{equation}
  \label{eq:Kxc}
\begin{split}
  K^\text{xc}_{ia,jb} = \iint & \psi_i(\rr) \psi^*_a(\rr)\,
  \frac{\delta^2 \Exc}{\delta n_\sigma(\rr)\, \delta n_{\sigma'}(\rr')} \\
  & \times \psi^*_j(\rr') \psi_b(\rr')
  \,\mathrm{d}^3 r \,\mathrm{d}^3 r',
\end{split}
\end{equation}
which are likewise given in the spin-orbital basis: $\psi_i$
denotes the spatial part of the spin orbital $i$, and $\sigma$ and
$\sigma'$ are the spins of the orbital pairs $ia$ and $jb$.
Closed-shell spin adaptation produces the singlet and
triplet kernel combinations discussed in \cref{sec:response}.

For the
semi-local functionals of \cref{eq:Exc}, the second functional
derivative only contributes at $\rr = \rr'$, and \cref{eq:Kxc}
collapses to a single quadrature sum.
In the LDA, for example,
\begin{equation}
  \label{eq:KxcLDA}
\begin{split}
  K^\text{xc}_{ia,jb} = \sum_g & w_g\, \psi_i(\rr_g)\, \psi^*_a(\rr_g) \\
  & \times
  \frac{\partial^2 \exc}{\partial n_\sigma\, \partial n_{\sigma'}}(\rr_g)\,
  \psi^*_j(\rr_g)\, \psi_b(\rr_g).
\end{split}
\end{equation}
This is a
contraction of the molecular orbital values $\psi_p(\rr_g) =
\sum_\alpha C_{\alpha p} \chi_\alpha(\rr_g)$---that is, of the molecular-orbital
coefficients and the atomic basis-function values---against the
pointwise second derivative $\partial^2 \exc / \partial n_\sigma\,
\partial n_{\sigma'}$, the LDA form of $\fxc$. For GGAs and mGGAs, the chain rule
brings in the derivatives of $\exc$ with respect to $\gamma$,
$\nabla^2 n$, and $\tau$, together with basis-function gradients.
Instead of the single pointwise factor of \cref{eq:KxcLDA}, the
quadrature sum then contains one term for each second derivative of
$\exc$: for instance, 28 for a spin-polarized $\tau$-mGGA, as counted in
\cref{sec:intro}. Each term is contracted with its own combination
of orbital values and orbital gradients; the fully expanded matrix
elements for the spin-polarized $\tau$-mGGA case are given in
\cref{app:tmgga}.

The orbital-rotation Hessian is also the engine of \emph{direct
minimization} of the total energy, in which the orbitals are
parameterized by exponential rotations $\mathbf{C}(\boldsymbol{\kappa}) =
\mathbf{C} \exp(\boldsymbol{\kappa})$\footnote{Two conventions
  coexist in the literature: the rotations are parametrized either as
  $\exp(\boldsymbol{\kappa})$, as here, or as
  $\exp(-\boldsymbol{\kappa})$. The choice flips the sign of the
  orbital gradient---and of any odd-order orbital derivative---while
  the Hessian at the expansion point is invariant, being even in
  $\boldsymbol{\kappa}$. As the convention is often left unstated,
  \xck{} carries the sign as an explicit parameter.} and the energy is minimized in the
rotation parameters $\boldsymbol{\kappa}$. We have recently described a
reusable open-source implementation of trust-region orbital
optimization in Ref.~\onlinecite{Greiner2026_JCTC_881}.
The second-order xc contribution to the orbital-rotation Hessian
consists of the same $\mathbf{A}$ and $\mathbf{B}$ matrices of
\cref{eq:Amat,eq:Bmat}. Using this solver in a DFT
calculation will therefore require exactly the second-order xc
kernels discussed in this work: the kernels that power the
stability analysis also enable the use of quadratically convergent
solvers.

\subsection{Time-dependent DFT}
\label{sec:tddft}

In time-dependent DFT (TD-DFT),\cite{Runge1984_PRL_997} the
excitation energies $\Omega$
are obtained from the poles of the linear response of the ground
state. Together with the transition amplitudes $(\mathbf{X},
\mathbf{Y})$, they follow from the generalized eigenvalue problem
known as the Casida equation\cite{Casida1995__155}
\begin{equation}
  \label{eq:casida}
  \begin{pmatrix} \mathbf{A} & \mathbf{B} \\
    \mathbf{B} & \mathbf{A} \end{pmatrix}
  \begin{pmatrix} \mathbf{X} \\ \mathbf{Y} \end{pmatrix}
  = \Omega
  \begin{pmatrix} \mathbf{1} & \mathbf{0} \\
    \mathbf{0} & -\mathbf{1} \end{pmatrix}
  \begin{pmatrix} \mathbf{X} \\ \mathbf{Y} \end{pmatrix}.
\end{equation}
\Cref{eq:casida} is built from the same matrices $\mathbf{A}$ and
$\mathbf{B}$ of \cref{eq:Amat,eq:Bmat} as the stability
analysis,\cite{Bauernschmitt1996_CPL_454,
Bauernschmitt1996_JCP_9047, Bauernschmitt1997_CPL_573}
likewise assuming integer occupations. The full
eigenvalue problem of \cref{eq:casida} is also known as the
random-phase approximation (RPA). Hirata and
Head-Gordon introduced the Tamm--Dancoff approximation (TDA) to
TD-DFT in 1999, obtained by setting $\mathbf{B} = \mathbf{0}$ in
\cref{eq:casida}.\cite{Hirata1999_CPL_291}

The use of the ground-state kernel of
\cref{eq:Kxc} in \cref{eq:casida} constitutes the adiabatic
approximation: the xc
kernel of the exact linear response theory is frequency
dependent,\cite{Gross1985_PRL_2850} but it is replaced by the
frequency-independent second functional derivative of the
ground-state xc energy. The adiabatic approximation neglects memory
effects, and it misses certain classes of excitations, such as
states of double-excitation character.\cite{Maitra2016_JCP_220901}
It is nonetheless the standard choice both in molecular
calculations and in the solid state.
All xc kernels discussed in this work are adiabatic.

Although the discussion above is written in terms of occupied and
virtual orbitals, and is thereby limited to integer occupations,
Casida's original formulation of
\cref{eq:casida} already anticipated fractional occupation
numbers: the matrix elements were weighted by the
occupation-number differences of the orbital
pairs.\cite{Casida1995__155}

Real-time TD-DFT offers an alternative to the linear-response
formulation: the time-dependent Kohn--Sham state is propagated in
time, which only requires Hamiltonian builds, that is,
\cref{eq:vxcmat} evaluated at every time
step.\cite{Yabana1996_PRB_4484} Fractional
occupations are feasible also in this approach, where they
enter as the fixed weights of the propagated orbitals in the
density.

Excitation energies are not the only quantities the response
equations yield. Excited-state
nuclear gradients---first geometric derivatives of the excitation
energies---were first obtained by Van Caillie and
Amos,\cite{VanCaillie1999_CPL_249, VanCaillie2000_CPL_159} and were
cast in the general variational, spin-unrestricted form used here by
Furche and Ahlrichs in 2002.\cite{Furche2002_JCP_7433} Because the excitation energy
depends on the ground-state orbitals, its gradient requires the
orbital response, which is obtained from a Z-vector
equation.\cite{Handy1984_JCP_5031, Furche2002_JCP_7433} The right-hand side of this
equation contains the third xc derivative $\kxc$ contracted with two
transition density matrices. In analogy to \cref{eq:Kxc}, the
corresponding matrix elements read
\begin{equation}
  \label{eq:Kxc3}
\begin{split}
  K^{\text{xc},(3)}_{ia,jb,kc} = \iiint &
  \psi_i(\rr) \psi^*_a(\rr)\,
  \psi^*_j(\rr') \psi_b(\rr') \psi^*_k(\rr'') \psi_c(\rr'') \\
  & \times
  \frac{\delta^3 \Exc}{\delta n_\sigma(\rr)\, \delta n_{\sigma'}(\rr')\,
    \delta n_{\sigma''}(\rr'')} \\
  & \times \mathrm{d}^3 r \,\mathrm{d}^3 r' \,\mathrm{d}^3 r'',
\end{split}
\end{equation}
which again collapse to single quadrature sums for semilocal
functionals.

\subsection{Magnetic properties}
\label{sec:magnetic}

Nuclear
magnetic resonance (NMR) shieldings are mixed second derivatives of
the energy,
\begin{equation}
  \label{eq:shielding}
  \sigma^I_{st} = \frac{\mathrm{d}^2 E}{\mathrm{d} B_s\,
    \mathrm{d} m^I_t},
\end{equation}
with respect to the external magnetic field $\mathbf{B}$ and the
magnetic moment $\mathbf{m}^I$ of nucleus $I$; the magnetizability
\begin{equation}
  \label{eq:magnetizability}
  \xi_{st} = -\frac{\mathrm{d}^2 E}{\mathrm{d} B_s\, \mathrm{d} B_t}
\end{equation}
is the corresponding second derivative with respect to the field
alone. The field enters the electronic Hamiltonian by minimal
substitution,
\begin{equation}
  \label{eq:hmag}
  \hat{h}(\mathbf{B}) = \frac{1}{2} \left[ -i\nabla +
    \mathbf{A}(\rr) \right]^2 + V(\rr),
\end{equation}
where the vector potential of the uniform field,
\begin{equation}
  \label{eq:Afield}
  \mathbf{A}(\rr) = \frac{1}{2c}\, \mathbf{B} \times (\rr -
  \mathbf{R}_O),
\end{equation}
depends on an arbitrary gauge origin $\mathbf{R}_O$. The nuclear
magnetic moments enter analogously through their own vector
potentials,
\begin{equation}
  \label{eq:Anuc}
  \mathbf{A}^I(\rr) = \frac{1}{c}\,
  \frac{\mathbf{m}^I \times (\rr - \mathbf{R}_I)}
       {|\rr - \mathbf{R}_I|^3},
\end{equation}
which are centered on the nuclei and thereby carry no gauge
ambiguity.

A challenge to notice here is
the gauge-origin dependence of the vector potential in
\cref{eq:Afield}: although exact results are independent of the
gauge origin, results in finite basis sets are not. In the
LCAO context, the dependence on the gauge origin can be eliminated
to first order with the field-dependent London orbitals, also
known as gauge-including atomic orbitals
(GIAOs),\cite{London1937_JPlR_397, Ditchfield1974_MP_789,
Wolinski1990_JACS_8251}
\begin{equation}
  \label{eq:london}
  \chi_\alpha^{\mathbf{B}}(\rr) = \exp\left[ -\frac{i}{2c}\,
    (\mathbf{B} \times \mathbf{R}_\alpha) \cdot \rr \right]
  \chi_\alpha(\rr),
\end{equation}
where $\mathbf{R}_\alpha$ is the center of the basis function and
$c$ is the speed of light in atomic units. The London orbitals add
an explicit field dependence of the basis functions on top of that
of the Hamiltonian. DFT implementations of GIAOs appeared in the
mid-1990s,\cite{Schreckenbach1995_JPC_606, Lee1995_JCP_10095,
Cheeseman1996_JCP_5497} shortly after the first DFT calculations of
NMR shieldings, which employed the individual gauge for localized
orbitals instead.\cite{Malkin1993_CPL_87, Malkin1994_JACS_5898}

Following \cref{eq:d2E}, both the shielding and the
magnetizability require the response of the density matrix to the
field, which is obtained from the CPKS equations. For a static perturbation $x$, these take the form
\begin{equation}
  \label{eq:cpks}
  (\mathbf{A} \pm \mathbf{B})\, \mathbf{U}^x = \mathbf{b}^x,
\end{equation}
where $\mathbf{A}$ and $\mathbf{B}$ are the matrices of
\cref{eq:Amat,eq:Bmat}, again in the spin-orbital basis, the $+$
sign applies to real and the $-$ sign to imaginary perturbations,
$\mathbf{U}^x$ collects the occupied--virtual response
coefficients, and the right-hand side
$\mathbf{b}^x$ is assembled from the derivative Fock matrix. The Hartree--Fock
counterpart of \cref{eq:cpks} was derived in finite basis sets by
\citet{Stevens1963_JCP_550}, who considered both real and
imaginary perturbations and applied the equations to the magnetic
susceptibility and shieldings of LiH. The modern matrix
formulation is due to \citet{Gerratt1968_JCP_1719}, which was
generalized to
spin orbitals by \citet{Pople1979_IJQC_225}.

The magnetic perturbation
is imaginary. Because the kernel matrix elements of \cref{eq:Kxc}
enter \cref{eq:Amat,eq:Bmat} identically, they cancel in
$\mathbf{A} - \mathbf{B}$. Thus, the magnetic response involves
no semilocal $\fxc$ contribution at all.

The vanishing of the exchange--correlation part of the magnetic Hessian for functionals
of the density alone was the motivation for current-density functional
theory,\cite{Vignale1987_PRL_2360, Vignale1989_PRL_115} in which
the xc energy is a functional of the density and the vorticity
$\bm\nu = \nabla \times (\mathbf{j}_p / n)$. The vorticity dependence enters as an additive
term in the local approximation,
$\Exc[n, \bm\nu] = \Exc[n] + \int g(n)\, |\bm\nu(\rr)|^2
\,\mathrm{d}^3 r$, so that the magnetic response acquires an xc
contribution already at the LDA level, where no kinetic energy
density exists. \Citet{Lee1994_CPL_225} used this route to report
the first magnetizabilities from a CPKS calculation.

The xc terms do always enter through the field derivatives of the
London orbitals of \cref{eq:london}: wherever the contraction
features basis functions tabulated at the quadrature points, these
are replaced by their field-perturbed counterparts. The
explicit field derivatives of the xc Fock matrix with London
orbitals are kernels generated by \xck{} in the catalog of
\cref{sec:implementation}, in both the restricted and the
unrestricted cases---the restricted kernels serve the shieldings
and magnetizabilities of closed-shell molecules, while the
unrestricted ones are needed, for example, for the
shieldings and $g$-tensors of open-shell species---and they are
validated against finite differences with numerically phased
basis functions.

As \cref{eq:shielding} is a mixed second derivative, the response
equations can be solved for either perturbation, which is known as
the interchange theorem of double perturbation
theory.\cite{Dalgarno1958_PRSAMPES_245} Differentiating
first with respect to the field requires CPKS solutions for only
the three Cartesian components of $\mathbf{B}$, whereupon the
shieldings of all nuclei follow at once, and the magnetizability
of \cref{eq:magnetizability} can be evaluated from the same
response densities; this is the standard
route of molecular implementations.\cite{Ditchfield1974_MP_789,
Wolinski1990_JACS_8251}

Differentiating first with respect to the
nuclear moments instead requires three CPKS solutions per nucleus,
but the perturbations of \cref{eq:Anuc} are localized around the
nuclei, whereas the field perturbation is global. This reversal
was pioneered in the solid state as the converse
approach;\cite{Thonhauser2009_JCP_101101} in the molecular
context, the locality can be exploited through screening to
compute the shieldings of selected nuclei in large systems at
sublinear cost.\cite{Beer2011_JCP_74102}

The same freedom appears in other mixed responses: optical rotation, for example,
can be computed from either the electric or the magnetic
response, with time-periodic magnetic-field-dependent basis
functions making both routes gauge-origin
independent.\cite{Krykunov2005_JCP_114103}

Indirect spin--spin couplings are mixed second derivatives with
respect to two nuclear magnetic moments, and involve real triplet
perturbations, namely the Fermi-contact and spin-dipole
operators.\cite{Helgaker1999_CR_293, Sychrovsky2000_JCP_3530,
Helgaker2000_JCP_9402} Hyperfine couplings are first derivatives
with respect to a single nuclear magnetic moment, and are obtained in
the spin-unrestricted formalism as expectation values over the
converged spin density,\cite{Arbuznikov2002_PCCP_5467} so no response
equation enters. The spin--spin CPKS equations of
\cref{eq:cpks} carry the triplet kernel, i.e., the
$f^{\uparrow\uparrow} - f^{\uparrow\downarrow}$ combination
discussed in \cref{sec:response}. The
magnetic response of density functionals moreover introduces the
paramagnetic current density as a further
ingredient,\cite{Lee1995_JCP_10095, Tellgren2014_JCP_34101} since the
kinetic energy density built from the canonical momentum is not gauge
invariant in a magnetic field, whereby a $\tau$-dependent functional
would yield gauge-origin dependent results. Gauge
consistency is restored for such functionals
through the current-corrected kinetic energy
density.\cite{Dobson1993_JCP_8870, Maximoff2004_CPL_408} Current-density kernels
will be discussed in \cref{sec:conclusions}.

\subsection{Higher orders of response}
\label{sec:higherorders}
The CPKS equation~\eqref{eq:cpks} is the first member of a
hierarchy.\cite{Olsen1985_JCP_3235, Helgaker2012_CR_543} At every
order of perturbation theory the operator on the left-hand side is
the same---the orbital-rotation Hessian $\mathbf{A} \pm
\mathbf{B}$, whose xc content is the $\fxc$ contraction---and only
the right-hand side changes,
\begin{equation}
  \label{eq:cpksn}
  (\mathbf{A} \pm \mathbf{B})\, \mathbf{U}^{x_1 \cdots x_n}
  = \mathbf{b}^{x_1 \cdots x_n}.
\end{equation}
The right-hand sides collect exclusively known
quantities:\cite{Christiansen1998_IJQC_1,
Thorvaldsen2008_JCP_214108} perturbed integrals, products of
lower-order responses, and higher functional derivatives contracted
with the already-computed lower-order perturbed densities. Writing the xc part of
$\mathbf{b}$ at second order explicitly,
\begin{equation}
  \label{eq:rhsxc}
\begin{split}
  b^{\text{xc},xy}_{\alpha\beta} = {}&
  \sum_{\gamma\delta} \sum_{\varepsilon\zeta}
  \frac{\partial^3 \Exc}
       {\partial P_{\alpha\beta}\, \partial P_{\gamma\delta}\,
        \partial P_{\varepsilon\zeta}}\,
  P^{x}_{\gamma\delta}\, P^{y}_{\varepsilon\zeta} \\
  & + \sum_{\gamma\delta}
  \frac{\partial^2 \Exc}
       {\partial P_{\alpha\beta}\, \partial P_{\gamma\delta}}\,
  P^{xy}_{\gamma\delta},
\end{split}
\end{equation}
that is, the $\kxc$ contraction over two first-order perturbed
density matrices, plus the $\fxc$ contraction over the second-order
one. The third-order equations repeat the structure with the
$\lxc$ contraction over three first-order densities, and so on to
any order; geometric perturbations add the perturbed-kernel seeds
that arise when the kernel is evaluated over displaced basis
functions and quadrature points.

A code with a working first-order CPKS solver
therefore already contains all of the machinery required for
quadratic and cubic response---hyperpolarizabilities and other
properties via the $(2n+1)$ rule, and the higher-order responses
themselves where the asymmetric routes are
preferable\cite{Helgaker2012_CR_543}---except for the seed
contractions, every one of which is generated by \xck{} at
arbitrary order. Extending a host from linear to higher-order
response thereby reduces from deriving and implementing new
operator expressions to assembling right-hand sides from generated
kernels. The generated kernels are exact expressions evaluated on
whatever discretization the host supplies, so the discretization
error remains the host's to control; the grid sensitivity observed
for r$^2$SCAN in \cref{sec:results} illustrates this at low order.

This division of labor is not hypothetical. The open-ended
response code of \citet{Ringholm2014_JCC_622} uses recursion over the
density-matrix quasienergy formulation of
\citet{Thorvaldsen2008_JCP_214108} in combination with time- and
perturbation-dependent basis sets to assemble any response property
at the Hartree--Fock and Kohn--Sham levels, the latter restricted to
GGA and hybrid functionals.\cite{Ringholm2014_JCP_34103} Ringholm \emph{et al.}
identified the availability of external routines for integral
derivatives and for the exchange--correlation kernel contributions
(in particular for high-order geometric derivatives) as factors
limiting its applicability. The
companion demonstration of that machinery---the analytic cubic and
quartic force fields of \citet{Ringholm2014_JCP_34103}---correspondingly
employed a combination of automatic differentiation and hand-written
code to supply the functional derivatives (see \cref{sec:xcfuncomparison}
for discussion), while multiphoton absorption matrix elements, which involve only
one-electron perturbations that leave the basis set unperturbed, have
been computed to arbitrary order.\cite{Friese2015_JCTC_1129} In contrast, \xck{} is able to
generate any of these expressions in full, at any level of DFT.

A practical caveat accompanies high-order response. The perturbed
densities entering such properties describe the response of the
diffuse outer valence region to external fields, and are therefore
considerably more extended than the ground-state density. This is felt throughout the pipeline. Atom-centered quadrature
grids are at their sparsest far from the nuclei, so the quadrature
requirements of a given functional grow with the order of the
derivative. The same holds for the basis set: an
atom-centered expansion has to be augmented with diffuse
functions\cite{Woon1994_JCP_2975} to represent the extended
perturbed densities, while a real-space discretization needs a
correspondingly larger simulation cell. Grids and basis sets
validated for energies and gradients should therefore not be
assumed sufficient for high-order response properties. The same
caution should be extended to all other parameters of a calculation,
as well, such as screening and convergence thresholds, auxiliary
basis sets, and pseudopotentials or projector augmentation, for
example, since they are usually set up once against ground-state or
low-lying excited-state benchmarks and then reused without further
consideration.

\subsection{The real-space potential}
\label{sec:realspace}

In the real-space form of the generalized Kohn--Sham (gKS)
approach,\cite{Seidl1996_PRB_3764} the xc potential of a
$\tau$-dependent functional is an operator that acts on the
orbitals. Functional differentiation of the xc
energy with the commonly used chain rule and integration by parts
yields
\begin{equation}
  \label{eq:gks}
\begin{split}
  \hat{v}_\text{xc}\, \psi_p(\rr) ={}& \bigg[
  \frac{\partial \exc}{\partial n}
  - \nabla \cdot \left( 2\, \frac{\partial \exc}{\partial \gamma}\,
  \nabla n \right) \\
  & \quad + \nabla^2\, \frac{\partial \exc}{\partial (\nabla^2 n)}
  \bigg]\, \psi_p(\rr) \\
  & - \frac{1}{2} \nabla \cdot \left[
    \frac{\partial \exc}{\partial \tau}\, \nabla \psi_p(\rr)
    \right].
\end{split}
\end{equation}

\Cref{eq:gks} illustrates how the complexity of the xc terms
escalates with the rung of the functional. For an LDA, only the
first term survives, and the potential is a simple local function
of the density. For a GGA, the second term is already nontrivial:
expanding the divergence with the chain rule brings in the second
derivatives of $\exc$ and the second derivatives of the
density. Continuing to the higher rungs, $\tau$-dependent
mGGAs require second derivatives of the xc energy density, the
density, and the orbitals, while Laplacian dependence pushes to
third derivatives of $\exc$ and fourth derivatives of $n$. This
complexity is likely the
reason why implementations that require the potential as a
function on a real-space grid traditionally stop at the GGA level. \xck{} is designed to derive and generate
exactly such complicated, expanded expressions.

In a compact basis, the potential is instead evaluated through the
matrix elements of \cref{eq:vxcmat}, which are obtained by direct
differentiation. The $\tau$ term is
unambiguous in that form, and, in contrast to the expanded forms
of \cref{eq:gks}, only the \emph{first} derivatives of
$\exc$ appear. The real-space and finite-basis forms coincide when
the integrals are evaluated exactly, but they differ on a finite
quadrature grid; the directly differentiated form of
\cref{eq:vxcmat} is the one that is variationally consistent with
the computed energy. This form is likewise generated by \xck{}, the
xc Fock matrix being the first member of the derivative tower of
\cref{sec:implementation}.

\subsection{Further properties and ingredient extensions}
\label{sec:extensions}

The remaining properties of interest combine the threads developed
above, and are collected in \cref{tab:properties} to augment our
discussion of the potential applications of \xck{}. They differ only
in which perturbations are mixed.

Vibrational spectroscopies mix the geometric response with a field.
Raman intensities are geometric derivatives of the
polarizability,\cite{Komornicki1979_JCP_2014, Frisch1986_JCP_531,
Rappoport2007_JCP_201104} i.e., the third
derivatives
\begin{equation}
  \label{eq:raman}
  \frac{\mathrm{d}^3 E}{\mathrm{d}\mathcal{E}_s\,
    \mathrm{d}\mathcal{E}_t\, \mathrm{d}\mathbf{R}_I}
\end{equation}
with respect to two components of an external electric field
$\boldsymbol{\mathcal{E}}$ and the nuclear coordinates
$\mathbf{R}_I$. The
required field and geometric responses are obtained with CPKS,
i.e., \cref{eq:cpks}; following the discussion of \cref{eq:d3E}, the
assembly then combines the $\kxc$ kernel with the geometric
collocation operands.

The chiroptical vibrational spectroscopies mix in a magnetic
perturbation as well, and thereby also require the field-differentiated
London orbitals of \cref{eq:london}. Vibrational circular dichroism
requires the atomic axial tensors,\cite{Stephens1985_JPC_748,
Cheeseman1996_CPL_211, Stephens1994_JPC_11623} which couple the
geometric and magnetic responses $\partial \mathbf{P} /
\partial \mathbf{R}_I$ and $\partial \mathbf{P} / \partial B_s$ of
\cref{eq:cpks}, while Raman optical activity adds the geometric
derivatives of the mixed electric-dipole--magnetic-dipole
polarizability.\cite{Barron1992_SAAMBS_1051, Barron1992_SAAMBS_1193,
Polavarapu1993_JACS_7736, Helgaker1994_FD_165, Ruud2002_JPCA_7448}

The nonlinear optical properties---first and second
hyperpolarizabilities and multi-photon absorption cross
sections---mix electric perturbations alone, and are governed by the
$\kxc$ and $\lxc$ kernels as discussed at the beginning of
\cref{sec:theory}.

Finally, several
methods extend the ingredient set of \cref{eq:Exc}. As discussed in
\cref{sec:magnetic}, current-density DFT adds the paramagnetic
current density, either through the vorticity or through the
gauge-corrected kinetic energy
density.\cite{Vignale1987_PRL_2360, Vignale1989_PRL_115,
Dobson1993_JCP_8870, Tellgren2014_JCP_34101} Spin-flip
TD-DFT, whose collinear formulation couples the spin-flip block
through exact exchange alone,\cite{Shao2003_JCP_4807} is instead
built on a noncollinear spin density.\cite{Wang2004_JCP_12191,
Bast2009_IJQC_2091} Multiconfiguration pair-density functional
theory adds an on-top pair density.\cite{LiManni2014_JCTC_3669,
Carlson2015_JCTC_82, Carlson2015_JCTC_4077} Local hybrid functionals
add a dependence on the exact-exchange energy density, and possibly
also on the reduced density
Hessian.\cite{Jaramillo2003_JCP_1068, Maier2019_WIRCMS_1378,
Schattenberg2021_JPCA_2697}

Once the new ingredients and their
density-matrix seeds have been defined, the corresponding kernels
follow mechanically in the present framework, as we demonstrate for
the gauge-corrected kinetic energy density and for the density
Hessian of local hybrids in \cref{sec:conclusions}. Some of these extensions also require new derivatives from the
functional library itself: the vorticity route calls for derivatives
with respect to $\bm\nu$, and the noncollinear case for derivatives
with respect to the magnetization. Neither is yet supported in
Libxc, but they may be added in a future release.

\begin{table*}
\caption{An overview of further relevant molecular properties,
which have been taken into account in the design of \xck{}.}
\label{tab:properties}
\footnotesize
\begin{ruledtabular}
\begin{tabular}{p{0.33\textwidth}lll}
Property & kernel order & parity / symmetry & comment \\
\hline
harmonic frequencies (CPKS Hessian)\cite{Johnson1993_CPL_133, Deglmann2002_CPL_511} & $\fxc$ & singlet & further operands: $\partial\chi/\partial\mathbf{R}$, $\partial w_g/\partial\mathbf{R}$ \\
Raman intensities\cite{Komornicki1979_JCP_2014, Frisch1986_JCP_531, Rappoport2007_JCP_201104} & $\kxc$ & singlet & further operands: $\partial\chi/\partial\mathbf{R}$, $\partial w_g/\partial\mathbf{R}$ \\
Raman optical activity\cite{Barron1992_SAAMBS_1051,
  Barron1992_SAAMBS_1193, Polavarapu1993_JACS_7736,
  Helgaker1994_FD_165, Ruud2002_JPCA_7448} & $\kxc$ & imaginary
singlet $\times$ geometric & further operands:
$\partial\chi/\partial\mathbf{B}$;
$\partial\chi/\partial\mathbf{R}$, $\partial
w_g/\partial\mathbf{R}$ \\
quartic force fields (vibrational perturbation theory)\cite{Barone2004_JCP_14108, Ringholm2014_JCP_34103} & $\lxc$ & singlet & further operands: $\partial\chi/\partial\mathbf{R}$, $\partial w_g/\partial\mathbf{R}$ \\
vibrational circular dichroism\cite{Stephens1985_JPC_748, Cheeseman1996_CPL_211, Stephens1994_JPC_11623} & $\fxc$ & imaginary singlet $\times$ geometric & further operands: $\partial\chi/\partial\mathbf{B}$; $\partial\chi/\partial\mathbf{R}$, $\partial w_g/\partial\mathbf{R}$ \\
excited-state gradients\cite{VanCaillie1999_CPL_249, VanCaillie2000_CPL_159, Furche2002_JCP_7433} & $\kxc$ & singlet, triplet & further operands: $\partial\chi/\partial\mathbf{R}$, $\partial w_g/\partial\mathbf{R}$ \\
excited-state Hessians\cite{Liu2011_JCP_184111} & $\lxc$ & singlet, triplet & further operands: $\partial\chi/\partial\mathbf{R}$, $\partial w_g/\partial\mathbf{R}$ \\
first hyperpolarizability; second-harmonic generation; two-photon absorption\cite{Gisbergen1998_JCP_10644, Gisbergen1998_JCP_10657, Salek2002_JCP_9630, Norman2018_CR_7208} & $\kxc$ & singlet (mixed parities allowed) & --- \\
second hyperpolarizability; third-harmonic generation; three-photon absorption\cite{Gisbergen1998_JCP_10644, Jansik2005_JCP_54107, Norman2018_CR_7208} & $\lxc$ & singlet, mixed & --- \\
NMR shielding; magnetizability\cite{Schreckenbach1995_JPC_606, Cheeseman1996_JCP_5497, Helgaker1999_CR_293, Lee1994_CPL_225} & ($\fxc$)\footnotemark[1] & imaginary singlet & further operands: $\partial\chi/\partial\mathbf{B}$ \\
spin--spin coupling (Fermi contact, spin dipole)\cite{Dickson1996_JPC_5286, Helgaker1999_CR_293, Sychrovsky2000_JCP_3530, Helgaker2000_JCP_9402} & $\fxc$ & triplet & --- \\
current-DFT magnetic response ($\tau$ gauge)\cite{Dobson1993_JCP_8870, Lee1995_JCP_10095, Maximoff2004_CPL_408, Tellgren2014_JCP_34101} & $\fxc$ & imaginary singlet & ingredient: $\mathbf{j}_p$ (implemented) \\
spin-flip TD-DFT; exchange couplings\cite{Wang2004_JCP_12191, Bast2009_IJQC_2091} & $\fxc$ & noncollinear & ingredient: $(\rho, \mathbf{m})$ \\
multiconfiguration pair-density functional theory potentials and response\cite{LiManni2014_JCTC_3669, Carlson2015_JCTC_82, Carlson2015_JCTC_4077} & $\vxc$--$\fxc$ & singlet & ingredient: on-top $\Pi$ \\
local hybrid SCF and response\cite{Jaramillo2003_JCP_1068, Maier2019_WIRCMS_1378, Schattenberg2021_JPCA_2697} & $\vxc$--$\lxc$ & all & ingredients: $e_x(\rr)$; $\eta$ (implemented) \\
\end{tabular}
\end{ruledtabular}
\footnotetext[1]{The kernel contribution vanishes, as the
first-order density is imaginary; the xc terms enter through the
field-differentiated London orbitals instead. Current-dependent
functionals regain a kernel contribution through the current
ingredient (see the current-DFT row).}
\end{table*}

\section{Implementation}
\label{sec:implementation}

\subsection{Ingredients, response contractions, and pattern
  collapse}
\label{sec:response}

We now summarize the computational implementation. At the bottom
level, we find linear primitives of the form
\begin{equation}
  k(\rr) = \sum_{\alpha\beta} P_{\alpha\beta}\, Q_k\big(\chi^*_\alpha(\rr),
  \chi_\beta(\rr)\big),
  \label{eq:primitive}
\end{equation}
where $Q_k$ is a fixed sesquilinear kernel (bilinear in the case of
a real basis). The semi-local DFT ingredients are of this
form: the electron density $n$ of \cref{eq:nr} has
$Q_n = \chi^*_\alpha \chi_\beta$, the Cartesian components
$\partial_c n$ of the density gradient have
$Q_{\partial_c n} = \partial_c(\chi^*_\alpha \chi_\beta)$, the
density Laplacian $\nabla^2 n$ has
$Q_{\nabla^2 n} = \nabla^2(\chi^*_\alpha \chi_\beta)$, and the
kinetic energy density $\tau$ of \cref{eq:tau} has
$Q_\tau = \tfrac{1}{2} \nabla\chi^*_\alpha \cdot \nabla\chi_\beta$.
The reduced
gradient $\gamma$ of \cref{eq:gamma} is, in turn, an algebraic
function of the linear $\partial_c n$ primitives.

With these primitives in hand, the operation of the code is easy to
understand. Expressing \cref{eq:quad} symbolically in terms of the
primitives, the action of any derivative operator on the expression
can be computed via the chain rule. The derivative acts on $\exc$;
nuclear derivatives also act on the quadrature weights $w_g$, which
yields the grid-response terms. The important thing to notice is
that since $\gamma$ of \cref{eq:gamma} is quadratic in the density
matrix, one application of the chain rule produces terms that still
depend on the density matrix, which can then be acted on by the
next round of derivative operators.

\label{sec:geometric}%
The grid-response terms are important in LCAO calculations, because
the quadrature is built from atom-centered grids that move
with the atoms, as discussed after \cref{eq:quad}. Like
the functional derivatives of $\exc$, the grid-response terms are
fully generated by our engine as abstract symbolic expressions;
their values must be supplied by the quadrature code of the host
program. The basis-function, grid-point, and quadrature-weight
term classes are generated as separate kernels, so each emitter
plugin decides independently which classes its host evaluates: the
fixed-grid and grid-response gradients compared in
\cref{sec:geomresults} draw on the same set of generated kernels, and a host option
determines whether the grid-point and quadrature-weight kernels
are evaluated.

Many response quantities are formally high-rank tensors: the xc
kernel $g_{\alpha\beta,\gamma\delta} = \partial^2 \Exc /
\partial P_{\alpha\beta}\, \partial P_{\gamma\delta}$ carries four
basis-function indices, and every further derivative order adds two
more. Such tensors are never stored in memory, as their size would
be prohibitive; instead, they are contracted directly. The xc
interfaces of response solvers therefore take in (un)perturbed
density matrices and return Fock-like matrices. The response
equations can be formulated either with the density or with the
one-particle density matrix as the basic variable; we adopt the
latter, which goes back to the time-dependent Hartree--Fock-like
treatment of \citet{Bauernschmitt1996_CPL_454} that was placed on an
independent, systematic footing by
Furche.\cite{Furche2001_JCP_5982} The $m$-th order
response Fock matrix reads
\begin{equation}
  F^{X_1 \cdots X_m}_{\alpha\beta} = \sum_{\{\gamma_i \delta_i\}}
  \big(D^{m+1} \Exc\big)_{\alpha\beta, \gamma_1\delta_1, \ldots}
  \, D^{X_1}_{\gamma_1\delta_1} \cdots D^{X_m}_{\gamma_m\delta_m},
  \label{eq:respfock}
\end{equation}
where $D_{\gamma\delta} = \partial / \partial P_{\gamma\delta}$
denotes the density-matrix derivative, and the $\mathbf{D}^{X_i}$
are the perturbed density matrices.

Although \cref{eq:respfock} sums over $m$ pairs of basis-function
indices, it can be evaluated at low cost: the key is to contract
in the perturbed density matrices one at a time. Each such contraction collapses into per-point scalars: the
\emph{perturbed ingredients}
\begin{equation}
  k^{X}(\rr) = \sum_{\gamma\delta}
  \frac{\partial k}{\partial P_{\gamma\delta}}
  D^{X}_{\gamma\delta},
  \label{eq:pertingredient}
\end{equation}
where $\partial k / \partial P_{\gamma\delta} =
Q_k(\chi^*_\gamma, \chi_\delta)$ is the \emph{seed} of the
ingredient, cf.\ \cref{eq:primitive}. The perturbed reduced
gradient follows by the chain rule as $\gamma^{X} = 2\,
\nabla n \cdot \nabla n^{X}$. Because each contraction yields
per-point scalars, no high-rank intermediates ever arise, and
little memory is needed beyond the collocation matrices that an
ordinary xc potential build already requires.

If the
$\mathbf{D}^X$ were general matrices, the cost to assemble
\cref{eq:pertingredient} would be $\mathcal{O}(N^2
N_\text{grid})$, where $N$ is the number of basis functions.
However, response solvers build the perturbed density matrices
from occupied--virtual rotations, $\mathbf{D}^X =
\mathbf{C}_\text{o} \mathbf{X} \mathbf{C}_\text{v}^\dagger +
\text{h.c.}$, and this rank structure can be exploited:
collocating the orbital pairs directly reduces the cost of
\cref{eq:pertingredient} to $\mathcal{O}(N_\text{occ} N
N_\text{grid})$. When only the occupied--virtual block of the
output is needed, as in the sigma vector builds of iterative
solvers, this reduction applies to the assembly as well, as
we will discuss in \cref{sec:library}.

Once all of the
perturbed density matrices have been folded in, only one free
basis-function pair remains, and the Fock-like matrix is assembled
with a handful of matrix products, as we will discuss shortly. \Cref{eq:respfock} is multilinear in the
$\mathbf{D}^{X}$, so the higher-order perturbed densities of
quadratic and higher response enter as separate inputs supplied by
the solver.

A perturbed density matrix has independent spin blocks
$\mathbf{D}^{X,\uparrow}$ and $\mathbf{D}^{X,\downarrow}$, so each
spin density has its own response; the mixed reduced gradient
couples the two by the product rule,
$\gamma_{\sigma\sigma'}^{X} = \nabla n_\sigma^{X}\!\cdot\!\nabla n_{\sigma'} + \nabla
n_\sigma\!\cdot\!\nabla n_{\sigma'}^{X}$, where $\sigma$ and $\sigma'$
label the spin channels. Closed-shell spin adaptation is a
substitution: each perturbation is assigned a parity $p = \pm 1$
(singlet/triplet), and $\mathbf{D}^{X,\downarrow} = p\,
\mathbf{D}^{X,\uparrow}$ is set at the closed-shell reference. At
second order, this produces the familiar $f^{\uparrow\uparrow} \pm
f^{\uparrow\downarrow}$ combinations. At third order it reproduces as plain expression
arithmetic the closed-shell coefficient tables that have previously
been derived by hand.\cite{Salek2002_JCP_9630}

\label{sec:collapse}%
A straightforward application of the chain and Leibniz rules
leads to a rapid explosion in the number of terms generated as a
function of the number of ingredients and the order of the
response kernel. For example, evaluating a contraction in
\cref{eq:respfock} of the fourth density-matrix derivative of a
spin-resolved mGGA $\Exc$ that depends on both the kinetic
energy density and the density Laplacian, and thus on nine spin
ingredients, with three
perturbed density matrices leads to a total of 130\,566 terms.
However, each of these terms factorizes as a product of a
per-point scalar and two basis factors of indices $\alpha$ and
$\beta$. Only four patterns ($\chi\chi$, $\chi\partial_c\chi$,
$\partial_c\chi\,\partial_{c'}\chi$, and $\chi\nabla^2\chi$) arise
for fixed-basis response contractions \emph{of any order} for
functionals of the form of \cref{eq:Exc}.

Raising the derivative order does raise the order of the functional derivatives,
which are multiplied by further perturbed ingredients, but these
are all scalar values on the grid. The basis-function pair enters
only through the ingredient seeds, which contain at most the
Laplacian of the density. Even when the perturbation affects the
basis functions, as geometric derivatives do in LCAO calculations,
the pattern set stays compact: the differentiated basis functions
enter as new collocation factors alongside the original patterns,
and the geometric derivative of the $\tau$-mGGA Fock matrix,
for example, involves 30 patterns instead of the 10 of the
fixed-basis form.

The solution to the term explosion is therefore to collect terms by
pattern. This enables the
minimization of the number of expensive operations in the
computation of the matrix elements, which are the matrix
multiplications over the collocation data: the two factors are
basis functions in an LCAO Fock build, and orbital pairs when the
kernel is assembled directly in a transition-pair basis, but the
contraction is the same in either case.

Resolving the components and the order of the two basis factors,
the four patterns yield twelve component patterns: one $\chi\chi$;
the mixed patterns $\chi\partial_c\chi$ and $\partial_c\chi\,\chi$,
three of each; the three diagonal patterns
$\partial_c\chi\,\partial_c\chi$; and $\chi\nabla^2\chi$ and
$\nabla^2\chi\,\chi$. Component patterns that are transposes of
each other are evaluated together with a single matrix
multiplication and a transposed accumulation, provided their
coefficients are either identical or exactly opposite; the
resulting contribution is symmetric in the first case and
antisymmetric in the second, the latter arising for the
current-type ingredients of \cref{sec:conclusions}.

The patterns are then fused. Every pattern is a product of a
per-point coefficient with two basis factors. Patterns that share
a left basis factor therefore differ only in their
coefficient and their right factor, and their sum can be formed on
the grid before the matrix product is taken,
\begin{equation}
  \label{eq:fusion}
  \sum_i \sum_g w_g\, c^{(i)}_g\, \chi_{\alpha g}\, f^{(i)}_{\beta g}
  = \sum_g \chi_{\alpha g} \bigg[ w_g \sum_i c^{(i)}_g\,
    f^{(i)}_{\beta g} \bigg],
\end{equation}
where the $f^{(i)}$ are the differing right-hand basis factors and
the $c^{(i)}$ their coefficients. The bracket is one array of
the dimensions of the collocation matrix, assembled with elementwise
work, so a group of patterns costs a single matrix multiplication
instead of one each. Patterns sharing a transpose sign
are fused, so the group is accumulated as
$\mathbf{t} \pm \mathbf{t}^\mathsf{T}$ exactly as a single pair
would be. A pattern whose two basis factors are equal is its own
transpose partner, and would therefore be counted twice; it joins a
symmetric group with half its coefficient, the transposed
accumulation supplying the other half. For a complex basis the
transposes are Hermitian throughout, as the per-point coefficients
remain real. For a GGA this fuses all
seven patterns into one matrix multiplication, which is the form
that hand-written implementations use;\cite{Pople1992_CPL_557} for
an mGGA, four remain, as the three diagonal patterns
$\partial_c\chi\,\partial_c\chi$ carry the Cartesian index on both
sides and share no left factor. The twelve component patterns of
the 130\,566-monomial kernel thus require only four matrix
multiplications, and the same four suffice for the thirty patterns
of the density-Hessian kernels, whose additional patterns differ
only in their right-hand factor.
\subsection{The \xck{} library}
\label{sec:library}

\xck{} is a small Python library built on
SymPy\cite{Meurer2017_PCS_103} for the symbolic layer. The library
proceeds in stages. The integrand of the requested kernel is first
assembled symbolically from the ingredient definitions. The
derivative operators are then applied to it in a monomial
representation, and the resulting terms are collapsed by pattern,
as described in \cref{sec:response}. In the final step, the collapsed
expressions are emitted as source code. We now discuss key aspects
of the implementation.

\begin{description}
\item[Monomial representation] All derivative operators are applied in a
dedicated monomial representation (coefficient dictionaries keyed by
power tuples), term by term with hash-map accumulation. Generic
symbolic manipulation of the full expressions would be intractable
beyond third order, whereas the monomial representation yields
typical kernels in seconds and even the largest kernels of the
catalog in about a minute.

\item[Code emission] The spin components are resolved into
concrete array layouts only at the emission stage: the emitted
code follows Libxc's component
packing, in which the \texttt{v2rho2} array, for example, stores
the $\uparrow\uparrow$, $\uparrow\downarrow$, and
$\downarrow\downarrow$ components in this order. The packing is an
emitter convention: a backend
could equally well address the array layout of another
provider, such as XCFun.\cite{Ekstroem2010_JCTC_1971}

\item[Emitter backends] The emitter backend is extendable, and the
collapsed expressions can be written out in any desired form. We
currently provide the following backends. First, a NumPy backend produces
runtime-compiled \texttt{einsum} functions. Second, a C backend
emits the monomials as static coefficient/factor tables. While these
tables can be large (up to hundreds of megabytes), they are data that
is walked by a fixed evaluator of a few dozen lines. The compiler
never has to see these large expressions, which could otherwise make
compilation intractable. Third,
host-idiom backends rewrite the collapsed kernels into the adopting
code's native contraction style, such as machine-generated C++
include files for Psi4\cite{Smith2020_JCP_184108} and a
machine-generated NumPy module for
GPAW,\cite{Mortensen2024_JCP_92503} which are demonstrated in
\cref{sec:results}.

\item[Kernel catalog] A catalog generator enumerates the complete
set of kernels; the present 169 kernels are listed in full in
\cref{tab:catalog}. The
catalog spans the energy, Fock, and response contractions
through fourth order; seven functional families; and the unpolarized,
unrestricted, and spin-adapted cases, which are available for all
seven families. The
catalog is complete through fourth order for every family: no
derivative order is out of reach. While going to higher orders leads
to larger and larger emitted code due to the explosion in the number
of terms, the derivation does not become more challenging. The extreme
case is the spin-resolved fourth-order density-Hessian contraction,
whose 6\,601\,743 monomials collapse onto the same 30 patterns and
4 matrix products as its lower orders, but amount to half a
gigabyte of source when
written out as explicit expressions; such kernels are the reason
for the table-driven backend of the preceding item that emits the
monomials as data. Each kernel is
accompanied by a machine-readable manifest that declares its
operands, the Libxc arrays it consumes by name, and its term
ownership; the kernels contain only xc terms, while the Coulomb and
exact-exchange terms remain host-owned. Conventions that differ between hosts---the sign of
$\boldsymbol{\kappa}$ in the orbital rotation
$e^{\pm\boldsymbol{\kappa}}$,\cite{Note1} occupation-factor placement,
spin-component packing,
singlet/triplet parities---are explicit parameters or documented
constants throughout.
\end{description}

The emitters also cover the complex cases discussed in
\cref{sec:theory}: complex orbital coefficients
and complex basis functions. Complex orbital
coefficients over a real basis
require no new generated code. With a real basis, every ingredient
seed is a bilinear with a definite symmetry in the basis-function
pair: the symmetric seeds of the density-type ingredients pick out
the real part of the Hermitian density matrix, while the
antisymmetric, current-type seeds pick out the imaginary part (see
the discussion of the paramagnetic current density in
\cref{sec:conclusions}). The
runtime layer therefore splits each density matrix into these two
real parts, evaluates the generated real kernels on them, and
reassembles the complex Fock matrix from the general real output.

Complex basis functions, in contrast, change the expressions
themselves, as each basis-function product then carries a complex
conjugate on the bra side. They are served by a sesquilinear
emission mode, in which the conjugated basis values enter the
contractions as operands of their own. The
per-point coefficients are unaffected, as all the input ingredients
of a Hermitian density matrix remain real also for a complex
basis. For this reason, the transposed accumulation of the
pattern collapse generalizes to the Hermitian transpose, and the
sesquilinear kernels require no additional matrix
multiplications.
Both paths are validated against finite differences in the real and
imaginary parts of the density matrix.

Although our expressions have featured density
matrices, this has merely been a device for the formal derivation.
DFT implementations in a verbose basis, such as plane waves, never
form the density matrix, as its dimension would be prohibitive;
instead, they compute the electron density and other ingredients
one orbital at a time, using the factorized form of \cref{eq:dm}.
Such matrix-free hosts are served by \xck{}'s two-sided emission
mode, in which the two free indices of the output are contracted
with two independently chosen sets of function values on the
grid.

Supplying the values of the occupied orbitals on one side
and those of the virtual orbitals on the other yields the response
matrix directly in the compact molecular-orbital basis, providing
what is needed for the host-side iterative response solvers, such as
Davidson-type eigensolvers\cite{Davidson1975_JCP_87} of TD-DFT and
linear solvers for CPKS. The perturbed
ingredients are likewise computed from the orbital values with a
handful of matrix products.

In contrast to the LCAO
mode, which assembles Fock-like matrices in the atomic-orbital
basis, the two-sided mode never forms a matrix of the basis-set
dimension. Both emission routes evaluate the same collapsed
expressions and agree to machine precision. Alternatively, such
hosts can consume only the pointwise coefficient stage of the
present library and keep their own operator application.

The generated code is of the same quality as hand-written code.
The higher-order expressions routinely generated by \xck{} are in
any case outside the reach of a manual implementation. The emitted
expressions are tensor contractions
over batches of grid points. When the expressions are emitted in
full as Einstein sums, the contraction engine is free to choose
the optimal evaluation order for the whole expression.

The pattern-collapsed form of \cref{sec:collapse} makes the cost
structure explicit. However many terms the chain rule produces,
they only ever combine grid-point data---the collocated basis
functions, the (perturbed) ingredients formed from the density
matrices, and the functional derivatives from the functional
library---into a handful of flat arrays of the grid dimension: one
array per pattern.

These flat arrays are then sandwiched between
the basis-function values, or the molecular-orbital values in the
two-sided mode, which turns the contraction into a matrix
product. The millions of monomials of \cref{tab:collapse} are thus
elementwise work on grid-sized vectors, while the only operations
that scale with the basis-set dimension are the few
matrix-matrix multiplications of the patterns, which run at the
speed of the underlying BLAS library.

This structure also suits hardware accelerators such as graphics
processing units (GPUs): the
coefficient assembly consists of a large number of independent
arithmetic operations on grid-point data, with no communication
between grid points, while the pattern contractions map onto
vendor-tuned matrix multiplication. The only requirement is that
the monomial tables fit in device memory, which will be seen to be
the case even for the largest kernels (see \cref{sec:collapsecount}).

\begin{table*}
\caption{The complete catalog of the 169 generated kernels. The
kernel names follow the pattern
\texttt{xck\_<family>\_<case>\_o<order>[\_<parities>]}. The family
is \texttt{lda}, \texttt{gga}, \texttt{mgga\_tau} ($\tau$-only
mGGA), \texttt{mgga\_lapl} (Laplacian-only mGGA),
\texttt{mgga} (full mGGA, which depends on both
$\nabla^2 n$ and $\tau$, as needed, e.g., for the
Becke--Roussel functional\cite{Becke1989_PRA_3761}),
\texttt{cmgga\_tau} (current-corrected mGGA, i.e., a
$\tau$-only mGGA in which $\tau$ is replaced by the
gauge-corrected $\tilde\tau$ of \cref{eq:ctau},
discussed in \cref{sec:conclusions}), or
\texttt{hmgga} (mGGA extended by the density-Hessian
ingredient $\eta$ of the local-hybrid calibration functions,
likewise discussed in \cref{sec:conclusions}). The case is restricted (\texttt{r}), the unrestricted
output spin channels (\texttt{ua}, \texttt{ub}), or closed-shell
spin-adapted (\texttt{st}). Order 0 denotes the collocated energy
density, and order $n \geq 1$ the $n$-th density-matrix derivative
contracted with $n - 1$ perturbed density matrices, i.e.,
\cref{eq:respfock} with $m = n - 1$; the trailing parities mark
each perturbation as singlet (\texttt{p}) or triplet (\texttt{m}).
The \texttt{giao} kernels are the explicit magnetic-field
derivatives of the Fock matrix with London orbitals, emitted as the
real factor $\mathbf{K}_s$ of $\partial F^\text{xc}_{\alpha\beta} /
\partial B_s = (i/2c)\, (K_s)_{\alpha\beta}$ at a real reference.
The energy kernel is only emitted once, in the restricted set: the
contraction $\sum_g w_g \rho_g \text{zk}_g$ of the quadrature
weights, the total density, and Libxc's energy density per particle
zk is identical in the two cases, as the spin resolution resides in
the host-side Libxc call that yields zk.}
\label{tab:catalog}
\footnotesize
\begin{ruledtabular}
\begin{tabular}{llp{0.62\textwidth}}
family & case & kernels \\
\hline
LDA & restricted &
\texttt{xck\_lda\_r\_o0},
\texttt{xck\_lda\_r\_o1},
\texttt{xck\_lda\_r\_o2},
\texttt{xck\_lda\_r\_o3},
\texttt{xck\_lda\_r\_o4},
\texttt{xck\_lda\_r\_giao} \\
 & unrestricted &
\texttt{xck\_lda\_ua\_o1},
\texttt{xck\_lda\_ua\_o2},
\texttt{xck\_lda\_ua\_o3},
\texttt{xck\_lda\_ua\_o4},
\texttt{xck\_lda\_ub\_o1},
\texttt{xck\_lda\_ub\_o2},
\texttt{xck\_lda\_ub\_o3},
\texttt{xck\_lda\_ub\_o4},
\texttt{xck\_lda\_ua\_giao},
\texttt{xck\_lda\_ub\_giao} \\
 & spin-adapted &
\texttt{xck\_lda\_st\_o2\_p},
\texttt{xck\_lda\_st\_o2\_m},
\texttt{xck\_lda\_st\_o3\_pp},
\texttt{xck\_lda\_st\_o3\_pm},
\texttt{xck\_lda\_st\_o3\_mm},
\texttt{xck\_lda\_st\_o4\_ppp},
\texttt{xck\_lda\_st\_o4\_ppm},
\texttt{xck\_lda\_st\_o4\_pmm},
\texttt{xck\_lda\_st\_o4\_mmm} \\
GGA & restricted &
\texttt{xck\_gga\_r\_o0},
\texttt{xck\_gga\_r\_o1},
\texttt{xck\_gga\_r\_o2},
\texttt{xck\_gga\_r\_o3},
\texttt{xck\_gga\_r\_o4},
\texttt{xck\_gga\_r\_giao} \\
 & unrestricted &
\texttt{xck\_gga\_ua\_o1},
\texttt{xck\_gga\_ua\_o2},
\texttt{xck\_gga\_ua\_o3},
\texttt{xck\_gga\_ua\_o4},
\texttt{xck\_gga\_ub\_o1},
\texttt{xck\_gga\_ub\_o2},
\texttt{xck\_gga\_ub\_o3},
\texttt{xck\_gga\_ub\_o4},
\texttt{xck\_gga\_ua\_giao},
\texttt{xck\_gga\_ub\_giao} \\
 & spin-adapted &
\texttt{xck\_gga\_st\_o2\_p},
\texttt{xck\_gga\_st\_o2\_m},
\texttt{xck\_gga\_st\_o3\_pp},
\texttt{xck\_gga\_st\_o3\_pm},
\texttt{xck\_gga\_st\_o3\_mm},
\texttt{xck\_gga\_st\_o4\_ppp},
\texttt{xck\_gga\_st\_o4\_ppm},
\texttt{xck\_gga\_st\_o4\_pmm},
\texttt{xck\_gga\_st\_o4\_mmm} \\
$\tau$ mGGA & restricted &
\texttt{xck\_mgga\_tau\_r\_o0},
\texttt{xck\_mgga\_tau\_r\_o1},
\texttt{xck\_mgga\_tau\_r\_o2},
\texttt{xck\_mgga\_tau\_r\_o3},
\texttt{xck\_mgga\_tau\_r\_o4},
\texttt{xck\_mgga\_tau\_r\_giao} \\
 & unrestricted &
\texttt{xck\_mgga\_tau\_ua\_o1},
\texttt{xck\_mgga\_tau\_ua\_o2},
\texttt{xck\_mgga\_tau\_ua\_o3},
\texttt{xck\_mgga\_tau\_ua\_o4},
\texttt{xck\_mgga\_tau\_ub\_o1},
\texttt{xck\_mgga\_tau\_ub\_o2},
\texttt{xck\_mgga\_tau\_ub\_o3},
\texttt{xck\_mgga\_tau\_ub\_o4},
\texttt{xck\_mgga\_tau\_ua\_giao},
\texttt{xck\_mgga\_tau\_ub\_giao} \\
 & spin-adapted &
\texttt{xck\_mgga\_tau\_st\_o2\_p},
\texttt{xck\_mgga\_tau\_st\_o2\_m},
\texttt{xck\_mgga\_tau\_st\_o3\_pp},
\texttt{xck\_mgga\_tau\_st\_o3\_pm},
\texttt{xck\_mgga\_tau\_st\_o3\_mm},
\texttt{xck\_mgga\_tau\_st\_o4\_ppp},
\texttt{xck\_mgga\_tau\_st\_o4\_ppm},
\texttt{xck\_mgga\_tau\_st\_o4\_pmm},
\texttt{xck\_mgga\_tau\_st\_o4\_mmm} \\
Laplacian mGGA & restricted &
\texttt{xck\_mgga\_lapl\_r\_o0},
\texttt{xck\_mgga\_lapl\_r\_o1},
\texttt{xck\_mgga\_lapl\_r\_o2},
\texttt{xck\_mgga\_lapl\_r\_o3},
\texttt{xck\_mgga\_lapl\_r\_o4},
\texttt{xck\_mgga\_lapl\_r\_giao} \\
 & unrestricted &
\texttt{xck\_mgga\_lapl\_ua\_o1},
\texttt{xck\_mgga\_lapl\_ua\_o2},
\texttt{xck\_mgga\_lapl\_ua\_o3},
\texttt{xck\_mgga\_lapl\_ua\_o4},
\texttt{xck\_mgga\_lapl\_ub\_o1},
\texttt{xck\_mgga\_lapl\_ub\_o2},
\texttt{xck\_mgga\_lapl\_ub\_o3},
\texttt{xck\_mgga\_lapl\_ub\_o4},
\texttt{xck\_mgga\_lapl\_ua\_giao},
\texttt{xck\_mgga\_lapl\_ub\_giao} \\
 & spin-adapted &
\texttt{xck\_mgga\_lapl\_st\_o2\_p},
\texttt{xck\_mgga\_lapl\_st\_o2\_m},
\texttt{xck\_mgga\_lapl\_st\_o3\_pp},
\texttt{xck\_mgga\_lapl\_st\_o3\_pm},
\texttt{xck\_mgga\_lapl\_st\_o3\_mm},
\texttt{xck\_mgga\_lapl\_st\_o4\_ppp},
\texttt{xck\_mgga\_lapl\_st\_o4\_ppm},
\texttt{xck\_mgga\_lapl\_st\_o4\_pmm},
\texttt{xck\_mgga\_lapl\_st\_o4\_mmm} \\
full mGGA & restricted &
\texttt{xck\_mgga\_r\_o0},
\texttt{xck\_mgga\_r\_o1},
\texttt{xck\_mgga\_r\_o2},
\texttt{xck\_mgga\_r\_o3},
\texttt{xck\_mgga\_r\_o4},
\texttt{xck\_mgga\_r\_giao} \\
 & unrestricted &
\texttt{xck\_mgga\_ua\_o1},
\texttt{xck\_mgga\_ua\_o2},
\texttt{xck\_mgga\_ua\_o3},
\texttt{xck\_mgga\_ua\_o4},
\texttt{xck\_mgga\_ub\_o1},
\texttt{xck\_mgga\_ub\_o2},
\texttt{xck\_mgga\_ub\_o3},
\texttt{xck\_mgga\_ub\_o4},
\texttt{xck\_mgga\_ua\_giao},
\texttt{xck\_mgga\_ub\_giao} \\
 & spin-adapted &
\texttt{xck\_mgga\_st\_o2\_p},
\texttt{xck\_mgga\_st\_o2\_m},
\texttt{xck\_mgga\_st\_o3\_pp},
\texttt{xck\_mgga\_st\_o3\_pm},
\texttt{xck\_mgga\_st\_o3\_mm},
\texttt{xck\_mgga\_st\_o4\_ppp},
\texttt{xck\_mgga\_st\_o4\_ppm},
\texttt{xck\_mgga\_st\_o4\_pmm},
\texttt{xck\_mgga\_st\_o4\_mmm} \\
current-corrected & restricted &
\texttt{xck\_cmgga\_tau\_r\_o0},
\texttt{xck\_cmgga\_tau\_r\_o1},
\texttt{xck\_cmgga\_tau\_r\_o2},
\texttt{xck\_cmgga\_tau\_r\_o3},
\texttt{xck\_cmgga\_tau\_r\_o4} \\
 & unrestricted &
\texttt{xck\_cmgga\_tau\_ua\_o1},
\texttt{xck\_cmgga\_tau\_ua\_o2},
\texttt{xck\_cmgga\_tau\_ua\_o3},
\texttt{xck\_cmgga\_tau\_ua\_o4},
\texttt{xck\_cmgga\_tau\_ub\_o1},
\texttt{xck\_cmgga\_tau\_ub\_o2},
\texttt{xck\_cmgga\_tau\_ub\_o3},
\texttt{xck\_cmgga\_tau\_ub\_o4} \\
 & spin-adapted &
\texttt{xck\_cmgga\_tau\_st\_o2\_p},
\texttt{xck\_cmgga\_tau\_st\_o2\_m},
\texttt{xck\_cmgga\_tau\_st\_o3\_pp},
\texttt{xck\_cmgga\_tau\_st\_o3\_pm},
\texttt{xck\_cmgga\_tau\_st\_o3\_mm},
\texttt{xck\_cmgga\_tau\_st\_o4\_ppp},
\texttt{xck\_cmgga\_tau\_st\_o4\_ppm},
\texttt{xck\_cmgga\_tau\_st\_o4\_pmm},
\texttt{xck\_cmgga\_tau\_st\_o4\_mmm} \\
density-Hessian & restricted &
\texttt{xck\_hmgga\_r\_o0},
\texttt{xck\_hmgga\_r\_o1},
\texttt{xck\_hmgga\_r\_o2},
\texttt{xck\_hmgga\_r\_o3},
\texttt{xck\_hmgga\_r\_o4} \\
 & unrestricted &
\texttt{xck\_hmgga\_ua\_o1},
\texttt{xck\_hmgga\_ua\_o2},
\texttt{xck\_hmgga\_ua\_o3},
\texttt{xck\_hmgga\_ua\_o4},
\texttt{xck\_hmgga\_ub\_o1},
\texttt{xck\_hmgga\_ub\_o2},
\texttt{xck\_hmgga\_ub\_o3},
\texttt{xck\_hmgga\_ub\_o4} \\
 & spin-adapted &
\texttt{xck\_hmgga\_st\_o2\_p},
\texttt{xck\_hmgga\_st\_o2\_m},
\texttt{xck\_hmgga\_st\_o3\_pp},
\texttt{xck\_hmgga\_st\_o3\_pm},
\texttt{xck\_hmgga\_st\_o3\_mm},
\texttt{xck\_hmgga\_st\_o4\_ppp},
\texttt{xck\_hmgga\_st\_o4\_ppm},
\texttt{xck\_hmgga\_st\_o4\_pmm},
\texttt{xck\_hmgga\_st\_o4\_mmm} \\
\end{tabular}
\end{ruledtabular}
\end{table*}

\section{Computational details}
\label{sec:compdet}

\subsection{Gaussian-basis calculations: Psi4 and PySCF}
\label{sec:compgauss}

We ran calculations with an early development version of
Psi4\cite{Smith2020_JCP_184108} version 1.12. We also performed
comparison calculations with PySCF\cite{Sun2018_WIRCMS_1340} version
2.13.1.  Libxc\cite{Lehtola2018_S_1} version 7.1.2 was employed to
evaluate the density functionals in both programs, ensuring a fair
comparison.\cite{Lehtola2023_JCP_114116} The results agree to within
the convergence thresholds of the programs, as detailed below. Both
programs employ the radial quadrature scheme of Treutler and
Ahlrichs,\cite{Treutler1995_JCP_346} and pruning of the quadrature
grid was disabled in both.

We demonstrate the code on H$_2$O (spin-restricted) and triplet CH$_2$
(spin-unrestricted). The systems were chosen to exercise both of the
generated code paths: a closed-shell reference and an open-shell
reference with two unpaired electrons.  The geometries of the studied
molecules are given in \cref{tab:geometries}. The geometries
correspond to experimental structures and were not optimized at the
levels of theory used; the fixed geometries were used in all of
the calculations in both programs. The use of non-equilibrium
geometries also makes the validation of the nuclear derivatives more
stringent, as the gradients do not vanish. The input and output files
of all of the calculations are available as described in the data
availability statement.

All Psi4 calculations that include the quadrature-grid response were
run with the \texttt{DFT\_GRID\_RESPONSE=true},
\texttt{DFT\_BLOCK\_SCHEME=atomic}, and
\texttt{DFT\_GRID\_ORIENTATION=false} settings. The atomic blocking
scheme assigns every batch of quadrature points to its parent atom, as
is required by the analytic derivatives of the quadrature
weights. Disabling the use of the standard grid orientation fixes
the orientation of the
atomic grids in the laboratory frame, since the grid-response
expressions assume that quadrature points translate rigidly with their
parent atoms. If the grid is instead allowed to reorient as the
geometry changes (which is the default), the finite-difference
gradients no longer match the exact analytic gradient, since they
then also sample the unmodeled reorientation of the
quadrature grid; in such a misconfigured calculation we observed
spurious deviations of the order of $10^{-3}\,E_h/a_0^2$ in the
finite-difference Hessians on the coarse grid specified below.

\begin{table}
\caption{The employed Cartesian geometries (in \AA{}) for the
molecules in the demonstration calculations.}
\label{tab:geometries}
\begin{ruledtabular}
\begin{tabular}{llrrr}
Molecule & Atom & $x$ & $y$ & $z$ \\
\hline
H$_2$O & O & 0.0000 & 0.0000 & 0.0000 \\
       & H & 0.0000 & 0.7570 & $-$0.5858 \\
       & H & 0.0000 & $-$0.7570 & $-$0.5858 \\
CH$_2$ & C & 0.0000 & 0.0000 & 0.0000 \\
       & H & 0.0000 & 0.9914 & $-$0.4208 \\
       & H & 0.0000 & $-$0.9914 & $-$0.4208 \\
\end{tabular}
\end{ruledtabular}
\end{table}

The calculations employ the SPW92,\cite{Bloch1929_ZP_545,
  Dirac1930_MPCPS_376, Perdew1992_PRB_13244}
PBE,\cite{Perdew1996_PRL_3865} TPSS,\cite{Tao2003_PRL_146401} and
r$^2$SCAN\cite{Furness2020_JPCL_8208} functionals, which span the LDA,
GGA, and mGGA rungs.
Calculations for excitation energies and stability analysis use the
aug-cc-pVTZ basis set,\cite{Dunning1989_JCP_1007, Kendall1992_JCP_6796} and are validated
against the independent implementation in PySCF. Calculations of
geometric derivatives, in contrast, are validated against high-order
finite differences. This requires many displaced calculations, and
thus we chose to use the small split-valence pcseg-0 basis
set,\cite{Jensen2001_JCP_9113, Jensen2014_JCTC_1074} whose segmented
contraction suits Psi4's integral code.

Although the studied systems are tiny and the pcseg-0 basis set is too
small for predictive accuracy in meaningful applications, the whole
point of our demonstrations is to show that the \xck{} implementation
affords correct analytic expressions. The restriction to small systems
and basis sets is useful for this purpose, since the implementation
can be verified to high precision. Applications to larger, chemically
relevant systems and basis sets are possible with the generated code,
but they are not necessary to establish the correctness of the present
approach.

\subsection{Finite-difference and plane-wave calculations: GPAW}
\label{sec:compgpaw}

The plane-wave and finite-difference demonstrations of
\cref{sec:gpaw} were run with a development version of
GPAW\cite{Mortensen2005_PRB_35109, Enkovaara2010_JPCM_253202,
Mortensen2024_JCP_92503} 26.7.1b1 carrying the generated kernels,
the standard GPAW projector augmented-wave
(PAW)\cite{Bloechl1994_PRB_17953}
datasets (version 24.11.0), and Libxc\cite{Lehtola2018_S_1} version
7.1.2 for the functional-derivative arrays; following standard GPAW
practice, the mGGA calculations employ the PBE PAW datasets.
Each demonstration of \cref{sec:gpaw} has its own setup: the stress
and dielectric-response demonstrations concern bulk silicon in the
plane-wave mode, while the Casida demonstration uses the water
molecule in the real-space finite-difference mode, thereby
exercising the generated kernels on both grid representations.

The stress calculations used the two-atom diamond cell of silicon ($a
= 5.43$ \AA{}) sheared by 2\%, a 340 eV plane-wave cutoff, a $2 \times
2 \times 2$ Monkhorst--Pack\cite{Monkhorst1976_PRB_5188}
$\mathbf{k}$-point grid, densities converged to $10^{-8}$, and
symmetrized strain steps of $10^{-3}$ for the finite-difference
reference.  The dielectric-response calculations used a PBE ground
state with a 400 eV plane-wave cutoff, an $8 \times 8 \times 8$
Monkhorst--Pack $\mathbf{k}$-point grid, and 60 bands; the response
was evaluated with a 50 eV plane-wave cutoff, local-field effects, and
a broadening of $\eta$ = 0.25 eV (0.001 eV for the static dielectric
constants).

The Casida calculations used the water
molecule in a box with a grid spacing of
0.22 \AA{} and 3.5 \AA{} of vacuum, with all couplings among the four
valence orbitals and the six lowest unoccupied states, and the
implementation's default two-point differentiation step of
$10^{-5}$ for the reference kernels; this small box suffices, as
the analytic and finite-difference kernels are compared within the
same discretization, whose incompleteness cancels identically. The
cross-code comparison of the excitation energies instead requires
converged absolute energies, and the lowest excitations of water
have partial Rydberg character that a small box confines; it
therefore used a larger box (7 \AA{} of vacuum, grid spacing 0.18
\AA{}, ten converged unoccupied states) on the GPAW side.

\section{Results}
\label{sec:results}

\subsection{Term explosion, pattern collapse, and fusion}
\label{sec:collapsecount}

Before turning to the demonstration calculations, we quantify the
effect of the pattern collapse and fusion of \cref{sec:collapse}.
\Cref{tab:collapse} lists the number of terms produced for the
spin-restricted kernels of the catalog of \cref{tab:catalog} by a
straightforward application of the chain and Leibniz rules, the
number of patterns that remain after collapse, and the number of
matrix products left after fusion. The term count grows by
a factor of 5--20 per order, depending on the family, reaching, at fourth order,
862 terms for a GGA, 4452 for a full mGGA, 32\,542 for a
current-corrected mGGA, and 518\,139 for a density-Hessian
mGGA, whereas the number of patterns is a constant
of the functional family: 1 for the LDA, 7 for a GGA, 10 for the
$\tau$-mGGA and its current-corrected variant, 12 for the
Laplacian-dependent and full mGGAs, and 30 for the
density-Hessian mGGA, whose cubic
ingredient brings in higher basis-function derivatives. Fusing the operations, i.e., taking the right-hand operand of the
matrix product as a common factor, then further reduces this number
to 1 for LDAs and GGAs, and 4 for every mGGA family, regardless of
the order of the response kernel.

\Cref{tab:collapse} counts distinct monomials (no two terms of the
expansion coincide), which must be evaluated pointwise on the
quadrature grid. The reason the pattern and matrix-product counts do
not grow with the order is that raising the derivative order raises
the order of the functional derivatives and multiplies in further
perturbed ingredients, but these are per-point scalars. The
basis-function pair enters only through the ingredient seeds, whose
form does not depend on the order. Going to higher order only adds
more terms in the contractions.

The monomials are the reason for the size of the emitted source.
Each monomial is a numeric coefficient together with the operands it
multiplies and the power of each. Therefore, the monomials can be
evaluated by a fixed evaluator employing an intermediate
representation, instead of having to be implemented in purpose-built
compiled code. Specifically, a response kernel can be expressed as a
table of monomial entries, plus a fixed evaluator of a few dozen
lines. An entry such as $4\, \partial_x n^{X}\, (\partial_x n)^2\,
(\partial^2 \exc/\partial\gamma^2)\, w_g$ is stored as the number
4 and the list of its four factors. The table-driven C backend of
\cref{sec:library} is implemented in this fashion.

Casting the kernels as data is also what
makes them well suited to accelerators, and increasingly so with the
derivative order. The input to a kernel does grow with the order,
since each further order brings in another perturbed density matrix
and thereby another set of perturbed ingredient fields; but it grows
only linearly, while the number of monomials to be formed from those
fields grows by the factor of 5--20 per order quantified above. For
a GGA, going from the Fock matrix to the fourth derivative multiplies
the number of input arrays by three and the number of monomials by
more than a hundred, so the arithmetic performed per unit of input
data rises steeply with the order of the response. High-order kernels are therefore compute-bound---the regime in which accelerators are most
effective---while the table itself is read-only and shared by every
grid point.

\begin{table}
\caption{Number of terms generated by the chain and Leibniz rules,
the number of patterns remaining after the collapse of
\cref{sec:collapse}, and the number of matrix products they fuse
into, for the spin-restricted kernels of order $n$
(the $n$-th density-matrix derivative). Both the number of patterns
$N_\text{pat}$ and the number of matrix products $N_\text{mul}$ are
independent of the order.}
\label{tab:collapse}
\begin{ruledtabular}
\begin{tabular}{lrrrrrr}
 & \multicolumn{4}{c}{terms at order} & & \\
\cline{2-5}
family & 1 & 2 & 3 & 4 & $N_\text{pat}$ & $N_\text{mul}$ \\
\hline
LDA & 1 & 1 & 1 & 1 & 1 & 1 \\
GGA & 7 & 34 & 169 & 862 & 7 & 1 \\
$\tau$ mGGA & 10 & 56 & 328 & 1970 & 10 & 4 \\
Laplacian mGGA & 12 & 66 & 384 & 2298 & 12 & 4 \\
full mGGA & 15 & 96 & 645 & 4452 & 12 & 4 \\
current-corrected mGGA & 19 & 227 & 2729 & 32\,542 & 10 & 4 \\
density-Hessian mGGA & 54 & 1173 & 25\,029 & 518\,139 & 30 & 4 \\
\end{tabular}
\end{ruledtabular}
\end{table}

\subsection{Comparison to the DIRAC generator}
\label{sec:diraccomparison}

The generator in
DIRAC\cite{Saue2020_JCP_204104} discussed in \cref{sec:intro}
produces the scalar and vector coefficients $u$ and
$\mathbf{v}$ of
\begin{equation}
  \label{eq:uvform}
  K^\text{xc}_{\alpha\beta} = \sum_g w_g \Big[
  u\, \chi^*_\alpha \chi_\beta
  + 2\, \mathbf{v} \cdot \nabla\big(\chi^*_\alpha \chi_\beta\big)
  \Big]_{\rr_g},
\end{equation}
which is \cref{eq:vxcmat} truncated at the GGA level, with
$u = \partial \exc / \partial n$ and
$\mathbf{v} = (\partial \exc / \partial \gamma)\, \nabla n$ at the
unperturbed reference. Although written here for the unperturbed
reference, \cref{eq:uvform} is the form used at every order: the
generator emits one routine per order to obtain the perturbed $u$ and
$\mathbf{v}$, and a single contraction routine assembles all of them.
This is the order-independence of \cref{sec:collapsecount} seen from
the other side. The basis-function patterns are fixed by the
ingredient seeds and do not change with the order; only the
coefficients multiplying them do, which allows the two
patterns of \cref{eq:uvform} to be hard-coded once. The generator's four emitted routines
contain 5, 10, 21, and 35 terms at the first, second, third, and
fourth orders, respectively; these are the monomials that make up
$u$ and $\mathbf{v}$. Writing $b$ for a
perturbation and $u^b$, $n^b$ for the corresponding first-order
perturbed quantities, the routine for the first-order response
assembles the two monomials of
\begin{equation}
  \label{eq:diracub}
  u^b = \exc^{(2,0)} n^b + \exc^{(1,1)} Z_{0,b}
\end{equation}
and the three of
\begin{equation}
  \label{eq:diracvb}
  \mathbf{v}^b = \big(\exc^{(1,1)} n^b
  + \exc^{(0,2)} Z_{0,b}\big) \nabla n
  + \exc^{(0,1)} \nabla n^b,
\end{equation}
five monomials in all. Here the superscripts on $\exc$ count
differentiations with respect to $n$ and to
$Z_{a,b} = 2\nabla n^a \cdot \nabla n^b$, so that $Z_{0,b}$ pairs
the unperturbed density gradient with the perturbed one.

The contraction of these coefficients against the basis functions
is performed in DIRAC by hand-written Fortran, in which the two patterns
$\chi^*_\alpha \chi_\beta$ and
$\nabla(\chi^*_\alpha \chi_\beta)$ of \cref{eq:uvform} are
hard-coded. That hand-written layer confines the generator to
functionals whose matrix elements have those two patterns:
extending it to the kinetic energy density, to the density
Laplacian, or to perturbations that differentiate the basis
functions would require writing new contraction code by hand for
each new pattern.

In the present work these patterns are generated
together with the coefficients: resolving
$\nabla(\chi^*_\alpha \chi_\beta)$ by the product rule and
resolving the Cartesian components gives the seven GGA patterns
$\chi\chi$, the three $\chi\,\partial_c\chi$, and the three
$\partial_c\chi\,\chi$ of \cref{tab:collapse}. The mixed patterns are transpose
partners and are evaluated together. The fusion of
\cref{eq:fusion} then collects all of them onto the single left
factor $\chi$, so one matrix multiplication and a transposed
accumulation deliver the whole GGA kernel. Resolving the patterns
explicitly and fusing them afterwards therefore costs nothing
relative to keeping $\nabla(\chi^*_\alpha\chi_\beta)$ intact, and
it makes the higher rungs reachable, since the kinetic
energy density and the density Laplacian introduce patterns that
\cref{eq:uvform} cannot express at all.

The two implementations agree on the arithmetic but differ in how it
is organized. DIRAC's hand-written layer performs the same fold as
\cref{eq:fusion}, forming the combination $u \chi_\beta + 2
\mathbf{v}\cdot\nabla\chi_\beta$ before the contraction, so
neither implementation performs more contractions than the other,
and both are optimal in that sense.

There is a small difference in the two implementations, though.
DIRAC accumulates the result one quadrature point at a time, as a
rank-one update of the Fock matrix for each grid point, whereas the
kernels generated by \xck{} contract a whole batch of grid points in
a single matrix multiplication. DIRAC's rank-one update is limited by memory
bandwidth, while \xck{}'s batched contraction is a level-3 BLAS
operation that reaches a much larger fraction of the peak
floating-point throughput.
\subsection{Comparison to runtime automatic differentiation}
\label{sec:xcfuncomparison}

The other existing alternative approach mentioned in
\cref{sec:intro} is that of \citet{Ringholm2014_JCP_34103}, where
the perturbed densities are fed into
XCFun\cite{Ekstroem2010_JCTC_1971} as the coefficients of a
generalized density Taylor series using its \emph{contracted} mode,
in which the caller never receives a derivative array at all. Its
arithmetic is carried out over $n$
nilpotent perturbation parameters $\epsilon_1, \ldots, \epsilon_n$,
\begin{equation}
  \label{eq:nilpotent}
  \epsilon_a \epsilon_b = \epsilon_b \epsilon_a, \qquad
  \epsilon_a^2 = 0,
\end{equation}
so that a quantity carrying such parameters is a multilinear
polynomial with $2^n$ coefficients, one per subset of the
perturbations. The caller supplies every ingredient in this form,
\begin{equation}
  \label{eq:genden}
  k(\epsilon) = \sum_{S \subseteq \{1,\ldots,n\}} k^{S}
  \prod_{a \in S} \epsilon_a,
\end{equation}
where $k^{S}$ is the ingredient differentiated with respect to the
perturbations in $S$: $k^{\emptyset}$ is the unperturbed value,
$k^{\{a\}}$ the first-order perturbed one, and so on.

When $\exc$ is evaluated on such arguments, the nilpotency truncates
the expansion by itself. For a single perturbation, $k(\epsilon) =
k^{\emptyset} + k^{\{1\}} \epsilon_1$, every term beyond the
first order carries $\epsilon_1^2$ and vanishes, leaving
\begin{equation}
  \label{eq:nilpfirst}
  \exc\big(k(\epsilon)\big) = \exc\big(k^{\emptyset}\big)
  + \bigg[ \sum_k \frac{\partial \exc}{\partial k}\,
    k^{\{1\}} \bigg] \epsilon_1 ,
\end{equation}
whose $\epsilon_1$ coefficient is the first-order perturbed energy
density, which has already been contracted with the perturbed
ingredients. The same
happens at every order: the coefficient of
$\epsilon_1 \epsilon_2 \cdots \epsilon_n$ collects the
products in which each perturbation appears exactly once, which is the
$n$-th order chain rule, while all other products are annihilated.

The functional derivatives $\partial \exc / \partial k$ arise
inside this arithmetic as intermediate values, which are computed
point by point using automatic differentiation. The calculation
yields directly the contracted energy densities as an array of $2^n$
numbers indexed by the subsets of the perturbations. For example,
the contracted mode returns four numbers for a spin-polarized
$\tau$-mGGA at second order, while a $\tau$-mGGA functional has 28
symmetry-unique second derivatives with respect to its parameters
that are thus all included in the calculation. This means that
differentiation and contraction are performed in a single pass in
the contracted mode of XCFun.

The contracted mode automates the functional side of the chain rule:
given the perturbed ingredients, it returns the contracted energy
density. However, the basis-function side is still necessary, as the
perturbed ingredients $k^{S}$ of
\cref{eq:genden} have to be formed in the first place from the
perturbed density matrices and the collocated basis functions, that
is, from the seeds of \cref{eq:primitive}. Moreover, the computation of actual matrix elements requires
carrying the contracted result back onto basis-function pairs. While in the implementation of
\citet{Ringholm2014_JCP_34103} the input and output evaluation is
still written by hand, \xck{} automates these steps as well, as was
demonstrated in \cref{tab:collapse}.

Contracting at run time therefore carries three costs relative to
generating the contraction layer. The $2^n$-coefficient arithmetic
is repeated at every quadrature point of every evaluation. No
intermediate expression survives, so neither the pattern collapse of
\cref{sec:collapse} nor the fusion of \cref{eq:fusion} has anything
to act on. And the basis-function side still has to be written by
hand for every new ingredient and every new type of perturbation,
which is why high-order response implementations have been scarce. \xck{} removes all of these three issues.

\subsection{Demonstrations in Psi4}
\label{sec:psi4demo}

\subsubsection{New functionality}
\label{sec:psi4new}

We demonstrate \xck{} in practice by implementing a number of
missing features in Psi4,\cite{Smith2020_JCP_184108} which is a
popular open-source quantum chemistry program that has been used
to build a number of quantum mechanical datasets for training
machine learning potentials, see
Refs.~\onlinecite{Ballesteros2021_JCP_154104, Isert2022_SD_273,
Khrabrov2022_PCCP_25853, Eastman2023_SD_11, Neeser2023_CDC_101040,
Spronk2023_SD_619, Ullah2024_MLST_41001, Khan2025_SD_1551,
Kuryla2025_JCP_224313, Zeng2025_SD_693}, for example. We implemented TD-DFT,
CPKS, and stability analysis for mGGA functionals, which were
previously unavailable in Psi4,\cite{Smith2020_JCP_184108} using code generated by \xck{}. We also implemented analytic
nuclear Hessians for GGA and mGGA functionals, as well as the
response of the quadrature grid to nuclear displacements, which
was previously missing altogether, since we believe these features
might be interesting for future machine learning applications.

While the kernels of the catalog can also be emitted into a
low-level compiled library, which is written in C++ but exposes C
and Fortran interfaces, and which is generated and compiled on
demand with a configure-time choice of the functional families and
derivative orders, we expect
that most downstream packages will instead choose to interface with
the plugin-based emission system. As the kernel code only needs to
be generated once, it is usually easiest to bundle the generated
code in the host program's idiom as part of each program package;
this may also be the path of least effort, as writing a custom
emitter is not difficult, while it avoids the potential
difficulties of additional external dependencies and also allows
requirement-based customization of the used kernels.

Because the emitted source is idiomatic,
the approach is moreover not tied to any specific data format,
application programming interface, or parallelization scheme of
the host program. It works with any host design, as the interface
defined by the equations is simple: quadrature weights, collocated
basis-function data, ground-state and perturbed fields, and Libxc
derivative arrays. The sources generated in this way are
indistinguishable from hand-written kernels.

The drawback of such static bundling is that improvements to the \xck{}
generator that further optimize the expressions do not propagate
automatically. However, despite its novelty, \xck{} already emits
implementations that are on par with hand-optimized expressions;
moreover, any improvements can still be adopted by rerunning the
code generator. Following this approach, we implemented the new
features in Psi4's\cite{Smith2020_JCP_184108} own C++ idiom in its \texttt{libfock} module.

\subsubsection{Excitation energies and stability analysis}
\label{sec:psi4exc}

We start out by computing TD-DFT excitation energies and performing
stability analysis on H$_2$O and CH$_2$ with mGGA functionals in
Psi4,\cite{Smith2020_JCP_184108} demonstrating the newly unlocked
capabilities. Reference calculations are performed with
PySCF,\cite{Sun2018_WIRCMS_1340} which already implements this
feature.
The excitation energies and stability eigenvalues were computed
with a dense quadrature grid of 150 radial and 974 angular
points, which is much larger than the Psi4\cite{Smith2020_JCP_184108} default of 75 radial
and 302 angular points.

\Cref{tab:tdscf} shows TDA and RPA
excitation energies of H$_2$O for both the singlet and the triplet
manifold, computed with the SPW92, PBE, TPSS and r$^2$SCAN
functionals, which span the LDA, GGA and mGGA rungs; the mGGA
calculations exercise the generated $\fxc$ contractions, including
the $\tau$ second-derivative terms, while the SPW92 and PBE rows
employ Psi4's pre-existing LDA and GGA kernels. Every triplet
excitation lies below its singlet partner, as it must.

The Psi4\cite{Smith2020_JCP_184108} SPW92, PBE and TPSS values agree
with the independent PySCF reference values to within
$2\times10^{-7}\,E_h$, while the r$^2$SCAN values deviate by up to
$7\times10^{-6}\,E_h$. The two programs have differences in the way
they discard points with small weights, as well as in the details of
the atomic partitioning, and the larger r$^2$SCAN deviations reflect
that functional's numerical sensitivity to such grid-level
differences.\cite{Sitkiewicz2022_JPCL_5963,
Lehtola2022_JCP_174114, Lehtola2023_JCP_114116,
Sitkiewicz2024_JCTC_3144}

\begin{table}
\caption{Lowest three TDA and RPA excitation energies of H$_2$O (in
$E_h$) computed with Psi4\cite{Smith2020_JCP_184108} in the
aug-cc-pVTZ basis set; the mGGA rows employ the kernels
generated in this work. Values from an independent
PySCF\cite{Sun2018_WIRCMS_1340} calculation are given for comparison
on the following line of each entry. S and T denote the singlet and
triplet manifolds.}
\label{tab:tdscf}
\begin{ruledtabular}
\begin{tabular}{llccc}
 & & State 1 & State 2 & State 3 \\
\hline
SPW92, TDA, S & Psi4 & 0.2422473 & 0.2901472 & 0.3183025 \\
 & PySCF & 0.2422473 & 0.2901472 & 0.3183025 \\
SPW92, TDA, T & Psi4 & 0.2324788 & 0.2886215 & 0.3051647 \\
 & PySCF & 0.2324788 & 0.2886215 & 0.3051647 \\
SPW92, RPA, S & Psi4 & 0.2417355 & 0.2900713 & 0.3175074 \\
 & PySCF & 0.2417355 & 0.2900713 & 0.3175074 \\
SPW92, RPA, T & Psi4 & 0.2321030 & 0.2885041 & 0.3046183 \\
 & PySCF & 0.2321031 & 0.2885041 & 0.3046183 \\
PBE, TDA, S & Psi4 & 0.2361604 & 0.2826427 & 0.3151239 \\
 & PySCF & 0.2361605 & 0.2826429 & 0.3151241 \\
PBE, TDA, T & Psi4 & 0.2239190 & 0.2789412 & 0.2989918 \\
 & PySCF & 0.2239192 & 0.2789414 & 0.2989920 \\
PBE, RPA, S & Psi4 & 0.2357512 & 0.2825899 & 0.3144289 \\
 & PySCF & 0.2357514 & 0.2825901 & 0.3144291 \\
PBE, RPA, T & Psi4 & 0.2232595 & 0.2786641 & 0.2980051 \\
 & PySCF & 0.2232597 & 0.2786643 & 0.2980053 \\
TPSS, TDA, S & Psi4 & 0.2429825 & 0.2875300 & 0.3231382 \\
 & PySCF & 0.2429826 & 0.2875301 & 0.3231383 \\
TPSS, TDA, T & Psi4 & 0.2324641 & 0.2845176 & 0.3085792 \\
 & PySCF & 0.2324642 & 0.2845178 & 0.3085793 \\
TPSS, RPA, S & Psi4 & 0.2427067 & 0.2874906 & 0.3225789 \\
 & PySCF & 0.2427068 & 0.2874908 & 0.3225790 \\
TPSS, RPA, T & Psi4 & 0.2316178 & 0.2841616 & 0.3071799 \\
 & PySCF & 0.2316179 & 0.2841617 & 0.3071800 \\
r$^2$SCAN, TDA, S & Psi4 & 0.2692902 & 0.3120062 & 0.3492015 \\
 & PySCF & 0.2692902 & 0.3120063 & 0.3492023 \\
r$^2$SCAN, TDA, T & Psi4 & 0.2553411 & 0.3093165 & 0.3300395 \\
 & PySCF & 0.2553411 & 0.3093165 & 0.3300441 \\
r$^2$SCAN, RPA, S & Psi4 & 0.2689389 & 0.3119177 & 0.3485056 \\
 & PySCF & 0.2689389 & 0.3119178 & 0.3485065 \\
r$^2$SCAN, RPA, T & Psi4 & 0.2528721 & 0.3086198 & 0.3268353 \\
 & PySCF & 0.2528721 & 0.3086198 & 0.3268420 \\
\end{tabular}
\end{ruledtabular}
\end{table}

The spin-resolved kernels are demonstrated on triplet CH$_2$ in
the spin-unrestricted formalism. \Cref{tab:tdscfuks} collects TDA
and RPA excitation energies computed with SPW92, PBE, TPSS, and
r$^2$SCAN, spanning the LDA, GGA, and mGGA rungs. The SPW92, PBE, and TPSS
energies reproduce the PySCF references to within
$3\times10^{-8}\,E_h$; the r$^2$SCAN energies deviate by
up to $7\times10^{-7}\,E_h$,
again reflecting the grid sensitivity of this functional.

The TD-DFT kernels
also power the stability analysis of the unrestricted solutions:
the lowest eigenvalue of the electronic Hessian at the TPSS
solution is $0.1963084\,E_h$ (PBE: $0.1891941\,E_h$), which matches
the exact diagonalization of the corresponding electronic Hessian
in PySCF to within $10^{-7}\,E_h$.
We additionally sanity-checked the stability analysis on H$_2$ at
an internuclear distance of 1.8~\AA{}, which is in the
dissociation regime. The default initial guess lands on the spin-restricted
solution where the gerade orbital is doubly occupied, which places
electron density between the nuclei and leads to a high energy. The stability analysis
correctly detects that this solution is internally unstable, as
perturbing the orbitals to break spatial and spin symmetry leads
to a significantly lower-energy solution that corresponds to a
qualitatively correct description of the molecule at
dissociation.

\begin{table}
\caption{Lowest three TDA and RPA excitation energies of triplet
CH$_2$ (in $E_h$) computed with the generated spin-resolved kernels
in Psi4,\cite{Smith2020_JCP_184108} in the spin-unrestricted
formalism and the aug-cc-pVTZ basis set.
Values from an independent PySCF\cite{Sun2018_WIRCMS_1340}
calculation are given for comparison on the following line of each
entry.}
\label{tab:tdscfuks}
\begin{ruledtabular}
\begin{tabular}{llccc}
 & & State 1 & State 2 & State 3 \\
\hline
SPW92, TDA & Psi4 & 0.1960895 & 0.2282731 & 0.2284930 \\
 & PySCF & 0.1960895 & 0.2282731 & 0.2284930 \\
SPW92, RPA & Psi4 & 0.1958425 & 0.2281975 & 0.2282846 \\
 & PySCF & 0.1958425 & 0.2281975 & 0.2282846 \\
PBE, TDA & Psi4 & 0.1919292 & 0.2240969 & 0.2251764 \\
 & PySCF & 0.1919292 & 0.2240969 & 0.2251764 \\
PBE, RPA & Psi4 & 0.1916825 & 0.2239879 & 0.2249526 \\
 & PySCF & 0.1916825 & 0.2239879 & 0.2249526 \\
TPSS, TDA & Psi4 & 0.1993673 & 0.2300366 & 0.2337437 \\
 & PySCF & 0.1993673 & 0.2300366 & 0.2337437 \\
TPSS, RPA & Psi4 & 0.1991147 & 0.2299035 & 0.2334941 \\
 & PySCF & 0.1991147 & 0.2299035 & 0.2334940 \\
r$^2$SCAN, TDA & Psi4 & 0.2275469 & 0.2573908 & 0.2616146 \\
 & PySCF & 0.2275474 & 0.2573914 & 0.2616148 \\
r$^2$SCAN, RPA & Psi4 & 0.2267315 & 0.2571005 & 0.2608766 \\
 & PySCF & 0.2267320 & 0.2571012 & 0.2608769 \\
\end{tabular}
\end{ruledtabular}
\end{table}

\subsubsection{Geometric derivatives}
\label{sec:geomresults}

The geometric derivatives are demonstrated by end-to-end force and
Hessian calculations on H$_2$O (spin-restricted) and CH$_2$
(spin-unrestricted) with the SPW92, PBE, TPSS, and r$^2$SCAN
functionals. As
the analytic Hessian requires the solution of the CPKS equations,
these calculations test those kernels as well.
\Cref{tab:geometric} collects the maximum absolute deviations of
the analytic derivatives from central finite differences with 5-,
7-, and 9-point stencils on a deliberately coarse quadrature grid
of 30 radial and 110 angular points. Of the quantities collected in
\cref{tab:geometric}, only the gradients without the grid response
could be computed with Psi4\cite{Smith2020_JCP_184108} before the
functionality added in this work.

As the quadrature grid approaches completeness, the grid
response terms become negligible. This is why we have chosen a grid
that is too small, to make the grid response
noticeable: relative to the dense grid of 150 radial and 974
angular points, the coarse grid incurs energy errors of
$7.5\times10^{-5}$ (SPW92), $1.2\times10^{-5}$ (PBE),
$1.9\times10^{-5}$ (TPSS), and $1.1\times10^{-3}\,E_h$
(r$^2$SCAN) for H$_2$O, and of $2.1\times10^{-5}$,
$4.1\times10^{-5}$, $6.4\times10^{-5}$, and
$2.0\times10^{-4}\,E_h$, respectively, for CH$_2$.

With the grid response
enabled, the analytic gradient reproduces the finite-difference
gradient of the energy to $10^{-11}$--$10^{-10}\,E_h/a_0$; the
deviation shrinks with increasing stencil order until it saturates
at the numerical noise floor, which confirms that the analytic
gradient is the exact derivative of the computed energy. In
contrast, the gradient without the grid response deviates from the
same reference by up to $10^{-3}\,E_h/a_0$
($10^{-2}\,E_h/a_0$ for r$^2$SCAN), and this deviation is
independent of the stencil order: it is a true error of the
fixed-grid gradient expression.

The analytic Hessians, which include the grid response, match
five-point finite differences of the analytic gradients, which
include it likewise, to
$5\times10^{-9}\,E_h/a_0^2$ or better for SPW92, PBE, and TPSS.
With the 7- and 9-point stencils the deviations drop by up to two
orders of magnitude and then stop improving, which shows that the
five-point results are limited by the truncation error of the
stencil while the higher-order ones saturate at
$10^{-11}$--$10^{-10}\,E_h/a_0^2$. This saturation reflects the
accumulated numerical noise of the calculations: unlike the
fixed-grid gradient error discussed above, the deviation
\emph{decreases} with the stencil order before reaching the floor,
which an error in the derivative expressions could not do. The larger r$^2$SCAN Hessian
deviations of ${\sim}10^{-8}\,E_h/a_0^2$ reflect the
numerical noise of this functional. The grid
response thus removes the grid-convergence caveat from the analytic
derivatives, as discussed in \cref{sec:geometric}.

\begin{table}
\caption{Maximum absolute deviations of the analytic nuclear
derivatives from central finite differences with 5-, 7-, and
9-point stencils, for H$_2$O (spin-restricted) and CH$_2$
(spin-unrestricted) in the pcseg-0 basis on a deliberately coarse
quadrature grid of 30 radial and 110 angular points. The gradients
(in $E_h/a_0$) are compared against finite differences of the
energy, with the quadrature-grid response excluded from (w/o) and
included in (w/) the analytic gradient; the Hessians (in
$E_h/a_0^2$), which include the grid response, are compared
against finite differences of the analytic gradients, which also
include the grid response.}
\label{tab:geometric}
\begin{ruledtabular}
\begin{tabular}{llccc}
 & & \multicolumn{2}{c}{Gradient} & Hessian \\
\cline{3-4}
 & & w/o response & w/ response & \\
\hline
\multicolumn{5}{l}{(a) 5-point stencil} \\
\hline
SPW92 & H$_2$O & $1\times10^{-4}$ & $7\times10^{-11}$ & $8\times10^{-10}$ \\
     & CH$_2$ & $1\times10^{-4}$ & $4\times10^{-11}$ & $1\times10^{-9}$ \\
PBE & H$_2$O & $2\times10^{-4}$ & $6\times10^{-11}$ & $2\times10^{-9}$ \\
     & CH$_2$ & $1\times10^{-4}$ & $4\times10^{-11}$ & $7\times10^{-10}$ \\
TPSS & H$_2$O & $5\times10^{-4}$ & $1\times10^{-10}$ & $3\times10^{-9}$ \\
     & CH$_2$ & $6\times10^{-4}$ & $2\times10^{-10}$ & $4\times10^{-9}$ \\
r$^2$SCAN & H$_2$O & $9\times10^{-3}$ & $8\times10^{-11}$ & $5\times10^{-8}$ \\
     & CH$_2$ & $8\times10^{-3}$ & $5\times10^{-10}$ & $2\times10^{-8}$ \\
\hline
\multicolumn{5}{l}{(b) 7-point stencil} \\
\hline
SPW92 & H$_2$O & $1\times10^{-4}$ & $5\times10^{-11}$ & $6\times10^{-10}$ \\
     & CH$_2$ & $1\times10^{-4}$ & $1\times10^{-11}$ & $1\times10^{-10}$ \\
PBE & H$_2$O & $2\times10^{-4}$ & $2\times10^{-11}$ & $4\times10^{-10}$ \\
     & CH$_2$ & $1\times10^{-4}$ & $1\times10^{-11}$ & $4\times10^{-10}$ \\
TPSS & H$_2$O & $5\times10^{-4}$ & $3\times10^{-11}$ & $2\times10^{-11}$ \\
     & CH$_2$ & $6\times10^{-4}$ & $1\times10^{-11}$ & $2\times10^{-10}$ \\
r$^2$SCAN & H$_2$O & $9\times10^{-3}$ & $2\times10^{-11}$ & $3\times10^{-8}$ \\
     & CH$_2$ & $8\times10^{-3}$ & $1\times10^{-11}$ & $1\times10^{-8}$ \\
\hline
\multicolumn{5}{l}{(c) 9-point stencil} \\
\hline
SPW92 & H$_2$O & $1\times10^{-4}$ & $5\times10^{-11}$ & $6\times10^{-10}$ \\
     & CH$_2$ & $1\times10^{-4}$ & $1\times10^{-11}$ & $1\times10^{-10}$ \\
PBE & H$_2$O & $2\times10^{-4}$ & $2\times10^{-11}$ & $4\times10^{-10}$ \\
     & CH$_2$ & $1\times10^{-4}$ & $1\times10^{-11}$ & $4\times10^{-10}$ \\
TPSS & H$_2$O & $5\times10^{-4}$ & $3\times10^{-11}$ & $3\times10^{-11}$ \\
     & CH$_2$ & $6\times10^{-4}$ & $1\times10^{-11}$ & $2\times10^{-10}$ \\
r$^2$SCAN & H$_2$O & $9\times10^{-3}$ & $2\times10^{-11}$ & $2\times10^{-8}$ \\
     & CH$_2$ & $8\times10^{-3}$ & $2\times10^{-11}$ & $9\times10^{-9}$ \\
\end{tabular}
\end{ruledtabular}
\end{table}

\subsection{Demonstrations in GPAW}
\label{sec:gpaw}

To demonstrate that the generated kernels are indifferent to the
discretization, we have also piloted \xck{} in the
GPAW\cite{Mortensen2005_PRB_35109, Enkovaara2010_JPCM_253202,
Mortensen2024_JCP_92503} program, whose plane-wave and
finite-difference modes both carry the density and the potential on
uniform real-space grids, and whose response machinery was hitherto limited to
adiabatic LDA kernels at every entry point.

\subsubsection{Stress tensors}
\label{sec:gpawstress}
The counterpart of the quadrature-grid response of
\cref{sec:geometric} on a uniform grid is the explicit
cell-deformation (strain) derivative. Analytic strain derivatives are notoriously laborious to derive
and implement by hand. Yet, the stress theorem\cite{Nielsen1985_PRB_3780}, the
metric-tensor formulation of strain in density-functional
perturbation theory\cite{Hamann2005_PRB_35117}, and the
all-electron strain derivatives of numeric atom-centered
frameworks\cite{Knuth2015_CPC_33} can all be mechanically derived
by differentiation.

The cell's deformation can be written as $\mathbf{r} \to
(\mathbf{1} + \bm\epsilon) \mathbf{r}$ at fixed orbital
expansion coefficients. The coefficients respond to the deformation
as well, but that response is a separate Pulay-type contribution, like
the orthonormality term of \cref{eq:force}, and is
the host's to supply; in a plane-wave basis it surfaces as the
finite-cutoff Pulay stress discussed below. Now, every ingredient transforms by a single master law,
built from $\det(\mathbf{1} + \bm\epsilon)^{-1}$ together with
the metric $(\mathbf{1} + \bm\epsilon)^{-\mathsf{T}}$ acting on
the gradient indices; the quadrature weight, being the volume
element, carries the opposite power $\det(\mathbf{1} +
\bm\epsilon)$. Differentiating this law to the requested order generates
every strain seed. The only operand beyond those already discussed
is the kinetic-energy-density tensor
$\tau_{ab} = \tfrac12 \sum_i \partial_a \psi_i \partial_b \psi_i$.

Periodic LCAO discretizations obey a different but equally
mechanical strain law: atom-centered basis functions are
shape-rigid, so under strain at fixed fractional coordinates every
orbital center and its periodic images displace by
$\bm\epsilon (\mathbf{R}_\mu + \mathbf{T})$, while the Bloch phases
are strain-invariant at fixed reduced $\mathbf{k}$ (the wave vector
transforms contragradiently to the lattice vectors). The explicit
LCAO strain derivative is therefore a position-weighted lattice sum
over the nuclear-displacement derivative classes of
\cref{sec:geomderiv}, requiring no new generated kernels; the two
seed families thus map one-to-one onto the plane-wave stress
theorem\cite{Nielsen1985_PRB_3792} and the atom-centered strain
derivatives of Ref.~\onlinecite{Knuth2015_CPC_33}.

The generated first strain derivative reproduces GPAW's
hand-written semilocal exchange-correlation stress for a sheared
silicon crystal to machine precision ($3\times10^{-15}$ relative) for LDA, PBE, and TPSS,
regenerating an independent production implementation as demonstrated above for the LCAO codes. The regeneration also
holds at the level of complete calculations: substituting the
generated expression for GPAW's exchange-correlation stress term
leaves the total stress tensor of the sheared
crystal unchanged to floating-point roundoff
(\cref{tab:gpawstress}), and the analytic tensors agree with central
finite differences of the self-consistent total energy with respect
to the applied strain. The finite-difference residuals themselves
follow the expected pattern: the shear components, which are
insensitive to the basis-set incompleteness, agree to
$10^{-6}$--$10^{-5}$ eV/\AA$^3$, while the diagonal components carry
the $\sim\!2\times10^{-4}$ eV/\AA$^3$ Pulay contribution of the
finite plane-wave cutoff, which the analytic stress omits by
construction.

\begin{table}
\caption{Total stress tensor of a silicon crystal sheared by 2\%,
computed in GPAW's plane-wave mode with the exchange-correlation
stress term replaced by the \xck{}-generated strain expression.
Columns: maximum relative deviation of the full tensor from stock
GPAW, and absolute deviations (eV/\AA$^3$) of the $xx$ and $xy$
components from central finite differences of the self-consistent
energy.}
\label{tab:gpawstress}
\begin{ruledtabular}
\begin{tabular}{lccc}
 & vs.\ stock GPAW & $\sigma_{xx}$ vs.\ FD & $\sigma_{xy}$ vs.\ FD \\
\hline
PBE & $4\times10^{-14}$ & $1\times10^{-4}$ & $1\times10^{-5}$ \\
TPSS & $2\times10^{-14}$ & $2\times10^{-4}$ & $5\times10^{-6}$ \\
r$^2$SCAN & $2\times10^{-14}$ & $2\times10^{-4}$ & $4\times10^{-7}$ \\
\end{tabular}
\end{ruledtabular}
\end{table}

\subsubsection{Excitation energies}
\label{sec:gpawexc}
For linear response, GPAW has hitherto estimated the
exchange-correlation kernel by finite differences: the coupling
matrix element is formed by displacing the ground-state density
along one pair density and differencing the resulting
potential,\cite{Walter2008_JCP_244101}
\begin{equation}
  \label{eq:gpawfd}
  K^\text{xc}_{ia,jb} = \int n_{jb}(\rr)\,
  \frac{\vxc[n + h n_{ia}](\rr) - \vxc[n - h n_{ia}](\rr)}{2h}
  \,\mathrm{d}^3 r ,
\end{equation}
where $n_{ia} = \psi_i \psi^*_a$ is the pair density and $h$ is a
numerical displacement parameter. The generated $\fxc$ contraction
replaces \cref{eq:gpawfd}.
For PBE, the analytic route reproduces the finite-difference
transition energies to $10^{-7}$ eV, the truncation floor of the
two-point differentiation.

However, $\tau$-mGGAs fall outside the scope of \cref{eq:gpawfd}, as
they also depend on the kinetic energy density as an additional
independent ingredient. The construction can be extended by displacing
$\tau$ along the pair kinetic-energy density $\tau_{ia}$ alongside
the density; we build such a reference in the supplementary
material to validate the generated kernels. It is not,
however, what GPAW's Casida module implements, and it inherits
every drawback discussed below. A further difficulty is that TPSS-type functionals
are only piecewise smooth in the iso-orbital indicator
$z = \tau_W/\tau$, so that near-iso-orbital regions (bonds, lone
pairs and density tails) place derivative discontinuities under
any finite-difference stencil.

The more fundamental issue with \cref{eq:gpawfd} is that it is an
approximation. The displaced density $n \pm h n_{ia}$ (and, in an
extended form, $\tau \pm h \tau_{ia}$) has to stay within the
range in which the functional is well behaved, which is not
guaranteed: pair densities change sign, so the displacement can
drive the density negative in the tails, and $\tau$ is bounded from
below by the von Weizs\"acker value. The step $h$ must simultaneously be small enough for the
$O(h^2)$ truncation error of the central difference of
\cref{eq:gpawfd} to be negligible and large enough to avoid
roundoff, and a single global value cannot
meet both conditions everywhere: the density spans many orders of
magnitude across the grid, and so therefore does the relative
displacement $h n_{ia}/n$. Uniform accuracy would call for a
point-dependent step $h(\rr)$. The analytic derivatives of this
work are evaluated at the ground-state density itself, and are
one-sided evaluations of the Libxc derivatives wherever the
functional is only piecewise smooth; they are therefore free of all
of these concerns, and furnish what is to our knowledge the first
mGGA Casida calculation in GPAW.

The closed-shell spin-adapted kernels extend the Casida machinery
to triplet excitations, which GPAW did not support at
any rung: the spin-parity substitution of \cref{sec:collapse} yields
the spin-flip (triplet) kernel combinations mechanically, with the
polarized functional-derivative arrays evaluated at the
spin-compensated density.
The triplet coupling elements were
validated against second differences of the exchange-correlation
energy under antisymmetric spin-channel perturbations, with the
density, its gradient and the kinetic energy density
all displaced along the corresponding pair fields. The agreement
is $4\times10^{-7}$ for PBE over the full simulation cell. For
TPSS it reaches $2\times10^{-7}$ once the low-density and
near-iso-orbital regions are masked identically on both routes,
and then only at a displacement of $2.5\times10^{-4}$; an eightfold
larger step degrades it to $3\times10^{-4}$. The masking and the
step tuning are only required by the finite-difference reference
calculation.

The five lowest singlet and triplet states for PBE
and TPSS are shown in \cref{tab:gpawtddft}; the triplet states
again lie below their singlet partners. \Cref{tab:gpawtddft} also
reproduces some data from \cref{tab:tdscf}: the Psi4 PBE RPA rows
provide a cross-code check against an implementation that is fully
independent of the present work (we did not modify the pre-existing GGA
Casida functionality in Psi4), whereas both sides of the TPSS
comparison rely on analytic mGGA kernels from \xck{},
demonstrating cross-discretization consistency. The two lowest states of each multiplicity agree
to $0.06$ eV or better for PBE and to $0.09$ eV for TPSS, while the
singlet--triplet splitting of the lowest pair agrees to $0.02$ eV
for PBE and to $0.04$ eV for TPSS. TDA data are not reported in
\cref{tab:gpawtddft}, since GPAW's Casida module does not offer
such an option.

\begin{table}
\caption{Lowest five excitation energies (eV) of H$_2$O from the
generated kernels in two discretizations: Psi4 in the
aug-cc-pVTZ basis, all-electron, and GPAW in finite-difference
mode with a grid spacing of 0.18~\AA{}, 7~\AA{} of vacuum, and
PBE-generated PAW setups. The Psi4 values are the RPA results of
\cref{tab:tdscf} in eV, extended from three states to five. The two
lowest states of each multiplicity agree to within $0.03$--$0.09$
eV, while the higher
states diverge as they enter the Rydberg regime, where a finite
simulation box and a Gaussian basis describe diffuse states
differently.}
\label{tab:gpawtddft}
\begin{ruledtabular}
\begin{tabular}{llccccc}
 & & \multicolumn{5}{c}{state} \\
\cline{3-7}
 & & 1 & 2 & 3 & 4 & 5 \\
\hline
\multicolumn{7}{l}{Singlet} \\
PBE & Psi4 & 6.41 & 7.69 & 8.56 & 9.16 & 9.81 \\
 & GPAW & 6.46 & 7.66 & 7.97 & 8.10 & 8.44 \\
TPSS & Psi4 & 6.60 & 7.82 & 8.78 & 9.32 & 9.98 \\
 & GPAW & 6.65 & 7.79 & 8.08 & 8.20 & 8.55 \\
\hline
\multicolumn{7}{l}{Triplet} \\
PBE & Psi4 & 6.08 & 7.58 & 8.11 & 8.98 & 9.16 \\
 & GPAW & 6.13 & 7.60 & 7.85 & 7.94 & 8.18 \\
TPSS & Psi4 & 6.30 & 7.73 & 8.36 & 9.14 & 9.41 \\
 & GPAW & 6.39 & 7.75 & 7.99 & 8.06 & 8.46 \\
\end{tabular}
\end{ruledtabular}
\end{table}

The differences observed in \cref{tab:gpawtddft} have two sources. The Psi4 calculations are
all-electron, whereas GPAW employs the PAW approximation with
PBE-generated setups; the TPSS numbers therefore additionally
carry the mismatch between the functional and the setups it is
evaluated with, which possibly explains why TPSS agrees less
closely than PBE.
The one-particle discretizations differ as well---a Gaussian basis
against a real-space grid in a finite simulation box---and this
difference dominates the higher states which are in the Rydberg
regime. Rydberg states are straightforward to capture on the
real-space grid, but are trickier for atom-centered basis
expansions, where they may require doubly or triply augmented
basis sets\cite{Woon1994_JCP_2975} or dedicated Rydberg
functions;\cite{Kaufmann1989_JPBAMOP_2223} we have not attempted
either here. The Gaussian-basis excitation energies are therefore
expected to lie above the grid values for these states, which is
what \cref{tab:gpawtddft} shows: a basis that cannot extend far
enough leaves the Rydberg states variationally too high, and the
discrepancy grows with the state index.

\subsubsection{Dielectric response}
\label{sec:gpawdiel}
Before now, GPAW's reciprocal-space kernel
matrix\cite{Yan2011_PRB_245122} for periodic dielectric response
was restricted to kernels diagonal in real space, which were stored
as $\fxc(\mathbf{G} - \mathbf{G}')$. This representation cannot
accommodate the gradient structure of a GGA kernel. The
semilocal kernel operator is, however, represented exactly by a
symmetric matrix of per-point coefficient fields $c_{ab}(\rr)$ over
the derivative slots $(1, \partial_x, \partial_y, \partial_z)$,
whence
\begin{equation}
  \label{eq:KGG}
  K_\text{xc}(\mathbf{G}, \mathbf{G}') = \frac 1 {V_0} \sum_{ab}
  f_a^*(\mathbf{q} + \mathbf{G})\,
  \tilde c_{ab}(\mathbf{G} - \mathbf{G}')\,
  f_b(\mathbf{q} + \mathbf{G}'),
\end{equation}
with $f_1 = 1$ and $f_i = \mathrm{i} (\mathbf{q} + \mathbf{G})_i$,
and $\tilde c_{ab}$ the Fourier transforms of the generated
coefficient fields. The assembled kernel matrix agrees with
brute-force real-space matrix elements of the operator to machine
precision ($10^{-16}$ relative), and adiabatic GGA kernels become
available through GPAW's standard dielectric-function interface.
In contrast to the rest of the manuscript, this section does not
present any results at the mGGA level, as this would require
kinetic pair-density response functions to be included in GPAW's
Dyson equation solver, which is host-side plumbing outside the
kernel layer.

\begin{figure}
\includegraphics[width=\columnwidth]{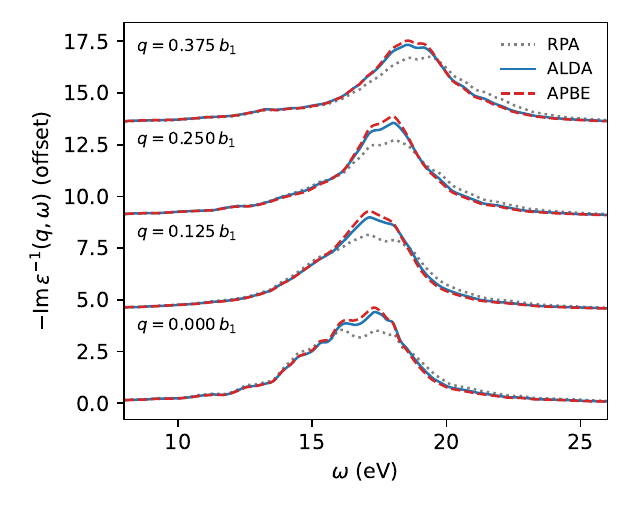}
\caption{Loss function $-\mathrm{Im}\,\epsilon^{-1}(\mathbf{q},
\omega)$ of silicon at four momenta along $\langle 111 \rangle$ (offset
vertically), computed in GPAW without an exchange--correlation
kernel (RPA), with the adiabatic LDA kernel, and with the
machine-generated adiabatic PBE kernel of \cref{eq:KGG}.}
\label{fig:plasmon}
\end{figure}

As a demonstration, \cref{fig:plasmon} shows the loss function
$-\mathrm{Im}\,\epsilon^{-1}(\mathbf{q},\omega)$ of silicon at four
momenta along the $\langle 111 \rangle$ direction with no
exchange--correlation kernel (RPA, in the sense standard for the
dielectric response, $\fxc = 0$; not to be confused with the RPA of
\cref{sec:tddft}), the ALDA, and adiabatic PBE. The exchange-correlation kernels shift the long-wavelength
plasmon from 16.1 (RPA) to 17.3 eV, bracketing the
measured 16.7 eV.\cite{Stiebling1978_PRL_1293} The plasmon dispersion, that is, the variation of the plasmon
energy $\hbar\omega_p$ with the momentum transfer $q$, is upward and
quadratic across the momenta sampled here, all of which lie below
the critical wave vector $q_c \approx 1.1$~\AA$^{-1}$ beyond which
the measured dispersion flattens; the experiments also
resolve a pronounced anisotropy between the
$\langle 100 \rangle$ and $\langle 111 \rangle$
directions,\cite{Stiebling1978_PRL_1293} so the present
momenta run along the latter.

Fitting
$\hbar\omega_p(q) = \hbar\omega_p(0) + \alpha \hbar^2 q^2 / m$ to the
computed finite-momentum peaks gives dispersion coefficients $\alpha = 0.57$ (RPA),
0.35 (ALDA) and 0.37 (adiabatic PBE), against the measured
$\alpha_{\langle 111 \rangle} = 0.32 \pm
0.02$:\cite{Stiebling1978_PRL_1293} the exchange--correlation kernel
removes most of the RPA overestimate, and the generated
gradient-corrected kernel is as close to experiment as ALDA. The
gradient corrections are otherwise small for
this observable: peak positions within 0.05 eV of ALDA (this is the
frequency resolution of these calculations), and lineshape changes
below 7\%. Indeed, this is the physically
expected result for the homogeneous-like valence electrons of
silicon, and it serves as a useful sanity check on the generated
kernel.

\section{Summary and discussion}
\label{sec:conclusions}

In the last decade, Libxc has eliminated the need for each code to
maintain its
own hand-differentiated functional implementations, yet each code
has still needed to connect the xc functional derivatives computed
by Libxc to build the actual response kernels. The response
capabilities of many production codes have been limited to LDA or
GGA functionals, since the challenge of implementing this
contraction stack grows rapidly with the number of ingredients of
the density functional.

We have discussed in this work the various ways the community has
tackled this problem. Hand-written implementations duplicate the
same derivations across codes, each in its own notation and
conventions. They do not scale to higher orders and are hard to
maintain. The generator in the source tree of DIRAC
(\cref{sec:diraccomparison}) is limited to GGAs, performs neither
pattern collapse nor deduplication, and still relies on
hand-written code. Runtime automatic differentiation
(\cref{sec:xcfuncomparison}) is slow and likewise still requires
hand-written code. Finite differences (\cref{sec:gpaw}) carry no
guarantee on the achieved precision.

In this work, we have eliminated the need to hand-develop
duplicate implementations of response codes by tying the
machine-generated functional derivatives computed by Libxc
together with an automatic application of the chain rule for the
various perturbations. Symbolic differentiation of the basis-set
ingredients combined with opaque handling of the xc derivatives
allows the production of any xc response matrix element to any
derivative order, and all of this is now available in \xck{}: a
free and open-source library under the BSD-3-Clause license.

Combined with pattern deduplication and the use
of (complex-conjugation) symmetry, \xck{} produces implementations
that construct xc kernels with the same algorithms as the best
hand-optimized codes, replacing weeks, months, or even years of
developer effort in deriving, implementing, and debugging
expressions term by term
with a few seconds to a few hours of code generation. The generated expressions are expressible as
Einstein sums, and they can be emitted in an arbitrary format by a
customizable plugin interface. This also enables drop-in support
for GPUs, either by dispatching the
Einstein sums to CuPy at run time, or by emitting low-level CUDA.
Such implementations should be especially attractive at the higher
response orders, where the number of terms to be evaluated grows by
one to two orders of magnitude per order while the data entering
and leaving the calculation remain compact. The inputs are the
basis functions on the quadrature grid and the (un)relaxed density
matrices, and the output is the kernel matrix itself. Everything in
between---the Libxc derivative arrays, which can already be
evaluated on the GPU, and their contractions with the basis and
density data---is intermediate, and need never leave the device.

The original motivation and initial impulse for this work came several
years ago, thinking how to implement response properties and
state-of-the-art orbital optimizers in
ERKALE,\cite{Lehtola2012_JCC_1572}
HelFEM,\cite{Lehtola2019_IJQC_25945, Lehtola2019_IJQC_25944,
  Lehtola2020_MP_1597989, Lehtola2020_PRA_12516,
  Lehtola2023_JCTC_2502, Lehtola2023_JCTC_4033, Lehtola2023_JPCA_4180}
Psi4,\cite{Smith2020_JCP_184108} and
PySCF.\cite{Sun2020_JCP_24109} There are many other programs in
addition to Psi4 and PySCF that already have implemented a quantum
chemistry pipeline for molecular properties, such as the solution of
the CPKS equations to compute perturbed density matrices, but that
have only implemented the necessary xc kernels for LDA or GGA density
functionals, due to the significant amount of development work
required for fully-fledged support of mGGA functionals, for
example. We believe \xck{} will be of great help to implementing
these features to various programs.

The automatic code generation enabled by \xck{} eliminates the
mechanical programming barrier hitherto hindering the use of
state-of-the-art density functionals across codes, enabling their
extension to the maximal capability of their theoretical
frameworks. For example, since the left-hand side of the response
equations does not change with the order of perturbation theory, and
the right-hand side consists of functional derivatives contracted with
perturbed densities, \xck{} appears to unlock higher-order response
theory (\cref{sec:higherorders}) across various codes as it is able to
generate the complicated right-hand side.

We exemplified \xck{} and its capabilities in this work in two host
programs of very different discretizations: the molecular Psi4
program\cite{Smith2020_JCP_184108} which employs Gaussian basis sets,
and the GPAW program\cite{Mortensen2024_JCP_92503} which is based on
grid representations and the PAW approach. We introduced generated
kernels in Psi4 providing mGGA response kernels, GGA and mGGA analytic
Hessians, and exact quadrature grid response, while the generated
kernels added to GPAW provide mGGA and triplet Casida couplings (the
latter were previously absent at any rung), and
gradient-corrected kernels for the periodic dielectric response, where
only the (renormalized) adiabatic LDA kernel and the bootstrap
approximation\cite{Sharma2011_PRL_186401} were previously available. We also regenerated GPAW's hand-written stress
implementation and found agreement at machine precision.

While most of the present discussion was in the scope of LCAO
calculations for illustrative purposes, we want to stress here that
solid-state codes beyond GPAW are also a highly likely field of
application of the present approach. The expressions of this work were
written for complex basis functions and complex orbital coefficients,
so they apply as such also to the plane-wave and Bloch-state
setting. Like molecular DFT codes, solid-state implementations of
density-functional perturbation theory\cite{Baroni2001_RMP_515,
  Gonze1997_PRB_10337} often support only LDA or GGA functionals, even
though the underlying solver structure is largely independent of the
type of the semi-local density functional. The exception concerns the
kinetic energy density: when the response equations are formulated
for the density alone, as in the Dyson equation of the dielectric
response, a $\tau$-dependent kernel also calls for the corresponding
kinetic pair-density response functions
(\cref{sec:gpawdiel}). The generated kernels declare this
requirement through their manifests, but supplying such response
functions remains host-side work. The kernels are likewise
indifferent to the fractional occupations (occupation smearing) that
are often used in the solid state: the kernels contract whatever
perturbed density matrices the host supplies, while the occupation and
Fermi-level responses of metallic density-functional perturbation
theory\cite{Gironcoli1995_PRB_6773} reside in the host's response
equations.  The only step needed to generalize these programs to the
full mGGA ladder is to implement the missing contraction kernels,
which is now possible thanks to \xck{}, as demonstrated on GPAW in
\cref{sec:gpaw}.

We wish to mention in this context the complementary route taken by
the DFTK code,\cite{Herbst2021_JP_69} whose Julia implementation is
designed to compose with algorithmic differentiation.
\citet{Schmitz2025_nCM_6} recently combined algorithmic
differentiation with the exact structures of density-functional
perturbation theory for computing derivatives for plane-wave DFT.  The
two approaches meet at the kernel layer: \xck{} could be used to
generate the symbolic expressions for the contractions, making DFTK
both a natural cross-validation partner for the generated kernels and
a straightforward integration target.

\subsection{New ingredients and future outlook}
\label{sec:newingredients}

Another motivation for this work has been to facilitate the
development of novel types of functionals in Libxc and in
quantum chemistry codes, where one is faced with the chicken-and-egg
problem: new density functionals must be tested before they are
included in Libxc, but testing is only possible if the novel type of
density functional is supported in an actual electronic structure
program.

Specifically, local hybrid functionals\cite{Jaramillo2003_JCP_1068,
  Maier2019_WIRCMS_1378} introduce a new ingredient: the local
exact-exchange energy density $e_\text{x}(\rr)$, which also means that
the SCF and response equations gain novel terms. Furthermore, the
gauge ambiguity of $e_\text{x}(\rr)$ has been addressed in some
functionals with calibration
functions.\cite{Arbuznikov2014_JCP_204101} Second-generation
calibration functions\cite{Schattenberg2021_JPCA_2697,
  Maier2016_PCCP_21133} introduce further ingredients, such as the
gradient-projected density Hessian
\begin{equation}
  \label{eq:eta}
  \eta_\sigma(\rr) = \nabla n_\sigma(\rr)^{\mathsf{T}}\,
  \big(\nabla\nabla^{\mathsf{T}} n_\sigma(\rr)\big)\, \nabla n_\sigma(\rr)
\end{equation}
which again leads to novel contributions to the SCF procedure. As is clear from \cref{eq:eta}, the ingredient is
\emph{cubic} in the density matrix, and it leads to the
complicated-looking Fock matrix contribution
\begin{equation}
  \label{eq:etafock}
\begin{split}
  F^{\eta,\sigma}_{\alpha\beta} = \sum_g w_g\,
  \frac{\partial \exc}{\partial \eta_\sigma} \Big[ &
  2\, \nabla(\chi^*_\alpha \chi_\beta)^{\mathsf{T}}\,
  (\nabla\nabla^{\mathsf{T}} n_\sigma)\, \nabla n_\sigma \\
  &+ \nabla n_\sigma^{\mathsf{T}}\,
  \nabla\nabla^{\mathsf{T}}(\chi^*_\alpha \chi_\beta)\,
  \nabla n_\sigma \Big]_{\rr_g},
\end{split}
\end{equation}
which is automatically generated by \xck{}. To our knowledge, the
only published account of these additional contributions to the
Fock matrix is the spin-resolved expression stated, without
derivation, in the appendix of
Ref.~\onlinecite{Maier2016_PCCP_21133}.

While the manual implementation of such xc kernels is already
unattractive for normal mGGAs, these new ingredients further increase
the development effort for any novel type of functionals. This is
reflected in the literature of local hybrids: the implementation of
local-hybrid self-consistency, TD-DFT kernels, and excitation-energy
gradients is a series of per-code, per-kernel derivation
papers.\cite{Bahmann2015_JCTC_1540, Maier2015_JCTC_4226,
  Grotjahn2019_JCTC_5508, Fuerst2023_JCTC_488} The hand derivation of
just the linear magnetic-response Fock contribution of $\eta$ spans
several display equations.\cite{Schattenberg2021_JPCA_2697}

In the present framework, expressions can be rapidly generated to any
order from a single few-line ingredient definition, and the
complicated hand-derivations are now made completely unnecessary by
\xck{}. In the \xck{} framework, a new ingredient is defined
\emph{once} through its value and its density-matrix seed; the entire
tower of matrix elements and response contractions for functionals
built on it then follows mechanically, enabling rapid development.

Since \xck{} is written in Python, it could even be used as a
runtime backend for automatic xc differentiation. Given a functional's
input ingredients, the library can be used to generate optimized
expressions for its xc kernels. The host program can then parse the
operand lists of the generated kernels to determine what quantities
each kernel requires. For example, computing the forces of local
hybrids will require the exact-exchange energy density and its
geometric derivative. The host can then dispatch the evaluation of
these operands to its own machinery to form the necessary arrays,
after which the contractions can be carried out as Einstein sums with
full evaluation-order optimization.

While the low-order response kernels generate in seconds, generating
the most complex high-order kernels does take time, which is why
pregeneration of static implementations in program packages will still
likely be the most attractive route.

In the following, we will discuss three future directions where we see
great promise in \xck{}: current-density DFT (\cref{sec:outlookcdft}),
non-collinear and relativistic functionals (\cref{sec:outlooknc}), and
the use of curvilinear coordinate systems (\cref{sec:outlookcurv}).

\subsubsection{Current-density DFT}
\label{sec:outlookcdft}

In gauge-corrected mGGA theory, the paramagnetic current density
\begin{equation}
  \label{eq:jp}
  \mathbf{j}_p(\rr) = \frac{1}{2i} \sum_{\alpha\beta} P_{\alpha\beta}
  \left[ \chi^*_\alpha(\rr)\, \nabla \chi_\beta(\rr)
    - \chi_\beta(\rr)\, \nabla \chi^*_\alpha(\rr) \right]
\end{equation}
is used to gauge-correct the local kinetic
energy as\cite{Dobson1993_JCP_8870, Becke2002_JCP_6935}
\begin{equation}
  \label{eq:ctau}
  \tilde\tau(\rr) = \tau(\rr) - \frac{j_p^2(\rr)}{2n(\rr)}.
\end{equation}
Importantly, while this route fits within the scope of normal mGGAs,
the modification of $\tau$ leads to changes in all of the working
equations, which until now have had to be derived manually. With the
help of \xck{}, any standard mGGA in Libxc can be converted into a
current-corrected functional, with kernels automatically generated to
arbitrary order.

The current density enters the ingredient layer of \cref{eq:primitive}
on an equal footing with the semi-local ingredients, with the
sesquilinear kernel $Q_{\mathbf{j}_p} = (2i)^{-1} (\chi^*_\alpha
\nabla\chi_\beta - \chi_\beta \nabla\chi^*_\alpha)$; unlike the
density-type ingredients, it is antisymmetric in the basis-function
pair.

For example, relative to the mGGA Fock matrix of \cref{eq:vxcmat}, the
current dependence adds the contribution
\begin{equation}
  \label{eq:jfock}
\begin{split}
  F^{\mathbf{j}}_{\alpha\beta} = \sum_g w_g\,
  \frac{\partial \exc}{\partial \tilde\tau} \bigg[ &
  \frac{j_p^2}{2 n^2}\, \chi^*_\alpha \chi_\beta \\
  & - \frac{\mathbf{j}_p}{n} \cdot
  \frac{\chi^*_\alpha \nabla \chi_\beta
    - \chi_\beta \nabla \chi^*_\alpha}{2i} \bigg]_{\rr_g},
\end{split}
\end{equation}
which follows from the chain rule through $\tilde\tau$.

The current density $\mathbf{j}_p$ is an anti-Hermitian form in the
density matrix, in contrast to the Hermitian forms of the density-type
ingredients.  It thus picks up the imaginary, antisymmetric part of
the density matrix, which, as discussed in \cref{sec:intro}, usually
does not contribute to the semi-local energy of \cref{eq:Exc}.

The gauge-corrected $\tilde\tau$ is not the only way to make the xc
energy depend on the current. The original current-density functional
theory of \citet{Vignale1987_PRL_2360} instead employs the vorticity
of \cref{sec:magnetic}, which enters as a separate functional
variable.  The two routes have
complementary scopes: the $\tilde\tau$ correction applies only to
$\tau$-dependent mGGA functionals without changes to the
functional's base form, whereas the vorticity route applies at any
rung (including the LDA), but requires genuinely new functional
parametrizations for the $\boldsymbol{\nu}$ dependence.

Both flavors of current-density DFT fit within the present framework:
$\mathbf{j}_p$ is an admissible ingredient in either case, and the
vorticity is an algebraic function of $\mathbf{j}_p$ and $n$ in the
same way that $\gamma$ is an algebraic function of $\nabla n$.
Generating the vorticity kernels will, however, require the
corresponding functional derivatives $\partial \exc / \partial
\bm\nu$, which are not currently available in Libxc.

\subsubsection{Noncollinear and relativistic functionals}
\label{sec:outlooknc}
Noncollinear functionals are another natural direction for future
work. The components of the spin magnetization $\mathbf{m}$ are
sesquilinear forms in the two-component basis functions, and they
thereby fit the ingredient layer of the present framework. Should
support for noncollinear functionals be added in Libxc, the
corresponding response kernels that are needed in noncollinear
spin-flip TD-DFT,\cite{Wang2004_JCP_12191, Bast2009_IJQC_2091} for
example, could be generated with \xck{} as well.

The same reasoning extends to the relativistic regime: in the recent
four-component Pauli-quaternion formulation of Bersson \emph{et
al.},\cite{Bersson2026_PCCP_} the charge and magnetization densities
and their gradients are sesquilinear forms in the large- and
small-component blocks of the density matrix, with the small-component
operands following mechanically from the restricted kinetic balance
condition, and the exchange--correlation potential is assembled by
hand-derived instances of exactly the chain rule that \xck{}
mechanizes. Indeed, \xck{} already generates the corresponding
relativistic kernels at the LDA and GGA levels: differentiating the
locally collinear map once reproduces the hand-derived potential
expressions of Ref.~\onlinecite{Bersson2026_PCCP_}, and
differentiating it twice yields the noncollinear $\fxc$ kernel of
four- and two-component TD-DFT, which to our knowledge had not been
derived before. The locally collinear map assigns the spin channels
\begin{equation}
  \label{eq:ncmap}
  n_\uparrow = \tfrac{1}{2}\left(n + m\right), \qquad
  n_\downarrow = \tfrac{1}{2}\left(n - m\right),
\end{equation}
with $m = |\mathbf{m}|$ the magnitude of the spin magnetization and
$\hat{\mathbf{m}} = \mathbf{m}/m$ its direction. Differentiating
\cref{eq:ncmap} twice gives, at the LDA level, the compact result
\begin{align}
  \label{eq:ncfxc}
  f^{nn} ={}& \tfrac{1}{4}\left(f^{\uparrow\uparrow}
    + 2 f^{\uparrow\downarrow} + f^{\downarrow\downarrow}\right), \\
  \label{eq:ncfxcnm}
  f^{n m_i} ={}& \tfrac{1}{4}\, \hat{m}_i
    \left(f^{\uparrow\uparrow} - f^{\downarrow\downarrow}\right), \\
  \label{eq:ncfxcmm}
  f^{m_i m_j} ={}& \tfrac{1}{4}\, \hat{m}_i \hat{m}_j
    \left(f^{\uparrow\uparrow} - 2 f^{\uparrow\downarrow}
      + f^{\downarrow\downarrow}\right) \nonumber \\
  & + \frac{\delta_{ij} - \hat{m}_i \hat{m}_j}{2m}
    \left(v^{\uparrow} - v^{\downarrow}\right),
\end{align}
where $f^{\sigma\sigma'}$ and $v^{\sigma}$ are the collinear second
and first derivatives supplied by Libxc. The first term of
\cref{eq:ncfxcmm} is the longitudinal response, which reduces to the
familiar collinear combinations, while the second is the transverse
(spin-flip) response projected onto the plane perpendicular to
$\hat{\mathbf{m}}$. The latter has no collinear counterpart and is
governed by the potential difference alone; its $1/m$ prefactor is
why the spin-compensated limit requires care. The GGA
counterpart follows from the same differentiation, but is far less
compact: the $16\times16$ coefficient matrix over the density and
magnetization components and their gradients carries 11\,070
distinct terms, and is best left to the generator. The generated kernels reproduce the collinear and
transverse (spin-flip) limits exactly, and are validated against
finite differences of the four-component energy over an explicit
restricted-kinetic-balance spinor basis, including the small-component
contributions; the mGGA rung awaits a definition of the noncollinear
kinetic energy density.

\subsubsection{Curvilinear coordinates}
\label{sec:outlookcurv}

Finally, some programs require the use of curvilinear coordinates,
such as spherical coordinates for atoms, and prolate spheroidal
coordinates for diatomic molecules.\cite{Lehtola2019_IJQC_25968}
Adaptive curvilinear coordinates have likewise been used in
solid-state calculations to concentrate the resolution where the
density varies most rapidly.\cite{Gygi1993_PRB_11692,
Hamann1995_PRB_7337} In these cases, the ingredients are built from derivatives taken along
the curvilinear axes, scaled by the corresponding Lam\'e scale factors
$h_i$. The chain rule itself is unaffected by the coordinate system:
written in the physical (orthonormal) components $g_i = h_i^{-1}
\partial_i n$, the reduced gradient of \cref{eq:gamma} is again
$\gamma = \sum_i g_i^2$, and every expression generated in this work
carries over unchanged. The metric enters solely through how an
ingredient is built from the basis functions, so that supporting a new
coordinate system amounts to declaring its scale factors.

Support for spherical and prolate spheroidal coordinates, as well as
for the two reductions in which an angular coordinate has been
integrated out analytically, leaving a residual operator on a block
index of the density matrix is already available in \xck{}: in the
spherically averaged atom, the density matrix is blocked by angular
momentum and block $l$ contributes $l(l+1) n_l(r)/r^2$ to the kinetic
energy density of \cref{eq:tau},\cite{Lehtola2023_JCTC_2502} and in
the cylindrically symmetric diatomic molecule block $m$ analogously
contributes $m^2 n_m/h_\varphi^2$. These four geometries are those
already employed in the finite-element program
HelFEM,\cite{Lehtola2019_IJQC_25944, Lehtola2019_IJQC_25945,
  Lehtola2020_MP_1597989, Lehtola2020_PRA_12516,
  Lehtola2023_JCTC_2502, Lehtola2023_JCTC_4033,
  Lehtola2023_JPCA_4180} for which we have
generated expressions for the second-order response kernels in the
spherically averaged atomic case, which we have furthermore verified
against finite differences of the energy.

The sole ingredient that does not survive the change of coordinates is
the density Laplacian, which becomes the Laplace--Beltrami operator
and brings in derivatives of the scale factors. The Laplacian is
therefore refused by \xck{} in curvilinear coordinates.

\subsection{Concluding remarks}
\label{sec:concluding}
In summary, while \xck{} is a small Python library, it has a
tremendous number of potential applications, both in enabling
existing codes to catch up with the state of the art and in
unlocking and accelerating the development of new types of density
functionals. Together with Libxc, \xck{} enables fully automatic
implementations of density-functional theory, from the symbolic
definition of the functional in Libxc to efficient implementations
of arbitrary response kernels in downstream codes. We warmly
invite developers of electronic structure codes to try out \xck{}
to automatically generate expressions for their programs;
contributions of new emitter plugins and functional/ingredient
definitions are welcome in the repository.

\appendix
\crefalias{section}{appendix}
\crefalias{subsection}{appendix}

\section{Kernel matrix elements for a
  \texorpdfstring{$\tau$}{tau}-mGGA}
\label{app:tmgga}

Here, we illustrate the verbosity of the hand-derived expressions
with the kernel matrix elements of \cref{eq:Kxc} for a
spin-polarized $\tau$-mGGA, $\exc = \exc(n_\uparrow,
n_\downarrow, \gamma_{\uparrow\uparrow},
\gamma_{\uparrow\downarrow}, \gamma_{\downarrow\downarrow},
\tau_\uparrow, \tau_\downarrow)$. These matrix elements are needed
to build the $\mathbf{A}$ and $\mathbf{B}$ matrices of
\cref{eq:Amat,eq:Bmat}. The structure of these matrix elements is standard; see, for
example, the density-matrix response formalism of
\citet{Furche2001_JCP_5982} and the mGGA TD-DFT implementation of
Bates and Furche.\cite{Bates2012_JCP_164105} We restate them in the
notation of the present work. As in the main text, $\sigma$ and $\sigma'$
denote the spins of the orbital pairs $ia$ and $jb$. The
density-matrix seeds of the ingredients for the pair $ia$ are
\begin{align}
  q^{n_\varsigma}_{ia} ={}& \delta_{\varsigma\sigma}\,
  \psi_i \psi^*_a, \label{eq:seedn} \\
  q^{\gamma_{\varsigma\varsigma'}}_{ia} ={}&
  \left( \delta_{\varsigma\sigma}\, \nabla n_{\varsigma'}
  + \delta_{\varsigma'\sigma}\, \nabla n_{\varsigma} \right)
  \cdot \nabla(\psi_i \psi^*_a), \label{eq:seedg} \\
  q^{\tau_\varsigma}_{ia} ={}& \tfrac{1}{2}\,
  \delta_{\varsigma\sigma}\, \nabla \psi_i \cdot \nabla \psi^*_a,
  \label{eq:seedt}
\end{align}
where $\varsigma$ and $\varsigma'$ label the spin channels of the
ingredients. The kernel matrix elements then read
\begin{equation}
  \label{eq:Kxctmgga}
\begin{split}
  K^\text{xc}_{ia,jb} = \sum_g w_g \bigg[ &
  \sum_{Q, Q'} \frac{\partial^2 \exc}{\partial Q\, \partial Q'}\,
  q^{Q}_{ia}\, \big(q^{Q'}_{jb}\big)^* \\
  & + \bigg( 2\, \delta_{\sigma\sigma'}\,
  \frac{\partial \exc}{\partial \gamma_{\sigma\sigma}}
  + (1 - \delta_{\sigma\sigma'})\,
  \frac{\partial \exc}{\partial \gamma_{\uparrow\downarrow}} \bigg)
  \\
  & \quad \times
  \nabla(\psi_i \psi^*_a) \cdot \nabla(\psi_j \psi^*_b)^*
  \bigg]_{\rr_g},
\end{split}
\end{equation}
where $Q$ and $Q'$ run over the seven ingredients, and all factors
are evaluated at the grid points. The double sum contains the 28
symmetry-unique second derivatives of $\exc$ that were counted in
\cref{sec:intro}; in Libxc's naming convention, these are the three
components of \texttt{v2rho2}, the six of \texttt{v2rhosigma}, the
four of \texttt{v2rhotau}, the six of \texttt{v2sigma2}, the six of
\texttt{v2sigmatau}, and the three of \texttt{v2tau2}. Each second
derivative is multiplied by its own combination of orbital values
and orbital gradients through the seeds of
\cref{eq:seedn,eq:seedg,eq:seedt}. The last term of
\cref{eq:Kxctmgga} arises from the curvature of $\gamma$, i.e., from
its second derivative with respect to the density matrix. For real
orbitals, the same $K^\text{xc}_{ia,jb}$ enters both $\mathbf{A}$
and $\mathbf{B}$. \xck{} generates \cref{eq:Kxctmgga}---and its
third- and fourth-order counterparts, which are far too long to
print---automatically.

\begin{acknowledgments}
The author thanks the Academy of Finland for financial support under
project no.~350282 and 353749.
\end{acknowledgments}

\section*{Artificial intelligence use}
The \xck{} library and this manuscript were developed with
substantial assistance from a large language model (Claude,
Anthropic), which was used as an interactive programming and
editing tool. All generated code is validated against independent
reference implementations and against finite differences, as
described in \cref{sec:results}. The author has
reviewed the code, the results, and the text, and assumes full
responsibility for the content of this work.

\section*{Data availability}
The \xck{} library, including all validation suites and the
kernel-catalog generator, is freely and openly available at
\url{https://github.com/susilehtola/libxckernel}. The Psi4\cite{Smith2020_JCP_184108}
implementation of the new features is freely and openly available in a pull
request at
\url{https://github.com/psi4/psi4/pull/3458}, and the
GPAW\cite{Mortensen2024_JCP_92503} implementation of
\cref{sec:gpaw} in a merge request at
\url{https://gitlab.com/gpaw/gpaw/-/merge_requests/3425}. The input
and output files of the calculations reported in this work are
included in the supplementary material, together with the run and
analysis scripts that reproduce every table and figure of
\cref{sec:results} from scratch through a single entry point, the
exact program versions, and the branches of the host programs that
carry the generated kernels.

\bibliography{citations,custom}

\begin{thebibliography}{210}%
\makeatletter
\providecommand \@ifxundefined [1]{%
 \@ifx{#1\undefined}
}%
\providecommand \@ifnum [1]{%
 \ifnum #1\expandafter \@firstoftwo
 \else \expandafter \@secondoftwo
 \fi
}%
\providecommand \@ifx [1]{%
 \ifx #1\expandafter \@firstoftwo
 \else \expandafter \@secondoftwo
 \fi
}%
\providecommand \natexlab [1]{#1}%
\providecommand \enquote  [1]{``#1''}%
\providecommand \bibnamefont  [1]{#1}%
\providecommand \bibfnamefont [1]{#1}%
\providecommand \citenamefont [1]{#1}%
\providecommand \href@noop [0]{\@secondoftwo}%
\providecommand \href [0]{\begingroup \@sanitize@url \@href}%
\providecommand \@href[1]{\@@startlink{#1}\@@href}%
\providecommand \@@href[1]{\endgroup#1\@@endlink}%
\providecommand \@sanitize@url [0]{\catcode `\\12\catcode `\$12\catcode
  `\&12\catcode `\#12\catcode `\^12\catcode `\_12\catcode `\%12\relax}%
\providecommand \@@startlink[1]{}%
\providecommand \@@endlink[0]{}%
\providecommand \url  [0]{\begingroup\@sanitize@url \@url }%
\providecommand \@url [1]{\endgroup\@href {#1}{\urlprefix }}%
\providecommand \urlprefix  [0]{URL }%
\providecommand \Eprint [0]{\href }%
\providecommand \doibase [0]{https://doi.org/}%
\providecommand \selectlanguage [0]{\@gobble}%
\providecommand \bibinfo  [0]{\@secondoftwo}%
\providecommand \bibfield  [0]{\@secondoftwo}%
\providecommand \translation [1]{[#1]}%
\providecommand \BibitemOpen [0]{}%
\providecommand \bibitemStop [0]{}%
\providecommand \bibitemNoStop [0]{.\EOS\space}%
\providecommand \EOS [0]{\spacefactor3000\relax}%
\providecommand \BibitemShut  [1]{\csname bibitem#1\endcsname}%
\let\auto@bib@innerbib\@empty
\bibitem [{\citenamefont {Casida}(1995)}]{Casida1995__155}%
  \BibitemOpen
  \bibfield  {author} {\bibinfo {author} {\bibfnamefont {M.~E.}\ \bibnamefont
  {Casida}},\ }\enquote {\bibinfo {title} {Time-dependent density functional
  response theory for molecules},}\ in\ \href
  {https://doi.org/10.1142/9789812830586_0005} {\emph {\bibinfo {booktitle}
  {Recent Advances in Computational Chemistry}}}\ (\bibinfo  {publisher} {WORLD
  SCIENTIFIC},\ \bibinfo {year} {1995})\ pp.\ \bibinfo {pages}
  {155--192}\BibitemShut {NoStop}%
\bibitem [{\citenamefont {Olsen}\ and\ \citenamefont
  {J{\o}rgensen}(1985)}]{Olsen1985_JCP_3235}%
  \BibitemOpen
  \bibfield  {author} {\bibinfo {author} {\bibfnamefont {J.}~\bibnamefont
  {Olsen}}\ and\ \bibinfo {author} {\bibfnamefont {P.}~\bibnamefont
  {J{\o}rgensen}},\ }\bibfield  {title} {\enquote {\bibinfo {title} {Linear and
  nonlinear response functions for an exact state and for an {MCSCF} state},}\
  }\href {https://doi.org/10.1063/1.448223} {\bibfield  {journal} {\bibinfo
  {journal} {J. Chem. Phys.}\ }\textbf {\bibinfo {volume} {82}},\ \bibinfo
  {pages} {3235--3264} (\bibinfo {year} {1985})}\BibitemShut {NoStop}%
\bibitem [{\citenamefont {Helgaker}\ \emph {et~al.}(2012)\citenamefont
  {Helgaker}, \citenamefont {Coriani}, \citenamefont {J{\o}rgensen},
  \citenamefont {Kristensen}, \citenamefont {Olsen},\ and\ \citenamefont
  {Ruud}}]{Helgaker2012_CR_543}%
  \BibitemOpen
  \bibfield  {author} {\bibinfo {author} {\bibfnamefont {T.}~\bibnamefont
  {Helgaker}}, \bibinfo {author} {\bibfnamefont {S.}~\bibnamefont {Coriani}},
  \bibinfo {author} {\bibfnamefont {P.}~\bibnamefont {J{\o}rgensen}}, \bibinfo
  {author} {\bibfnamefont {K.}~\bibnamefont {Kristensen}}, \bibinfo {author}
  {\bibfnamefont {J.}~\bibnamefont {Olsen}},\ and\ \bibinfo {author}
  {\bibfnamefont {K.}~\bibnamefont {Ruud}},\ }\bibfield  {title} {\enquote
  {\bibinfo {title} {{Recent advances in wave function-based methods of
  molecular-property calculations}},}\ }\href
  {https://doi.org/10.1021/cr2002239} {\bibfield  {journal} {\bibinfo
  {journal} {Chem. Rev.}\ }\textbf {\bibinfo {volume} {112}},\ \bibinfo {pages}
  {543--631} (\bibinfo {year} {2012})}\BibitemShut {NoStop}%
\bibitem [{\citenamefont {Norman}\ and\ \citenamefont
  {Dreuw}(2018)}]{Norman2018_CR_7208}%
  \BibitemOpen
  \bibfield  {author} {\bibinfo {author} {\bibfnamefont {P.}~\bibnamefont
  {Norman}}\ and\ \bibinfo {author} {\bibfnamefont {A.}~\bibnamefont {Dreuw}},\
  }\bibfield  {title} {\enquote {\bibinfo {title} {Simulating x-ray
  spectroscopies and calculating core-excited states of molecules},}\ }\href
  {https://doi.org/10.1021/acs.chemrev.8b00156} {\bibfield  {journal} {\bibinfo
   {journal} {Chem. Rev.}\ }\textbf {\bibinfo {volume} {118}},\ \bibinfo
  {pages} {7208--7248} (\bibinfo {year} {2018})}\BibitemShut {NoStop}%
\bibitem [{\citenamefont {Hohenberg}\ and\ \citenamefont
  {Kohn}(1964)}]{Hohenberg1964_PR_864}%
  \BibitemOpen
  \bibfield  {author} {\bibinfo {author} {\bibfnamefont {P.}~\bibnamefont
  {Hohenberg}}\ and\ \bibinfo {author} {\bibfnamefont {W.}~\bibnamefont
  {Kohn}},\ }\bibfield  {title} {\enquote {\bibinfo {title} {Inhomogeneous
  electron gas},}\ }\href {https://doi.org/10.1103/PhysRev.136.B864} {\bibfield
   {journal} {\bibinfo  {journal} {Phys. Rev.}\ }\textbf {\bibinfo {volume}
  {136}},\ \bibinfo {pages} {B864--B871} (\bibinfo {year} {1964})}\BibitemShut
  {NoStop}%
\bibitem [{\citenamefont {Kohn}\ and\ \citenamefont
  {Sham}(1965)}]{Kohn1965_PR_1133}%
  \BibitemOpen
  \bibfield  {author} {\bibinfo {author} {\bibfnamefont {W.}~\bibnamefont
  {Kohn}}\ and\ \bibinfo {author} {\bibfnamefont {L.~J.}\ \bibnamefont
  {Sham}},\ }\bibfield  {title} {\enquote {\bibinfo {title} {Self-consistent
  equations including exchange and correlation effects},}\ }\href
  {https://doi.org/10.1103/PhysRev.140.A1133} {\bibfield  {journal} {\bibinfo
  {journal} {Phys. Rev.}\ }\textbf {\bibinfo {volume} {140}},\ \bibinfo {pages}
  {A1133--A1138} (\bibinfo {year} {1965})}\BibitemShut {NoStop}%
\bibitem [{\citenamefont
  {Lehtola}(2019{\natexlab{a}})}]{Lehtola2019_IJQC_25945}%
  \BibitemOpen
  \bibfield  {author} {\bibinfo {author} {\bibfnamefont {S.}~\bibnamefont
  {Lehtola}},\ }\bibfield  {title} {\enquote {\bibinfo {title} {Fully numerical
  {Hartree}--{Fock} and density functional calculations. {I}. {Atoms}},}\
  }\href {https://doi.org/10.1002/qua.25945} {\bibfield  {journal} {\bibinfo
  {journal} {Int. J. Quantum Chem.}\ }\textbf {\bibinfo {volume} {119}},\
  \bibinfo {pages} {e25945} (\bibinfo {year} {2019}{\natexlab{a}})},\ \Eprint
  {https://arxiv.org/abs/1810.11651} {arXiv:1810.11651} \BibitemShut {NoStop}%
\bibitem [{\citenamefont
  {Lehtola}(2019{\natexlab{b}})}]{Lehtola2019_IJQC_25944}%
  \BibitemOpen
  \bibfield  {author} {\bibinfo {author} {\bibfnamefont {S.}~\bibnamefont
  {Lehtola}},\ }\bibfield  {title} {\enquote {\bibinfo {title} {Fully numerical
  {Hartree}--{Fock} and density functional calculations. {II}. {Diatomic}
  molecules},}\ }\href {https://doi.org/10.1002/qua.25944} {\bibfield
  {journal} {\bibinfo  {journal} {Int. J. Quantum Chem.}\ }\textbf {\bibinfo
  {volume} {119}},\ \bibinfo {pages} {e25944} (\bibinfo {year}
  {2019}{\natexlab{b}})},\ \Eprint {https://arxiv.org/abs/1810.11653}
  {arXiv:1810.11653} \BibitemShut {NoStop}%
\bibitem [{\citenamefont {Becke}(1993{\natexlab{a}})}]{Becke1993_JCP_1372}%
  \BibitemOpen
  \bibfield  {author} {\bibinfo {author} {\bibfnamefont {A.~D.}\ \bibnamefont
  {Becke}},\ }\bibfield  {title} {\enquote {\bibinfo {title} {A new mixing of
  {Hartree}--{Fock} and local density-functional theories},}\ }\href
  {https://doi.org/10.1063/1.464304} {\bibfield  {journal} {\bibinfo  {journal}
  {J. Chem. Phys.}\ }\textbf {\bibinfo {volume} {98}},\ \bibinfo {pages}
  {1372--1377} (\bibinfo {year} {1993}{\natexlab{a}})}\BibitemShut {NoStop}%
\bibitem [{\citenamefont {Becke}(1993{\natexlab{b}})}]{Becke1993_JCP_5648}%
  \BibitemOpen
  \bibfield  {author} {\bibinfo {author} {\bibfnamefont {A.~D.}\ \bibnamefont
  {Becke}},\ }\bibfield  {title} {\enquote {\bibinfo {title}
  {Density-functional thermochemistry. {III}. {The} role of exact exchange},}\
  }\href {https://doi.org/10.1063/1.464913} {\bibfield  {journal} {\bibinfo
  {journal} {J. Chem. Phys.}\ }\textbf {\bibinfo {volume} {98}},\ \bibinfo
  {pages} {5648--5652} (\bibinfo {year} {1993}{\natexlab{b}})}\BibitemShut
  {NoStop}%
\bibitem [{\citenamefont {Stephens}\ \emph {et~al.}(1994)\citenamefont
  {Stephens}, \citenamefont {Devlin}, \citenamefont {Chabalowski},\ and\
  \citenamefont {Frisch}}]{Stephens1994_JPC_11623}%
  \BibitemOpen
  \bibfield  {author} {\bibinfo {author} {\bibfnamefont {P.~J.}\ \bibnamefont
  {Stephens}}, \bibinfo {author} {\bibfnamefont {F.~J.}\ \bibnamefont
  {Devlin}}, \bibinfo {author} {\bibfnamefont {C.~F.}\ \bibnamefont
  {Chabalowski}},\ and\ \bibinfo {author} {\bibfnamefont {M.~J.}\ \bibnamefont
  {Frisch}},\ }\bibfield  {title} {\enquote {\bibinfo {title} {Ab initio
  calculation of vibrational absorption and circular dichroism spectra using
  density functional force fields},}\ }\href
  {https://doi.org/10.1021/j100096a001} {\bibfield  {journal} {\bibinfo
  {journal} {J. Phys. Chem.}\ }\textbf {\bibinfo {volume} {98}},\ \bibinfo
  {pages} {11623--11627} (\bibinfo {year} {1994})}\BibitemShut {NoStop}%
\bibitem [{\citenamefont {Savin}\ and\ \citenamefont
  {Flad}(1995)}]{Savin1995_IJQC_327}%
  \BibitemOpen
  \bibfield  {author} {\bibinfo {author} {\bibfnamefont {A.}~\bibnamefont
  {Savin}}\ and\ \bibinfo {author} {\bibfnamefont {H.-J.}\ \bibnamefont
  {Flad}},\ }\bibfield  {title} {\enquote {\bibinfo {title} {{Density
  functionals for the Yukawa electron-electron interaction}},}\ }\href
  {https://doi.org/10.1002/qua.560560417} {\bibfield  {journal} {\bibinfo
  {journal} {Int. J. Quantum Chem.}\ }\textbf {\bibinfo {volume} {56}},\
  \bibinfo {pages} {327--332} (\bibinfo {year} {1995})}\BibitemShut {NoStop}%
\bibitem [{\citenamefont {Leininger}\ \emph {et~al.}(1997)\citenamefont
  {Leininger}, \citenamefont {Stoll}, \citenamefont {Werner},\ and\
  \citenamefont {Savin}}]{Leininger1997_CPL_151}%
  \BibitemOpen
  \bibfield  {author} {\bibinfo {author} {\bibfnamefont {T.}~\bibnamefont
  {Leininger}}, \bibinfo {author} {\bibfnamefont {H.}~\bibnamefont {Stoll}},
  \bibinfo {author} {\bibfnamefont {H.-J.}\ \bibnamefont {Werner}},\ and\
  \bibinfo {author} {\bibfnamefont {A.}~\bibnamefont {Savin}},\ }\bibfield
  {title} {\enquote {\bibinfo {title} {Combining long-range configuration
  interaction with short-range density functionals},}\ }\href
  {https://doi.org/10.1016/S0009-2614(97)00758-6} {\bibfield  {journal}
  {\bibinfo  {journal} {Chem. Phys. Lett.}\ }\textbf {\bibinfo {volume}
  {275}},\ \bibinfo {pages} {151--160} (\bibinfo {year} {1997})}\BibitemShut
  {NoStop}%
\bibitem [{\citenamefont {Iikura}\ \emph {et~al.}(2001)\citenamefont {Iikura},
  \citenamefont {Tsuneda}, \citenamefont {Yanai},\ and\ \citenamefont
  {Hirao}}]{Iikura2001_JCP_3540}%
  \BibitemOpen
  \bibfield  {author} {\bibinfo {author} {\bibfnamefont {H.}~\bibnamefont
  {Iikura}}, \bibinfo {author} {\bibfnamefont {T.}~\bibnamefont {Tsuneda}},
  \bibinfo {author} {\bibfnamefont {T.}~\bibnamefont {Yanai}},\ and\ \bibinfo
  {author} {\bibfnamefont {K.}~\bibnamefont {Hirao}},\ }\bibfield  {title}
  {\enquote {\bibinfo {title} {{A long-range correction scheme for
  generalized-gradient-approximation exchange functionals}},}\ }\href
  {https://doi.org/10.1063/1.1383587} {\bibfield  {journal} {\bibinfo
  {journal} {J. Chem. Phys.}\ }\textbf {\bibinfo {volume} {115}},\ \bibinfo
  {pages} {3540} (\bibinfo {year} {2001})}\BibitemShut {NoStop}%
\bibitem [{\citenamefont {Yanai}, \citenamefont {Tew},\ and\ \citenamefont
  {Handy}(2004)}]{Yanai2004_CPL_51}%
  \BibitemOpen
  \bibfield  {author} {\bibinfo {author} {\bibfnamefont {T.}~\bibnamefont
  {Yanai}}, \bibinfo {author} {\bibfnamefont {D.~P.}\ \bibnamefont {Tew}},\
  and\ \bibinfo {author} {\bibfnamefont {N.~C.}\ \bibnamefont {Handy}},\
  }\bibfield  {title} {\enquote {\bibinfo {title} {{A new hybrid
  exchange--correlation functional using the Coulomb-attenuating method
  (CAM-B3LYP)}},}\ }\href {https://doi.org/10.1016/j.cplett.2004.06.011}
  {\bibfield  {journal} {\bibinfo  {journal} {Chem. Phys. Lett.}\ }\textbf
  {\bibinfo {volume} {393}},\ \bibinfo {pages} {51--57} (\bibinfo {year}
  {2004})}\BibitemShut {NoStop}%
\bibitem [{\citenamefont {Heyd}, \citenamefont {Scuseria},\ and\ \citenamefont
  {Ernzerhof}(2003)}]{Heyd2003_JCP_8207}%
  \BibitemOpen
  \bibfield  {author} {\bibinfo {author} {\bibfnamefont {J.}~\bibnamefont
  {Heyd}}, \bibinfo {author} {\bibfnamefont {G.~E.}\ \bibnamefont {Scuseria}},\
  and\ \bibinfo {author} {\bibfnamefont {M.}~\bibnamefont {Ernzerhof}},\
  }\bibfield  {title} {\enquote {\bibinfo {title} {Hybrid functionals based on
  a screened {Coulomb} potential},}\ }\href {https://doi.org/10.1063/1.1564060}
  {\bibfield  {journal} {\bibinfo  {journal} {J. Chem. Phys.}\ }\textbf
  {\bibinfo {volume} {118}},\ \bibinfo {pages} {8207--8215} (\bibinfo {year}
  {2003})}\BibitemShut {NoStop}%
\bibitem [{\citenamefont {Heyd}, \citenamefont {Scuseria},\ and\ \citenamefont
  {Ernzerhof}(2006)}]{Heyd2006_JCP_219906}%
  \BibitemOpen
  \bibfield  {author} {\bibinfo {author} {\bibfnamefont {J.}~\bibnamefont
  {Heyd}}, \bibinfo {author} {\bibfnamefont {G.~E.}\ \bibnamefont {Scuseria}},\
  and\ \bibinfo {author} {\bibfnamefont {M.}~\bibnamefont {Ernzerhof}},\
  }\bibfield  {title} {\enquote {\bibinfo {title} {Erratum: "{Hybrid}
  functionals based on a screened {Coulomb} potential" [{J}. {Chem}. {Phys}.
  118, 8207 (2003)]},}\ }\href {https://doi.org/10.1063/1.2204597} {\bibfield
  {journal} {\bibinfo  {journal} {J. Chem. Phys.}\ }\textbf {\bibinfo {volume}
  {124}},\ \bibinfo {pages} {219906} (\bibinfo {year} {2006})}\BibitemShut
  {NoStop}%
\bibitem [{\citenamefont {Mardirossian}\ and\ \citenamefont
  {Head-Gordon}(2016)}]{Mardirossian2016_JCP_214110}%
  \BibitemOpen
  \bibfield  {author} {\bibinfo {author} {\bibfnamefont {N.}~\bibnamefont
  {Mardirossian}}\ and\ \bibinfo {author} {\bibfnamefont {M.}~\bibnamefont
  {Head-Gordon}},\ }\bibfield  {title} {\enquote {\bibinfo {title}
  {$\omega${B97M-V}: A combinatorially optimized, range-separated hybrid,
  meta-{GGA} density functional with {VV10} nonlocal correlation},}\ }\href
  {https://doi.org/10.1063/1.4952647} {\bibfield  {journal} {\bibinfo
  {journal} {J. Chem. Phys.}\ }\textbf {\bibinfo {volume} {144}},\ \bibinfo
  {pages} {214110} (\bibinfo {year} {2016})}\BibitemShut {NoStop}%
\bibitem [{\citenamefont {Becke}(1988)}]{Becke1988_JCP_2547}%
  \BibitemOpen
  \bibfield  {author} {\bibinfo {author} {\bibfnamefont {A.~D.}\ \bibnamefont
  {Becke}},\ }\bibfield  {title} {\enquote {\bibinfo {title} {A multicenter
  numerical integration scheme for polyatomic molecules},}\ }\href
  {https://doi.org/10.1063/1.454033} {\bibfield  {journal} {\bibinfo  {journal}
  {J. Chem. Phys.}\ }\textbf {\bibinfo {volume} {88}},\ \bibinfo {pages}
  {2547--2553} (\bibinfo {year} {1988})}\BibitemShut {NoStop}%
\bibitem [{\citenamefont {Murray}, \citenamefont {Handy},\ and\ \citenamefont
  {Laming}(1993)}]{Murray1993_MP_997}%
  \BibitemOpen
  \bibfield  {author} {\bibinfo {author} {\bibfnamefont {C.~W.}\ \bibnamefont
  {Murray}}, \bibinfo {author} {\bibfnamefont {N.~C.}\ \bibnamefont {Handy}},\
  and\ \bibinfo {author} {\bibfnamefont {G.~J.}\ \bibnamefont {Laming}},\
  }\bibfield  {title} {\enquote {\bibinfo {title} {{Quadrature schemes for
  integrals of density functional theory}},}\ }\href
  {https://doi.org/10.1080/00268979300100651} {\bibfield  {journal} {\bibinfo
  {journal} {Mol. Phys.}\ }\textbf {\bibinfo {volume} {78}},\ \bibinfo {pages}
  {997--1014} (\bibinfo {year} {1993})}\BibitemShut {NoStop}%
\bibitem [{\citenamefont {Gill}, \citenamefont {Johnson},\ and\ \citenamefont
  {Pople}(1993)}]{Gill1993_CPL_506}%
  \BibitemOpen
  \bibfield  {author} {\bibinfo {author} {\bibfnamefont {P.~M.~W.}\
  \bibnamefont {Gill}}, \bibinfo {author} {\bibfnamefont {B.~G.}\ \bibnamefont
  {Johnson}},\ and\ \bibinfo {author} {\bibfnamefont {J.~A.}\ \bibnamefont
  {Pople}},\ }\bibfield  {title} {\enquote {\bibinfo {title} {{A standard grid
  for density functional calculations}},}\ }\href
  {https://doi.org/10.1016/0009-2614(93)80125-9} {\bibfield  {journal}
  {\bibinfo  {journal} {Chem. Phys. Lett.}\ }\textbf {\bibinfo {volume}
  {209}},\ \bibinfo {pages} {506--512} (\bibinfo {year} {1993})}\BibitemShut
  {NoStop}%
\bibitem [{\citenamefont {Treutler}\ and\ \citenamefont
  {Ahlrichs}(1995)}]{Treutler1995_JCP_346}%
  \BibitemOpen
  \bibfield  {author} {\bibinfo {author} {\bibfnamefont {O.}~\bibnamefont
  {Treutler}}\ and\ \bibinfo {author} {\bibfnamefont {R.}~\bibnamefont
  {Ahlrichs}},\ }\bibfield  {title} {\enquote {\bibinfo {title} {{Efficient
  molecular numerical integration schemes}},}\ }\href
  {https://doi.org/10.1063/1.469408} {\bibfield  {journal} {\bibinfo  {journal}
  {J. Chem. Phys.}\ }\textbf {\bibinfo {volume} {102}},\ \bibinfo {pages} {346}
  (\bibinfo {year} {1995})}\BibitemShut {NoStop}%
\bibitem [{\citenamefont {Stratmann}, \citenamefont {Scuseria},\ and\
  \citenamefont {Frisch}(1996)}]{Stratmann1996_CPL_213}%
  \BibitemOpen
  \bibfield  {author} {\bibinfo {author} {\bibfnamefont {R.~E.}\ \bibnamefont
  {Stratmann}}, \bibinfo {author} {\bibfnamefont {G.~E.}\ \bibnamefont
  {Scuseria}},\ and\ \bibinfo {author} {\bibfnamefont {M.~J.}\ \bibnamefont
  {Frisch}},\ }\bibfield  {title} {\enquote {\bibinfo {title} {{Achieving
  linear scaling in exchange-correlation density functional quadratures}},}\
  }\href {https://doi.org/10.1016/0009-2614(96)00600-8} {\bibfield  {journal}
  {\bibinfo  {journal} {Chem. Phys. Lett.}\ }\textbf {\bibinfo {volume}
  {257}},\ \bibinfo {pages} {213--223} (\bibinfo {year} {1996})}\BibitemShut
  {NoStop}%
\bibitem [{\citenamefont {Towler}, \citenamefont {Zupan},\ and\ \citenamefont
  {Caus{\`{a}}}(1996)}]{Towler1996_CPC_181}%
  \BibitemOpen
  \bibfield  {author} {\bibinfo {author} {\bibfnamefont {M.~D.}\ \bibnamefont
  {Towler}}, \bibinfo {author} {\bibfnamefont {A.}~\bibnamefont {Zupan}},\ and\
  \bibinfo {author} {\bibfnamefont {M.}~\bibnamefont {Caus{\`{a}}}},\
  }\bibfield  {title} {\enquote {\bibinfo {title} {Density functional theory in
  periodic systems using local {Gaussian} basis sets},}\ }\href
  {https://doi.org/10.1016/0010-4655(96)00078-1} {\bibfield  {journal}
  {\bibinfo  {journal} {Comput. Phys. Commun.}\ }\textbf {\bibinfo {volume}
  {98}},\ \bibinfo {pages} {181--205} (\bibinfo {year} {1996})}\BibitemShut
  {NoStop}%
\bibitem [{\citenamefont {Boerrigter}, \citenamefont {{Te Velde}},\ and\
  \citenamefont {Baerends}(1988)}]{Boerrigter1988_IJQC_87}%
  \BibitemOpen
  \bibfield  {author} {\bibinfo {author} {\bibfnamefont {P.~M.}\ \bibnamefont
  {Boerrigter}}, \bibinfo {author} {\bibfnamefont {G.}~\bibnamefont {{Te
  Velde}}},\ and\ \bibinfo {author} {\bibfnamefont {J.~E.}\ \bibnamefont
  {Baerends}},\ }\bibfield  {title} {\enquote {\bibinfo {title}
  {{Three-dimensional numerical integration for electronic structure
  calculations}},}\ }\href {https://doi.org/10.1002/qua.560330204} {\bibfield
  {journal} {\bibinfo  {journal} {Int. J. Quantum Chem.}\ }\textbf {\bibinfo
  {volume} {33}},\ \bibinfo {pages} {87--113} (\bibinfo {year}
  {1988})}\BibitemShut {NoStop}%
\bibitem [{\citenamefont {te~Velde}\ and\ \citenamefont
  {Baerends}(1991)}]{Velde1991_PRB_7888}%
  \BibitemOpen
  \bibfield  {author} {\bibinfo {author} {\bibfnamefont {G.}~\bibnamefont
  {te~Velde}}\ and\ \bibinfo {author} {\bibfnamefont {E.~J.}\ \bibnamefont
  {Baerends}},\ }\bibfield  {title} {\enquote {\bibinfo {title} {{Precise
  density-functional method for periodic structures}},}\ }\href
  {https://doi.org/10.1103/PhysRevB.44.7888} {\bibfield  {journal} {\bibinfo
  {journal} {Phys. Rev. B}\ }\textbf {\bibinfo {volume} {44}},\ \bibinfo
  {pages} {7888--7903} (\bibinfo {year} {1991})}\BibitemShut {NoStop}%
\bibitem [{\citenamefont {te~Velde}\ and\ \citenamefont
  {Baerends}(1992)}]{Velde1992_JCP_84}%
  \BibitemOpen
  \bibfield  {author} {\bibinfo {author} {\bibfnamefont {G.}~\bibnamefont
  {te~Velde}}\ and\ \bibinfo {author} {\bibfnamefont {E.~J.}\ \bibnamefont
  {Baerends}},\ }\bibfield  {title} {\enquote {\bibinfo {title} {Numerical
  integration for polyatomic systems},}\ }\href
  {https://doi.org/10.1016/0021-9991(92)90277-6} {\bibfield  {journal}
  {\bibinfo  {journal} {J. Comput. Phys.}\ }\textbf {\bibinfo {volume} {99}},\
  \bibinfo {pages} {84--98} (\bibinfo {year} {1992})}\BibitemShut {NoStop}%
\bibitem [{\citenamefont {Franchini}, \citenamefont {Philipsen},\ and\
  \citenamefont {Visscher}(2013)}]{Franchini2013_JCC_1819}%
  \BibitemOpen
  \bibfield  {author} {\bibinfo {author} {\bibfnamefont {M.}~\bibnamefont
  {Franchini}}, \bibinfo {author} {\bibfnamefont {P.~H.~T.}\ \bibnamefont
  {Philipsen}},\ and\ \bibinfo {author} {\bibfnamefont {L.}~\bibnamefont
  {Visscher}},\ }\bibfield  {title} {\enquote {\bibinfo {title} {The {Becke}
  fuzzy cells integration scheme in the amsterdam density functional program
  suite},}\ }\href {https://doi.org/10.1002/jcc.23323} {\bibfield  {journal}
  {\bibinfo  {journal} {J. Comput. Chem.}\ }\textbf {\bibinfo {volume} {34}},\
  \bibinfo {pages} {1819--1827} (\bibinfo {year} {2013})}\BibitemShut {NoStop}%
\bibitem [{\citenamefont {Ihm}, \citenamefont {Zunger},\ and\ \citenamefont
  {Cohen}(1979)}]{Ihm1979_JPCSSP_4409}%
  \BibitemOpen
  \bibfield  {author} {\bibinfo {author} {\bibfnamefont {J.}~\bibnamefont
  {Ihm}}, \bibinfo {author} {\bibfnamefont {A.}~\bibnamefont {Zunger}},\ and\
  \bibinfo {author} {\bibfnamefont {M.~L.}\ \bibnamefont {Cohen}},\ }\bibfield
  {title} {\enquote {\bibinfo {title} {Momentum-space formalism for the total
  energy of solids},}\ }\href {https://doi.org/10.1088/0022-3719/12/21/009}
  {\bibfield  {journal} {\bibinfo  {journal} {J. Phys. C Solid State Phys.}\
  }\textbf {\bibinfo {volume} {12}},\ \bibinfo {pages} {4409--4422} (\bibinfo
  {year} {1979})}\BibitemShut {NoStop}%
\bibitem [{\citenamefont {Payne}\ \emph {et~al.}(1992)\citenamefont {Payne},
  \citenamefont {Teter}, \citenamefont {Allan}, \citenamefont {Arias},\ and\
  \citenamefont {Joannopoulos}}]{Payne1992_RMP_1045}%
  \BibitemOpen
  \bibfield  {author} {\bibinfo {author} {\bibfnamefont {M.~C.}\ \bibnamefont
  {Payne}}, \bibinfo {author} {\bibfnamefont {M.~P.}\ \bibnamefont {Teter}},
  \bibinfo {author} {\bibfnamefont {D.~C.}\ \bibnamefont {Allan}}, \bibinfo
  {author} {\bibfnamefont {T.~A.}\ \bibnamefont {Arias}},\ and\ \bibinfo
  {author} {\bibfnamefont {J.~D.}\ \bibnamefont {Joannopoulos}},\ }\bibfield
  {title} {\enquote {\bibinfo {title} {{Iterative minimization techniques for
  ab initio total-energy calculations: Molecular dynamics and conjugate
  gradients}},}\ }\href {https://doi.org/10.1103/RevModPhys.64.1045} {\bibfield
   {journal} {\bibinfo  {journal} {Rev. Mod. Phys.}\ }\textbf {\bibinfo
  {volume} {64}},\ \bibinfo {pages} {1045--1097} (\bibinfo {year}
  {1992})}\BibitemShut {NoStop}%
\bibitem [{\citenamefont {Roothaan}(1951)}]{Roothaan1951_RMP_69}%
  \BibitemOpen
  \bibfield  {author} {\bibinfo {author} {\bibfnamefont {C.~C.~J.}\
  \bibnamefont {Roothaan}},\ }\bibfield  {title} {\enquote {\bibinfo {title}
  {New developments in molecular orbital theory},}\ }\href
  {https://doi.org/10.1103/RevModPhys.23.69} {\bibfield  {journal} {\bibinfo
  {journal} {Rev. Mod. Phys.}\ }\textbf {\bibinfo {volume} {23}},\ \bibinfo
  {pages} {69--89} (\bibinfo {year} {1951})}\BibitemShut {NoStop}%
\bibitem [{\citenamefont {Hall}(1951)}]{Hall1951_PRSAMPES_541}%
  \BibitemOpen
  \bibfield  {author} {\bibinfo {author} {\bibfnamefont {G.~G.}\ \bibnamefont
  {Hall}},\ }\bibfield  {title} {\enquote {\bibinfo {title} {The molecular
  orbital theory of chemical valency. {VIII}. {A} method of calculating
  ionization potentials},}\ }\href {https://doi.org/10.1098/rspa.1951.0048}
  {\bibfield  {journal} {\bibinfo  {journal} {Proc. R. Soc. A Math. Phys. Eng.
  Sci.}\ }\textbf {\bibinfo {volume} {205}},\ \bibinfo {pages} {541--552}
  (\bibinfo {year} {1951})}\BibitemShut {NoStop}%
\bibitem [{\citenamefont {Lehtola}, \citenamefont {Blockhuys},\ and\
  \citenamefont {{Van Alsenoy}}(2020)}]{Lehtola2020_M_1218}%
  \BibitemOpen
  \bibfield  {author} {\bibinfo {author} {\bibfnamefont {S.}~\bibnamefont
  {Lehtola}}, \bibinfo {author} {\bibfnamefont {F.}~\bibnamefont {Blockhuys}},\
  and\ \bibinfo {author} {\bibfnamefont {C.}~\bibnamefont {{Van Alsenoy}}},\
  }\bibfield  {title} {\enquote {\bibinfo {title} {An overview of
  self-consistent field calculations within finite basis sets},}\ }\href
  {https://doi.org/10.3390/molecules25051218} {\bibfield  {journal} {\bibinfo
  {journal} {Molecules}\ }\textbf {\bibinfo {volume} {25}},\ \bibinfo {pages}
  {1218} (\bibinfo {year} {2020})},\ \Eprint {https://arxiv.org/abs/1912.12029}
  {arXiv:1912.12029} \BibitemShut {NoStop}%
\bibitem [{\citenamefont {Lehtola}\ and\ \citenamefont
  {Burns}(2025)}]{Lehtola2025_JPCA_5651}%
  \BibitemOpen
  \bibfield  {author} {\bibinfo {author} {\bibfnamefont {S.}~\bibnamefont
  {Lehtola}}\ and\ \bibinfo {author} {\bibfnamefont {L.~A.}\ \bibnamefont
  {Burns}},\ }\bibfield  {title} {\enquote {\bibinfo {title}
  {{OpenOrbitalOptimizer}---a reusable open source library for self-consistent
  field calculations},}\ }\href {https://doi.org/10.1021/acs.jpca.5c02110}
  {\bibfield  {journal} {\bibinfo  {journal} {J. Phys. Chem. A}\ }\textbf
  {\bibinfo {volume} {129}},\ \bibinfo {pages} {5651--5664} (\bibinfo {year}
  {2025})},\ \Eprint {https://arxiv.org/abs/2503.23034} {2503.23034
  [physics.comp-ph]} \BibitemShut {NoStop}%
\bibitem [{\citenamefont {Baerends}, \citenamefont {Ellis},\ and\ \citenamefont
  {Ros}(1973)}]{Baerends1973_CP_41}%
  \BibitemOpen
  \bibfield  {author} {\bibinfo {author} {\bibfnamefont {E.~J.}\ \bibnamefont
  {Baerends}}, \bibinfo {author} {\bibfnamefont {D.~E.}\ \bibnamefont
  {Ellis}},\ and\ \bibinfo {author} {\bibfnamefont {P.}~\bibnamefont {Ros}},\
  }\bibfield  {title} {\enquote {\bibinfo {title} {{Self-consistent molecular
  Hartree--Fock--Slater calculations I. The computational procedure}},}\ }\href
  {https://doi.org/10.1016/0301-0104(73)80059-X} {\bibfield  {journal}
  {\bibinfo  {journal} {Chem. Phys.}\ }\textbf {\bibinfo {volume} {2}},\
  \bibinfo {pages} {41--51} (\bibinfo {year} {1973})}\BibitemShut {NoStop}%
\bibitem [{\citenamefont {Kobayashi}\ \emph {et~al.}(1991)\citenamefont
  {Kobayashi}, \citenamefont {Kurita}, \citenamefont {Kumahora},\ and\
  \citenamefont {Tago}}]{Kobayashi1991_PRA_5810}%
  \BibitemOpen
  \bibfield  {author} {\bibinfo {author} {\bibfnamefont {K.}~\bibnamefont
  {Kobayashi}}, \bibinfo {author} {\bibfnamefont {N.}~\bibnamefont {Kurita}},
  \bibinfo {author} {\bibfnamefont {H.}~\bibnamefont {Kumahora}},\ and\
  \bibinfo {author} {\bibfnamefont {K.}~\bibnamefont {Tago}},\ }\bibfield
  {title} {\enquote {\bibinfo {title} {Bond-energy calculations of \ce{Cu2},
  \ce{Ag2}, and cuag with the generalized gradient approximation},}\ }\href
  {https://doi.org/10.1103/PhysRevA.43.5810} {\bibfield  {journal} {\bibinfo
  {journal} {Phys. Rev. A}\ }\textbf {\bibinfo {volume} {43}},\ \bibinfo
  {pages} {5810--5813} (\bibinfo {year} {1991})}\BibitemShut {NoStop}%
\bibitem [{\citenamefont {Pople}, \citenamefont {Gill},\ and\ \citenamefont
  {Johnson}(1992)}]{Pople1992_CPL_557}%
  \BibitemOpen
  \bibfield  {author} {\bibinfo {author} {\bibfnamefont {J.~A.}\ \bibnamefont
  {Pople}}, \bibinfo {author} {\bibfnamefont {P.~M.~W.}\ \bibnamefont {Gill}},\
  and\ \bibinfo {author} {\bibfnamefont {B.~G.}\ \bibnamefont {Johnson}},\
  }\bibfield  {title} {\enquote {\bibinfo {title} {{Kohn--Sham
  density-functional theory within a finite basis set}},}\ }\href
  {https://doi.org/10.1016/0009-2614(92)85009-Y} {\bibfield  {journal}
  {\bibinfo  {journal} {Chem. Phys. Lett.}\ }\textbf {\bibinfo {volume}
  {199}},\ \bibinfo {pages} {557--560} (\bibinfo {year} {1992})}\BibitemShut
  {NoStop}%
\bibitem [{\citenamefont {Neumann}, \citenamefont {Nobes},\ and\ \citenamefont
  {Handy}(1996)}]{Neumann1996_MP_1}%
  \BibitemOpen
  \bibfield  {author} {\bibinfo {author} {\bibfnamefont {R.}~\bibnamefont
  {Neumann}}, \bibinfo {author} {\bibfnamefont {R.~H.}\ \bibnamefont {Nobes}},\
  and\ \bibinfo {author} {\bibfnamefont {N.~C.}\ \bibnamefont {Handy}},\
  }\bibfield  {title} {\enquote {\bibinfo {title} {{Exchange functionals and
  potentials}},}\ }\href {https://doi.org/10.1080/00268979650027630} {\bibfield
   {journal} {\bibinfo  {journal} {Mol. Phys.}\ }\textbf {\bibinfo {volume}
  {87}},\ \bibinfo {pages} {1--36} (\bibinfo {year} {1996})}\BibitemShut
  {NoStop}%
\bibitem [{\citenamefont {Furness}\ \emph
  {et~al.}(2020{\natexlab{a}})\citenamefont {Furness}, \citenamefont {Kaplan},
  \citenamefont {Ning}, \citenamefont {Perdew},\ and\ \citenamefont
  {Sun}}]{Furness2020_JPCL_8208}%
  \BibitemOpen
  \bibfield  {author} {\bibinfo {author} {\bibfnamefont {J.~W.}\ \bibnamefont
  {Furness}}, \bibinfo {author} {\bibfnamefont {A.~D.}\ \bibnamefont {Kaplan}},
  \bibinfo {author} {\bibfnamefont {J.}~\bibnamefont {Ning}}, \bibinfo {author}
  {\bibfnamefont {J.~P.}\ \bibnamefont {Perdew}},\ and\ \bibinfo {author}
  {\bibfnamefont {J.}~\bibnamefont {Sun}},\ }\bibfield  {title} {\enquote
  {\bibinfo {title} {Accurate and numerically efficient r$^2${SCAN}
  meta-generalized gradient approximation},}\ }\href
  {https://doi.org/10.1021/acs.jpclett.0c02405} {\bibfield  {journal} {\bibinfo
   {journal} {J. Phys. Chem. Lett.}\ }\textbf {\bibinfo {volume} {11}},\
  \bibinfo {pages} {8208--8215} (\bibinfo {year}
  {2020}{\natexlab{a}})}\BibitemShut {NoStop}%
\bibitem [{\citenamefont {Furness}\ \emph
  {et~al.}(2020{\natexlab{b}})\citenamefont {Furness}, \citenamefont {Kaplan},
  \citenamefont {Ning}, \citenamefont {Perdew},\ and\ \citenamefont
  {Sun}}]{Furness2020_JPCL_9248}%
  \BibitemOpen
  \bibfield  {author} {\bibinfo {author} {\bibfnamefont {J.~W.}\ \bibnamefont
  {Furness}}, \bibinfo {author} {\bibfnamefont {A.~D.}\ \bibnamefont {Kaplan}},
  \bibinfo {author} {\bibfnamefont {J.}~\bibnamefont {Ning}}, \bibinfo {author}
  {\bibfnamefont {J.~P.}\ \bibnamefont {Perdew}},\ and\ \bibinfo {author}
  {\bibfnamefont {J.}~\bibnamefont {Sun}},\ }\bibfield  {title} {\enquote
  {\bibinfo {title} {Correction to "{Accurate} and numerically efficient
  r$^2${SCAN} meta-generalized gradient approximation"},}\ }\href
  {https://doi.org/10.1021/acs.jpclett.0c03077} {\bibfield  {journal} {\bibinfo
   {journal} {J. Phys. Chem. Lett.}\ }\textbf {\bibinfo {volume} {11}},\
  \bibinfo {pages} {9248--9248} (\bibinfo {year}
  {2020}{\natexlab{b}})}\BibitemShut {NoStop}%
\bibitem [{\citenamefont {Lehtola}\ \emph {et~al.}(2018)\citenamefont
  {Lehtola}, \citenamefont {Steigemann}, \citenamefont {Oliveira},\ and\
  \citenamefont {Marques}}]{Lehtola2018_S_1}%
  \BibitemOpen
  \bibfield  {author} {\bibinfo {author} {\bibfnamefont {S.}~\bibnamefont
  {Lehtola}}, \bibinfo {author} {\bibfnamefont {C.}~\bibnamefont {Steigemann}},
  \bibinfo {author} {\bibfnamefont {M.~J.~T.}\ \bibnamefont {Oliveira}},\ and\
  \bibinfo {author} {\bibfnamefont {M.~A.~L.}\ \bibnamefont {Marques}},\
  }\bibfield  {title} {\enquote {\bibinfo {title} {Recent developments in
  {LIBXC}---a comprehensive library of functionals for density functional
  theory},}\ }\href {https://doi.org/10.1016/j.softx.2017.11.002} {\bibfield
  {journal} {\bibinfo  {journal} {SoftwareX}\ }\textbf {\bibinfo {volume}
  {7}},\ \bibinfo {pages} {1--5} (\bibinfo {year} {2018})}\BibitemShut
  {NoStop}%
\bibitem [{\citenamefont {{Libxc developers}}(2026)}]{LibxcWeb}%
  \BibitemOpen
  \bibfield  {author} {\bibinfo {author} {\bibnamefont {{Libxc developers}}},\
  }\href@noop {} {\enquote {\bibinfo {title} {{Libxc --- in which programs is
  it used?}}}\ }\bibinfo {howpublished} {\url{https://libxc.gitlab.io/}}
  (\bibinfo {year} {2026}),\ \bibinfo {note} {accessed 10 August
  2026}\BibitemShut {NoStop}%
\bibitem [{\citenamefont {Marques}, \citenamefont {Oliveira},\ and\
  \citenamefont {Burnus}(2012)}]{Marques2012_CPC_2272}%
  \BibitemOpen
  \bibfield  {author} {\bibinfo {author} {\bibfnamefont {M.~A.~L.}\
  \bibnamefont {Marques}}, \bibinfo {author} {\bibfnamefont {M.~J.~T.}\
  \bibnamefont {Oliveira}},\ and\ \bibinfo {author} {\bibfnamefont
  {T.}~\bibnamefont {Burnus}},\ }\bibfield  {title} {\enquote {\bibinfo {title}
  {{Libxc: A library of exchange and correlation functionals for density
  functional theory}},}\ }\href {https://doi.org/10.1016/j.cpc.2012.05.007}
  {\bibfield  {journal} {\bibinfo  {journal} {Comput. Phys. Commun.}\ }\textbf
  {\bibinfo {volume} {183}},\ \bibinfo {pages} {2272--2281} (\bibinfo {year}
  {2012})}\BibitemShut {NoStop}%
\bibitem [{\citenamefont {Lehtola}\ and\ \citenamefont
  {Karttunen}(2022)}]{Lehtola2022_WIRCMS_1610}%
  \BibitemOpen
  \bibfield  {author} {\bibinfo {author} {\bibfnamefont {S.}~\bibnamefont
  {Lehtola}}\ and\ \bibinfo {author} {\bibfnamefont {A.~J.}\ \bibnamefont
  {Karttunen}},\ }\bibfield  {title} {\enquote {\bibinfo {title} {Free and open
  source software for computational chemistry education},}\ }\href
  {https://doi.org/10.1002/wcms.1610} {\bibfield  {journal} {\bibinfo
  {journal} {Wiley Interdiscip. Rev. Comput. Mol. Sci.}\ }\textbf {\bibinfo
  {volume} {12}},\ \bibinfo {pages} {e1610} (\bibinfo {year}
  {2022})}\BibitemShut {NoStop}%
\bibitem [{\citenamefont
  {Lehtola}(2023{\natexlab{a}})}]{Lehtola2023_JCP_180901}%
  \BibitemOpen
  \bibfield  {author} {\bibinfo {author} {\bibfnamefont {S.}~\bibnamefont
  {Lehtola}},\ }\bibfield  {title} {\enquote {\bibinfo {title} {A call to arms:
  Making the case for more reusable libraries},}\ }\href
  {https://doi.org/10.1063/5.0175165} {\bibfield  {journal} {\bibinfo
  {journal} {J. Chem. Phys.}\ }\textbf {\bibinfo {volume} {159}},\ \bibinfo
  {pages} {180901} (\bibinfo {year} {2023}{\natexlab{a}})}\BibitemShut
  {NoStop}%
\bibitem [{\citenamefont {Ekstr{\"{o}}m}\ \emph {et~al.}(2010)\citenamefont
  {Ekstr{\"{o}}m}, \citenamefont {Visscher}, \citenamefont {Bast},
  \citenamefont {Thorvaldsen},\ and\ \citenamefont
  {Ruud}}]{Ekstroem2010_JCTC_1971}%
  \BibitemOpen
  \bibfield  {author} {\bibinfo {author} {\bibfnamefont {U.}~\bibnamefont
  {Ekstr{\"{o}}m}}, \bibinfo {author} {\bibfnamefont {L.}~\bibnamefont
  {Visscher}}, \bibinfo {author} {\bibfnamefont {R.}~\bibnamefont {Bast}},
  \bibinfo {author} {\bibfnamefont {A.~J.}\ \bibnamefont {Thorvaldsen}},\ and\
  \bibinfo {author} {\bibfnamefont {K.}~\bibnamefont {Ruud}},\ }\bibfield
  {title} {\enquote {\bibinfo {title} {Arbitrary-order density functional
  response theory from automatic differentiation},}\ }\href
  {https://doi.org/10.1021/ct100117s} {\bibfield  {journal} {\bibinfo
  {journal} {J. Chem. Theory Comput.}\ }\textbf {\bibinfo {volume} {6}},\
  \bibinfo {pages} {1971--1980} (\bibinfo {year} {2010})}\BibitemShut {NoStop}%
\bibitem [{\citenamefont {Saue}\ \emph {et~al.}(2020)\citenamefont {Saue},
  \citenamefont {Bast}, \citenamefont {Gomes}, \citenamefont {Jensen},
  \citenamefont {Visscher}, \citenamefont {Aucar}, \citenamefont {{Di
  Remigio}}, \citenamefont {Dyall}, \citenamefont {Eliav}, \citenamefont
  {Fasshauer}, \citenamefont {Fleig}, \citenamefont {Halbert}, \citenamefont
  {Hedeg{\aa}rd}, \citenamefont {Helmich-Paris}, \citenamefont {Ilia{\v{s}}},
  \citenamefont {Jacob}, \citenamefont {Knecht}, \citenamefont {Laerdahl},
  \citenamefont {Vidal}, \citenamefont {Nayak}, \citenamefont {Olejniczak},
  \citenamefont {Olsen}, \citenamefont {Pernpointner}, \citenamefont {Senjean},
  \citenamefont {Shee}, \citenamefont {Sunaga},\ and\ \citenamefont {van
  Stralen}}]{Saue2020_JCP_204104}%
  \BibitemOpen
  \bibfield  {author} {\bibinfo {author} {\bibfnamefont {T.}~\bibnamefont
  {Saue}}, \bibinfo {author} {\bibfnamefont {R.}~\bibnamefont {Bast}}, \bibinfo
  {author} {\bibfnamefont {A.~S.~P.}\ \bibnamefont {Gomes}}, \bibinfo {author}
  {\bibfnamefont {H.~J.~A.}\ \bibnamefont {Jensen}}, \bibinfo {author}
  {\bibfnamefont {L.}~\bibnamefont {Visscher}}, \bibinfo {author}
  {\bibfnamefont {I.~A.}\ \bibnamefont {Aucar}}, \bibinfo {author}
  {\bibfnamefont {R.}~\bibnamefont {{Di Remigio}}}, \bibinfo {author}
  {\bibfnamefont {K.~G.}\ \bibnamefont {Dyall}}, \bibinfo {author}
  {\bibfnamefont {E.}~\bibnamefont {Eliav}}, \bibinfo {author} {\bibfnamefont
  {E.}~\bibnamefont {Fasshauer}}, \bibinfo {author} {\bibfnamefont
  {T.}~\bibnamefont {Fleig}}, \bibinfo {author} {\bibfnamefont
  {L.}~\bibnamefont {Halbert}}, \bibinfo {author} {\bibfnamefont {E.~D.}\
  \bibnamefont {Hedeg{\aa}rd}}, \bibinfo {author} {\bibfnamefont
  {B.}~\bibnamefont {Helmich-Paris}}, \bibinfo {author} {\bibfnamefont
  {M.}~\bibnamefont {Ilia{\v{s}}}}, \bibinfo {author} {\bibfnamefont {C.~R.}\
  \bibnamefont {Jacob}}, \bibinfo {author} {\bibfnamefont {S.}~\bibnamefont
  {Knecht}}, \bibinfo {author} {\bibfnamefont {J.~K.}\ \bibnamefont
  {Laerdahl}}, \bibinfo {author} {\bibfnamefont {M.~L.}\ \bibnamefont {Vidal}},
  \bibinfo {author} {\bibfnamefont {M.~K.}\ \bibnamefont {Nayak}}, \bibinfo
  {author} {\bibfnamefont {M.}~\bibnamefont {Olejniczak}}, \bibinfo {author}
  {\bibfnamefont {J.~M.~H.}\ \bibnamefont {Olsen}}, \bibinfo {author}
  {\bibfnamefont {M.}~\bibnamefont {Pernpointner}}, \bibinfo {author}
  {\bibfnamefont {B.}~\bibnamefont {Senjean}}, \bibinfo {author} {\bibfnamefont
  {A.}~\bibnamefont {Shee}}, \bibinfo {author} {\bibfnamefont {A.}~\bibnamefont
  {Sunaga}},\ and\ \bibinfo {author} {\bibfnamefont {J.~N.~P.}\ \bibnamefont
  {van Stralen}},\ }\bibfield  {title} {\enquote {\bibinfo {title} {{The DIRAC
  code for relativistic molecular calculations}},}\ }\href
  {https://doi.org/10.1063/5.0004844} {\bibfield  {journal} {\bibinfo
  {journal} {J. Chem. Phys.}\ }\textbf {\bibinfo {volume} {152}},\ \bibinfo
  {pages} {204104} (\bibinfo {year} {2020})},\ \Eprint
  {https://arxiv.org/abs/2002.06121} {arXiv:2002.06121} \BibitemShut {NoStop}%
\bibitem [{\citenamefont {Bauernschmitt}\ and\ \citenamefont
  {Ahlrichs}(1996{\natexlab{a}})}]{Bauernschmitt1996_CPL_454}%
  \BibitemOpen
  \bibfield  {author} {\bibinfo {author} {\bibfnamefont {R.}~\bibnamefont
  {Bauernschmitt}}\ and\ \bibinfo {author} {\bibfnamefont {R.}~\bibnamefont
  {Ahlrichs}},\ }\bibfield  {title} {\enquote {\bibinfo {title} {{Treatment of
  electronic excitations within the adiabatic approximation of time dependent
  density functional theory}},}\ }\href
  {https://doi.org/10.1016/0009-2614(96)00440-X} {\bibfield  {journal}
  {\bibinfo  {journal} {Chem. Phys. Lett.}\ }\textbf {\bibinfo {volume}
  {256}},\ \bibinfo {pages} {454--464} (\bibinfo {year}
  {1996}{\natexlab{a}})}\BibitemShut {NoStop}%
\bibitem [{\citenamefont {Jamorski}, \citenamefont {Casida},\ and\
  \citenamefont {Salahub}(1996)}]{Jamorski1996_JCP_5134}%
  \BibitemOpen
  \bibfield  {author} {\bibinfo {author} {\bibfnamefont {C.}~\bibnamefont
  {Jamorski}}, \bibinfo {author} {\bibfnamefont {M.~E.}\ \bibnamefont
  {Casida}},\ and\ \bibinfo {author} {\bibfnamefont {D.~R.}\ \bibnamefont
  {Salahub}},\ }\bibfield  {title} {\enquote {\bibinfo {title} {{Dynamic
  polarizabilities and excitation spectra from a molecular implementation of
  time-dependent density-functional response theory: N2 as a case study}},}\
  }\href {https://doi.org/10.1063/1.471140} {\bibfield  {journal} {\bibinfo
  {journal} {J. Chem. Phys.}\ }\textbf {\bibinfo {volume} {104}},\ \bibinfo
  {pages} {5134} (\bibinfo {year} {1996})}\BibitemShut {NoStop}%
\bibitem [{\citenamefont {Petersilka}, \citenamefont {Gossmann},\ and\
  \citenamefont {Gross}(1996)}]{Petersilka1996_PRL_1212}%
  \BibitemOpen
  \bibfield  {author} {\bibinfo {author} {\bibfnamefont {M.}~\bibnamefont
  {Petersilka}}, \bibinfo {author} {\bibfnamefont {U.~J.}\ \bibnamefont
  {Gossmann}},\ and\ \bibinfo {author} {\bibfnamefont {E.~K.~U.}\ \bibnamefont
  {Gross}},\ }\bibfield  {title} {\enquote {\bibinfo {title} {Excitation
  energies from time-dependent density-functional theory},}\ }\href
  {https://doi.org/10.1103/PhysRevLett.76.1212} {\bibfield  {journal} {\bibinfo
   {journal} {Phys. Rev. Lett.}\ }\textbf {\bibinfo {volume} {76}},\ \bibinfo
  {pages} {1212--1215} (\bibinfo {year} {1996})}\BibitemShut {NoStop}%
\bibitem [{\citenamefont {Stratmann}, \citenamefont {Scuseria},\ and\
  \citenamefont {Frisch}(1998)}]{Stratmann1998_JCP_8218}%
  \BibitemOpen
  \bibfield  {author} {\bibinfo {author} {\bibfnamefont {R.~E.}\ \bibnamefont
  {Stratmann}}, \bibinfo {author} {\bibfnamefont {G.~E.}\ \bibnamefont
  {Scuseria}},\ and\ \bibinfo {author} {\bibfnamefont {M.~J.}\ \bibnamefont
  {Frisch}},\ }\bibfield  {title} {\enquote {\bibinfo {title} {{An efficient
  implementation of time-dependent density-functional theory for the
  calculation of excitation energies of large molecules}},}\ }\href
  {https://doi.org/10.1063/1.477483} {\bibfield  {journal} {\bibinfo  {journal}
  {J. Chem. Phys.}\ }\textbf {\bibinfo {volume} {109}},\ \bibinfo {pages}
  {8218} (\bibinfo {year} {1998})}\BibitemShut {NoStop}%
\bibitem [{\citenamefont {Tozer}\ and\ \citenamefont
  {Handy}(1998)}]{Tozer1998_JCP_10180}%
  \BibitemOpen
  \bibfield  {author} {\bibinfo {author} {\bibfnamefont {D.~J.}\ \bibnamefont
  {Tozer}}\ and\ \bibinfo {author} {\bibfnamefont {N.~C.}\ \bibnamefont
  {Handy}},\ }\bibfield  {title} {\enquote {\bibinfo {title} {{Improving
  virtual Kohn--Sham orbitals and eigenvalues: Application to excitation
  energies and static polarizabilities}},}\ }\href
  {https://doi.org/10.1063/1.477711} {\bibfield  {journal} {\bibinfo  {journal}
  {J. Chem. Phys.}\ }\textbf {\bibinfo {volume} {109}},\ \bibinfo {pages}
  {10180--10189} (\bibinfo {year} {1998})},\ \Eprint
  {https://arxiv.org/abs/S0021-9606~98!30946-0} {arXiv:S0021-9606~98!30946-0}
  \BibitemShut {NoStop}%
\bibitem [{\citenamefont {Hirata}\ and\ \citenamefont
  {Head-Gordon}(1999{\natexlab{a}})}]{Hirata1999_CPL_375}%
  \BibitemOpen
  \bibfield  {author} {\bibinfo {author} {\bibfnamefont {S.}~\bibnamefont
  {Hirata}}\ and\ \bibinfo {author} {\bibfnamefont {M.}~\bibnamefont
  {Head-Gordon}},\ }\bibfield  {title} {\enquote {\bibinfo {title}
  {Time-dependent density functional theory for radicals},}\ }\href
  {https://doi.org/10.1016/S0009-2614(99)00137-2} {\bibfield  {journal}
  {\bibinfo  {journal} {Chem. Phys. Lett.}\ }\textbf {\bibinfo {volume}
  {302}},\ \bibinfo {pages} {375--382} (\bibinfo {year}
  {1999}{\natexlab{a}})}\BibitemShut {NoStop}%
\bibitem [{\citenamefont {G{\"o}rling}\ \emph {et~al.}(1999)\citenamefont
  {G{\"o}rling}, \citenamefont {Heinze}, \citenamefont {Ruzankin},
  \citenamefont {Staufer},\ and\ \citenamefont
  {R{\"o}sch}}]{Goerling1999_JCP_2785}%
  \BibitemOpen
  \bibfield  {author} {\bibinfo {author} {\bibfnamefont {A.}~\bibnamefont
  {G{\"o}rling}}, \bibinfo {author} {\bibfnamefont {H.~H.}\ \bibnamefont
  {Heinze}}, \bibinfo {author} {\bibfnamefont {S.~P.}\ \bibnamefont
  {Ruzankin}}, \bibinfo {author} {\bibfnamefont {M.}~\bibnamefont {Staufer}},\
  and\ \bibinfo {author} {\bibfnamefont {N.}~\bibnamefont {R{\"o}sch}},\
  }\bibfield  {title} {\enquote {\bibinfo {title} {Density- and
  density-matrix-based coupled {Kohn}--{Sham} methods for dynamic
  polarizabilities and excitation energies of molecules},}\ }\href
  {https://doi.org/10.1063/1.477922} {\bibfield  {journal} {\bibinfo  {journal}
  {J. Chem. Phys.}\ }\textbf {\bibinfo {volume} {110}},\ \bibinfo {pages}
  {2785--2799} (\bibinfo {year} {1999})}\BibitemShut {NoStop}%
\bibitem [{\citenamefont {van Gisbergen}, \citenamefont {Snijders},\ and\
  \citenamefont {Baerends}(1999)}]{Gisbergen1999_CPC_119}%
  \BibitemOpen
  \bibfield  {author} {\bibinfo {author} {\bibfnamefont {S.}~\bibnamefont {van
  Gisbergen}}, \bibinfo {author} {\bibfnamefont {J.}~\bibnamefont {Snijders}},\
  and\ \bibinfo {author} {\bibfnamefont {E.}~\bibnamefont {Baerends}},\
  }\bibfield  {title} {\enquote {\bibinfo {title} {Implementation of
  time-dependent density functional response equations},}\ }\href
  {https://doi.org/10.1016/S0010-4655(99)00187-3} {\bibfield  {journal}
  {\bibinfo  {journal} {Comput. Phys. Commun.}\ }\textbf {\bibinfo {volume}
  {118}},\ \bibinfo {pages} {119--138} (\bibinfo {year} {1999})}\BibitemShut
  {NoStop}%
\bibitem [{\citenamefont {Rinkevicius}\ \emph {et~al.}(2003)\citenamefont
  {Rinkevicius}, \citenamefont {Tunell}, \citenamefont {Sa{\l}ek},
  \citenamefont {Vahtras},\ and\ \citenamefont
  {{\AA}gren}}]{Rinkevicius2003_JCP_34}%
  \BibitemOpen
  \bibfield  {author} {\bibinfo {author} {\bibfnamefont {Z.}~\bibnamefont
  {Rinkevicius}}, \bibinfo {author} {\bibfnamefont {I.}~\bibnamefont {Tunell}},
  \bibinfo {author} {\bibfnamefont {P.}~\bibnamefont {Sa{\l}ek}}, \bibinfo
  {author} {\bibfnamefont {O.}~\bibnamefont {Vahtras}},\ and\ \bibinfo {author}
  {\bibfnamefont {H.}~\bibnamefont {{\AA}gren}},\ }\bibfield  {title} {\enquote
  {\bibinfo {title} {Restricted density functional theory of linear
  time-dependent properties in open-shell molecules},}\ }\href
  {https://doi.org/10.1063/1.1577329} {\bibfield  {journal} {\bibinfo
  {journal} {J. Chem. Phys.}\ }\textbf {\bibinfo {volume} {119}},\ \bibinfo
  {pages} {34--46} (\bibinfo {year} {2003})}\BibitemShut {NoStop}%
\bibitem [{\citenamefont {Ringholm}\ \emph {et~al.}(2014)\citenamefont
  {Ringholm}, \citenamefont {Jonsson}, \citenamefont {Bast}, \citenamefont
  {Gao}, \citenamefont {Thorvaldsen}, \citenamefont {Ekstr{\"{o}}m},
  \citenamefont {Helgaker},\ and\ \citenamefont
  {Ruud}}]{Ringholm2014_JCP_34103}%
  \BibitemOpen
  \bibfield  {author} {\bibinfo {author} {\bibfnamefont {M.}~\bibnamefont
  {Ringholm}}, \bibinfo {author} {\bibfnamefont {D.}~\bibnamefont {Jonsson}},
  \bibinfo {author} {\bibfnamefont {R.}~\bibnamefont {Bast}}, \bibinfo {author}
  {\bibfnamefont {B.}~\bibnamefont {Gao}}, \bibinfo {author} {\bibfnamefont
  {A.~J.}\ \bibnamefont {Thorvaldsen}}, \bibinfo {author} {\bibfnamefont
  {U.}~\bibnamefont {Ekstr{\"{o}}m}}, \bibinfo {author} {\bibfnamefont
  {T.}~\bibnamefont {Helgaker}},\ and\ \bibinfo {author} {\bibfnamefont
  {K.}~\bibnamefont {Ruud}},\ }\bibfield  {title} {\enquote {\bibinfo {title}
  {{Analytic cubic and quartic force fields using density-functional
  theory}},}\ }\href {https://doi.org/10.1063/1.4861003} {\bibfield  {journal}
  {\bibinfo  {journal} {J. Chem. Phys.}\ }\textbf {\bibinfo {volume} {140}},\
  \bibinfo {pages} {034103} (\bibinfo {year} {2014})}\BibitemShut {NoStop}%
\bibitem [{\citenamefont {Mortensen}\ \emph {et~al.}(2024)\citenamefont
  {Mortensen}, \citenamefont {Larsen}, \citenamefont {Kuisma}, \citenamefont
  {Ivanov}, \citenamefont {Taghizadeh}, \citenamefont {Peterson}, \citenamefont
  {Haldar}, \citenamefont {Dohn}, \citenamefont {Sch{\"{a}}fer}, \citenamefont
  {J{\'{o}}nsson}, \citenamefont {Hermes}, \citenamefont {Nilsson},
  \citenamefont {Kastlunger}, \citenamefont {Levi}, \citenamefont
  {J{\'{o}}nsson}, \citenamefont {H{\"{a}}kkinen}, \citenamefont {Fojt},
  \citenamefont {Kangsabanik}, \citenamefont {S{\o}dequist}, \citenamefont
  {Lehtom{\"{a}}ki}, \citenamefont {Heske}, \citenamefont {Enkovaara},
  \citenamefont {Winther}, \citenamefont {Dulak}, \citenamefont {Melander},
  \citenamefont {Ovesen}, \citenamefont {Louhivuori}, \citenamefont {Walter},
  \citenamefont {Gjerding}, \citenamefont {Lopez-Acevedo}, \citenamefont
  {Erhart}, \citenamefont {Warmbier}, \citenamefont {W{\"{u}}rdemann},
  \citenamefont {Kaappa}, \citenamefont {Latini}, \citenamefont {Boland},
  \citenamefont {Bligaard}, \citenamefont {Skovhus}, \citenamefont {Susi},
  \citenamefont {Maxson}, \citenamefont {Rossi}, \citenamefont {Chen},
  \citenamefont {Schmerwitz}, \citenamefont {Schi{\o}tz}, \citenamefont
  {Olsen}, \citenamefont {Jacobsen},\ and\ \citenamefont
  {Thygesen}}]{Mortensen2024_JCP_92503}%
  \BibitemOpen
  \bibfield  {author} {\bibinfo {author} {\bibfnamefont {J.~J.}\ \bibnamefont
  {Mortensen}}, \bibinfo {author} {\bibfnamefont {A.~H.}\ \bibnamefont
  {Larsen}}, \bibinfo {author} {\bibfnamefont {M.}~\bibnamefont {Kuisma}},
  \bibinfo {author} {\bibfnamefont {A.~V.}\ \bibnamefont {Ivanov}}, \bibinfo
  {author} {\bibfnamefont {A.}~\bibnamefont {Taghizadeh}}, \bibinfo {author}
  {\bibfnamefont {A.}~\bibnamefont {Peterson}}, \bibinfo {author}
  {\bibfnamefont {A.}~\bibnamefont {Haldar}}, \bibinfo {author} {\bibfnamefont
  {A.~O.}\ \bibnamefont {Dohn}}, \bibinfo {author} {\bibfnamefont
  {C.}~\bibnamefont {Sch{\"{a}}fer}}, \bibinfo {author} {\bibfnamefont
  {E.~{\"{O}}.}\ \bibnamefont {J{\'{o}}nsson}}, \bibinfo {author}
  {\bibfnamefont {E.~D.}\ \bibnamefont {Hermes}}, \bibinfo {author}
  {\bibfnamefont {F.~A.}\ \bibnamefont {Nilsson}}, \bibinfo {author}
  {\bibfnamefont {G.}~\bibnamefont {Kastlunger}}, \bibinfo {author}
  {\bibfnamefont {G.}~\bibnamefont {Levi}}, \bibinfo {author} {\bibfnamefont
  {H.}~\bibnamefont {J{\'{o}}nsson}}, \bibinfo {author} {\bibfnamefont
  {H.}~\bibnamefont {H{\"{a}}kkinen}}, \bibinfo {author} {\bibfnamefont
  {J.}~\bibnamefont {Fojt}}, \bibinfo {author} {\bibfnamefont {J.}~\bibnamefont
  {Kangsabanik}}, \bibinfo {author} {\bibfnamefont {J.}~\bibnamefont
  {S{\o}dequist}}, \bibinfo {author} {\bibfnamefont {J.}~\bibnamefont
  {Lehtom{\"{a}}ki}}, \bibinfo {author} {\bibfnamefont {J.}~\bibnamefont
  {Heske}}, \bibinfo {author} {\bibfnamefont {J.}~\bibnamefont {Enkovaara}},
  \bibinfo {author} {\bibfnamefont {K.~T.}\ \bibnamefont {Winther}}, \bibinfo
  {author} {\bibfnamefont {M.}~\bibnamefont {Dulak}}, \bibinfo {author}
  {\bibfnamefont {M.~M.}\ \bibnamefont {Melander}}, \bibinfo {author}
  {\bibfnamefont {M.}~\bibnamefont {Ovesen}}, \bibinfo {author} {\bibfnamefont
  {M.}~\bibnamefont {Louhivuori}}, \bibinfo {author} {\bibfnamefont
  {M.}~\bibnamefont {Walter}}, \bibinfo {author} {\bibfnamefont
  {M.}~\bibnamefont {Gjerding}}, \bibinfo {author} {\bibfnamefont
  {O.}~\bibnamefont {Lopez-Acevedo}}, \bibinfo {author} {\bibfnamefont
  {P.}~\bibnamefont {Erhart}}, \bibinfo {author} {\bibfnamefont
  {R.}~\bibnamefont {Warmbier}}, \bibinfo {author} {\bibfnamefont
  {R.}~\bibnamefont {W{\"{u}}rdemann}}, \bibinfo {author} {\bibfnamefont
  {S.}~\bibnamefont {Kaappa}}, \bibinfo {author} {\bibfnamefont
  {S.}~\bibnamefont {Latini}}, \bibinfo {author} {\bibfnamefont {T.~M.}\
  \bibnamefont {Boland}}, \bibinfo {author} {\bibfnamefont {T.}~\bibnamefont
  {Bligaard}}, \bibinfo {author} {\bibfnamefont {T.}~\bibnamefont {Skovhus}},
  \bibinfo {author} {\bibfnamefont {T.}~\bibnamefont {Susi}}, \bibinfo {author}
  {\bibfnamefont {T.}~\bibnamefont {Maxson}}, \bibinfo {author} {\bibfnamefont
  {T.}~\bibnamefont {Rossi}}, \bibinfo {author} {\bibfnamefont
  {X.}~\bibnamefont {Chen}}, \bibinfo {author} {\bibfnamefont {Y.~L.~A.}\
  \bibnamefont {Schmerwitz}}, \bibinfo {author} {\bibfnamefont
  {J.}~\bibnamefont {Schi{\o}tz}}, \bibinfo {author} {\bibfnamefont
  {T.}~\bibnamefont {Olsen}}, \bibinfo {author} {\bibfnamefont {K.~W.}\
  \bibnamefont {Jacobsen}},\ and\ \bibinfo {author} {\bibfnamefont {K.~S.}\
  \bibnamefont {Thygesen}},\ }\bibfield  {title} {\enquote {\bibinfo {title}
  {{GPAW}: An open {Python} package for electronic structure calculations},}\
  }\href {https://doi.org/10.1063/5.0182685} {\bibfield  {journal} {\bibinfo
  {journal} {J. Chem. Phys.}\ }\textbf {\bibinfo {volume} {160}},\ \bibinfo
  {pages} {092503} (\bibinfo {year} {2024})}\BibitemShut {NoStop}%
\bibitem [{\citenamefont {Smith}\ \emph {et~al.}(2020)\citenamefont {Smith},
  \citenamefont {Burns}, \citenamefont {Simmonett}, \citenamefont {Parrish},
  \citenamefont {Schieber}, \citenamefont {Galvelis}, \citenamefont {Kraus},
  \citenamefont {Kruse}, \citenamefont {{Di Remigio}}, \citenamefont
  {Alenaizan}, \citenamefont {James}, \citenamefont {Lehtola}, \citenamefont
  {Misiewicz}, \citenamefont {Scheurer}, \citenamefont {Shaw}, \citenamefont
  {Schriber}, \citenamefont {Xie}, \citenamefont {Glick}, \citenamefont
  {Sirianni}, \citenamefont {O'Brien}, \citenamefont {Waldrop}, \citenamefont
  {Kumar}, \citenamefont {Hohenstein}, \citenamefont {Pritchard}, \citenamefont
  {Brooks}, \citenamefont {Schaefer}, \citenamefont {Sokolov}, \citenamefont
  {Patkowski}, \citenamefont {DePrince}, \citenamefont {Bozkaya}, \citenamefont
  {King}, \citenamefont {Evangelista}, \citenamefont {Turney}, \citenamefont
  {Crawford},\ and\ \citenamefont {Sherrill}}]{Smith2020_JCP_184108}%
  \BibitemOpen
  \bibfield  {author} {\bibinfo {author} {\bibfnamefont {D.~G.~A.}\
  \bibnamefont {Smith}}, \bibinfo {author} {\bibfnamefont {L.~A.}\ \bibnamefont
  {Burns}}, \bibinfo {author} {\bibfnamefont {A.~C.}\ \bibnamefont
  {Simmonett}}, \bibinfo {author} {\bibfnamefont {R.~M.}\ \bibnamefont
  {Parrish}}, \bibinfo {author} {\bibfnamefont {M.~C.}\ \bibnamefont
  {Schieber}}, \bibinfo {author} {\bibfnamefont {R.}~\bibnamefont {Galvelis}},
  \bibinfo {author} {\bibfnamefont {P.}~\bibnamefont {Kraus}}, \bibinfo
  {author} {\bibfnamefont {H.}~\bibnamefont {Kruse}}, \bibinfo {author}
  {\bibfnamefont {R.}~\bibnamefont {{Di Remigio}}}, \bibinfo {author}
  {\bibfnamefont {A.}~\bibnamefont {Alenaizan}}, \bibinfo {author}
  {\bibfnamefont {A.~M.}\ \bibnamefont {James}}, \bibinfo {author}
  {\bibfnamefont {S.}~\bibnamefont {Lehtola}}, \bibinfo {author} {\bibfnamefont
  {J.~P.}\ \bibnamefont {Misiewicz}}, \bibinfo {author} {\bibfnamefont
  {M.}~\bibnamefont {Scheurer}}, \bibinfo {author} {\bibfnamefont {R.~A.}\
  \bibnamefont {Shaw}}, \bibinfo {author} {\bibfnamefont {J.~B.}\ \bibnamefont
  {Schriber}}, \bibinfo {author} {\bibfnamefont {Y.}~\bibnamefont {Xie}},
  \bibinfo {author} {\bibfnamefont {Z.~L.}\ \bibnamefont {Glick}}, \bibinfo
  {author} {\bibfnamefont {D.~A.}\ \bibnamefont {Sirianni}}, \bibinfo {author}
  {\bibfnamefont {J.~S.}\ \bibnamefont {O'Brien}}, \bibinfo {author}
  {\bibfnamefont {J.~M.}\ \bibnamefont {Waldrop}}, \bibinfo {author}
  {\bibfnamefont {A.}~\bibnamefont {Kumar}}, \bibinfo {author} {\bibfnamefont
  {E.~G.}\ \bibnamefont {Hohenstein}}, \bibinfo {author} {\bibfnamefont
  {B.~P.}\ \bibnamefont {Pritchard}}, \bibinfo {author} {\bibfnamefont {B.~R.}\
  \bibnamefont {Brooks}}, \bibinfo {author} {\bibfnamefont {H.~F.}\
  \bibnamefont {Schaefer}}, \bibinfo {author} {\bibfnamefont {A.~Y.}\
  \bibnamefont {Sokolov}}, \bibinfo {author} {\bibfnamefont {K.}~\bibnamefont
  {Patkowski}}, \bibinfo {author} {\bibfnamefont {A.~E.}\ \bibnamefont
  {DePrince}}, \bibinfo {author} {\bibfnamefont {U.}~\bibnamefont {Bozkaya}},
  \bibinfo {author} {\bibfnamefont {R.~A.}\ \bibnamefont {King}}, \bibinfo
  {author} {\bibfnamefont {F.~A.}\ \bibnamefont {Evangelista}}, \bibinfo
  {author} {\bibfnamefont {J.~M.}\ \bibnamefont {Turney}}, \bibinfo {author}
  {\bibfnamefont {T.~D.}\ \bibnamefont {Crawford}},\ and\ \bibinfo {author}
  {\bibfnamefont {C.~D.}\ \bibnamefont {Sherrill}},\ }\bibfield  {title}
  {\enquote {\bibinfo {title} {\textsc{Psi4} 1.4: Open-source software for
  high-throughput quantum chemistry},}\ }\href
  {https://doi.org/10.1063/5.0006002} {\bibfield  {journal} {\bibinfo
  {journal} {J. Chem. Phys.}\ }\textbf {\bibinfo {volume} {152}},\ \bibinfo
  {pages} {184108} (\bibinfo {year} {2020})}\BibitemShut {NoStop}%
\bibitem [{\citenamefont {Peng}(1941)}]{Peng1941_PRSAMPES_499}%
  \BibitemOpen
  \bibfield  {author} {\bibinfo {author} {\bibfnamefont {H.~W.}\ \bibnamefont
  {Peng}},\ }\bibfield  {title} {\enquote {\bibinfo {title} {Perturbation
  theory for the self-consistent field},}\ }\href
  {https://doi.org/10.1098/rspa.1941.0071} {\bibfield  {journal} {\bibinfo
  {journal} {Proc. R. Soc. A: Math. Phys. Eng. Sci.}\ }\textbf {\bibinfo
  {volume} {178}},\ \bibinfo {pages} {499--505} (\bibinfo {year}
  {1941})}\BibitemShut {NoStop}%
\bibitem [{\citenamefont {Dalgarno}(1959)}]{Dalgarno1959_PRSAMPES_282}%
  \BibitemOpen
  \bibfield  {author} {\bibinfo {author} {\bibfnamefont {A.}~\bibnamefont
  {Dalgarno}},\ }\bibfield  {title} {\enquote {\bibinfo {title} {Perturbation
  theory for atomic systems},}\ }\href {https://doi.org/10.1098/rspa.1959.0108}
  {\bibfield  {journal} {\bibinfo  {journal} {Proc. R. Soc. A: Math. Phys. Eng.
  Sci.}\ }\textbf {\bibinfo {volume} {251}},\ \bibinfo {pages} {282--290}
  (\bibinfo {year} {1959})}\BibitemShut {NoStop}%
\bibitem [{\citenamefont {van Gisbergen}, \citenamefont {Snijders},\ and\
  \citenamefont {Baerends}(1998{\natexlab{a}})}]{Gisbergen1998_JCP_10644}%
  \BibitemOpen
  \bibfield  {author} {\bibinfo {author} {\bibfnamefont {S.~J.~A.}\
  \bibnamefont {van Gisbergen}}, \bibinfo {author} {\bibfnamefont {J.~G.}\
  \bibnamefont {Snijders}},\ and\ \bibinfo {author} {\bibfnamefont {E.~J.}\
  \bibnamefont {Baerends}},\ }\bibfield  {title} {\enquote {\bibinfo {title}
  {Calculating frequency-dependent hyperpolarizabilities using time-dependent
  density functional theory},}\ }\href {https://doi.org/10.1063/1.477762}
  {\bibfield  {journal} {\bibinfo  {journal} {J. Chem. Phys.}\ }\textbf
  {\bibinfo {volume} {109}},\ \bibinfo {pages} {10644--10656} (\bibinfo {year}
  {1998}{\natexlab{a}})}\BibitemShut {NoStop}%
\bibitem [{\citenamefont {Sa{\l}ek}\ \emph {et~al.}(2002)\citenamefont
  {Sa{\l}ek}, \citenamefont {Vahtras}, \citenamefont {Helgaker},\ and\
  \citenamefont {{\AA}gren}}]{Salek2002_JCP_9630}%
  \BibitemOpen
  \bibfield  {author} {\bibinfo {author} {\bibfnamefont {P.}~\bibnamefont
  {Sa{\l}ek}}, \bibinfo {author} {\bibfnamefont {O.}~\bibnamefont {Vahtras}},
  \bibinfo {author} {\bibfnamefont {T.}~\bibnamefont {Helgaker}},\ and\
  \bibinfo {author} {\bibfnamefont {H.}~\bibnamefont {{\AA}gren}},\ }\bibfield
  {title} {\enquote {\bibinfo {title} {Density-functional theory of linear and
  nonlinear time-dependent molecular properties},}\ }\href
  {https://doi.org/10.1063/1.1516805} {\bibfield  {journal} {\bibinfo
  {journal} {J. Chem. Phys.}\ }\textbf {\bibinfo {volume} {117}},\ \bibinfo
  {pages} {9630--9645} (\bibinfo {year} {2002})}\BibitemShut {NoStop}%
\bibitem [{\citenamefont {Furche}\ and\ \citenamefont
  {Ahlrichs}(2002)}]{Furche2002_JCP_7433}%
  \BibitemOpen
  \bibfield  {author} {\bibinfo {author} {\bibfnamefont {F.}~\bibnamefont
  {Furche}}\ and\ \bibinfo {author} {\bibfnamefont {R.}~\bibnamefont
  {Ahlrichs}},\ }\bibfield  {title} {\enquote {\bibinfo {title} {{Adiabatic
  time-dependent density functional methods for excited state properties}},}\
  }\href {https://doi.org/10.1063/1.1508368} {\bibfield  {journal} {\bibinfo
  {journal} {J. Chem. Phys.}\ }\textbf {\bibinfo {volume} {117}},\ \bibinfo
  {pages} {7433} (\bibinfo {year} {2002})}\BibitemShut {NoStop}%
\bibitem [{\citenamefont {Hylleraas}(1930)}]{Hylleraas1930_ZP_209}%
  \BibitemOpen
  \bibfield  {author} {\bibinfo {author} {\bibfnamefont {E.~A.}\ \bibnamefont
  {Hylleraas}},\ }\bibfield  {title} {\enquote {\bibinfo {title} {{\"U}ber den
  {Grundterm} der {Zweielektronenprobleme} von \ce{H-}, {He}, \ce{Li+},
  \ce{Be++} usw.}}\ }\href {https://doi.org/10.1007/bf01397032} {\bibfield
  {journal} {\bibinfo  {journal} {Z. Phys.}\ }\textbf {\bibinfo {volume}
  {65}},\ \bibinfo {pages} {209--225} (\bibinfo {year} {1930})}\BibitemShut
  {NoStop}%
\bibitem [{\citenamefont {Jansik}\ \emph {et~al.}(2005)\citenamefont {Jansik},
  \citenamefont {Sa{\l}ek}, \citenamefont {Jonsson}, \citenamefont {Vahtras},\
  and\ \citenamefont {{\AA}gren}}]{Jansik2005_JCP_54107}%
  \BibitemOpen
  \bibfield  {author} {\bibinfo {author} {\bibfnamefont {B.}~\bibnamefont
  {Jansik}}, \bibinfo {author} {\bibfnamefont {P.}~\bibnamefont {Sa{\l}ek}},
  \bibinfo {author} {\bibfnamefont {D.}~\bibnamefont {Jonsson}}, \bibinfo
  {author} {\bibfnamefont {O.}~\bibnamefont {Vahtras}},\ and\ \bibinfo {author}
  {\bibfnamefont {H.}~\bibnamefont {{\AA}gren}},\ }\bibfield  {title} {\enquote
  {\bibinfo {title} {Cubic response functions in time-dependent density
  functional theory},}\ }\href {https://doi.org/10.1063/1.1811605} {\bibfield
  {journal} {\bibinfo  {journal} {J. Chem. Phys.}\ }\textbf {\bibinfo {volume}
  {122}},\ \bibinfo {pages} {054107} (\bibinfo {year} {2005})}\BibitemShut
  {NoStop}%
\bibitem [{\citenamefont {Liu}\ and\ \citenamefont
  {Liang}(2011)}]{Liu2011_JCP_184111}%
  \BibitemOpen
  \bibfield  {author} {\bibinfo {author} {\bibfnamefont {J.}~\bibnamefont
  {Liu}}\ and\ \bibinfo {author} {\bibfnamefont {W.}~\bibnamefont {Liang}},\
  }\bibfield  {title} {\enquote {\bibinfo {title} {Analytical approach for the
  excited-state {Hessian} in time-dependent density functional theory:
  Formalism, implementation, and performance},}\ }\href
  {https://doi.org/10.1063/1.3659312} {\bibfield  {journal} {\bibinfo
  {journal} {J. Chem. Phys.}\ }\textbf {\bibinfo {volume} {135}},\ \bibinfo
  {pages} {184111} (\bibinfo {year} {2011})}\BibitemShut {NoStop}%
\bibitem [{\citenamefont {Pulay}(1969)}]{Pulay1969_MP_197}%
  \BibitemOpen
  \bibfield  {author} {\bibinfo {author} {\bibfnamefont {P.}~\bibnamefont
  {Pulay}},\ }\bibfield  {title} {\enquote {\bibinfo {title} {Ab initio
  calculation of force constants and equilibrium geometries in polyatomic
  molecules},}\ }\href {https://doi.org/10.1080/00268976900100941} {\bibfield
  {journal} {\bibinfo  {journal} {Mol. Phys.}\ }\textbf {\bibinfo {volume}
  {17}},\ \bibinfo {pages} {197--204} (\bibinfo {year} {1969})}\BibitemShut
  {NoStop}%
\bibitem [{\citenamefont {Satoko}(1981)}]{Satoko1981_CPL_111}%
  \BibitemOpen
  \bibfield  {author} {\bibinfo {author} {\bibfnamefont {C.}~\bibnamefont
  {Satoko}},\ }\bibfield  {title} {\enquote {\bibinfo {title} {Direct force
  calculation in the ${X}_\alpha$ method and its application to chemisorption
  of an oxygen atom on the {Al}(111) surface},}\ }\href
  {https://doi.org/10.1016/0009-2614(81)80300-4} {\bibfield  {journal}
  {\bibinfo  {journal} {Chem. Phys. Lett.}\ }\textbf {\bibinfo {volume} {83}},\
  \bibinfo {pages} {111--115} (\bibinfo {year} {1981})}\BibitemShut {NoStop}%
\bibitem [{\citenamefont {Versluis}\ and\ \citenamefont
  {Ziegler}(1988)}]{Versluis1988_JCP_322}%
  \BibitemOpen
  \bibfield  {author} {\bibinfo {author} {\bibfnamefont {L.}~\bibnamefont
  {Versluis}}\ and\ \bibinfo {author} {\bibfnamefont {T.}~\bibnamefont
  {Ziegler}},\ }\bibfield  {title} {\enquote {\bibinfo {title} {The
  determination of molecular structures by density functional theory. {The}
  evaluation of analytical energy gradients by numerical integration},}\ }\href
  {https://doi.org/10.1063/1.454603} {\bibfield  {journal} {\bibinfo  {journal}
  {J. Chem. Phys.}\ }\textbf {\bibinfo {volume} {88}},\ \bibinfo {pages}
  {322--328} (\bibinfo {year} {1988})}\BibitemShut {NoStop}%
\bibitem [{\citenamefont {Fournier}, \citenamefont {Andzelm},\ and\
  \citenamefont {Salahub}(1989)}]{Fournier1989_JCP_6371}%
  \BibitemOpen
  \bibfield  {author} {\bibinfo {author} {\bibfnamefont {R.}~\bibnamefont
  {Fournier}}, \bibinfo {author} {\bibfnamefont {J.}~\bibnamefont {Andzelm}},\
  and\ \bibinfo {author} {\bibfnamefont {D.~R.}\ \bibnamefont {Salahub}},\
  }\bibfield  {title} {\enquote {\bibinfo {title} {Analytical gradient of the
  linear combination of {Gaussian}-type orbitals---local spin density
  energy},}\ }\href {https://doi.org/10.1063/1.456354} {\bibfield  {journal}
  {\bibinfo  {journal} {J. Chem. Phys.}\ }\textbf {\bibinfo {volume} {90}},\
  \bibinfo {pages} {6371--6377} (\bibinfo {year} {1989})}\BibitemShut {NoStop}%
\bibitem [{\citenamefont {Fournier}(1990)}]{Fournier1990_JCP_5422}%
  \BibitemOpen
  \bibfield  {author} {\bibinfo {author} {\bibfnamefont {R.}~\bibnamefont
  {Fournier}},\ }\bibfield  {title} {\enquote {\bibinfo {title} {Second and
  third derivatives of the linear combination of {Gaussian} type
  orbitals--local spin density energy},}\ }\href
  {https://doi.org/10.1063/1.458520} {\bibfield  {journal} {\bibinfo  {journal}
  {J. Chem. Phys.}\ }\textbf {\bibinfo {volume} {92}},\ \bibinfo {pages}
  {5422--5429} (\bibinfo {year} {1990})}\BibitemShut {NoStop}%
\bibitem [{\citenamefont {Delley}(1991)}]{Delley1991_JCP_7245}%
  \BibitemOpen
  \bibfield  {author} {\bibinfo {author} {\bibfnamefont {B.}~\bibnamefont
  {Delley}},\ }\bibfield  {title} {\enquote {\bibinfo {title} {Analytic energy
  derivatives in the numerical local-density-functional approach},}\ }\href
  {https://doi.org/10.1063/1.460208} {\bibfield  {journal} {\bibinfo  {journal}
  {J. Chem. Phys.}\ }\textbf {\bibinfo {volume} {94}},\ \bibinfo {pages}
  {7245--7250} (\bibinfo {year} {1991})}\BibitemShut {NoStop}%
\bibitem [{\citenamefont {Gerratt}\ and\ \citenamefont
  {Mills}(1968)}]{Gerratt1968_JCP_1719}%
  \BibitemOpen
  \bibfield  {author} {\bibinfo {author} {\bibfnamefont {J.}~\bibnamefont
  {Gerratt}}\ and\ \bibinfo {author} {\bibfnamefont {I.~M.}\ \bibnamefont
  {Mills}},\ }\bibfield  {title} {\enquote {\bibinfo {title} {Force constants
  and dipole-moment derivatives of molecules from perturbed {Hartree}--{Fock}
  calculations. {I}},}\ }\href {https://doi.org/10.1063/1.1670299} {\bibfield
  {journal} {\bibinfo  {journal} {J. Chem. Phys.}\ }\textbf {\bibinfo {volume}
  {49}},\ \bibinfo {pages} {1719} (\bibinfo {year} {1968})}\BibitemShut
  {NoStop}%
\bibitem [{\citenamefont {Pople}\ \emph {et~al.}(1979)\citenamefont {Pople},
  \citenamefont {Krishnan}, \citenamefont {Schlegel},\ and\ \citenamefont
  {Binkley}}]{Pople1979_IJQC_225}%
  \BibitemOpen
  \bibfield  {author} {\bibinfo {author} {\bibfnamefont {J.~A.}\ \bibnamefont
  {Pople}}, \bibinfo {author} {\bibfnamefont {R.}~\bibnamefont {Krishnan}},
  \bibinfo {author} {\bibfnamefont {H.~B.}\ \bibnamefont {Schlegel}},\ and\
  \bibinfo {author} {\bibfnamefont {J.~S.}\ \bibnamefont {Binkley}},\
  }\bibfield  {title} {\enquote {\bibinfo {title} {Derivative studies in
  {Hartree}--{Fock} and {M{\o}ller}--{Plesset} theories},}\ }\href
  {https://doi.org/10.1002/qua.560160825} {\bibfield  {journal} {\bibinfo
  {journal} {Int. J. Quantum Chem.}\ }\textbf {\bibinfo {volume} {16}},\
  \bibinfo {pages} {225--241} (\bibinfo {year} {1979})}\BibitemShut {NoStop}%
\bibitem [{\citenamefont {Komornicki}\ and\ \citenamefont
  {Fitzgerald}(1993)}]{Komornicki1993_JCP_1398}%
  \BibitemOpen
  \bibfield  {author} {\bibinfo {author} {\bibfnamefont {A.}~\bibnamefont
  {Komornicki}}\ and\ \bibinfo {author} {\bibfnamefont {G.}~\bibnamefont
  {Fitzgerald}},\ }\bibfield  {title} {\enquote {\bibinfo {title} {{Molecular
  gradients and hessians implemented in density functional theory}},}\ }\href
  {https://doi.org/10.1063/1.465054} {\bibfield  {journal} {\bibinfo  {journal}
  {J. Chem. Phys.}\ }\textbf {\bibinfo {volume} {98}},\ \bibinfo {pages}
  {1398--1421} (\bibinfo {year} {1993})}\BibitemShut {NoStop}%
\bibitem [{\citenamefont {Johnson}\ and\ \citenamefont
  {Frisch}(1993)}]{Johnson1993_CPL_133}%
  \BibitemOpen
  \bibfield  {author} {\bibinfo {author} {\bibfnamefont {B.~G.}\ \bibnamefont
  {Johnson}}\ and\ \bibinfo {author} {\bibfnamefont {M.~J.}\ \bibnamefont
  {Frisch}},\ }\bibfield  {title} {\enquote {\bibinfo {title} {Analytic second
  derivatives of the gradient-corrected density functional energy. {Effect} of
  quadrature weight derivatives},}\ }\href
  {https://doi.org/10.1016/0009-2614(93)E1238-C} {\bibfield  {journal}
  {\bibinfo  {journal} {Chem. Phys. Lett.}\ }\textbf {\bibinfo {volume}
  {216}},\ \bibinfo {pages} {133--140} (\bibinfo {year} {1993})}\BibitemShut
  {NoStop}%
\bibitem [{\citenamefont {Johnson}\ and\ \citenamefont
  {Fisch}(1994)}]{Johnson1994_JCP_7429}%
  \BibitemOpen
  \bibfield  {author} {\bibinfo {author} {\bibfnamefont {B.~G.}\ \bibnamefont
  {Johnson}}\ and\ \bibinfo {author} {\bibfnamefont {M.~J.}\ \bibnamefont
  {Fisch}},\ }\bibfield  {title} {\enquote {\bibinfo {title} {{An
  implementation of analytic second derivatives of the gradient-corrected
  density functional energy}},}\ }\href {https://doi.org/10.1063/1.466887}
  {\bibfield  {journal} {\bibinfo  {journal} {J. Chem. Phys.}\ }\textbf
  {\bibinfo {volume} {100}},\ \bibinfo {pages} {7429--7442} (\bibinfo {year}
  {1994})}\BibitemShut {NoStop}%
\bibitem [{\citenamefont {Barone}(2004)}]{Barone2004_JCP_14108}%
  \BibitemOpen
  \bibfield  {author} {\bibinfo {author} {\bibfnamefont {V.}~\bibnamefont
  {Barone}},\ }\bibfield  {title} {\enquote {\bibinfo {title} {Anharmonic
  vibrational properties by a fully automated second-order perturbative
  approach},}\ }\href {https://doi.org/10.1063/1.1824881} {\bibfield  {journal}
  {\bibinfo  {journal} {J. Chem. Phys.}\ }\textbf {\bibinfo {volume} {122}},\
  \bibinfo {pages} {014108} (\bibinfo {year} {2004})}\BibitemShut {NoStop}%
\bibitem [{\citenamefont {Thouless}(1960)}]{Thouless1960_NP_225}%
  \BibitemOpen
  \bibfield  {author} {\bibinfo {author} {\bibfnamefont {D.~J.}\ \bibnamefont
  {Thouless}},\ }\bibfield  {title} {\enquote {\bibinfo {title} {{Stability
  conditions and nuclear rotations in the Hartree--Fock theory}},}\ }\href
  {https://doi.org/10.1016/0029-5582(60)90048-1} {\bibfield  {journal}
  {\bibinfo  {journal} {Nucl. Phys.}\ }\textbf {\bibinfo {volume} {21}},\
  \bibinfo {pages} {225--232} (\bibinfo {year} {1960})}\BibitemShut {NoStop}%
\bibitem [{\citenamefont {{\v{C}}{\'i}{\v{z}}ek}\ and\ \citenamefont
  {Paldus}(1967)}]{Cizek1967_JCP_3976}%
  \BibitemOpen
  \bibfield  {author} {\bibinfo {author} {\bibfnamefont {J.}~\bibnamefont
  {{\v{C}}{\'i}{\v{z}}ek}}\ and\ \bibinfo {author} {\bibfnamefont
  {J.}~\bibnamefont {Paldus}},\ }\bibfield  {title} {\enquote {\bibinfo {title}
  {Stability conditions for the solutions of the {Hartree}--{Fock} equations
  for atomic and molecular systems. {Application} to the pi-electron model of
  cyclic polyenes},}\ }\href {https://doi.org/10.1063/1.1701562} {\bibfield
  {journal} {\bibinfo  {journal} {J. Chem. Phys.}\ }\textbf {\bibinfo {volume}
  {47}},\ \bibinfo {pages} {3976--3985} (\bibinfo {year} {1967})}\BibitemShut
  {NoStop}%
\bibitem [{\citenamefont {Seeger}\ and\ \citenamefont
  {Pople}(1977)}]{Seeger1977_JCP_3045}%
  \BibitemOpen
  \bibfield  {author} {\bibinfo {author} {\bibfnamefont {R.}~\bibnamefont
  {Seeger}}\ and\ \bibinfo {author} {\bibfnamefont {J.~A.}\ \bibnamefont
  {Pople}},\ }\bibfield  {title} {\enquote {\bibinfo {title} {{Self-consistent
  molecular orbital methods. XVIII. Constraints and stability in Hartree--Fock
  theory}},}\ }\href {https://doi.org/10.1063/1.434318} {\bibfield  {journal}
  {\bibinfo  {journal} {J. Chem. Phys.}\ }\textbf {\bibinfo {volume} {66}},\
  \bibinfo {pages} {3045--3050} (\bibinfo {year} {1977})}\BibitemShut {NoStop}%
\bibitem [{\citenamefont {Bauernschmitt}\ and\ \citenamefont
  {Ahlrichs}(1996{\natexlab{b}})}]{Bauernschmitt1996_JCP_9047}%
  \BibitemOpen
  \bibfield  {author} {\bibinfo {author} {\bibfnamefont {R.}~\bibnamefont
  {Bauernschmitt}}\ and\ \bibinfo {author} {\bibfnamefont {R.}~\bibnamefont
  {Ahlrichs}},\ }\bibfield  {title} {\enquote {\bibinfo {title} {{Stability
  analysis for solutions of the closed shell Kohn--Sham equation}},}\ }\href
  {https://doi.org/10.1063/1.471637} {\bibfield  {journal} {\bibinfo  {journal}
  {J. Chem. Phys.}\ }\textbf {\bibinfo {volume} {104}},\ \bibinfo {pages}
  {9047} (\bibinfo {year} {1996}{\natexlab{b}})}\BibitemShut {NoStop}%
\bibitem [{Note1()}]{Note1}%
  \BibitemOpen
  \bibinfo {note} {Two conventions coexist in the literature: the rotations are
  parametrized either as $\exp (\protect \bm {\kappa })$, as here, or as $\exp
  (-\protect \bm {\kappa })$. The choice flips the sign of the orbital
  gradient---and of any odd-order orbital derivative---while the Hessian at the
  expansion point is invariant, being even in $\protect \bm {\kappa }$. As the
  convention is often left unstated, \protect \textsc {libxckernel}{} carries
  the sign as an explicit parameter.}\BibitemShut {Stop}%
\bibitem [{\citenamefont {Greiner}\ \emph {et~al.}(2026)\citenamefont
  {Greiner}, \citenamefont {H{\o}yvik}, \citenamefont {Lehtola},\ and\
  \citenamefont {Eriksen}}]{Greiner2026_JCTC_881}%
  \BibitemOpen
  \bibfield  {author} {\bibinfo {author} {\bibfnamefont {J.}~\bibnamefont
  {Greiner}}, \bibinfo {author} {\bibfnamefont {I.-M.}\ \bibnamefont
  {H{\o}yvik}}, \bibinfo {author} {\bibfnamefont {S.}~\bibnamefont {Lehtola}},\
  and\ \bibinfo {author} {\bibfnamefont {J.~J.}\ \bibnamefont {Eriksen}},\
  }\bibfield  {title} {\enquote {\bibinfo {title} {A reusable library for
  second-order orbital optimization using the trust region method},}\ }\href
  {https://doi.org/10.1021/acs.jctc.5c01576} {\bibfield  {journal} {\bibinfo
  {journal} {J. Chem. Theory Comput.}\ }\textbf {\bibinfo {volume} {22}},\
  \bibinfo {pages} {881--895} (\bibinfo {year} {2026})}\BibitemShut {NoStop}%
\bibitem [{\citenamefont {Runge}\ and\ \citenamefont
  {Gross}(1984)}]{Runge1984_PRL_997}%
  \BibitemOpen
  \bibfield  {author} {\bibinfo {author} {\bibfnamefont {E.}~\bibnamefont
  {Runge}}\ and\ \bibinfo {author} {\bibfnamefont {E.~K.~U.}\ \bibnamefont
  {Gross}},\ }\bibfield  {title} {\enquote {\bibinfo {title}
  {Density-functional theory for time-dependent systems},}\ }\href
  {https://doi.org/10.1103/PhysRevLett.52.997} {\bibfield  {journal} {\bibinfo
  {journal} {Phys. Rev. Lett.}\ }\textbf {\bibinfo {volume} {52}},\ \bibinfo
  {pages} {997--1000} (\bibinfo {year} {1984})}\BibitemShut {NoStop}%
\bibitem [{\citenamefont {Bauernschmitt}\ \emph {et~al.}(1997)\citenamefont
  {Bauernschmitt}, \citenamefont {H{\"{a}}ser}, \citenamefont {Treutler},\ and\
  \citenamefont {Ahlrichs}}]{Bauernschmitt1997_CPL_573}%
  \BibitemOpen
  \bibfield  {author} {\bibinfo {author} {\bibfnamefont {R.}~\bibnamefont
  {Bauernschmitt}}, \bibinfo {author} {\bibfnamefont {M.}~\bibnamefont
  {H{\"{a}}ser}}, \bibinfo {author} {\bibfnamefont {O.}~\bibnamefont
  {Treutler}},\ and\ \bibinfo {author} {\bibfnamefont {R.}~\bibnamefont
  {Ahlrichs}},\ }\bibfield  {title} {\enquote {\bibinfo {title} {{Calculation
  of excitation energies within time-dependent density functional theory using
  auxiliary basis set expansions}},}\ }\href
  {https://doi.org/10.1016/S0009-2614(96)01343-7} {\bibfield  {journal}
  {\bibinfo  {journal} {Chem. Phys. Lett.}\ }\textbf {\bibinfo {volume}
  {264}},\ \bibinfo {pages} {573--578} (\bibinfo {year} {1997})}\BibitemShut
  {NoStop}%
\bibitem [{\citenamefont {Hirata}\ and\ \citenamefont
  {Head-Gordon}(1999{\natexlab{b}})}]{Hirata1999_CPL_291}%
  \BibitemOpen
  \bibfield  {author} {\bibinfo {author} {\bibfnamefont {S.}~\bibnamefont
  {Hirata}}\ and\ \bibinfo {author} {\bibfnamefont {M.}~\bibnamefont
  {Head-Gordon}},\ }\bibfield  {title} {\enquote {\bibinfo {title}
  {Time-dependent density functional theory within the {Tamm}--{Dancoff}
  approximation},}\ }\href {https://doi.org/10.1016/S0009-2614(99)01149-5}
  {\bibfield  {journal} {\bibinfo  {journal} {Chem. Phys. Lett.}\ }\textbf
  {\bibinfo {volume} {314}},\ \bibinfo {pages} {291--299} (\bibinfo {year}
  {1999}{\natexlab{b}})}\BibitemShut {NoStop}%
\bibitem [{\citenamefont {Gross}\ and\ \citenamefont
  {Kohn}(1985)}]{Gross1985_PRL_2850}%
  \BibitemOpen
  \bibfield  {author} {\bibinfo {author} {\bibfnamefont {E.~K.~U.}\
  \bibnamefont {Gross}}\ and\ \bibinfo {author} {\bibfnamefont
  {W.}~\bibnamefont {Kohn}},\ }\bibfield  {title} {\enquote {\bibinfo {title}
  {Local density-functional theory of frequency-dependent linear response},}\
  }\href {https://doi.org/10.1103/PhysRevLett.55.2850} {\bibfield  {journal}
  {\bibinfo  {journal} {Phys. Rev. Lett.}\ }\textbf {\bibinfo {volume} {55}},\
  \bibinfo {pages} {2850--2852} (\bibinfo {year} {1985})}\BibitemShut {NoStop}%
\bibitem [{\citenamefont {Maitra}(2016)}]{Maitra2016_JCP_220901}%
  \BibitemOpen
  \bibfield  {author} {\bibinfo {author} {\bibfnamefont {N.~T.}\ \bibnamefont
  {Maitra}},\ }\bibfield  {title} {\enquote {\bibinfo {title} {{Perspective:
  Fundamental aspects of time-dependent density functional theory}},}\ }\href
  {https://doi.org/10.1063/1.4953039} {\bibfield  {journal} {\bibinfo
  {journal} {J. Chem. Phys.}\ }\textbf {\bibinfo {volume} {144}},\ \bibinfo
  {pages} {220901} (\bibinfo {year} {2016})}\BibitemShut {NoStop}%
\bibitem [{\citenamefont {Yabana}\ and\ \citenamefont
  {Bertsch}(1996)}]{Yabana1996_PRB_4484}%
  \BibitemOpen
  \bibfield  {author} {\bibinfo {author} {\bibfnamefont {K.}~\bibnamefont
  {Yabana}}\ and\ \bibinfo {author} {\bibfnamefont {G.~F.}\ \bibnamefont
  {Bertsch}},\ }\bibfield  {title} {\enquote {\bibinfo {title} {Time-dependent
  local-density approximation in real time},}\ }\href
  {https://doi.org/10.1103/PhysRevB.54.4484} {\bibfield  {journal} {\bibinfo
  {journal} {Phys. Rev. B}\ }\textbf {\bibinfo {volume} {54}},\ \bibinfo
  {pages} {4484--4487} (\bibinfo {year} {1996})}\BibitemShut {NoStop}%
\bibitem [{\citenamefont {{Van Caillie}}\ and\ \citenamefont
  {Amos}(1999)}]{VanCaillie1999_CPL_249}%
  \BibitemOpen
  \bibfield  {author} {\bibinfo {author} {\bibfnamefont {C.}~\bibnamefont {{Van
  Caillie}}}\ and\ \bibinfo {author} {\bibfnamefont {R.~D.}\ \bibnamefont
  {Amos}},\ }\bibfield  {title} {\enquote {\bibinfo {title} {{Geometric
  derivatives of excitation energies using SCF and DFT}},}\ }\href
  {https://doi.org/10.1016/S0009-2614(99)00646-6} {\bibfield  {journal}
  {\bibinfo  {journal} {Chem. Phys. Lett.}\ }\textbf {\bibinfo {volume}
  {308}},\ \bibinfo {pages} {249--255} (\bibinfo {year} {1999})}\BibitemShut
  {NoStop}%
\bibitem [{\citenamefont {{Van Caillie}}\ and\ \citenamefont
  {Amos}(2000)}]{VanCaillie2000_CPL_159}%
  \BibitemOpen
  \bibfield  {author} {\bibinfo {author} {\bibfnamefont {C.}~\bibnamefont {{Van
  Caillie}}}\ and\ \bibinfo {author} {\bibfnamefont {R.~D.}\ \bibnamefont
  {Amos}},\ }\bibfield  {title} {\enquote {\bibinfo {title} {{Geometric
  derivatives of density functional theory excitation energies using
  gradient-corrected functionals}},}\ }\href
  {https://doi.org/10.1016/S0009-2614(99)01346-9} {\bibfield  {journal}
  {\bibinfo  {journal} {Chem. Phys. Lett.}\ }\textbf {\bibinfo {volume}
  {317}},\ \bibinfo {pages} {159--164} (\bibinfo {year} {2000})}\BibitemShut
  {NoStop}%
\bibitem [{\citenamefont {Handy}\ and\ \citenamefont
  {Schaefer}(1984)}]{Handy1984_JCP_5031}%
  \BibitemOpen
  \bibfield  {author} {\bibinfo {author} {\bibfnamefont {N.~C.}\ \bibnamefont
  {Handy}}\ and\ \bibinfo {author} {\bibfnamefont {H.~F.}\ \bibnamefont
  {Schaefer}},\ }\bibfield  {title} {\enquote {\bibinfo {title} {{On the
  evaluation of analytic energy derivatives for correlated wave functions}},}\
  }\href {https://doi.org/10.1063/1.447489} {\bibfield  {journal} {\bibinfo
  {journal} {J. Chem. Phys.}\ }\textbf {\bibinfo {volume} {81}},\ \bibinfo
  {pages} {5031} (\bibinfo {year} {1984})}\BibitemShut {NoStop}%
\bibitem [{\citenamefont {London}(1937)}]{London1937_JPlR_397}%
  \BibitemOpen
  \bibfield  {author} {\bibinfo {author} {\bibfnamefont {F.}~\bibnamefont
  {London}},\ }\bibfield  {title} {\enquote {\bibinfo {title} {{Th{\'{e}}orie
  quantique des courants interatomiques dans les combinaisons aromatiques}},}\
  }\href {https://doi.org/10.1051/jphysrad:01937008010039700} {\bibfield
  {journal} {\bibinfo  {journal} {J. Phys. le Radium}\ }\textbf {\bibinfo
  {volume} {8}},\ \bibinfo {pages} {397--409} (\bibinfo {year}
  {1937})}\BibitemShut {NoStop}%
\bibitem [{\citenamefont {Ditchfield}(1974)}]{Ditchfield1974_MP_789}%
  \BibitemOpen
  \bibfield  {author} {\bibinfo {author} {\bibfnamefont {R.}~\bibnamefont
  {Ditchfield}},\ }\bibfield  {title} {\enquote {\bibinfo {title}
  {{Self-consistent perturbation theory of diamagnetism. I. A gauge-invariant
  LCAO method for N.M.R. chemical shifts}},}\ }\href
  {https://doi.org/10.1080/00268977400100711} {\bibfield  {journal} {\bibinfo
  {journal} {Mol. Phys.}\ }\textbf {\bibinfo {volume} {27}},\ \bibinfo {pages}
  {789--807} (\bibinfo {year} {1974})}\BibitemShut {NoStop}%
\bibitem [{\citenamefont {Wolinski}, \citenamefont {Hinton},\ and\
  \citenamefont {Pulay}(1990)}]{Wolinski1990_JACS_8251}%
  \BibitemOpen
  \bibfield  {author} {\bibinfo {author} {\bibfnamefont {K.}~\bibnamefont
  {Wolinski}}, \bibinfo {author} {\bibfnamefont {J.~F.}\ \bibnamefont
  {Hinton}},\ and\ \bibinfo {author} {\bibfnamefont {P.}~\bibnamefont
  {Pulay}},\ }\bibfield  {title} {\enquote {\bibinfo {title} {Efficient
  implementation of the gauge-independent atomic orbital method for {NMR}
  chemical shift calculations},}\ }\href {https://doi.org/10.1021/ja00179a005}
  {\bibfield  {journal} {\bibinfo  {journal} {J. Am. Chem. Soc.}\ }\textbf
  {\bibinfo {volume} {112}},\ \bibinfo {pages} {8251--8260} (\bibinfo {year}
  {1990})}\BibitemShut {NoStop}%
\bibitem [{\citenamefont {Schreckenbach}\ and\ \citenamefont
  {Ziegler}(1995)}]{Schreckenbach1995_JPC_606}%
  \BibitemOpen
  \bibfield  {author} {\bibinfo {author} {\bibfnamefont {G.}~\bibnamefont
  {Schreckenbach}}\ and\ \bibinfo {author} {\bibfnamefont {T.}~\bibnamefont
  {Ziegler}},\ }\bibfield  {title} {\enquote {\bibinfo {title} {Calculation of
  {NMR} shielding tensors using gauge-including atomic orbitals and modern
  density functional theory},}\ }\href {https://doi.org/10.1021/j100002a024}
  {\bibfield  {journal} {\bibinfo  {journal} {J. Phys. Chem.}\ }\textbf
  {\bibinfo {volume} {99}},\ \bibinfo {pages} {606--611} (\bibinfo {year}
  {1995})}\BibitemShut {NoStop}%
\bibitem [{\citenamefont {Lee}, \citenamefont {Handy},\ and\ \citenamefont
  {Colwell}(1995)}]{Lee1995_JCP_10095}%
  \BibitemOpen
  \bibfield  {author} {\bibinfo {author} {\bibfnamefont {A.~M.}\ \bibnamefont
  {Lee}}, \bibinfo {author} {\bibfnamefont {N.~C.}\ \bibnamefont {Handy}},\
  and\ \bibinfo {author} {\bibfnamefont {S.~M.}\ \bibnamefont {Colwell}},\
  }\bibfield  {title} {\enquote {\bibinfo {title} {{The density functional
  calculation of nuclear shielding constants using London atomic orbitals}},}\
  }\href {https://doi.org/10.1063/1.469912} {\bibfield  {journal} {\bibinfo
  {journal} {J. Chem. Phys.}\ }\textbf {\bibinfo {volume} {103}},\ \bibinfo
  {pages} {10095--10109} (\bibinfo {year} {1995})}\BibitemShut {NoStop}%
\bibitem [{\citenamefont {Cheeseman}\ \emph
  {et~al.}(1996{\natexlab{a}})\citenamefont {Cheeseman}, \citenamefont
  {Trucks}, \citenamefont {Keith},\ and\ \citenamefont
  {Frisch}}]{Cheeseman1996_JCP_5497}%
  \BibitemOpen
  \bibfield  {author} {\bibinfo {author} {\bibfnamefont {J.~R.}\ \bibnamefont
  {Cheeseman}}, \bibinfo {author} {\bibfnamefont {G.~W.}\ \bibnamefont
  {Trucks}}, \bibinfo {author} {\bibfnamefont {T.~A.}\ \bibnamefont {Keith}},\
  and\ \bibinfo {author} {\bibfnamefont {M.~J.}\ \bibnamefont {Frisch}},\
  }\bibfield  {title} {\enquote {\bibinfo {title} {{A comparison of models for
  calculating nuclear magnetic resonance shielding tensors}},}\ }\href
  {https://doi.org/10.1063/1.471789} {\bibfield  {journal} {\bibinfo  {journal}
  {J. Chem. Phys.}\ }\textbf {\bibinfo {volume} {104}},\ \bibinfo {pages}
  {5497} (\bibinfo {year} {1996}{\natexlab{a}})}\BibitemShut {NoStop}%
\bibitem [{\citenamefont {Malkin}, \citenamefont {Malkina},\ and\ \citenamefont
  {Salahub}(1993)}]{Malkin1993_CPL_87}%
  \BibitemOpen
  \bibfield  {author} {\bibinfo {author} {\bibfnamefont {V.~G.}\ \bibnamefont
  {Malkin}}, \bibinfo {author} {\bibfnamefont {O.~L.}\ \bibnamefont
  {Malkina}},\ and\ \bibinfo {author} {\bibfnamefont {D.~R.}\ \bibnamefont
  {Salahub}},\ }\bibfield  {title} {\enquote {\bibinfo {title} {Calculations of
  {NMR} shielding constants beyond uncoupled density functional theory. {IGLO}
  approach},}\ }\href {https://doi.org/10.1016/0009-2614(93)85609-r} {\bibfield
   {journal} {\bibinfo  {journal} {Chem. Phys. Lett.}\ }\textbf {\bibinfo
  {volume} {204}},\ \bibinfo {pages} {87--95} (\bibinfo {year}
  {1993})}\BibitemShut {NoStop}%
\bibitem [{\citenamefont {Malkin}\ \emph {et~al.}(1994)\citenamefont {Malkin},
  \citenamefont {Malkina}, \citenamefont {Casida},\ and\ \citenamefont
  {Salahub}}]{Malkin1994_JACS_5898}%
  \BibitemOpen
  \bibfield  {author} {\bibinfo {author} {\bibfnamefont {V.~G.}\ \bibnamefont
  {Malkin}}, \bibinfo {author} {\bibfnamefont {O.~L.}\ \bibnamefont {Malkina}},
  \bibinfo {author} {\bibfnamefont {M.~E.}\ \bibnamefont {Casida}},\ and\
  \bibinfo {author} {\bibfnamefont {D.~R.}\ \bibnamefont {Salahub}},\
  }\bibfield  {title} {\enquote {\bibinfo {title} {Nuclear magnetic resonance
  shielding tensors calculated with a sum-over-states density functional
  perturbation theory},}\ }\href {https://doi.org/10.1021/ja00092a046}
  {\bibfield  {journal} {\bibinfo  {journal} {J. Am. Chem. Soc.}\ }\textbf
  {\bibinfo {volume} {116}},\ \bibinfo {pages} {5898--5908} (\bibinfo {year}
  {1994})}\BibitemShut {NoStop}%
\bibitem [{\citenamefont {Stevens}, \citenamefont {Pitzer},\ and\ \citenamefont
  {Lipscomb}(1963)}]{Stevens1963_JCP_550}%
  \BibitemOpen
  \bibfield  {author} {\bibinfo {author} {\bibfnamefont {R.~M.}\ \bibnamefont
  {Stevens}}, \bibinfo {author} {\bibfnamefont {R.~M.}\ \bibnamefont
  {Pitzer}},\ and\ \bibinfo {author} {\bibfnamefont {W.~N.}\ \bibnamefont
  {Lipscomb}},\ }\bibfield  {title} {\enquote {\bibinfo {title} {Perturbed
  {Hartree}--{Fock} calculations. {I}. {Magnetic} susceptibility and shielding
  in the {LiH} molecule},}\ }\href {https://doi.org/10.1063/1.1733693}
  {\bibfield  {journal} {\bibinfo  {journal} {J. Chem. Phys.}\ }\textbf
  {\bibinfo {volume} {38}},\ \bibinfo {pages} {550--560} (\bibinfo {year}
  {1963})}\BibitemShut {NoStop}%
\bibitem [{\citenamefont {Vignale}\ and\ \citenamefont
  {Rasolt}(1987)}]{Vignale1987_PRL_2360}%
  \BibitemOpen
  \bibfield  {author} {\bibinfo {author} {\bibfnamefont {G.}~\bibnamefont
  {Vignale}}\ and\ \bibinfo {author} {\bibfnamefont {M.}~\bibnamefont
  {Rasolt}},\ }\bibfield  {title} {\enquote {\bibinfo {title}
  {Density-functional theory in strong magnetic fields},}\ }\href
  {https://doi.org/10.1103/PhysRevLett.59.2360} {\bibfield  {journal} {\bibinfo
   {journal} {Phys. Rev. Lett.}\ }\textbf {\bibinfo {volume} {59}},\ \bibinfo
  {pages} {2360--2363} (\bibinfo {year} {1987})}\BibitemShut {NoStop}%
\bibitem [{\citenamefont {Vignale}\ and\ \citenamefont
  {Rasolt}(1989)}]{Vignale1989_PRL_115}%
  \BibitemOpen
  \bibfield  {author} {\bibinfo {author} {\bibfnamefont {G.}~\bibnamefont
  {Vignale}}\ and\ \bibinfo {author} {\bibfnamefont {M.}~\bibnamefont
  {Rasolt}},\ }\bibfield  {title} {\enquote {\bibinfo {title}
  {Density-functional theory in strong magnetic fields},}\ }\href
  {https://doi.org/10.1103/PhysRevLett.62.115} {\bibfield  {journal} {\bibinfo
  {journal} {Phys. Rev. Lett.}\ }\textbf {\bibinfo {volume} {62}},\ \bibinfo
  {pages} {115--115} (\bibinfo {year} {1989})}\BibitemShut {NoStop}%
\bibitem [{\citenamefont {Lee}, \citenamefont {Colwell},\ and\ \citenamefont
  {Handy}(1994)}]{Lee1994_CPL_225}%
  \BibitemOpen
  \bibfield  {author} {\bibinfo {author} {\bibfnamefont {A.~M.}\ \bibnamefont
  {Lee}}, \bibinfo {author} {\bibfnamefont {S.~M.}\ \bibnamefont {Colwell}},\
  and\ \bibinfo {author} {\bibfnamefont {N.~C.}\ \bibnamefont {Handy}},\
  }\bibfield  {title} {\enquote {\bibinfo {title} {The calculation of
  magnetisabilities using current density functional theory},}\ }\href
  {https://doi.org/10.1016/0009-2614(94)01009-9} {\bibfield  {journal}
  {\bibinfo  {journal} {Chem. Phys. Lett.}\ }\textbf {\bibinfo {volume}
  {229}},\ \bibinfo {pages} {225--232} (\bibinfo {year} {1994})}\BibitemShut
  {NoStop}%
\bibitem [{\citenamefont {Dalgarno}\ and\ \citenamefont
  {Stewart}(1958)}]{Dalgarno1958_PRSAMPES_245}%
  \BibitemOpen
  \bibfield  {author} {\bibinfo {author} {\bibfnamefont {A.}~\bibnamefont
  {Dalgarno}}\ and\ \bibinfo {author} {\bibfnamefont {A.~L.}\ \bibnamefont
  {Stewart}},\ }\bibfield  {title} {\enquote {\bibinfo {title} {A perturbation
  calculation of properties of the helium iso-electronic sequence},}\ }\href
  {https://doi.org/10.1098/rspa.1958.0182} {\bibfield  {journal} {\bibinfo
  {journal} {Proc. R. Soc. A: Math. Phys. Eng. Sci.}\ }\textbf {\bibinfo
  {volume} {247}},\ \bibinfo {pages} {245--259} (\bibinfo {year}
  {1958})}\BibitemShut {NoStop}%
\bibitem [{\citenamefont {Thonhauser}\ \emph {et~al.}(2009)\citenamefont
  {Thonhauser}, \citenamefont {Ceresoli}, \citenamefont {Mostofi},
  \citenamefont {Marzari}, \citenamefont {Resta},\ and\ \citenamefont
  {Vanderbilt}}]{Thonhauser2009_JCP_101101}%
  \BibitemOpen
  \bibfield  {author} {\bibinfo {author} {\bibfnamefont {T.}~\bibnamefont
  {Thonhauser}}, \bibinfo {author} {\bibfnamefont {D.}~\bibnamefont
  {Ceresoli}}, \bibinfo {author} {\bibfnamefont {A.~A.}\ \bibnamefont
  {Mostofi}}, \bibinfo {author} {\bibfnamefont {N.}~\bibnamefont {Marzari}},
  \bibinfo {author} {\bibfnamefont {R.}~\bibnamefont {Resta}},\ and\ \bibinfo
  {author} {\bibfnamefont {D.}~\bibnamefont {Vanderbilt}},\ }\bibfield  {title}
  {\enquote {\bibinfo {title} {A converse approach to the calculation of {NMR}
  shielding tensors},}\ }\href {https://doi.org/10.1063/1.3216028} {\bibfield
  {journal} {\bibinfo  {journal} {J. Chem. Phys.}\ }\textbf {\bibinfo {volume}
  {131}},\ \bibinfo {pages} {101101} (\bibinfo {year} {2009})}\BibitemShut
  {NoStop}%
\bibitem [{\citenamefont {Beer}, \citenamefont {Kussmann},\ and\ \citenamefont
  {Ochsenfeld}(2011)}]{Beer2011_JCP_74102}%
  \BibitemOpen
  \bibfield  {author} {\bibinfo {author} {\bibfnamefont {M.}~\bibnamefont
  {Beer}}, \bibinfo {author} {\bibfnamefont {J.}~\bibnamefont {Kussmann}},\
  and\ \bibinfo {author} {\bibfnamefont {C.}~\bibnamefont {Ochsenfeld}},\
  }\bibfield  {title} {\enquote {\bibinfo {title} {Nuclei-selected {NMR}
  shielding calculations: A sublinear-scaling quantum-chemical method},}\
  }\href {https://doi.org/10.1063/1.3526315} {\bibfield  {journal} {\bibinfo
  {journal} {J. Chem. Phys.}\ }\textbf {\bibinfo {volume} {134}},\ \bibinfo
  {pages} {074102} (\bibinfo {year} {2011})}\BibitemShut {NoStop}%
\bibitem [{\citenamefont {Krykunov}\ and\ \citenamefont
  {Autschbach}(2005)}]{Krykunov2005_JCP_114103}%
  \BibitemOpen
  \bibfield  {author} {\bibinfo {author} {\bibfnamefont {M.}~\bibnamefont
  {Krykunov}}\ and\ \bibinfo {author} {\bibfnamefont {J.}~\bibnamefont
  {Autschbach}},\ }\bibfield  {title} {\enquote {\bibinfo {title} {Calculation
  of optical rotation with time-periodic magnetic-field-dependent basis
  functions in approximate time-dependent density-functional theory},}\ }\href
  {https://doi.org/10.1063/1.2032428} {\bibfield  {journal} {\bibinfo
  {journal} {J. Chem. Phys.}\ }\textbf {\bibinfo {volume} {123}},\ \bibinfo
  {pages} {114103} (\bibinfo {year} {2005})}\BibitemShut {NoStop}%
\bibitem [{\citenamefont {Helgaker}, \citenamefont {Jaszu{\'{n}}ski},\ and\
  \citenamefont {Ruud}(1999)}]{Helgaker1999_CR_293}%
  \BibitemOpen
  \bibfield  {author} {\bibinfo {author} {\bibfnamefont {T.}~\bibnamefont
  {Helgaker}}, \bibinfo {author} {\bibfnamefont {M.}~\bibnamefont
  {Jaszu{\'{n}}ski}},\ and\ \bibinfo {author} {\bibfnamefont {K.}~\bibnamefont
  {Ruud}},\ }\bibfield  {title} {\enquote {\bibinfo {title} {Ab initio methods
  for the calculation of {NMR} shielding and indirect spin--spin coupling
  constants},}\ }\href {https://doi.org/10.1021/cr960017t} {\bibfield
  {journal} {\bibinfo  {journal} {Chem. Rev.}\ }\textbf {\bibinfo {volume}
  {99}},\ \bibinfo {pages} {293--352} (\bibinfo {year} {1999})}\BibitemShut
  {NoStop}%
\bibitem [{\citenamefont {Sychrovsk{\'y}}, \citenamefont {Gr{\"a}fenstein},\
  and\ \citenamefont {Cremer}(2000)}]{Sychrovsky2000_JCP_3530}%
  \BibitemOpen
  \bibfield  {author} {\bibinfo {author} {\bibfnamefont {V.}~\bibnamefont
  {Sychrovsk{\'y}}}, \bibinfo {author} {\bibfnamefont {J.}~\bibnamefont
  {Gr{\"a}fenstein}},\ and\ \bibinfo {author} {\bibfnamefont {D.}~\bibnamefont
  {Cremer}},\ }\bibfield  {title} {\enquote {\bibinfo {title} {Nuclear magnetic
  resonance spin--spin coupling constants from coupled perturbed density
  functional theory},}\ }\href {https://doi.org/10.1063/1.1286806} {\bibfield
  {journal} {\bibinfo  {journal} {J. Chem. Phys.}\ }\textbf {\bibinfo {volume}
  {113}},\ \bibinfo {pages} {3530--3547} (\bibinfo {year} {2000})}\BibitemShut
  {NoStop}%
\bibitem [{\citenamefont {Helgaker}, \citenamefont {Watson},\ and\
  \citenamefont {Handy}(2000)}]{Helgaker2000_JCP_9402}%
  \BibitemOpen
  \bibfield  {author} {\bibinfo {author} {\bibfnamefont {T.}~\bibnamefont
  {Helgaker}}, \bibinfo {author} {\bibfnamefont {M.}~\bibnamefont {Watson}},\
  and\ \bibinfo {author} {\bibfnamefont {N.~C.}\ \bibnamefont {Handy}},\
  }\bibfield  {title} {\enquote {\bibinfo {title} {Analytical calculation of
  nuclear magnetic resonance indirect spin--spin coupling constants at the
  generalized gradient approximation and hybrid levels of density-functional
  theory},}\ }\href {https://doi.org/10.1063/1.1321296} {\bibfield  {journal}
  {\bibinfo  {journal} {J. Chem. Phys.}\ }\textbf {\bibinfo {volume} {113}},\
  \bibinfo {pages} {9402--9409} (\bibinfo {year} {2000})}\BibitemShut {NoStop}%
\bibitem [{\citenamefont {Arbuznikov}\ \emph {et~al.}(2002)\citenamefont
  {Arbuznikov}, \citenamefont {Kaupp}, \citenamefont {Malkin}, \citenamefont
  {Reviakine},\ and\ \citenamefont {Malkina}}]{Arbuznikov2002_PCCP_5467}%
  \BibitemOpen
  \bibfield  {author} {\bibinfo {author} {\bibfnamefont {A.~V.}\ \bibnamefont
  {Arbuznikov}}, \bibinfo {author} {\bibfnamefont {M.}~\bibnamefont {Kaupp}},
  \bibinfo {author} {\bibfnamefont {V.~G.}\ \bibnamefont {Malkin}}, \bibinfo
  {author} {\bibfnamefont {R.}~\bibnamefont {Reviakine}},\ and\ \bibinfo
  {author} {\bibfnamefont {O.~L.}\ \bibnamefont {Malkina}},\ }\bibfield
  {title} {\enquote {\bibinfo {title} {{Validation study of meta-GGA
  functionals and of a model exchange-correlation potential in density
  functional calculations of EPR parameters}},}\ }\href
  {https://doi.org/10.1039/b207171a} {\bibfield  {journal} {\bibinfo  {journal}
  {Phys. Chem. Chem. Phys.}\ }\textbf {\bibinfo {volume} {4}},\ \bibinfo
  {pages} {5467--5474} (\bibinfo {year} {2002})}\BibitemShut {NoStop}%
\bibitem [{\citenamefont {Tellgren}\ \emph {et~al.}(2014)\citenamefont
  {Tellgren}, \citenamefont {Teale}, \citenamefont {Furness}, \citenamefont
  {Lange}, \citenamefont {Ekstr{\"{o}}m},\ and\ \citenamefont
  {Helgaker}}]{Tellgren2014_JCP_34101}%
  \BibitemOpen
  \bibfield  {author} {\bibinfo {author} {\bibfnamefont {E.~I.}\ \bibnamefont
  {Tellgren}}, \bibinfo {author} {\bibfnamefont {A.~M.}\ \bibnamefont {Teale}},
  \bibinfo {author} {\bibfnamefont {J.~W.}\ \bibnamefont {Furness}}, \bibinfo
  {author} {\bibfnamefont {K.~K.}\ \bibnamefont {Lange}}, \bibinfo {author}
  {\bibfnamefont {U.}~\bibnamefont {Ekstr{\"{o}}m}},\ and\ \bibinfo {author}
  {\bibfnamefont {T.}~\bibnamefont {Helgaker}},\ }\bibfield  {title} {\enquote
  {\bibinfo {title} {{Non-perturbative calculation of molecular magnetic
  properties within current-density functional theory}},}\ }\href
  {https://doi.org/10.1063/1.4861427} {\bibfield  {journal} {\bibinfo
  {journal} {J. Chem. Phys.}\ }\textbf {\bibinfo {volume} {140}},\ \bibinfo
  {pages} {034101} (\bibinfo {year} {2014})}\BibitemShut {NoStop}%
\bibitem [{\citenamefont {Dobson}(1993)}]{Dobson1993_JCP_8870}%
  \BibitemOpen
  \bibfield  {author} {\bibinfo {author} {\bibfnamefont {J.~F.}\ \bibnamefont
  {Dobson}},\ }\bibfield  {title} {\enquote {\bibinfo {title} {{Alternative
  expressions for the Fermi hole curvature}},}\ }\href
  {https://doi.org/10.1063/1.464444} {\bibfield  {journal} {\bibinfo  {journal}
  {J. Chem. Phys.}\ }\textbf {\bibinfo {volume} {98}},\ \bibinfo {pages}
  {8870--8872} (\bibinfo {year} {1993})}\BibitemShut {NoStop}%
\bibitem [{\citenamefont {Maximoff}\ and\ \citenamefont
  {Scuseria}(2004)}]{Maximoff2004_CPL_408}%
  \BibitemOpen
  \bibfield  {author} {\bibinfo {author} {\bibfnamefont {S.~N.}\ \bibnamefont
  {Maximoff}}\ and\ \bibinfo {author} {\bibfnamefont {G.~E.}\ \bibnamefont
  {Scuseria}},\ }\bibfield  {title} {\enquote {\bibinfo {title} {Nuclear
  magnetic resonance shielding tensors calculated with kinetic energy
  density-dependent exchange-correlation functionals},}\ }\href
  {https://doi.org/10.1016/j.cplett.2004.04.049} {\bibfield  {journal}
  {\bibinfo  {journal} {Chem. Phys. Lett.}\ }\textbf {\bibinfo {volume}
  {390}},\ \bibinfo {pages} {408--412} (\bibinfo {year} {2004})}\BibitemShut
  {NoStop}%
\bibitem [{\citenamefont {Christiansen}, \citenamefont {J{\o}rgensen},\ and\
  \citenamefont {H{\"{a}}ttig}(1998)}]{Christiansen1998_IJQC_1}%
  \BibitemOpen
  \bibfield  {author} {\bibinfo {author} {\bibfnamefont {O.}~\bibnamefont
  {Christiansen}}, \bibinfo {author} {\bibfnamefont {P.}~\bibnamefont
  {J{\o}rgensen}},\ and\ \bibinfo {author} {\bibfnamefont {C.}~\bibnamefont
  {H{\"{a}}ttig}},\ }\bibfield  {title} {\enquote {\bibinfo {title} {Response
  functions from {Fourier} component variational perturbation theory applied to
  a time-averaged quasienergy},}\ }\href
  {https://doi.org/10.1002/(SICI)1097-461X(1998)68:1<1::AID-QUA1>3.0.CO;2-Z}
  {\bibfield  {journal} {\bibinfo  {journal} {Int. J. Quantum Chem.}\ }\textbf
  {\bibinfo {volume} {68}},\ \bibinfo {pages} {1--52} (\bibinfo {year}
  {1998})}\BibitemShut {NoStop}%
\bibitem [{\citenamefont {Thorvaldsen}\ \emph {et~al.}(2008)\citenamefont
  {Thorvaldsen}, \citenamefont {Ruud}, \citenamefont {Kristensen},
  \citenamefont {J{\o}rgensen},\ and\ \citenamefont
  {Coriani}}]{Thorvaldsen2008_JCP_214108}%
  \BibitemOpen
  \bibfield  {author} {\bibinfo {author} {\bibfnamefont {A.~J.}\ \bibnamefont
  {Thorvaldsen}}, \bibinfo {author} {\bibfnamefont {K.}~\bibnamefont {Ruud}},
  \bibinfo {author} {\bibfnamefont {K.}~\bibnamefont {Kristensen}}, \bibinfo
  {author} {\bibfnamefont {P.}~\bibnamefont {J{\o}rgensen}},\ and\ \bibinfo
  {author} {\bibfnamefont {S.}~\bibnamefont {Coriani}},\ }\bibfield  {title}
  {\enquote {\bibinfo {title} {A density matrix-based quasienergy formulation
  of the {Kohn}--{Sham} density functional response theory using perturbation-
  and time-dependent basis sets},}\ }\href {https://doi.org/10.1063/1.2996351}
  {\bibfield  {journal} {\bibinfo  {journal} {J. Chem. Phys.}\ }\textbf
  {\bibinfo {volume} {129}},\ \bibinfo {pages} {214108} (\bibinfo {year}
  {2008})}\BibitemShut {NoStop}%
\bibitem [{\citenamefont {Ringholm}, \citenamefont {Jonsson},\ and\
  \citenamefont {Ruud}(2014)}]{Ringholm2014_JCC_622}%
  \BibitemOpen
  \bibfield  {author} {\bibinfo {author} {\bibfnamefont {M.}~\bibnamefont
  {Ringholm}}, \bibinfo {author} {\bibfnamefont {D.}~\bibnamefont {Jonsson}},\
  and\ \bibinfo {author} {\bibfnamefont {K.}~\bibnamefont {Ruud}},\ }\bibfield
  {title} {\enquote {\bibinfo {title} {A general, recursive, and open‐ended
  response code},}\ }\href {https://doi.org/10.1002/jcc.23533} {\bibfield
  {journal} {\bibinfo  {journal} {J. Comput. Chem.}\ }\textbf {\bibinfo
  {volume} {35}},\ \bibinfo {pages} {622--633} (\bibinfo {year}
  {2014})}\BibitemShut {NoStop}%
\bibitem [{\citenamefont {Friese}\ \emph {et~al.}(2015)\citenamefont {Friese},
  \citenamefont {Beerepoot}, \citenamefont {Ringholm},\ and\ \citenamefont
  {Ruud}}]{Friese2015_JCTC_1129}%
  \BibitemOpen
  \bibfield  {author} {\bibinfo {author} {\bibfnamefont {D.~H.}\ \bibnamefont
  {Friese}}, \bibinfo {author} {\bibfnamefont {M.~T.~P.}\ \bibnamefont
  {Beerepoot}}, \bibinfo {author} {\bibfnamefont {M.}~\bibnamefont
  {Ringholm}},\ and\ \bibinfo {author} {\bibfnamefont {K.}~\bibnamefont
  {Ruud}},\ }\bibfield  {title} {\enquote {\bibinfo {title} {Open-ended
  recursive approach for the calculation of multiphoton absorption matrix
  elements},}\ }\href {https://doi.org/10.1021/ct501113y} {\bibfield  {journal}
  {\bibinfo  {journal} {J. Chem. Theory Comput.}\ }\textbf {\bibinfo {volume}
  {11}},\ \bibinfo {pages} {1129--1144} (\bibinfo {year} {2015})}\BibitemShut
  {NoStop}%
\bibitem [{\citenamefont {Woon}\ and\ \citenamefont
  {Dunning}(1994)}]{Woon1994_JCP_2975}%
  \BibitemOpen
  \bibfield  {author} {\bibinfo {author} {\bibfnamefont {D.~E.}\ \bibnamefont
  {Woon}}\ and\ \bibinfo {author} {\bibfnamefont {T.~H.}\ \bibnamefont
  {Dunning}},\ }\bibfield  {title} {\enquote {\bibinfo {title} {{Gaussian basis
  sets for use in correlated molecular calculations. IV. Calculation of static
  electrical response properties}},}\ }\href {https://doi.org/10.1063/1.466439}
  {\bibfield  {journal} {\bibinfo  {journal} {J. Chem. Phys.}\ }\textbf
  {\bibinfo {volume} {100}},\ \bibinfo {pages} {2975} (\bibinfo {year}
  {1994})}\BibitemShut {NoStop}%
\bibitem [{\citenamefont {Seidl}\ \emph {et~al.}(1996)\citenamefont {Seidl},
  \citenamefont {G{\"{o}}rling}, \citenamefont {Vogl}, \citenamefont
  {Majewski},\ and\ \citenamefont {Levy}}]{Seidl1996_PRB_3764}%
  \BibitemOpen
  \bibfield  {author} {\bibinfo {author} {\bibfnamefont {A.}~\bibnamefont
  {Seidl}}, \bibinfo {author} {\bibfnamefont {A.}~\bibnamefont
  {G{\"{o}}rling}}, \bibinfo {author} {\bibfnamefont {P.}~\bibnamefont {Vogl}},
  \bibinfo {author} {\bibfnamefont {J.~A.}\ \bibnamefont {Majewski}},\ and\
  \bibinfo {author} {\bibfnamefont {M.}~\bibnamefont {Levy}},\ }\bibfield
  {title} {\enquote {\bibinfo {title} {{Generalized Kohn--Sham schemes and the
  band-gap problem}},}\ }\href {https://doi.org/10.1103/PhysRevB.53.3764}
  {\bibfield  {journal} {\bibinfo  {journal} {Phys. Rev. B}\ }\textbf {\bibinfo
  {volume} {53}},\ \bibinfo {pages} {3764--3774} (\bibinfo {year}
  {1996})}\BibitemShut {NoStop}%
\bibitem [{\citenamefont {Komornicki}\ and\ \citenamefont
  {McIver}(1979)}]{Komornicki1979_JCP_2014}%
  \BibitemOpen
  \bibfield  {author} {\bibinfo {author} {\bibfnamefont {A.}~\bibnamefont
  {Komornicki}}\ and\ \bibinfo {author} {\bibfnamefont {J.~W.}\ \bibnamefont
  {McIver}},\ }\bibfield  {title} {\enquote {\bibinfo {title} {An efficient ab
  initio method for computing infrared and {Raman} intensities: Application to
  ethylene},}\ }\href {https://doi.org/10.1063/1.437627} {\bibfield  {journal}
  {\bibinfo  {journal} {J. Chem. Phys.}\ }\textbf {\bibinfo {volume} {70}},\
  \bibinfo {pages} {2014--2016} (\bibinfo {year} {1979})}\BibitemShut {NoStop}%
\bibitem [{\citenamefont {Frisch}\ \emph {et~al.}(1986)\citenamefont {Frisch},
  \citenamefont {Yamaguchi}, \citenamefont {Gaw}, \citenamefont {Schaefer},\
  and\ \citenamefont {Binkley}}]{Frisch1986_JCP_531}%
  \BibitemOpen
  \bibfield  {author} {\bibinfo {author} {\bibfnamefont {M.~J.}\ \bibnamefont
  {Frisch}}, \bibinfo {author} {\bibfnamefont {Y.}~\bibnamefont {Yamaguchi}},
  \bibinfo {author} {\bibfnamefont {J.~F.}\ \bibnamefont {Gaw}}, \bibinfo
  {author} {\bibfnamefont {H.~F.}\ \bibnamefont {Schaefer}},\ and\ \bibinfo
  {author} {\bibfnamefont {J.~S.}\ \bibnamefont {Binkley}},\ }\bibfield
  {title} {\enquote {\bibinfo {title} {Analytic {Raman} intensities from
  molecular electronic wave functions},}\ }\href
  {https://doi.org/10.1063/1.450121} {\bibfield  {journal} {\bibinfo  {journal}
  {J. Chem. Phys.}\ }\textbf {\bibinfo {volume} {84}},\ \bibinfo {pages}
  {531--532} (\bibinfo {year} {1986})}\BibitemShut {NoStop}%
\bibitem [{\citenamefont {Rappoport}\ and\ \citenamefont
  {Furche}(2007)}]{Rappoport2007_JCP_201104}%
  \BibitemOpen
  \bibfield  {author} {\bibinfo {author} {\bibfnamefont {D.}~\bibnamefont
  {Rappoport}}\ and\ \bibinfo {author} {\bibfnamefont {F.}~\bibnamefont
  {Furche}},\ }\bibfield  {title} {\enquote {\bibinfo {title} {{Lagrangian
  approach to molecular vibrational Raman intensities using time-dependent
  hybrid density functional theory}},}\ }\href
  {https://doi.org/10.1063/1.2744026} {\bibfield  {journal} {\bibinfo
  {journal} {J. Chem. Phys.}\ }\textbf {\bibinfo {volume} {126}},\ \bibinfo
  {pages} {201104} (\bibinfo {year} {2007})}\BibitemShut {NoStop}%
\bibitem [{\citenamefont {Stephens}(1985)}]{Stephens1985_JPC_748}%
  \BibitemOpen
  \bibfield  {author} {\bibinfo {author} {\bibfnamefont {P.~J.}\ \bibnamefont
  {Stephens}},\ }\bibfield  {title} {\enquote {\bibinfo {title} {Theory of
  vibrational circular dichroism},}\ }\href
  {https://doi.org/10.1021/j100251a006} {\bibfield  {journal} {\bibinfo
  {journal} {J. Phys. Chem.}\ }\textbf {\bibinfo {volume} {89}},\ \bibinfo
  {pages} {748--752} (\bibinfo {year} {1985})}\BibitemShut {NoStop}%
\bibitem [{\citenamefont {Cheeseman}\ \emph
  {et~al.}(1996{\natexlab{b}})\citenamefont {Cheeseman}, \citenamefont
  {Frisch}, \citenamefont {Devlin},\ and\ \citenamefont
  {Stephens}}]{Cheeseman1996_CPL_211}%
  \BibitemOpen
  \bibfield  {author} {\bibinfo {author} {\bibfnamefont {J.~R.}\ \bibnamefont
  {Cheeseman}}, \bibinfo {author} {\bibfnamefont {M.~J.}\ \bibnamefont
  {Frisch}}, \bibinfo {author} {\bibfnamefont {F.~J.}\ \bibnamefont {Devlin}},\
  and\ \bibinfo {author} {\bibfnamefont {P.~J.}\ \bibnamefont {Stephens}},\
  }\bibfield  {title} {\enquote {\bibinfo {title} {Ab initio calculation of
  atomic axial tensors and vibrational rotational strengths using density
  functional theory},}\ }\href {https://doi.org/10.1016/0009-2614(96)00154-6}
  {\bibfield  {journal} {\bibinfo  {journal} {Chem. Phys. Lett.}\ }\textbf
  {\bibinfo {volume} {252}},\ \bibinfo {pages} {211--220} (\bibinfo {year}
  {1996}{\natexlab{b}})}\BibitemShut {NoStop}%
\bibitem [{\citenamefont {Barron}\ \emph {et~al.}(1992)\citenamefont {Barron},
  \citenamefont {Gargaro}, \citenamefont {Hecht}, \citenamefont {Polavarapu},\
  and\ \citenamefont {Sugeta}}]{Barron1992_SAAMBS_1051}%
  \BibitemOpen
  \bibfield  {author} {\bibinfo {author} {\bibfnamefont {L.~D.}\ \bibnamefont
  {Barron}}, \bibinfo {author} {\bibfnamefont {A.~R.}\ \bibnamefont {Gargaro}},
  \bibinfo {author} {\bibfnamefont {L.}~\bibnamefont {Hecht}}, \bibinfo
  {author} {\bibfnamefont {P.~L.}\ \bibnamefont {Polavarapu}},\ and\ \bibinfo
  {author} {\bibfnamefont {H.}~\bibnamefont {Sugeta}},\ }\bibfield  {title}
  {\enquote {\bibinfo {title} {Experimental and ab initio theoretical
  vibrational {Raman} optical activity of tartaric acid},}\ }\href
  {https://doi.org/10.1016/0584-8539(92)80115-d} {\bibfield  {journal}
  {\bibinfo  {journal} {Spectrochim. Acta - A: Mol. Biomol. Spectrosc.}\
  }\textbf {\bibinfo {volume} {48}},\ \bibinfo {pages} {1051--1066} (\bibinfo
  {year} {1992})}\BibitemShut {NoStop}%
\bibitem [{\citenamefont {Barron}, \citenamefont {Hecht},\ and\ \citenamefont
  {Polavarapu}(1992)}]{Barron1992_SAAMBS_1193}%
  \BibitemOpen
  \bibfield  {author} {\bibinfo {author} {\bibfnamefont {L.~D.}\ \bibnamefont
  {Barron}}, \bibinfo {author} {\bibfnamefont {L.}~\bibnamefont {Hecht}},\ and\
  \bibinfo {author} {\bibfnamefont {P.~L.}\ \bibnamefont {Polavarapu}},\
  }\bibfield  {title} {\enquote {\bibinfo {title} {Methyl torsion {Raman}
  optical activity in trans-2,3-dimethyloxirane and
  trans-2,3-dimethylthiirane},}\ }\href
  {https://doi.org/10.1016/0584-8539(92)80133-h} {\bibfield  {journal}
  {\bibinfo  {journal} {Spectrochim. Acta - A: Mol. Biomol. Spectrosc.}\
  }\textbf {\bibinfo {volume} {48}},\ \bibinfo {pages} {1193--1195} (\bibinfo
  {year} {1992})}\BibitemShut {NoStop}%
\bibitem [{\citenamefont {Polavarapu}\ \emph {et~al.}(1993)\citenamefont
  {Polavarapu}, \citenamefont {Black}, \citenamefont {Barron},\ and\
  \citenamefont {Hecht}}]{Polavarapu1993_JACS_7736}%
  \BibitemOpen
  \bibfield  {author} {\bibinfo {author} {\bibfnamefont {P.~L.}\ \bibnamefont
  {Polavarapu}}, \bibinfo {author} {\bibfnamefont {T.~M.}\ \bibnamefont
  {Black}}, \bibinfo {author} {\bibfnamefont {L.~D.}\ \bibnamefont {Barron}},\
  and\ \bibinfo {author} {\bibfnamefont {L.}~\bibnamefont {Hecht}},\ }\bibfield
   {title} {\enquote {\bibinfo {title} {Vibrational {Raman} optical activity in
  ({R})-(+)-3-methylcyclohexanone: experimental and ab initio theoretical
  studies and the origins of the unusual couplets},}\ }\href
  {https://doi.org/10.1021/ja00070a018} {\bibfield  {journal} {\bibinfo
  {journal} {J. Am. Chem. Soc.}\ }\textbf {\bibinfo {volume} {115}},\ \bibinfo
  {pages} {7736--7742} (\bibinfo {year} {1993})}\BibitemShut {NoStop}%
\bibitem [{\citenamefont {Helgaker}\ \emph {et~al.}(1994)\citenamefont
  {Helgaker}, \citenamefont {Ruud}, \citenamefont {Bak}, \citenamefont
  {J{\o}rgensen},\ and\ \citenamefont {Olsen}}]{Helgaker1994_FD_165}%
  \BibitemOpen
  \bibfield  {author} {\bibinfo {author} {\bibfnamefont {T.}~\bibnamefont
  {Helgaker}}, \bibinfo {author} {\bibfnamefont {K.}~\bibnamefont {Ruud}},
  \bibinfo {author} {\bibfnamefont {K.~L.}\ \bibnamefont {Bak}}, \bibinfo
  {author} {\bibfnamefont {P.}~\bibnamefont {J{\o}rgensen}},\ and\ \bibinfo
  {author} {\bibfnamefont {J.}~\bibnamefont {Olsen}},\ }\bibfield  {title}
  {\enquote {\bibinfo {title} {Vibrational {Raman} optical activity
  calculations using {London} atomic orbitals},}\ }\href
  {https://doi.org/10.1039/FD9949900165} {\bibfield  {journal} {\bibinfo
  {journal} {Faraday Discuss.}\ }\textbf {\bibinfo {volume} {99}},\ \bibinfo
  {pages} {165--180} (\bibinfo {year} {1994})}\BibitemShut {NoStop}%
\bibitem [{\citenamefont {Ruud}, \citenamefont {Helgaker},\ and\ \citenamefont
  {Bouř}(2002)}]{Ruud2002_JPCA_7448}%
  \BibitemOpen
  \bibfield  {author} {\bibinfo {author} {\bibfnamefont {K.}~\bibnamefont
  {Ruud}}, \bibinfo {author} {\bibfnamefont {T.}~\bibnamefont {Helgaker}},\
  and\ \bibinfo {author} {\bibfnamefont {P.}~\bibnamefont {Bouř}},\ }\bibfield
   {title} {\enquote {\bibinfo {title} {Gauge-origin independent
  density-functional theory calculations of vibrational {Raman} optical
  activity},}\ }\href {https://doi.org/10.1021/jp026037i} {\bibfield  {journal}
  {\bibinfo  {journal} {J. Phys. Chem. A}\ }\textbf {\bibinfo {volume} {106}},\
  \bibinfo {pages} {7448--7455} (\bibinfo {year} {2002})}\BibitemShut {NoStop}%
\bibitem [{\citenamefont {Shao}, \citenamefont {Head-Gordon},\ and\
  \citenamefont {Krylov}(2003)}]{Shao2003_JCP_4807}%
  \BibitemOpen
  \bibfield  {author} {\bibinfo {author} {\bibfnamefont {Y.}~\bibnamefont
  {Shao}}, \bibinfo {author} {\bibfnamefont {M.}~\bibnamefont {Head-Gordon}},\
  and\ \bibinfo {author} {\bibfnamefont {A.~I.}\ \bibnamefont {Krylov}},\
  }\bibfield  {title} {\enquote {\bibinfo {title} {{The spin--flip approach
  within time-dependent density functional theory: Theory and applications to
  diradicals}},}\ }\href {https://doi.org/10.1063/1.1545679} {\bibfield
  {journal} {\bibinfo  {journal} {J. Chem. Phys.}\ }\textbf {\bibinfo {volume}
  {118}},\ \bibinfo {pages} {4807} (\bibinfo {year} {2003})}\BibitemShut
  {NoStop}%
\bibitem [{\citenamefont {Wang}\ and\ \citenamefont
  {Ziegler}(2004)}]{Wang2004_JCP_12191}%
  \BibitemOpen
  \bibfield  {author} {\bibinfo {author} {\bibfnamefont {F.}~\bibnamefont
  {Wang}}\ and\ \bibinfo {author} {\bibfnamefont {T.}~\bibnamefont {Ziegler}},\
  }\bibfield  {title} {\enquote {\bibinfo {title} {Time-dependent density
  functional theory based on a noncollinear formulation of the
  exchange-correlation potential},}\ }\href {https://doi.org/10.1063/1.1821494}
  {\bibfield  {journal} {\bibinfo  {journal} {J. Chem. Phys.}\ }\textbf
  {\bibinfo {volume} {121}},\ \bibinfo {pages} {12191--12196} (\bibinfo {year}
  {2004})}\BibitemShut {NoStop}%
\bibitem [{\citenamefont {Bast}, \citenamefont {Jensen},\ and\ \citenamefont
  {Saue}(2009)}]{Bast2009_IJQC_2091}%
  \BibitemOpen
  \bibfield  {author} {\bibinfo {author} {\bibfnamefont {R.}~\bibnamefont
  {Bast}}, \bibinfo {author} {\bibfnamefont {H.~J.~A.}\ \bibnamefont
  {Jensen}},\ and\ \bibinfo {author} {\bibfnamefont {T.}~\bibnamefont {Saue}},\
  }\bibfield  {title} {\enquote {\bibinfo {title} {Relativistic adiabatic
  time-dependent density functional theory using hybrid functionals and
  noncollinear spin magnetization},}\ }\href
  {https://doi.org/10.1002/qua.22065} {\bibfield  {journal} {\bibinfo
  {journal} {Int. J. Quantum Chem.}\ }\textbf {\bibinfo {volume} {109}},\
  \bibinfo {pages} {2091--2112} (\bibinfo {year} {2009})}\BibitemShut {NoStop}%
\bibitem [{\citenamefont {{Li Manni}}\ \emph {et~al.}(2014)\citenamefont {{Li
  Manni}}, \citenamefont {Carlson}, \citenamefont {Luo}, \citenamefont {Ma},
  \citenamefont {Olsen}, \citenamefont {Truhlar},\ and\ \citenamefont
  {Gagliardi}}]{LiManni2014_JCTC_3669}%
  \BibitemOpen
  \bibfield  {author} {\bibinfo {author} {\bibfnamefont {G.}~\bibnamefont {{Li
  Manni}}}, \bibinfo {author} {\bibfnamefont {R.~K.}\ \bibnamefont {Carlson}},
  \bibinfo {author} {\bibfnamefont {S.}~\bibnamefont {Luo}}, \bibinfo {author}
  {\bibfnamefont {D.}~\bibnamefont {Ma}}, \bibinfo {author} {\bibfnamefont
  {J.}~\bibnamefont {Olsen}}, \bibinfo {author} {\bibfnamefont {D.~G.}\
  \bibnamefont {Truhlar}},\ and\ \bibinfo {author} {\bibfnamefont
  {L.}~\bibnamefont {Gagliardi}},\ }\bibfield  {title} {\enquote {\bibinfo
  {title} {Multiconfiguration pair-density functional theory},}\ }\href
  {https://doi.org/10.1021/ct500483t} {\bibfield  {journal} {\bibinfo
  {journal} {J. Chem. Theory Comput.}\ }\textbf {\bibinfo {volume} {10}},\
  \bibinfo {pages} {3669--3680} (\bibinfo {year} {2014})}\BibitemShut {NoStop}%
\bibitem [{\citenamefont {Carlson}\ \emph {et~al.}(2015)\citenamefont
  {Carlson}, \citenamefont {{Li Manni}}, \citenamefont {Sonnenberger},
  \citenamefont {Truhlar},\ and\ \citenamefont
  {Gagliardi}}]{Carlson2015_JCTC_82}%
  \BibitemOpen
  \bibfield  {author} {\bibinfo {author} {\bibfnamefont {R.~K.}\ \bibnamefont
  {Carlson}}, \bibinfo {author} {\bibfnamefont {G.}~\bibnamefont {{Li Manni}}},
  \bibinfo {author} {\bibfnamefont {A.~L.}\ \bibnamefont {Sonnenberger}},
  \bibinfo {author} {\bibfnamefont {D.~G.}\ \bibnamefont {Truhlar}},\ and\
  \bibinfo {author} {\bibfnamefont {L.}~\bibnamefont {Gagliardi}},\ }\bibfield
  {title} {\enquote {\bibinfo {title} {Multiconfiguration pair-density
  functional theory: Barrier heights and main group and transition metal
  energetics},}\ }\href {https://doi.org/10.1021/ct5008235} {\bibfield
  {journal} {\bibinfo  {journal} {J. Chem. Theory Comput.}\ }\textbf {\bibinfo
  {volume} {11}},\ \bibinfo {pages} {82--90} (\bibinfo {year}
  {2015})}\BibitemShut {NoStop}%
\bibitem [{\citenamefont {Carlson}, \citenamefont {Truhlar},\ and\
  \citenamefont {Gagliardi}(2015)}]{Carlson2015_JCTC_4077}%
  \BibitemOpen
  \bibfield  {author} {\bibinfo {author} {\bibfnamefont {R.~K.}\ \bibnamefont
  {Carlson}}, \bibinfo {author} {\bibfnamefont {D.~G.}\ \bibnamefont
  {Truhlar}},\ and\ \bibinfo {author} {\bibfnamefont {L.}~\bibnamefont
  {Gagliardi}},\ }\bibfield  {title} {\enquote {\bibinfo {title}
  {Multiconfiguration pair-density functional theory: A fully translated
  gradient approximation and its performance for transition metal dimers and
  the spectroscopy of \ce{Re2Cl8^{2-}}},}\ }\href
  {https://doi.org/10.1021/acs.jctc.5b00609} {\bibfield  {journal} {\bibinfo
  {journal} {J. Chem. Theory Comput.}\ }\textbf {\bibinfo {volume} {11}},\
  \bibinfo {pages} {4077--4085} (\bibinfo {year} {2015})}\BibitemShut {NoStop}%
\bibitem [{\citenamefont {Jaramillo}, \citenamefont {Scuseria},\ and\
  \citenamefont {Ernzerhof}(2003)}]{Jaramillo2003_JCP_1068}%
  \BibitemOpen
  \bibfield  {author} {\bibinfo {author} {\bibfnamefont {J.}~\bibnamefont
  {Jaramillo}}, \bibinfo {author} {\bibfnamefont {G.~E.}\ \bibnamefont
  {Scuseria}},\ and\ \bibinfo {author} {\bibfnamefont {M.}~\bibnamefont
  {Ernzerhof}},\ }\bibfield  {title} {\enquote {\bibinfo {title} {{Local hybrid
  functionals}},}\ }\href {https://doi.org/10.1063/1.1528936} {\bibfield
  {journal} {\bibinfo  {journal} {J. Chem. Phys.}\ }\textbf {\bibinfo {volume}
  {118}},\ \bibinfo {pages} {1068--1073} (\bibinfo {year} {2003})}\BibitemShut
  {NoStop}%
\bibitem [{\citenamefont {Maier}, \citenamefont {Arbuznikov},\ and\
  \citenamefont {Kaupp}(2019)}]{Maier2019_WIRCMS_1378}%
  \BibitemOpen
  \bibfield  {author} {\bibinfo {author} {\bibfnamefont {T.~M.}\ \bibnamefont
  {Maier}}, \bibinfo {author} {\bibfnamefont {A.~V.}\ \bibnamefont
  {Arbuznikov}},\ and\ \bibinfo {author} {\bibfnamefont {M.}~\bibnamefont
  {Kaupp}},\ }\bibfield  {title} {\enquote {\bibinfo {title} {{Local hybrid
  functionals: Theory, implementation, and performance of an emerging new tool
  in quantum chemistry and beyond}},}\ }\href
  {https://doi.org/10.1002/wcms.1378} {\bibfield  {journal} {\bibinfo
  {journal} {Wiley Interdiscip. Rev. Comput. Mol. Sci.}\ }\textbf {\bibinfo
  {volume} {9}},\ \bibinfo {pages} {e1378} (\bibinfo {year}
  {2019})}\BibitemShut {NoStop}%
\bibitem [{\citenamefont {Schattenberg}\ and\ \citenamefont
  {Kaupp}(2021)}]{Schattenberg2021_JPCA_2697}%
  \BibitemOpen
  \bibfield  {author} {\bibinfo {author} {\bibfnamefont {C.~J.}\ \bibnamefont
  {Schattenberg}}\ and\ \bibinfo {author} {\bibfnamefont {M.}~\bibnamefont
  {Kaupp}},\ }\bibfield  {title} {\enquote {\bibinfo {title} {Implementation
  and validation of local hybrid functionals with calibrated exchange-energy
  densities for nuclear shielding constants},}\ }\href
  {https://doi.org/10.1021/acs.jpca.1c01135} {\bibfield  {journal} {\bibinfo
  {journal} {J. Phys. Chem. A}\ }\textbf {\bibinfo {volume} {125}},\ \bibinfo
  {pages} {2697--2707} (\bibinfo {year} {2021})}\BibitemShut {NoStop}%
\bibitem [{\citenamefont {Deglmann}, \citenamefont {Furche},\ and\
  \citenamefont {Ahlrichs}(2002)}]{Deglmann2002_CPL_511}%
  \BibitemOpen
  \bibfield  {author} {\bibinfo {author} {\bibfnamefont {P.}~\bibnamefont
  {Deglmann}}, \bibinfo {author} {\bibfnamefont {F.}~\bibnamefont {Furche}},\
  and\ \bibinfo {author} {\bibfnamefont {R.}~\bibnamefont {Ahlrichs}},\
  }\bibfield  {title} {\enquote {\bibinfo {title} {{An efficient implementation
  of second analytical derivatives for density functional methods}},}\ }\href
  {https://doi.org/10.1016/S0009-2614(02)01084-9} {\bibfield  {journal}
  {\bibinfo  {journal} {Chem. Phys. Lett.}\ }\textbf {\bibinfo {volume}
  {362}},\ \bibinfo {pages} {511--518} (\bibinfo {year} {2002})}\BibitemShut
  {NoStop}%
\bibitem [{\citenamefont {van Gisbergen}, \citenamefont {Snijders},\ and\
  \citenamefont {Baerends}(1998{\natexlab{b}})}]{Gisbergen1998_JCP_10657}%
  \BibitemOpen
  \bibfield  {author} {\bibinfo {author} {\bibfnamefont {S.~J.~A.}\
  \bibnamefont {van Gisbergen}}, \bibinfo {author} {\bibfnamefont {J.~G.}\
  \bibnamefont {Snijders}},\ and\ \bibinfo {author} {\bibfnamefont {E.~J.}\
  \bibnamefont {Baerends}},\ }\bibfield  {title} {\enquote {\bibinfo {title}
  {Accurate density functional calculations on frequency-dependent
  hyperpolarizabilities of small molecules},}\ }\href
  {https://doi.org/10.1063/1.477763} {\bibfield  {journal} {\bibinfo  {journal}
  {J. Chem. Phys.}\ }\textbf {\bibinfo {volume} {109}},\ \bibinfo {pages}
  {10657--10668} (\bibinfo {year} {1998}{\natexlab{b}})}\BibitemShut {NoStop}%
\bibitem [{\citenamefont {Dickson}\ and\ \citenamefont
  {Ziegler}(1996)}]{Dickson1996_JPC_5286}%
  \BibitemOpen
  \bibfield  {author} {\bibinfo {author} {\bibfnamefont {R.~M.}\ \bibnamefont
  {Dickson}}\ and\ \bibinfo {author} {\bibfnamefont {T.}~\bibnamefont
  {Ziegler}},\ }\bibfield  {title} {\enquote {\bibinfo {title} {{NMR}
  spin--spin coupling constants from density functional theory with
  {Slater}-type basis functions},}\ }\href {https://doi.org/10.1021/jp951930l}
  {\bibfield  {journal} {\bibinfo  {journal} {J. Phys. Chem.}\ }\textbf
  {\bibinfo {volume} {100}},\ \bibinfo {pages} {5286--5290} (\bibinfo {year}
  {1996})}\BibitemShut {NoStop}%
\bibitem [{\citenamefont {Furche}(2001)}]{Furche2001_JCP_5982}%
  \BibitemOpen
  \bibfield  {author} {\bibinfo {author} {\bibfnamefont {F.}~\bibnamefont
  {Furche}},\ }\bibfield  {title} {\enquote {\bibinfo {title} {{On the density
  matrix based approach to time-dependent density functional response
  theory}},}\ }\href {https://doi.org/10.1063/1.1353585} {\bibfield  {journal}
  {\bibinfo  {journal} {J. Chem. Phys.}\ }\textbf {\bibinfo {volume} {114}},\
  \bibinfo {pages} {5982} (\bibinfo {year} {2001})}\BibitemShut {NoStop}%
\bibitem [{\citenamefont {Meurer}\ \emph {et~al.}(2017)\citenamefont {Meurer},
  \citenamefont {Smith}, \citenamefont {Paprocki}, \citenamefont
  {{\v{C}}ert{\'i}k}, \citenamefont {Kirpichev}, \citenamefont {Rocklin},
  \citenamefont {Kumar}, \citenamefont {Ivanov}, \citenamefont {Moore},
  \citenamefont {Singh}, \citenamefont {Rathnayake}, \citenamefont {Vig},
  \citenamefont {Granger}, \citenamefont {Muller}, \citenamefont {Bonazzi},
  \citenamefont {Gupta}, \citenamefont {Vats}, \citenamefont {Johansson},
  \citenamefont {Pedregosa}, \citenamefont {Curry}, \citenamefont {Terrel},
  \citenamefont {Rou{\v{c}}ka}, \citenamefont {Saboo}, \citenamefont
  {Fernando}, \citenamefont {Kulal}, \citenamefont {Cimrman},\ and\
  \citenamefont {Scopatz}}]{Meurer2017_PCS_103}%
  \BibitemOpen
  \bibfield  {author} {\bibinfo {author} {\bibfnamefont {A.}~\bibnamefont
  {Meurer}}, \bibinfo {author} {\bibfnamefont {C.~P.}\ \bibnamefont {Smith}},
  \bibinfo {author} {\bibfnamefont {M.}~\bibnamefont {Paprocki}}, \bibinfo
  {author} {\bibfnamefont {O.}~\bibnamefont {{\v{C}}ert{\'i}k}}, \bibinfo
  {author} {\bibfnamefont {S.~B.}\ \bibnamefont {Kirpichev}}, \bibinfo {author}
  {\bibfnamefont {M.}~\bibnamefont {Rocklin}}, \bibinfo {author} {\bibfnamefont
  {A.}~\bibnamefont {Kumar}}, \bibinfo {author} {\bibfnamefont
  {S.}~\bibnamefont {Ivanov}}, \bibinfo {author} {\bibfnamefont {J.~K.}\
  \bibnamefont {Moore}}, \bibinfo {author} {\bibfnamefont {S.}~\bibnamefont
  {Singh}}, \bibinfo {author} {\bibfnamefont {T.}~\bibnamefont {Rathnayake}},
  \bibinfo {author} {\bibfnamefont {S.}~\bibnamefont {Vig}}, \bibinfo {author}
  {\bibfnamefont {B.~E.}\ \bibnamefont {Granger}}, \bibinfo {author}
  {\bibfnamefont {R.~P.}\ \bibnamefont {Muller}}, \bibinfo {author}
  {\bibfnamefont {F.}~\bibnamefont {Bonazzi}}, \bibinfo {author} {\bibfnamefont
  {H.}~\bibnamefont {Gupta}}, \bibinfo {author} {\bibfnamefont
  {S.}~\bibnamefont {Vats}}, \bibinfo {author} {\bibfnamefont {F.}~\bibnamefont
  {Johansson}}, \bibinfo {author} {\bibfnamefont {F.}~\bibnamefont
  {Pedregosa}}, \bibinfo {author} {\bibfnamefont {M.~J.}\ \bibnamefont
  {Curry}}, \bibinfo {author} {\bibfnamefont {A.~R.}\ \bibnamefont {Terrel}},
  \bibinfo {author} {\bibfnamefont {{\v{S}}.}~\bibnamefont {Rou{\v{c}}ka}},
  \bibinfo {author} {\bibfnamefont {A.}~\bibnamefont {Saboo}}, \bibinfo
  {author} {\bibfnamefont {I.}~\bibnamefont {Fernando}}, \bibinfo {author}
  {\bibfnamefont {S.}~\bibnamefont {Kulal}}, \bibinfo {author} {\bibfnamefont
  {R.}~\bibnamefont {Cimrman}},\ and\ \bibinfo {author} {\bibfnamefont
  {A.}~\bibnamefont {Scopatz}},\ }\bibfield  {title} {\enquote {\bibinfo
  {title} {{SymPy}: symbolic computing in {Python}},}\ }\href
  {https://doi.org/10.7717/peerj-cs.103} {\bibfield  {journal} {\bibinfo
  {journal} {PeerJ Comput. Sci.}\ }\textbf {\bibinfo {volume} {3}},\ \bibinfo
  {pages} {e103} (\bibinfo {year} {2017})}\BibitemShut {NoStop}%
\bibitem [{\citenamefont {Davidson}(1975)}]{Davidson1975_JCP_87}%
  \BibitemOpen
  \bibfield  {author} {\bibinfo {author} {\bibfnamefont {E.~R.}\ \bibnamefont
  {Davidson}},\ }\bibfield  {title} {\enquote {\bibinfo {title} {The iterative
  calculation of a few of the lowest eigenvalues and corresponding eigenvectors
  of large real-symmetric matrices},}\ }\href
  {https://doi.org/10.1016/0021-9991(75)90065-0} {\bibfield  {journal}
  {\bibinfo  {journal} {J. Comput. Phys.}\ }\textbf {\bibinfo {volume} {17}},\
  \bibinfo {pages} {87--94} (\bibinfo {year} {1975})}\BibitemShut {NoStop}%
\bibitem [{\citenamefont {Becke}\ and\ \citenamefont
  {Roussel}(1989)}]{Becke1989_PRA_3761}%
  \BibitemOpen
  \bibfield  {author} {\bibinfo {author} {\bibfnamefont {A.~D.}\ \bibnamefont
  {Becke}}\ and\ \bibinfo {author} {\bibfnamefont {M.~R.}\ \bibnamefont
  {Roussel}},\ }\bibfield  {title} {\enquote {\bibinfo {title} {{Exchange holes
  in inhomogeneous systems: A coordinate-space model}},}\ }\href
  {https://doi.org/10.1103/PhysRevA.39.3761} {\bibfield  {journal} {\bibinfo
  {journal} {Phys. Rev. A}\ }\textbf {\bibinfo {volume} {39}},\ \bibinfo
  {pages} {3761--3767} (\bibinfo {year} {1989})}\BibitemShut {NoStop}%
\bibitem [{\citenamefont {Sun}\ \emph {et~al.}(2018)\citenamefont {Sun},
  \citenamefont {Berkelbach}, \citenamefont {Blunt}, \citenamefont {Booth},
  \citenamefont {Guo}, \citenamefont {Li}, \citenamefont {Liu}, \citenamefont
  {McClain}, \citenamefont {Sayfutyarova}, \citenamefont {Sharma},
  \citenamefont {Wouters},\ and\ \citenamefont {Chan}}]{Sun2018_WIRCMS_1340}%
  \BibitemOpen
  \bibfield  {author} {\bibinfo {author} {\bibfnamefont {Q.}~\bibnamefont
  {Sun}}, \bibinfo {author} {\bibfnamefont {T.~C.}\ \bibnamefont {Berkelbach}},
  \bibinfo {author} {\bibfnamefont {N.~S.}\ \bibnamefont {Blunt}}, \bibinfo
  {author} {\bibfnamefont {G.~H.}\ \bibnamefont {Booth}}, \bibinfo {author}
  {\bibfnamefont {S.}~\bibnamefont {Guo}}, \bibinfo {author} {\bibfnamefont
  {Z.}~\bibnamefont {Li}}, \bibinfo {author} {\bibfnamefont {J.}~\bibnamefont
  {Liu}}, \bibinfo {author} {\bibfnamefont {J.~D.}\ \bibnamefont {McClain}},
  \bibinfo {author} {\bibfnamefont {E.~R.}\ \bibnamefont {Sayfutyarova}},
  \bibinfo {author} {\bibfnamefont {S.}~\bibnamefont {Sharma}}, \bibinfo
  {author} {\bibfnamefont {S.}~\bibnamefont {Wouters}},\ and\ \bibinfo {author}
  {\bibfnamefont {G.~K.-L.}\ \bibnamefont {Chan}},\ }\bibfield  {title}
  {\enquote {\bibinfo {title} {{PySCF}: the {Python}-based simulations of
  chemistry framework},}\ }\href {https://doi.org/10.1002/wcms.1340} {\bibfield
   {journal} {\bibinfo  {journal} {Wiley Interdiscip. Rev. Comput. Mol. Sci.}\
  }\textbf {\bibinfo {volume} {8}},\ \bibinfo {pages} {e1340} (\bibinfo {year}
  {2018})},\ \Eprint {https://arxiv.org/abs/1701.08223} {arXiv:1701.08223}
  \BibitemShut {NoStop}%
\bibitem [{\citenamefont {Lehtola}\ and\ \citenamefont
  {Marques}(2023)}]{Lehtola2023_JCP_114116}%
  \BibitemOpen
  \bibfield  {author} {\bibinfo {author} {\bibfnamefont {S.}~\bibnamefont
  {Lehtola}}\ and\ \bibinfo {author} {\bibfnamefont {M.~A.~L.}\ \bibnamefont
  {Marques}},\ }\bibfield  {title} {\enquote {\bibinfo {title} {Reproducibility
  of density functional approximations: How new functionals should be
  reported},}\ }\href {https://doi.org/10.1063/5.0167763} {\bibfield  {journal}
  {\bibinfo  {journal} {J. Chem. Phys.}\ }\textbf {\bibinfo {volume} {159}},\
  \bibinfo {pages} {114116} (\bibinfo {year} {2023})}\BibitemShut {NoStop}%
\bibitem [{\citenamefont {Bloch}(1929)}]{Bloch1929_ZP_545}%
  \BibitemOpen
  \bibfield  {author} {\bibinfo {author} {\bibfnamefont {F.}~\bibnamefont
  {Bloch}},\ }\bibfield  {title} {\enquote {\bibinfo {title} {{Bemerkung} zur
  {Elektronentheorie} des {Ferromagnetismus} und der elektrischen
  {Leitf{\"{a}}higkeit}},}\ }\href {https://doi.org/10.1007/BF01340281}
  {\bibfield  {journal} {\bibinfo  {journal} {Z. Phys.}\ }\textbf {\bibinfo
  {volume} {57}},\ \bibinfo {pages} {545--555} (\bibinfo {year}
  {1929})}\BibitemShut {NoStop}%
\bibitem [{\citenamefont {Dirac}(1930)}]{Dirac1930_MPCPS_376}%
  \BibitemOpen
  \bibfield  {author} {\bibinfo {author} {\bibfnamefont {P.~A.~M.}\
  \bibnamefont {Dirac}},\ }\bibfield  {title} {\enquote {\bibinfo {title} {Note
  on exchange phenomena in the {Thomas} atom},}\ }\href
  {https://doi.org/10.1017/S0305004100016108} {\bibfield  {journal} {\bibinfo
  {journal} {Math. Proc. Cambridge Philos. Soc.}\ }\textbf {\bibinfo {volume}
  {26}},\ \bibinfo {pages} {376--385} (\bibinfo {year} {1930})}\BibitemShut
  {NoStop}%
\bibitem [{\citenamefont {Perdew}\ and\ \citenamefont
  {Wang}(1992)}]{Perdew1992_PRB_13244}%
  \BibitemOpen
  \bibfield  {author} {\bibinfo {author} {\bibfnamefont {J.~P.}\ \bibnamefont
  {Perdew}}\ and\ \bibinfo {author} {\bibfnamefont {Y.}~\bibnamefont {Wang}},\
  }\bibfield  {title} {\enquote {\bibinfo {title} {{Accurate and simple
  analytic representation of the electron-gas correlation energy}},}\ }\href
  {https://doi.org/10.1103/PhysRevB.45.13244} {\bibfield  {journal} {\bibinfo
  {journal} {Phys. Rev. B}\ }\textbf {\bibinfo {volume} {45}},\ \bibinfo
  {pages} {13244--13249} (\bibinfo {year} {1992})}\BibitemShut {NoStop}%
\bibitem [{\citenamefont {Perdew}, \citenamefont {Burke},\ and\ \citenamefont
  {Ernzerhof}(1996)}]{Perdew1996_PRL_3865}%
  \BibitemOpen
  \bibfield  {author} {\bibinfo {author} {\bibfnamefont {J.~P.}\ \bibnamefont
  {Perdew}}, \bibinfo {author} {\bibfnamefont {K.}~\bibnamefont {Burke}},\ and\
  \bibinfo {author} {\bibfnamefont {M.}~\bibnamefont {Ernzerhof}},\ }\bibfield
  {title} {\enquote {\bibinfo {title} {Generalized gradient approximation made
  simple},}\ }\href {https://doi.org/10.1103/PhysRevLett.77.3865} {\bibfield
  {journal} {\bibinfo  {journal} {Phys. Rev. Lett.}\ }\textbf {\bibinfo
  {volume} {77}},\ \bibinfo {pages} {3865--3868} (\bibinfo {year}
  {1996})}\BibitemShut {NoStop}%
\bibitem [{\citenamefont {Tao}\ \emph {et~al.}(2003)\citenamefont {Tao},
  \citenamefont {Perdew}, \citenamefont {Staroverov},\ and\ \citenamefont
  {Scuseria}}]{Tao2003_PRL_146401}%
  \BibitemOpen
  \bibfield  {author} {\bibinfo {author} {\bibfnamefont {J.}~\bibnamefont
  {Tao}}, \bibinfo {author} {\bibfnamefont {J.~P.}\ \bibnamefont {Perdew}},
  \bibinfo {author} {\bibfnamefont {V.~N.}\ \bibnamefont {Staroverov}},\ and\
  \bibinfo {author} {\bibfnamefont {G.~E.}\ \bibnamefont {Scuseria}},\
  }\bibfield  {title} {\enquote {\bibinfo {title} {Climbing the density
  functional ladder: Nonempirical meta-generalized gradient approximation
  designed for molecules and solids},}\ }\href
  {https://doi.org/10.1103/PhysRevLett.91.146401} {\bibfield  {journal}
  {\bibinfo  {journal} {Phys. Rev. Lett.}\ }\textbf {\bibinfo {volume} {91}},\
  \bibinfo {pages} {146401} (\bibinfo {year} {2003})}\BibitemShut {NoStop}%
\bibitem [{\citenamefont {Dunning}(1989)}]{Dunning1989_JCP_1007}%
  \BibitemOpen
  \bibfield  {author} {\bibinfo {author} {\bibfnamefont {T.~H.}\ \bibnamefont
  {Dunning}},\ }\bibfield  {title} {\enquote {\bibinfo {title} {{Gaussian basis
  sets for use in correlated molecular calculations. I. The atoms boron through
  neon and hydrogen}},}\ }\href {https://doi.org/10.1063/1.456153} {\bibfield
  {journal} {\bibinfo  {journal} {J. Chem. Phys.}\ }\textbf {\bibinfo {volume}
  {90}},\ \bibinfo {pages} {1007} (\bibinfo {year} {1989})}\BibitemShut
  {NoStop}%
\bibitem [{\citenamefont {Kendall}, \citenamefont {Dunning},\ and\
  \citenamefont {Harrison}(1992)}]{Kendall1992_JCP_6796}%
  \BibitemOpen
  \bibfield  {author} {\bibinfo {author} {\bibfnamefont {R.~A.}\ \bibnamefont
  {Kendall}}, \bibinfo {author} {\bibfnamefont {T.~H.}\ \bibnamefont
  {Dunning}},\ and\ \bibinfo {author} {\bibfnamefont {R.~J.}\ \bibnamefont
  {Harrison}},\ }\bibfield  {title} {\enquote {\bibinfo {title} {{Electron
  affinities of the first-row atoms revisited. Systematic basis sets and wave
  functions}},}\ }\href {https://doi.org/10.1063/1.462569} {\bibfield
  {journal} {\bibinfo  {journal} {J. Chem. Phys.}\ }\textbf {\bibinfo {volume}
  {96}},\ \bibinfo {pages} {6796} (\bibinfo {year} {1992})}\BibitemShut
  {NoStop}%
\bibitem [{\citenamefont {Jensen}(2001)}]{Jensen2001_JCP_9113}%
  \BibitemOpen
  \bibfield  {author} {\bibinfo {author} {\bibfnamefont {F.}~\bibnamefont
  {Jensen}},\ }\bibfield  {title} {\enquote {\bibinfo {title} {Polarization
  consistent basis sets: Principles},}\ }\href
  {https://doi.org/10.1063/1.1413524} {\bibfield  {journal} {\bibinfo
  {journal} {J. Chem. Phys.}\ }\textbf {\bibinfo {volume} {115}},\ \bibinfo
  {pages} {9113--9125} (\bibinfo {year} {2001})}\BibitemShut {NoStop}%
\bibitem [{\citenamefont {Jensen}(2014)}]{Jensen2014_JCTC_1074}%
  \BibitemOpen
  \bibfield  {author} {\bibinfo {author} {\bibfnamefont {F.}~\bibnamefont
  {Jensen}},\ }\bibfield  {title} {\enquote {\bibinfo {title} {Unifying general
  and segmented contracted basis sets. segmented polarization consistent basis
  sets},}\ }\href {https://doi.org/10.1021/ct401026a} {\bibfield  {journal}
  {\bibinfo  {journal} {J. Chem. Theory Comput.}\ }\textbf {\bibinfo {volume}
  {10}},\ \bibinfo {pages} {1074--1085} (\bibinfo {year} {2014})}\BibitemShut
  {NoStop}%
\bibitem [{\citenamefont {Mortensen}, \citenamefont {Hansen},\ and\
  \citenamefont {Jacobsen}(2005)}]{Mortensen2005_PRB_35109}%
  \BibitemOpen
  \bibfield  {author} {\bibinfo {author} {\bibfnamefont {J.~J.}\ \bibnamefont
  {Mortensen}}, \bibinfo {author} {\bibfnamefont {L.~B.}\ \bibnamefont
  {Hansen}},\ and\ \bibinfo {author} {\bibfnamefont {K.~W.}\ \bibnamefont
  {Jacobsen}},\ }\bibfield  {title} {\enquote {\bibinfo {title} {{Real-space
  grid implementation of the projector augmented wave method}},}\ }\href
  {https://doi.org/10.1103/PhysRevB.71.035109} {\bibfield  {journal} {\bibinfo
  {journal} {Phys. Rev. B}\ }\textbf {\bibinfo {volume} {71}},\ \bibinfo
  {pages} {035109} (\bibinfo {year} {2005})},\ \Eprint
  {https://arxiv.org/abs/0411218} {arXiv:0411218 [cond-mat]} \BibitemShut
  {NoStop}%
\bibitem [{\citenamefont {Enkovaara}\ \emph {et~al.}(2010)\citenamefont
  {Enkovaara}, \citenamefont {Rostgaard}, \citenamefont {Mortensen},
  \citenamefont {Chen}, \citenamefont {Du{\l}ak}, \citenamefont {Ferrighi},
  \citenamefont {Gavnholt}, \citenamefont {Glinsvad}, \citenamefont {Haikola},
  \citenamefont {Hansen}, \citenamefont {Kristoffersen}, \citenamefont
  {Kuisma}, \citenamefont {Larsen}, \citenamefont {Lehtovaara}, \citenamefont
  {Ljungberg}, \citenamefont {Lopez-Acevedo}, \citenamefont {Moses},
  \citenamefont {Ojanen}, \citenamefont {Olsen}, \citenamefont {Petzold},
  \citenamefont {Romero}, \citenamefont {Stausholm-M{\o}ller}, \citenamefont
  {Strange}, \citenamefont {Tritsaris}, \citenamefont {Vanin}, \citenamefont
  {Walter}, \citenamefont {Hammer}, \citenamefont {H{\"{a}}kkinen},
  \citenamefont {Madsen}, \citenamefont {Nieminen}, \citenamefont {N{\o}rskov},
  \citenamefont {Puska}, \citenamefont {Rantala}, \citenamefont {Schi{\o}tz},
  \citenamefont {Thygesen},\ and\ \citenamefont
  {Jacobsen}}]{Enkovaara2010_JPCM_253202}%
  \BibitemOpen
  \bibfield  {author} {\bibinfo {author} {\bibfnamefont {J.}~\bibnamefont
  {Enkovaara}}, \bibinfo {author} {\bibfnamefont {C.}~\bibnamefont
  {Rostgaard}}, \bibinfo {author} {\bibfnamefont {J.~J.}\ \bibnamefont
  {Mortensen}}, \bibinfo {author} {\bibfnamefont {J.}~\bibnamefont {Chen}},
  \bibinfo {author} {\bibfnamefont {M.}~\bibnamefont {Du{\l}ak}}, \bibinfo
  {author} {\bibfnamefont {L.}~\bibnamefont {Ferrighi}}, \bibinfo {author}
  {\bibfnamefont {J.}~\bibnamefont {Gavnholt}}, \bibinfo {author}
  {\bibfnamefont {C.}~\bibnamefont {Glinsvad}}, \bibinfo {author}
  {\bibfnamefont {V.}~\bibnamefont {Haikola}}, \bibinfo {author} {\bibfnamefont
  {H.~A.}\ \bibnamefont {Hansen}}, \bibinfo {author} {\bibfnamefont {H.~H.}\
  \bibnamefont {Kristoffersen}}, \bibinfo {author} {\bibfnamefont
  {M.}~\bibnamefont {Kuisma}}, \bibinfo {author} {\bibfnamefont {A.~H.}\
  \bibnamefont {Larsen}}, \bibinfo {author} {\bibfnamefont {L.}~\bibnamefont
  {Lehtovaara}}, \bibinfo {author} {\bibfnamefont {M.}~\bibnamefont
  {Ljungberg}}, \bibinfo {author} {\bibfnamefont {O.}~\bibnamefont
  {Lopez-Acevedo}}, \bibinfo {author} {\bibfnamefont {P.~G.}\ \bibnamefont
  {Moses}}, \bibinfo {author} {\bibfnamefont {J.}~\bibnamefont {Ojanen}},
  \bibinfo {author} {\bibfnamefont {T.}~\bibnamefont {Olsen}}, \bibinfo
  {author} {\bibfnamefont {V.}~\bibnamefont {Petzold}}, \bibinfo {author}
  {\bibfnamefont {N.~A.}\ \bibnamefont {Romero}}, \bibinfo {author}
  {\bibfnamefont {J.}~\bibnamefont {Stausholm-M{\o}ller}}, \bibinfo {author}
  {\bibfnamefont {M.}~\bibnamefont {Strange}}, \bibinfo {author} {\bibfnamefont
  {G.~A.}\ \bibnamefont {Tritsaris}}, \bibinfo {author} {\bibfnamefont
  {M.}~\bibnamefont {Vanin}}, \bibinfo {author} {\bibfnamefont
  {M.}~\bibnamefont {Walter}}, \bibinfo {author} {\bibfnamefont
  {B.}~\bibnamefont {Hammer}}, \bibinfo {author} {\bibfnamefont
  {H.}~\bibnamefont {H{\"{a}}kkinen}}, \bibinfo {author} {\bibfnamefont
  {G.~K.~H.}\ \bibnamefont {Madsen}}, \bibinfo {author} {\bibfnamefont {R.~M.}\
  \bibnamefont {Nieminen}}, \bibinfo {author} {\bibfnamefont {J.~K.}\
  \bibnamefont {N{\o}rskov}}, \bibinfo {author} {\bibfnamefont
  {M.}~\bibnamefont {Puska}}, \bibinfo {author} {\bibfnamefont {T.~T.}\
  \bibnamefont {Rantala}}, \bibinfo {author} {\bibfnamefont {J.}~\bibnamefont
  {Schi{\o}tz}}, \bibinfo {author} {\bibfnamefont {K.~S.}\ \bibnamefont
  {Thygesen}},\ and\ \bibinfo {author} {\bibfnamefont {K.~W.}\ \bibnamefont
  {Jacobsen}},\ }\bibfield  {title} {\enquote {\bibinfo {title} {{Electronic
  structure calculations with GPAW: a real-space implementation of the
  projector augmented-wave method.}}}\ }\href
  {https://doi.org/10.1088/0953-8984/22/25/253202} {\bibfield  {journal}
  {\bibinfo  {journal} {J. Phys. Condens. Matter}\ }\textbf {\bibinfo {volume}
  {22}},\ \bibinfo {pages} {253202} (\bibinfo {year} {2010})}\BibitemShut
  {NoStop}%
\bibitem [{\citenamefont {Bl{\"{o}}chl}(1994)}]{Bloechl1994_PRB_17953}%
  \BibitemOpen
  \bibfield  {author} {\bibinfo {author} {\bibfnamefont {P.~E.}\ \bibnamefont
  {Bl{\"{o}}chl}},\ }\bibfield  {title} {\enquote {\bibinfo {title} {{Projector
  augmented-wave method}},}\ }\href {https://doi.org/10.1103/PhysRevB.50.17953}
  {\bibfield  {journal} {\bibinfo  {journal} {Phys. Rev. B}\ }\textbf {\bibinfo
  {volume} {50}},\ \bibinfo {pages} {17953--17979} (\bibinfo {year}
  {1994})}\BibitemShut {NoStop}%
\bibitem [{\citenamefont {Monkhorst}\ and\ \citenamefont
  {Pack}(1976)}]{Monkhorst1976_PRB_5188}%
  \BibitemOpen
  \bibfield  {author} {\bibinfo {author} {\bibfnamefont {H.~J.}\ \bibnamefont
  {Monkhorst}}\ and\ \bibinfo {author} {\bibfnamefont {J.~D.}\ \bibnamefont
  {Pack}},\ }\bibfield  {title} {\enquote {\bibinfo {title} {{Special points
  for Brillouin-zone integrations}},}\ }\href
  {https://doi.org/10.1103/PhysRevB.13.5188} {\bibfield  {journal} {\bibinfo
  {journal} {Phys. Rev. B}\ }\textbf {\bibinfo {volume} {13}},\ \bibinfo
  {pages} {5188--5192} (\bibinfo {year} {1976})}\BibitemShut {NoStop}%
\bibitem [{\citenamefont {Ballesteros}, \citenamefont {Dunivan},\ and\
  \citenamefont {Lao}(2021)}]{Ballesteros2021_JCP_154104}%
  \BibitemOpen
  \bibfield  {author} {\bibinfo {author} {\bibfnamefont {F.}~\bibnamefont
  {Ballesteros}}, \bibinfo {author} {\bibfnamefont {S.}~\bibnamefont
  {Dunivan}},\ and\ \bibinfo {author} {\bibfnamefont {K.~U.}\ \bibnamefont
  {Lao}},\ }\bibfield  {title} {\enquote {\bibinfo {title} {Coupled cluster
  benchmarks of large noncovalent complexes: The l7 dataset as well as
  {DNA}{\textendash}ellipticine and buckycatcher{\textendash}fullerene},}\
  }\href {https://doi.org/10.1063/5.0042906} {\bibfield  {journal} {\bibinfo
  {journal} {J. Chem. Phys.}\ }\textbf {\bibinfo {volume} {154}},\ \bibinfo
  {pages} {154104} (\bibinfo {year} {2021})}\BibitemShut {NoStop}%
\bibitem [{\citenamefont {Isert}\ \emph {et~al.}(2022)\citenamefont {Isert},
  \citenamefont {Atz}, \citenamefont {Jim{\'{e}}nez-Luna},\ and\ \citenamefont
  {Schneider}}]{Isert2022_SD_273}%
  \BibitemOpen
  \bibfield  {author} {\bibinfo {author} {\bibfnamefont {C.}~\bibnamefont
  {Isert}}, \bibinfo {author} {\bibfnamefont {K.}~\bibnamefont {Atz}}, \bibinfo
  {author} {\bibfnamefont {J.}~\bibnamefont {Jim{\'{e}}nez-Luna}},\ and\
  \bibinfo {author} {\bibfnamefont {G.}~\bibnamefont {Schneider}},\ }\bibfield
  {title} {\enquote {\bibinfo {title} {{QMugs}, quantum mechanical properties
  of drug-like molecules},}\ }\href
  {https://doi.org/10.1038/s41597-022-01390-7} {\bibfield  {journal} {\bibinfo
  {journal} {Sci. Data}\ }\textbf {\bibinfo {volume} {9}},\ \bibinfo {pages}
  {273} (\bibinfo {year} {2022})}\BibitemShut {NoStop}%
\bibitem [{\citenamefont {Khrabrov}\ \emph {et~al.}(2022)\citenamefont
  {Khrabrov}, \citenamefont {Shenbin}, \citenamefont {Ryabov}, \citenamefont
  {Tsypin}, \citenamefont {Telepov}, \citenamefont {Alekseev}, \citenamefont
  {Grishin}, \citenamefont {Strashnov}, \citenamefont {Zhilyaev}, \citenamefont
  {Nikolenko},\ and\ \citenamefont {Kadurin}}]{Khrabrov2022_PCCP_25853}%
  \BibitemOpen
  \bibfield  {author} {\bibinfo {author} {\bibfnamefont {K.}~\bibnamefont
  {Khrabrov}}, \bibinfo {author} {\bibfnamefont {I.}~\bibnamefont {Shenbin}},
  \bibinfo {author} {\bibfnamefont {A.}~\bibnamefont {Ryabov}}, \bibinfo
  {author} {\bibfnamefont {A.}~\bibnamefont {Tsypin}}, \bibinfo {author}
  {\bibfnamefont {A.}~\bibnamefont {Telepov}}, \bibinfo {author} {\bibfnamefont
  {A.}~\bibnamefont {Alekseev}}, \bibinfo {author} {\bibfnamefont
  {A.}~\bibnamefont {Grishin}}, \bibinfo {author} {\bibfnamefont
  {P.}~\bibnamefont {Strashnov}}, \bibinfo {author} {\bibfnamefont
  {P.}~\bibnamefont {Zhilyaev}}, \bibinfo {author} {\bibfnamefont
  {S.}~\bibnamefont {Nikolenko}},\ and\ \bibinfo {author} {\bibfnamefont
  {A.}~\bibnamefont {Kadurin}},\ }\bibfield  {title} {\enquote {\bibinfo
  {title} {nabla{DFT}: Large-scale conformational energy and {Hamiltonian}
  prediction benchmark and dataset},}\ }\href
  {https://doi.org/10.1039/d2cp03966d} {\bibfield  {journal} {\bibinfo
  {journal} {Phys. Chem. Chem. Phys.}\ }\textbf {\bibinfo {volume} {24}},\
  \bibinfo {pages} {25853--25863} (\bibinfo {year} {2022})}\BibitemShut
  {NoStop}%
\bibitem [{\citenamefont {Eastman}\ \emph {et~al.}(2023)\citenamefont
  {Eastman}, \citenamefont {Behara}, \citenamefont {Dotson}, \citenamefont
  {Galvelis}, \citenamefont {Herr}, \citenamefont {Horton}, \citenamefont
  {Mao}, \citenamefont {Chodera}, \citenamefont {Pritchard}, \citenamefont
  {Wang}, \citenamefont {Fabritiis},\ and\ \citenamefont
  {Markland}}]{Eastman2023_SD_11}%
  \BibitemOpen
  \bibfield  {author} {\bibinfo {author} {\bibfnamefont {P.}~\bibnamefont
  {Eastman}}, \bibinfo {author} {\bibfnamefont {P.~K.}\ \bibnamefont {Behara}},
  \bibinfo {author} {\bibfnamefont {D.~L.}\ \bibnamefont {Dotson}}, \bibinfo
  {author} {\bibfnamefont {R.}~\bibnamefont {Galvelis}}, \bibinfo {author}
  {\bibfnamefont {J.~E.}\ \bibnamefont {Herr}}, \bibinfo {author}
  {\bibfnamefont {J.~T.}\ \bibnamefont {Horton}}, \bibinfo {author}
  {\bibfnamefont {Y.}~\bibnamefont {Mao}}, \bibinfo {author} {\bibfnamefont
  {J.~D.}\ \bibnamefont {Chodera}}, \bibinfo {author} {\bibfnamefont {B.~P.}\
  \bibnamefont {Pritchard}}, \bibinfo {author} {\bibfnamefont {Y.}~\bibnamefont
  {Wang}}, \bibinfo {author} {\bibfnamefont {G.~D.}\ \bibnamefont
  {Fabritiis}},\ and\ \bibinfo {author} {\bibfnamefont {T.~E.}\ \bibnamefont
  {Markland}},\ }\bibfield  {title} {\enquote {\bibinfo {title} {{SPICE}, a
  dataset of drug-like molecules and peptides for training machine learning
  potentials},}\ }\href {https://doi.org/10.1038/s41597-022-01882-6} {\bibfield
   {journal} {\bibinfo  {journal} {Sci. Data}\ }\textbf {\bibinfo {volume}
  {10}},\ \bibinfo {pages} {11} (\bibinfo {year} {2023})}\BibitemShut {NoStop}%
\bibitem [{\citenamefont {Neeser}\ \emph {et~al.}(2023)\citenamefont {Neeser},
  \citenamefont {Isert}, \citenamefont {Stuyver}, \citenamefont {Schneider},\
  and\ \citenamefont {Coley}}]{Neeser2023_CDC_101040}%
  \BibitemOpen
  \bibfield  {author} {\bibinfo {author} {\bibfnamefont {R.~M.}\ \bibnamefont
  {Neeser}}, \bibinfo {author} {\bibfnamefont {C.}~\bibnamefont {Isert}},
  \bibinfo {author} {\bibfnamefont {T.}~\bibnamefont {Stuyver}}, \bibinfo
  {author} {\bibfnamefont {G.}~\bibnamefont {Schneider}},\ and\ \bibinfo
  {author} {\bibfnamefont {C.~W.}\ \bibnamefont {Coley}},\ }\bibfield  {title}
  {\enquote {\bibinfo {title} {{QMugs} 1.1: Quantum mechanical properties of
  organic compounds commonly encountered in reactivity datasets},}\ }\href
  {https://doi.org/10.1016/j.cdc.2023.101040} {\bibfield  {journal} {\bibinfo
  {journal} {Chemical Data Collections}\ }\textbf {\bibinfo {volume} {46}},\
  \bibinfo {pages} {101040} (\bibinfo {year} {2023})}\BibitemShut {NoStop}%
\bibitem [{\citenamefont {Spronk}\ \emph {et~al.}(2023)\citenamefont {Spronk},
  \citenamefont {Glick}, \citenamefont {Metcalf}, \citenamefont {Sherrill},\
  and\ \citenamefont {Cheney}}]{Spronk2023_SD_619}%
  \BibitemOpen
  \bibfield  {author} {\bibinfo {author} {\bibfnamefont {S.~A.}\ \bibnamefont
  {Spronk}}, \bibinfo {author} {\bibfnamefont {Z.~L.}\ \bibnamefont {Glick}},
  \bibinfo {author} {\bibfnamefont {D.~P.}\ \bibnamefont {Metcalf}}, \bibinfo
  {author} {\bibfnamefont {C.~D.}\ \bibnamefont {Sherrill}},\ and\ \bibinfo
  {author} {\bibfnamefont {D.~L.}\ \bibnamefont {Cheney}},\ }\bibfield  {title}
  {\enquote {\bibinfo {title} {A quantum chemical interaction energy dataset
  for accurately modeling protein-ligand interactions},}\ }\href
  {https://doi.org/10.1038/s41597-023-02443-1} {\bibfield  {journal} {\bibinfo
  {journal} {Sci. Data}\ }\textbf {\bibinfo {volume} {10}},\ \bibinfo {pages}
  {619} (\bibinfo {year} {2023})}\BibitemShut {NoStop}%
\bibitem [{\citenamefont {Ullah}, \citenamefont {Chen},\ and\ \citenamefont
  {Dral}(2024)}]{Ullah2024_MLST_41001}%
  \BibitemOpen
  \bibfield  {author} {\bibinfo {author} {\bibfnamefont {A.}~\bibnamefont
  {Ullah}}, \bibinfo {author} {\bibfnamefont {Y.}~\bibnamefont {Chen}},\ and\
  \bibinfo {author} {\bibfnamefont {P.~O.}\ \bibnamefont {Dral}},\ }\bibfield
  {title} {\enquote {\bibinfo {title} {Molecular quantum chemical data sets and
  databases for machine learning potentials},}\ }\href
  {https://doi.org/10.1088/2632-2153/ad8f13} {\bibfield  {journal} {\bibinfo
  {journal} {Mach. Learn.: Sci. Technol.}\ }\textbf {\bibinfo {volume} {5}},\
  \bibinfo {pages} {041001} (\bibinfo {year} {2024})}\BibitemShut {NoStop}%
\bibitem [{\citenamefont {Khan}\ \emph {et~al.}(2025)\citenamefont {Khan},
  \citenamefont {Benali}, \citenamefont {Kim}, \citenamefont {von Rudorff},\
  and\ \citenamefont {von Lilienfeld}}]{Khan2025_SD_1551}%
  \BibitemOpen
  \bibfield  {author} {\bibinfo {author} {\bibfnamefont {D.}~\bibnamefont
  {Khan}}, \bibinfo {author} {\bibfnamefont {A.}~\bibnamefont {Benali}},
  \bibinfo {author} {\bibfnamefont {S.~Y.~H.}\ \bibnamefont {Kim}}, \bibinfo
  {author} {\bibfnamefont {G.~F.}\ \bibnamefont {von Rudorff}},\ and\ \bibinfo
  {author} {\bibfnamefont {O.~A.}\ \bibnamefont {von Lilienfeld}},\ }\bibfield
  {title} {\enquote {\bibinfo {title} {Quantum mechanical dataset of 836k
  neutral closed-shell molecules with up to 5 heavy atoms from {C}, {N}, {O},
  {F}, {Si}, {P}, {S}, {Cl}, {Br}},}\ }\href
  {https://doi.org/10.1038/s41597-025-05428-4} {\bibfield  {journal} {\bibinfo
  {journal} {Sci. Data}\ }\textbf {\bibinfo {volume} {12}},\ \bibinfo {pages}
  {1551} (\bibinfo {year} {2025})}\BibitemShut {NoStop}%
\bibitem [{\citenamefont {Kuryla}\ \emph {et~al.}(2025)\citenamefont {Kuryla},
  \citenamefont {Berger}, \citenamefont {Cs{\'a}nyi},\ and\ \citenamefont
  {Michaelides}}]{Kuryla2025_JCP_224313}%
  \BibitemOpen
  \bibfield  {author} {\bibinfo {author} {\bibfnamefont {D.}~\bibnamefont
  {Kuryla}}, \bibinfo {author} {\bibfnamefont {F.}~\bibnamefont {Berger}},
  \bibinfo {author} {\bibfnamefont {G.}~\bibnamefont {Cs{\'a}nyi}},\ and\
  \bibinfo {author} {\bibfnamefont {A.}~\bibnamefont {Michaelides}},\
  }\bibfield  {title} {\enquote {\bibinfo {title} {How accurate are {DFT}
  forces? {Unexpectedly} large uncertainties in molecular datasets},}\ }\href
  {https://doi.org/10.1063/5.0296997} {\bibfield  {journal} {\bibinfo
  {journal} {J. Chem. Phys.}\ }\textbf {\bibinfo {volume} {163}},\ \bibinfo
  {pages} {224313} (\bibinfo {year} {2025})}\BibitemShut {NoStop}%
\bibitem [{\citenamefont {Zeng}\ \emph {et~al.}(2025)\citenamefont {Zeng},
  \citenamefont {Giese}, \citenamefont {G{\"o}tz},\ and\ \citenamefont
  {York}}]{Zeng2025_SD_693}%
  \BibitemOpen
  \bibfield  {author} {\bibinfo {author} {\bibfnamefont {J.}~\bibnamefont
  {Zeng}}, \bibinfo {author} {\bibfnamefont {T.~J.}\ \bibnamefont {Giese}},
  \bibinfo {author} {\bibfnamefont {A.~W.}\ \bibnamefont {G{\"o}tz}},\ and\
  \bibinfo {author} {\bibfnamefont {D.~M.}\ \bibnamefont {York}},\ }\bibfield
  {title} {\enquote {\bibinfo {title} {The {QD}$\pi$ dataset, training data for
  drug-like molecules and biopolymer fragments and their interactions},}\
  }\href {https://doi.org/10.1038/s41597-025-04972-3} {\bibfield  {journal}
  {\bibinfo  {journal} {Sci. Data}\ }\textbf {\bibinfo {volume} {12}},\
  \bibinfo {pages} {693} (\bibinfo {year} {2025})}\BibitemShut {NoStop}%
\bibitem [{\citenamefont {Sitkiewicz}\ \emph {et~al.}(2022)\citenamefont
  {Sitkiewicz}, \citenamefont {Zale{\'{s}}ny}, \citenamefont {Ramos-Cordoba},
  \citenamefont {Luis},\ and\ \citenamefont
  {Matito}}]{Sitkiewicz2022_JPCL_5963}%
  \BibitemOpen
  \bibfield  {author} {\bibinfo {author} {\bibfnamefont {S.~P.}\ \bibnamefont
  {Sitkiewicz}}, \bibinfo {author} {\bibfnamefont {R.}~\bibnamefont
  {Zale{\'{s}}ny}}, \bibinfo {author} {\bibfnamefont {E.}~\bibnamefont
  {Ramos-Cordoba}}, \bibinfo {author} {\bibfnamefont {J.~M.}\ \bibnamefont
  {Luis}},\ and\ \bibinfo {author} {\bibfnamefont {E.}~\bibnamefont {Matito}},\
  }\bibfield  {title} {\enquote {\bibinfo {title} {How reliable are modern
  density functional approximations to simulate vibrational spectroscopies?}}\
  }\href {https://doi.org/10.1021/acs.jpclett.2c01278} {\bibfield  {journal}
  {\bibinfo  {journal} {J. Phys. Chem. Lett.}\ }\textbf {\bibinfo {volume}
  {13}},\ \bibinfo {pages} {5963--5968} (\bibinfo {year} {2022})}\BibitemShut
  {NoStop}%
\bibitem [{\citenamefont {Lehtola}\ and\ \citenamefont
  {Marques}(2022)}]{Lehtola2022_JCP_174114}%
  \BibitemOpen
  \bibfield  {author} {\bibinfo {author} {\bibfnamefont {S.}~\bibnamefont
  {Lehtola}}\ and\ \bibinfo {author} {\bibfnamefont {M.~A.~L.}\ \bibnamefont
  {Marques}},\ }\bibfield  {title} {\enquote {\bibinfo {title} {Many recent
  density functionals are numerically ill-behaved},}\ }\href
  {https://doi.org/10.1063/5.0121187} {\bibfield  {journal} {\bibinfo
  {journal} {J. Chem. Phys.}\ }\textbf {\bibinfo {volume} {157}},\ \bibinfo
  {pages} {174114} (\bibinfo {year} {2022})}\BibitemShut {NoStop}%
\bibitem [{\citenamefont {Sitkiewicz}\ \emph {et~al.}(2024)\citenamefont
  {Sitkiewicz}, \citenamefont {Ferrad{\'{a}}s}, \citenamefont {Ramos-Cordoba},
  \citenamefont {Zale{\'{s}}ny}, \citenamefont {Matito},\ and\ \citenamefont
  {Luis}}]{Sitkiewicz2024_JCTC_3144}%
  \BibitemOpen
  \bibfield  {author} {\bibinfo {author} {\bibfnamefont {S.~P.}\ \bibnamefont
  {Sitkiewicz}}, \bibinfo {author} {\bibfnamefont {R.~R.}\ \bibnamefont
  {Ferrad{\'{a}}s}}, \bibinfo {author} {\bibfnamefont {E.}~\bibnamefont
  {Ramos-Cordoba}}, \bibinfo {author} {\bibfnamefont {R.}~\bibnamefont
  {Zale{\'{s}}ny}}, \bibinfo {author} {\bibfnamefont {E.}~\bibnamefont
  {Matito}},\ and\ \bibinfo {author} {\bibfnamefont {J.~M.}\ \bibnamefont
  {Luis}},\ }\bibfield  {title} {\enquote {\bibinfo {title} {Spurious
  oscillations caused by density functional approximations: Who is to blame?
  exchange or correlation?}}\ }\href {https://doi.org/10.1021/acs.jctc.3c01339}
  {\bibfield  {journal} {\bibinfo  {journal} {J. Chem. Theory Comput.}\
  }\textbf {\bibinfo {volume} {20}},\ \bibinfo {pages} {3144--3153} (\bibinfo
  {year} {2024})}\BibitemShut {NoStop}%
\bibitem [{\citenamefont {Nielsen}\ and\ \citenamefont
  {Martin}(1985{\natexlab{a}})}]{Nielsen1985_PRB_3780}%
  \BibitemOpen
  \bibfield  {author} {\bibinfo {author} {\bibfnamefont {O.~H.}\ \bibnamefont
  {Nielsen}}\ and\ \bibinfo {author} {\bibfnamefont {R.~M.}\ \bibnamefont
  {Martin}},\ }\bibfield  {title} {\enquote {\bibinfo {title}
  {Quantum-mechanical theory of stress and force},}\ }\href
  {https://doi.org/10.1103/PhysRevB.32.3780} {\bibfield  {journal} {\bibinfo
  {journal} {Phys. Rev. B}\ }\textbf {\bibinfo {volume} {32}},\ \bibinfo
  {pages} {3780--3791} (\bibinfo {year} {1985}{\natexlab{a}})}\BibitemShut
  {NoStop}%
\bibitem [{\citenamefont {Hamann}\ \emph {et~al.}(2005)\citenamefont {Hamann},
  \citenamefont {Wu}, \citenamefont {Rabe},\ and\ \citenamefont
  {Vanderbilt}}]{Hamann2005_PRB_35117}%
  \BibitemOpen
  \bibfield  {author} {\bibinfo {author} {\bibfnamefont {D.~R.}\ \bibnamefont
  {Hamann}}, \bibinfo {author} {\bibfnamefont {X.}~\bibnamefont {Wu}}, \bibinfo
  {author} {\bibfnamefont {K.~M.}\ \bibnamefont {Rabe}},\ and\ \bibinfo
  {author} {\bibfnamefont {D.}~\bibnamefont {Vanderbilt}},\ }\bibfield  {title}
  {\enquote {\bibinfo {title} {Metric tensor formulation of strain in
  density-functional perturbation theory},}\ }\href
  {https://doi.org/10.1103/PhysRevB.71.035117} {\bibfield  {journal} {\bibinfo
  {journal} {Phys. Rev. B}\ }\textbf {\bibinfo {volume} {71}},\ \bibinfo
  {pages} {035117} (\bibinfo {year} {2005})}\BibitemShut {NoStop}%
\bibitem [{\citenamefont {Knuth}\ \emph {et~al.}(2015)\citenamefont {Knuth},
  \citenamefont {Carbogno}, \citenamefont {Atalla}, \citenamefont {Blum},\ and\
  \citenamefont {Scheffler}}]{Knuth2015_CPC_33}%
  \BibitemOpen
  \bibfield  {author} {\bibinfo {author} {\bibfnamefont {F.}~\bibnamefont
  {Knuth}}, \bibinfo {author} {\bibfnamefont {C.}~\bibnamefont {Carbogno}},
  \bibinfo {author} {\bibfnamefont {V.}~\bibnamefont {Atalla}}, \bibinfo
  {author} {\bibfnamefont {V.}~\bibnamefont {Blum}},\ and\ \bibinfo {author}
  {\bibfnamefont {M.}~\bibnamefont {Scheffler}},\ }\bibfield  {title} {\enquote
  {\bibinfo {title} {{All-electron formalism for total energy strain
  derivatives and stress tensor components for numeric atom-centered
  orbitals}},}\ }\href {https://doi.org/10.1016/j.cpc.2015.01.003} {\bibfield
  {journal} {\bibinfo  {journal} {Comput. Phys. Commun.}\ }\textbf {\bibinfo
  {volume} {190}},\ \bibinfo {pages} {33--50} (\bibinfo {year}
  {2015})}\BibitemShut {NoStop}%
\bibitem [{\citenamefont {Nielsen}\ and\ \citenamefont
  {Martin}(1985{\natexlab{b}})}]{Nielsen1985_PRB_3792}%
  \BibitemOpen
  \bibfield  {author} {\bibinfo {author} {\bibfnamefont {O.~H.}\ \bibnamefont
  {Nielsen}}\ and\ \bibinfo {author} {\bibfnamefont {R.~M.}\ \bibnamefont
  {Martin}},\ }\bibfield  {title} {\enquote {\bibinfo {title} {Stresses in
  semiconductors: Ab initio calculations on {Si}, {Ge}, and {GaAs}},}\ }\href
  {https://doi.org/10.1103/PhysRevB.32.3792} {\bibfield  {journal} {\bibinfo
  {journal} {Phys. Rev. B}\ }\textbf {\bibinfo {volume} {32}},\ \bibinfo
  {pages} {3792--3805} (\bibinfo {year} {1985}{\natexlab{b}})}\BibitemShut
  {NoStop}%
\bibitem [{\citenamefont {Walter}\ \emph {et~al.}(2008)\citenamefont {Walter},
  \citenamefont {H{\"a}kkinen}, \citenamefont {Lehtovaara}, \citenamefont
  {Puska}, \citenamefont {Enkovaara}, \citenamefont {Rostgaard},\ and\
  \citenamefont {Mortensen}}]{Walter2008_JCP_244101}%
  \BibitemOpen
  \bibfield  {author} {\bibinfo {author} {\bibfnamefont {M.}~\bibnamefont
  {Walter}}, \bibinfo {author} {\bibfnamefont {H.}~\bibnamefont
  {H{\"a}kkinen}}, \bibinfo {author} {\bibfnamefont {L.}~\bibnamefont
  {Lehtovaara}}, \bibinfo {author} {\bibfnamefont {M.}~\bibnamefont {Puska}},
  \bibinfo {author} {\bibfnamefont {J.}~\bibnamefont {Enkovaara}}, \bibinfo
  {author} {\bibfnamefont {C.}~\bibnamefont {Rostgaard}},\ and\ \bibinfo
  {author} {\bibfnamefont {J.~J.}\ \bibnamefont {Mortensen}},\ }\bibfield
  {title} {\enquote {\bibinfo {title} {Time-dependent density-functional theory
  in the projector augmented-wave method},}\ }\href
  {https://doi.org/10.1063/1.2943138} {\bibfield  {journal} {\bibinfo
  {journal} {J. Chem. Phys.}\ }\textbf {\bibinfo {volume} {128}},\ \bibinfo
  {pages} {244101} (\bibinfo {year} {2008})}\BibitemShut {NoStop}%
\bibitem [{\citenamefont {Kaufmann}, \citenamefont {Baumeister},\ and\
  \citenamefont {Jungen}(1989)}]{Kaufmann1989_JPBAMOP_2223}%
  \BibitemOpen
  \bibfield  {author} {\bibinfo {author} {\bibfnamefont {K.}~\bibnamefont
  {Kaufmann}}, \bibinfo {author} {\bibfnamefont {W.}~\bibnamefont
  {Baumeister}},\ and\ \bibinfo {author} {\bibfnamefont {M.}~\bibnamefont
  {Jungen}},\ }\bibfield  {title} {\enquote {\bibinfo {title} {{Universal
  Gaussian basis sets for an optimum representation of Rydberg and continuum
  wavefunctions}},}\ }\href {https://doi.org/10.1088/0953-4075/22/14/007}
  {\bibfield  {journal} {\bibinfo  {journal} {J. Phys. B: At. Mol. Opt. Phys.}\
  }\textbf {\bibinfo {volume} {22}},\ \bibinfo {pages} {2223--2240} (\bibinfo
  {year} {1989})}\BibitemShut {NoStop}%
\bibitem [{\citenamefont {Yan}\ \emph {et~al.}(2011)\citenamefont {Yan},
  \citenamefont {Mortensen}, \citenamefont {Jacobsen},\ and\ \citenamefont
  {Thygesen}}]{Yan2011_PRB_245122}%
  \BibitemOpen
  \bibfield  {author} {\bibinfo {author} {\bibfnamefont {J.}~\bibnamefont
  {Yan}}, \bibinfo {author} {\bibfnamefont {J.~J.}\ \bibnamefont {Mortensen}},
  \bibinfo {author} {\bibfnamefont {K.~W.}\ \bibnamefont {Jacobsen}},\ and\
  \bibinfo {author} {\bibfnamefont {K.~S.}\ \bibnamefont {Thygesen}},\
  }\bibfield  {title} {\enquote {\bibinfo {title} {Linear density response
  function in the projector augmented wave method: Applications to solids,
  surfaces, and interfaces},}\ }\href
  {https://doi.org/10.1103/PhysRevB.83.245122} {\bibfield  {journal} {\bibinfo
  {journal} {Phys. Rev. B}\ }\textbf {\bibinfo {volume} {83}},\ \bibinfo
  {pages} {245122} (\bibinfo {year} {2011})}\BibitemShut {NoStop}%
\bibitem [{\citenamefont {Stiebling}\ and\ \citenamefont
  {Raether}(1978)}]{Stiebling1978_PRL_1293}%
  \BibitemOpen
  \bibfield  {author} {\bibinfo {author} {\bibfnamefont {J.}~\bibnamefont
  {Stiebling}}\ and\ \bibinfo {author} {\bibfnamefont {H.}~\bibnamefont
  {Raether}},\ }\bibfield  {title} {\enquote {\bibinfo {title} {Dispersion of
  the volume plasmon of silicon (16.7 {eV}) at large wave vectors},}\ }\href
  {https://doi.org/10.1103/PhysRevLett.40.1293} {\bibfield  {journal} {\bibinfo
   {journal} {Phys. Rev. Lett.}\ }\textbf {\bibinfo {volume} {40}},\ \bibinfo
  {pages} {1293--1295} (\bibinfo {year} {1978})}\BibitemShut {NoStop}%
\bibitem [{\citenamefont {Lehtola}\ \emph {et~al.}(2012)\citenamefont
  {Lehtola}, \citenamefont {Hakala}, \citenamefont {Sakko},\ and\ \citenamefont
  {H{\"{a}}m{\"{a}}l{\"{a}}inen}}]{Lehtola2012_JCC_1572}%
  \BibitemOpen
  \bibfield  {author} {\bibinfo {author} {\bibfnamefont {J.}~\bibnamefont
  {Lehtola}}, \bibinfo {author} {\bibfnamefont {M.}~\bibnamefont {Hakala}},
  \bibinfo {author} {\bibfnamefont {A.}~\bibnamefont {Sakko}},\ and\ \bibinfo
  {author} {\bibfnamefont {K.}~\bibnamefont {H{\"{a}}m{\"{a}}l{\"{a}}inen}},\
  }\bibfield  {title} {\enquote {\bibinfo {title} {{ERKALE} -- a flexible
  program package for x-ray properties of atoms and molecules},}\ }\href
  {https://doi.org/10.1002/jcc.22987} {\bibfield  {journal} {\bibinfo
  {journal} {J. Comput. Chem.}\ }\textbf {\bibinfo {volume} {33}},\ \bibinfo
  {pages} {1572--1585} (\bibinfo {year} {2012})}\BibitemShut {NoStop}%
\bibitem [{\citenamefont {Lehtola}, \citenamefont {Dimitrova},\ and\
  \citenamefont {Sundholm}(2020)}]{Lehtola2020_MP_1597989}%
  \BibitemOpen
  \bibfield  {author} {\bibinfo {author} {\bibfnamefont {S.}~\bibnamefont
  {Lehtola}}, \bibinfo {author} {\bibfnamefont {M.}~\bibnamefont {Dimitrova}},\
  and\ \bibinfo {author} {\bibfnamefont {D.}~\bibnamefont {Sundholm}},\
  }\bibfield  {title} {\enquote {\bibinfo {title} {Fully numerical electronic
  structure calculations on diatomic molecules in weak to strong magnetic
  fields},}\ }\href {https://doi.org/10.1080/00268976.2019.1597989} {\bibfield
  {journal} {\bibinfo  {journal} {Mol. Phys.}\ }\textbf {\bibinfo {volume}
  {118}},\ \bibinfo {pages} {e1597989} (\bibinfo {year} {2020})},\ \Eprint
  {https://arxiv.org/abs/1812.06274} {arXiv:1812.06274} \BibitemShut {NoStop}%
\bibitem [{\citenamefont {Lehtola}(2020)}]{Lehtola2020_PRA_12516}%
  \BibitemOpen
  \bibfield  {author} {\bibinfo {author} {\bibfnamefont {S.}~\bibnamefont
  {Lehtola}},\ }\bibfield  {title} {\enquote {\bibinfo {title} {Fully numerical
  calculations on atoms with fractional occupations and range-separated
  exchange functionals},}\ }\href {https://doi.org/10.1103/PhysRevA.101.012516}
  {\bibfield  {journal} {\bibinfo  {journal} {Phys. Rev. A}\ }\textbf {\bibinfo
  {volume} {101}},\ \bibinfo {pages} {012516} (\bibinfo {year} {2020})},\
  \Eprint {https://arxiv.org/abs/1908.02528} {arXiv:1908.02528} \BibitemShut
  {NoStop}%
\bibitem [{\citenamefont
  {Lehtola}(2023{\natexlab{b}})}]{Lehtola2023_JCTC_2502}%
  \BibitemOpen
  \bibfield  {author} {\bibinfo {author} {\bibfnamefont {S.}~\bibnamefont
  {Lehtola}},\ }\bibfield  {title} {\enquote {\bibinfo {title} {Meta-{GGA}
  density functional calculations on atoms with spherically symmetric densities
  in the finite element formalism},}\ }\href
  {https://doi.org/10.1021/acs.jctc.3c00183} {\bibfield  {journal} {\bibinfo
  {journal} {J. Chem. Theory Comput.}\ }\textbf {\bibinfo {volume} {19}},\
  \bibinfo {pages} {2502--2517} (\bibinfo {year} {2023}{\natexlab{b}})},\
  \Eprint {https://arxiv.org/abs/2302.06284} {2302.06284} \BibitemShut
  {NoStop}%
\bibitem [{\citenamefont
  {Lehtola}(2023{\natexlab{c}})}]{Lehtola2023_JCTC_4033}%
  \BibitemOpen
  \bibfield  {author} {\bibinfo {author} {\bibfnamefont {S.}~\bibnamefont
  {Lehtola}},\ }\bibfield  {title} {\enquote {\bibinfo {title} {Accuracy of a
  recent regularized nuclear potential},}\ }\href
  {https://doi.org/10.1021/acs.jctc.3c00530} {\bibfield  {journal} {\bibinfo
  {journal} {J. Chem. Theory Comput.}\ }\textbf {\bibinfo {volume} {19}},\
  \bibinfo {pages} {4033--4039} (\bibinfo {year} {2023}{\natexlab{c}})},\
  \Eprint {https://arxiv.org/abs/2302.09557} {2302.09557} \BibitemShut
  {NoStop}%
\bibitem [{\citenamefont
  {Lehtola}(2023{\natexlab{d}})}]{Lehtola2023_JPCA_4180}%
  \BibitemOpen
  \bibfield  {author} {\bibinfo {author} {\bibfnamefont {S.}~\bibnamefont
  {Lehtola}},\ }\bibfield  {title} {\enquote {\bibinfo {title} {Atomic
  electronic structure calculations with {Hermite} interpolating
  polynomials},}\ }\href {https://doi.org/10.1021/acs.jpca.3c00729} {\bibfield
  {journal} {\bibinfo  {journal} {J. Phys. Chem. A}\ }\textbf {\bibinfo
  {volume} {127}},\ \bibinfo {pages} {4180--4193} (\bibinfo {year}
  {2023}{\natexlab{d}})},\ \Eprint {https://arxiv.org/abs/2302.00440}
  {2302.00440} \BibitemShut {NoStop}%
\bibitem [{\citenamefont {Sun}\ \emph {et~al.}(2020)\citenamefont {Sun},
  \citenamefont {Zhang}, \citenamefont {Banerjee}, \citenamefont {Bao},
  \citenamefont {Barbry}, \citenamefont {Blunt}, \citenamefont {Bogdanov},
  \citenamefont {Booth}, \citenamefont {Chen}, \citenamefont {Cui},
  \citenamefont {Eriksen}, \citenamefont {Gao}, \citenamefont {Guo},
  \citenamefont {Hermann}, \citenamefont {Hermes}, \citenamefont {Koh},
  \citenamefont {Koval}, \citenamefont {Lehtola}, \citenamefont {Li},
  \citenamefont {Liu}, \citenamefont {Mardirossian}, \citenamefont {McClain},
  \citenamefont {Motta}, \citenamefont {Mussard}, \citenamefont {Pham},
  \citenamefont {Pulkin}, \citenamefont {Purwanto}, \citenamefont {Robinson},
  \citenamefont {Ronca}, \citenamefont {Sayfutyarova}, \citenamefont
  {Scheurer}, \citenamefont {Schurkus}, \citenamefont {Smith}, \citenamefont
  {Sun}, \citenamefont {Sun}, \citenamefont {Upadhyay}, \citenamefont {Wagner},
  \citenamefont {Wang}, \citenamefont {White}, \citenamefont {Whitfield},
  \citenamefont {Williamson}, \citenamefont {Wouters}, \citenamefont {Yang},
  \citenamefont {Yu}, \citenamefont {Zhu}, \citenamefont {Berkelbach},
  \citenamefont {Sharma}, \citenamefont {Sokolov},\ and\ \citenamefont
  {Chan}}]{Sun2020_JCP_24109}%
  \BibitemOpen
  \bibfield  {author} {\bibinfo {author} {\bibfnamefont {Q.}~\bibnamefont
  {Sun}}, \bibinfo {author} {\bibfnamefont {X.}~\bibnamefont {Zhang}}, \bibinfo
  {author} {\bibfnamefont {S.}~\bibnamefont {Banerjee}}, \bibinfo {author}
  {\bibfnamefont {P.}~\bibnamefont {Bao}}, \bibinfo {author} {\bibfnamefont
  {M.}~\bibnamefont {Barbry}}, \bibinfo {author} {\bibfnamefont {N.~S.}\
  \bibnamefont {Blunt}}, \bibinfo {author} {\bibfnamefont {N.~A.}\ \bibnamefont
  {Bogdanov}}, \bibinfo {author} {\bibfnamefont {G.~H.}\ \bibnamefont {Booth}},
  \bibinfo {author} {\bibfnamefont {J.}~\bibnamefont {Chen}}, \bibinfo {author}
  {\bibfnamefont {Z.-H.}\ \bibnamefont {Cui}}, \bibinfo {author} {\bibfnamefont
  {J.~J.}\ \bibnamefont {Eriksen}}, \bibinfo {author} {\bibfnamefont
  {Y.}~\bibnamefont {Gao}}, \bibinfo {author} {\bibfnamefont {S.}~\bibnamefont
  {Guo}}, \bibinfo {author} {\bibfnamefont {J.}~\bibnamefont {Hermann}},
  \bibinfo {author} {\bibfnamefont {M.~R.}\ \bibnamefont {Hermes}}, \bibinfo
  {author} {\bibfnamefont {K.}~\bibnamefont {Koh}}, \bibinfo {author}
  {\bibfnamefont {P.}~\bibnamefont {Koval}}, \bibinfo {author} {\bibfnamefont
  {S.}~\bibnamefont {Lehtola}}, \bibinfo {author} {\bibfnamefont
  {Z.}~\bibnamefont {Li}}, \bibinfo {author} {\bibfnamefont {J.}~\bibnamefont
  {Liu}}, \bibinfo {author} {\bibfnamefont {N.}~\bibnamefont {Mardirossian}},
  \bibinfo {author} {\bibfnamefont {J.~D.}\ \bibnamefont {McClain}}, \bibinfo
  {author} {\bibfnamefont {M.}~\bibnamefont {Motta}}, \bibinfo {author}
  {\bibfnamefont {B.}~\bibnamefont {Mussard}}, \bibinfo {author} {\bibfnamefont
  {H.~Q.}\ \bibnamefont {Pham}}, \bibinfo {author} {\bibfnamefont
  {A.}~\bibnamefont {Pulkin}}, \bibinfo {author} {\bibfnamefont
  {W.}~\bibnamefont {Purwanto}}, \bibinfo {author} {\bibfnamefont {P.~J.}\
  \bibnamefont {Robinson}}, \bibinfo {author} {\bibfnamefont {E.}~\bibnamefont
  {Ronca}}, \bibinfo {author} {\bibfnamefont {E.~R.}\ \bibnamefont
  {Sayfutyarova}}, \bibinfo {author} {\bibfnamefont {M.}~\bibnamefont
  {Scheurer}}, \bibinfo {author} {\bibfnamefont {H.~F.}\ \bibnamefont
  {Schurkus}}, \bibinfo {author} {\bibfnamefont {J.~E.~T.}\ \bibnamefont
  {Smith}}, \bibinfo {author} {\bibfnamefont {C.}~\bibnamefont {Sun}}, \bibinfo
  {author} {\bibfnamefont {S.-N.}\ \bibnamefont {Sun}}, \bibinfo {author}
  {\bibfnamefont {S.}~\bibnamefont {Upadhyay}}, \bibinfo {author}
  {\bibfnamefont {L.~K.}\ \bibnamefont {Wagner}}, \bibinfo {author}
  {\bibfnamefont {X.}~\bibnamefont {Wang}}, \bibinfo {author} {\bibfnamefont
  {A.}~\bibnamefont {White}}, \bibinfo {author} {\bibfnamefont {J.~D.}\
  \bibnamefont {Whitfield}}, \bibinfo {author} {\bibfnamefont {M.~J.}\
  \bibnamefont {Williamson}}, \bibinfo {author} {\bibfnamefont
  {S.}~\bibnamefont {Wouters}}, \bibinfo {author} {\bibfnamefont
  {J.}~\bibnamefont {Yang}}, \bibinfo {author} {\bibfnamefont {J.~M.}\
  \bibnamefont {Yu}}, \bibinfo {author} {\bibfnamefont {T.}~\bibnamefont
  {Zhu}}, \bibinfo {author} {\bibfnamefont {T.~C.}\ \bibnamefont {Berkelbach}},
  \bibinfo {author} {\bibfnamefont {S.}~\bibnamefont {Sharma}}, \bibinfo
  {author} {\bibfnamefont {A.~Y.}\ \bibnamefont {Sokolov}},\ and\ \bibinfo
  {author} {\bibfnamefont {G.~K.-L.}\ \bibnamefont {Chan}},\ }\bibfield
  {title} {\enquote {\bibinfo {title} {Recent developments in the
  \textsc{PySCF} program package},}\ }\href {https://doi.org/10.1063/5.0006074}
  {\bibfield  {journal} {\bibinfo  {journal} {J. Chem. Phys.}\ }\textbf
  {\bibinfo {volume} {153}},\ \bibinfo {pages} {024109} (\bibinfo {year}
  {2020})},\ \Eprint {https://arxiv.org/abs/2002.12531} {arXiv:2002.12531}
  \BibitemShut {NoStop}%
\bibitem [{\citenamefont {Sharma}\ \emph {et~al.}(2011)\citenamefont {Sharma},
  \citenamefont {Dewhurst}, \citenamefont {Sanna},\ and\ \citenamefont
  {Gross}}]{Sharma2011_PRL_186401}%
  \BibitemOpen
  \bibfield  {author} {\bibinfo {author} {\bibfnamefont {S.}~\bibnamefont
  {Sharma}}, \bibinfo {author} {\bibfnamefont {J.~K.}\ \bibnamefont
  {Dewhurst}}, \bibinfo {author} {\bibfnamefont {A.}~\bibnamefont {Sanna}},\
  and\ \bibinfo {author} {\bibfnamefont {E.~K.~U.}\ \bibnamefont {Gross}},\
  }\bibfield  {title} {\enquote {\bibinfo {title} {Bootstrap approximation for
  the exchange-correlation kernel of time-dependent density-functional
  theory},}\ }\href {https://doi.org/10.1103/PhysRevLett.107.186401} {\bibfield
   {journal} {\bibinfo  {journal} {Phys. Rev. Lett.}\ }\textbf {\bibinfo
  {volume} {107}},\ \bibinfo {pages} {186401} (\bibinfo {year}
  {2011})}\BibitemShut {NoStop}%
\bibitem [{\citenamefont {Baroni}\ \emph {et~al.}(2001)\citenamefont {Baroni},
  \citenamefont {de~Gironcoli}, \citenamefont {Corso},\ and\ \citenamefont
  {Giannozzi}}]{Baroni2001_RMP_515}%
  \BibitemOpen
  \bibfield  {author} {\bibinfo {author} {\bibfnamefont {S.}~\bibnamefont
  {Baroni}}, \bibinfo {author} {\bibfnamefont {S.}~\bibnamefont
  {de~Gironcoli}}, \bibinfo {author} {\bibfnamefont {A.~D.}\ \bibnamefont
  {Corso}},\ and\ \bibinfo {author} {\bibfnamefont {P.}~\bibnamefont
  {Giannozzi}},\ }\bibfield  {title} {\enquote {\bibinfo {title} {Phonons and
  related crystal properties from density-functional perturbation theory},}\
  }\href {https://doi.org/10.1103/RevModPhys.73.515} {\bibfield  {journal}
  {\bibinfo  {journal} {Rev. Mod. Phys.}\ }\textbf {\bibinfo {volume} {73}},\
  \bibinfo {pages} {515--562} (\bibinfo {year} {2001})}\BibitemShut {NoStop}%
\bibitem [{\citenamefont {Gonze}(1997)}]{Gonze1997_PRB_10337}%
  \BibitemOpen
  \bibfield  {author} {\bibinfo {author} {\bibfnamefont {X.}~\bibnamefont
  {Gonze}},\ }\bibfield  {title} {\enquote {\bibinfo {title} {First-principles
  responses of solids to atomic displacements and homogeneous electric fields:
  Implementation of a conjugate-gradient algorithm},}\ }\href
  {https://doi.org/10.1103/PhysRevB.55.10337} {\bibfield  {journal} {\bibinfo
  {journal} {Phys. Rev. B}\ }\textbf {\bibinfo {volume} {55}},\ \bibinfo
  {pages} {10337--10354} (\bibinfo {year} {1997})}\BibitemShut {NoStop}%
\bibitem [{\citenamefont {de~Gironcoli}(1995)}]{Gironcoli1995_PRB_6773}%
  \BibitemOpen
  \bibfield  {author} {\bibinfo {author} {\bibfnamefont {S.}~\bibnamefont
  {de~Gironcoli}},\ }\bibfield  {title} {\enquote {\bibinfo {title} {Lattice
  dynamics of metals from density-functional perturbation theory},}\ }\href
  {https://doi.org/10.1103/PhysRevB.51.6773} {\bibfield  {journal} {\bibinfo
  {journal} {Phys. Rev. B}\ }\textbf {\bibinfo {volume} {51}},\ \bibinfo
  {pages} {6773--6776} (\bibinfo {year} {1995})}\BibitemShut {NoStop}%
\bibitem [{\citenamefont {Herbst}, \citenamefont {Levitt},\ and\ \citenamefont
  {Canc{\`{e}}s}(2021)}]{Herbst2021_JP_69}%
  \BibitemOpen
  \bibfield  {author} {\bibinfo {author} {\bibfnamefont {M.~F.}\ \bibnamefont
  {Herbst}}, \bibinfo {author} {\bibfnamefont {A.}~\bibnamefont {Levitt}},\
  and\ \bibinfo {author} {\bibfnamefont {E.}~\bibnamefont {Canc{\`{e}}s}},\
  }\bibfield  {title} {\enquote {\bibinfo {title} {{DFTK}: A {Julian} approach
  for simulating electrons in solids},}\ }\href
  {https://doi.org/10.21105/jcon.00069} {\bibfield  {journal} {\bibinfo
  {journal} {{JuliaCon} Proceedings}\ }\textbf {\bibinfo {volume} {3}},\
  \bibinfo {pages} {69} (\bibinfo {year} {2021})}\BibitemShut {NoStop}%
\bibitem [{\citenamefont {Schmitz}, \citenamefont {Ploumhans},\ and\
  \citenamefont {Herbst}(2025)}]{Schmitz2025_nCM_6}%
  \BibitemOpen
  \bibfield  {author} {\bibinfo {author} {\bibfnamefont {N.~F.}\ \bibnamefont
  {Schmitz}}, \bibinfo {author} {\bibfnamefont {B.}~\bibnamefont {Ploumhans}},\
  and\ \bibinfo {author} {\bibfnamefont {M.~F.}\ \bibnamefont {Herbst}},\
  }\bibfield  {title} {\enquote {\bibinfo {title} {Algorithmic differentiation
  for plane-wave {DFT}: materials design, error control and learning model
  parameters},}\ }\href {https://doi.org/10.1038/s41524-025-01880-3} {\bibfield
   {journal} {\bibinfo  {journal} {npj Comput. Mater.}\ }\textbf {\bibinfo
  {volume} {12}},\ \bibinfo {pages} {6} (\bibinfo {year} {2025})}\BibitemShut
  {NoStop}%
\bibitem [{\citenamefont {Arbuznikov}\ and\ \citenamefont
  {Kaupp}(2014)}]{Arbuznikov2014_JCP_204101}%
  \BibitemOpen
  \bibfield  {author} {\bibinfo {author} {\bibfnamefont {A.~V.}\ \bibnamefont
  {Arbuznikov}}\ and\ \bibinfo {author} {\bibfnamefont {M.}~\bibnamefont
  {Kaupp}},\ }\bibfield  {title} {\enquote {\bibinfo {title} {{Towards improved
  local hybrid functionals by calibration of exchange-energy densities}},}\
  }\href {https://doi.org/10.1063/1.4901238} {\bibfield  {journal} {\bibinfo
  {journal} {J. Chem. Phys.}\ }\textbf {\bibinfo {volume} {141}},\ \bibinfo
  {pages} {204101} (\bibinfo {year} {2014})}\BibitemShut {NoStop}%
\bibitem [{\citenamefont {Maier}\ \emph {et~al.}(2016)\citenamefont {Maier},
  \citenamefont {Haasler}, \citenamefont {Arbuznikov},\ and\ \citenamefont
  {Kaupp}}]{Maier2016_PCCP_21133}%
  \BibitemOpen
  \bibfield  {author} {\bibinfo {author} {\bibfnamefont {T.~M.}\ \bibnamefont
  {Maier}}, \bibinfo {author} {\bibfnamefont {M.}~\bibnamefont {Haasler}},
  \bibinfo {author} {\bibfnamefont {A.~V.}\ \bibnamefont {Arbuznikov}},\ and\
  \bibinfo {author} {\bibfnamefont {M.}~\bibnamefont {Kaupp}},\ }\bibfield
  {title} {\enquote {\bibinfo {title} {{New approaches for the calibration of
  exchange-energy densities in local hybrid functionals}},}\ }\href
  {https://doi.org/10.1039/C6CP00990E} {\bibfield  {journal} {\bibinfo
  {journal} {Phys. Chem. Chem. Phys.}\ }\textbf {\bibinfo {volume} {18}},\
  \bibinfo {pages} {21133--21144} (\bibinfo {year} {2016})}\BibitemShut
  {NoStop}%
\bibitem [{\citenamefont {Bahmann}\ and\ \citenamefont
  {Kaupp}(2015)}]{Bahmann2015_JCTC_1540}%
  \BibitemOpen
  \bibfield  {author} {\bibinfo {author} {\bibfnamefont {H.}~\bibnamefont
  {Bahmann}}\ and\ \bibinfo {author} {\bibfnamefont {M.}~\bibnamefont
  {Kaupp}},\ }\bibfield  {title} {\enquote {\bibinfo {title} {Efficient
  self-consistent implementation of local hybrid functionals},}\ }\href
  {https://doi.org/10.1021/ct501137x} {\bibfield  {journal} {\bibinfo
  {journal} {J. Chem. Theory Comput.}\ }\textbf {\bibinfo {volume} {11}},\
  \bibinfo {pages} {1540--1548} (\bibinfo {year} {2015})}\BibitemShut {NoStop}%
\bibitem [{\citenamefont {Maier}, \citenamefont {Bahmann},\ and\ \citenamefont
  {Kaupp}(2015)}]{Maier2015_JCTC_4226}%
  \BibitemOpen
  \bibfield  {author} {\bibinfo {author} {\bibfnamefont {T.~M.}\ \bibnamefont
  {Maier}}, \bibinfo {author} {\bibfnamefont {H.}~\bibnamefont {Bahmann}},\
  and\ \bibinfo {author} {\bibfnamefont {M.}~\bibnamefont {Kaupp}},\ }\bibfield
   {title} {\enquote {\bibinfo {title} {Efficient semi-numerical implementation
  of global and local hybrid functionals for time-dependent density functional
  theory},}\ }\href {https://doi.org/10.1021/acs.jctc.5b00624} {\bibfield
  {journal} {\bibinfo  {journal} {J. Chem. Theory Comput.}\ }\textbf {\bibinfo
  {volume} {11}},\ \bibinfo {pages} {4226--4237} (\bibinfo {year}
  {2015})}\BibitemShut {NoStop}%
\bibitem [{\citenamefont {Grotjahn}, \citenamefont {Furche},\ and\
  \citenamefont {Kaupp}(2019)}]{Grotjahn2019_JCTC_5508}%
  \BibitemOpen
  \bibfield  {author} {\bibinfo {author} {\bibfnamefont {R.}~\bibnamefont
  {Grotjahn}}, \bibinfo {author} {\bibfnamefont {F.}~\bibnamefont {Furche}},\
  and\ \bibinfo {author} {\bibfnamefont {M.}~\bibnamefont {Kaupp}},\ }\bibfield
   {title} {\enquote {\bibinfo {title} {Development and implementation of
  excited-state gradients for local hybrid functionals},}\ }\href
  {https://doi.org/10.1021/acs.jctc.9b00659} {\bibfield  {journal} {\bibinfo
  {journal} {J. Chem. Theory Comput.}\ }\textbf {\bibinfo {volume} {15}},\
  \bibinfo {pages} {5508--5522} (\bibinfo {year} {2019})}\BibitemShut {NoStop}%
\bibitem [{\citenamefont {F{\"u}rst}\ \emph {et~al.}(2023)\citenamefont
  {F{\"u}rst}, \citenamefont {Haasler}, \citenamefont {Grotjahn},\ and\
  \citenamefont {Kaupp}}]{Fuerst2023_JCTC_488}%
  \BibitemOpen
  \bibfield  {author} {\bibinfo {author} {\bibfnamefont {S.}~\bibnamefont
  {F{\"u}rst}}, \bibinfo {author} {\bibfnamefont {M.}~\bibnamefont {Haasler}},
  \bibinfo {author} {\bibfnamefont {R.}~\bibnamefont {Grotjahn}},\ and\
  \bibinfo {author} {\bibfnamefont {M.}~\bibnamefont {Kaupp}},\ }\bibfield
  {title} {\enquote {\bibinfo {title} {Full implementation, optimization, and
  evaluation of a range-separated local hybrid functional with wide accuracy
  for ground and excited states},}\ }\href
  {https://doi.org/10.1021/acs.jctc.2c00782} {\bibfield  {journal} {\bibinfo
  {journal} {J. Chem. Theory Comput.}\ }\textbf {\bibinfo {volume} {19}},\
  \bibinfo {pages} {488--502} (\bibinfo {year} {2023})}\BibitemShut {NoStop}%
\bibitem [{\citenamefont {Becke}(2002)}]{Becke2002_JCP_6935}%
  \BibitemOpen
  \bibfield  {author} {\bibinfo {author} {\bibfnamefont {A.~D.}\ \bibnamefont
  {Becke}},\ }\bibfield  {title} {\enquote {\bibinfo {title} {Current density
  in exchange-correlation functionals: Application to atomic states},}\ }\href
  {https://doi.org/10.1063/1.1503772} {\bibfield  {journal} {\bibinfo
  {journal} {J. Chem. Phys.}\ }\textbf {\bibinfo {volume} {117}},\ \bibinfo
  {pages} {6935--6938} (\bibinfo {year} {2002})}\BibitemShut {NoStop}%
\bibitem [{\citenamefont {Bersson}, \citenamefont {Kovtun},\ and\ \citenamefont
  {Li}(2026)}]{Bersson2026_PCCP_}%
  \BibitemOpen
  \bibfield  {author} {\bibinfo {author} {\bibfnamefont {J.~S.}\ \bibnamefont
  {Bersson}}, \bibinfo {author} {\bibfnamefont {M.}~\bibnamefont {Kovtun}},\
  and\ \bibinfo {author} {\bibfnamefont {X.}~\bibnamefont {Li}},\ }\bibfield
  {title} {\enquote {\bibinfo {title} {Four-component relativistic density
  functional theory: grid requirements and small-component contributions},}\
  }\href {https://doi.org/10.1039/d6cp02182d} {\bibfield  {journal} {\bibinfo
  {journal} {Phys. Chem. Chem. Phys.}\ } (\bibinfo {year} {2026}),\
  10.1039/d6cp02182d}\BibitemShut {NoStop}%
\bibitem [{\citenamefont
  {Lehtola}(2019{\natexlab{c}})}]{Lehtola2019_IJQC_25968}%
  \BibitemOpen
  \bibfield  {author} {\bibinfo {author} {\bibfnamefont {S.}~\bibnamefont
  {Lehtola}},\ }\bibfield  {title} {\enquote {\bibinfo {title} {A review on
  non-relativistic, fully numerical electronic structure calculations on atoms
  and diatomic molecules},}\ }\href {https://doi.org/10.1002/qua.25968}
  {\bibfield  {journal} {\bibinfo  {journal} {Int. J. Quantum Chem.}\ }\textbf
  {\bibinfo {volume} {119}},\ \bibinfo {pages} {e25968} (\bibinfo {year}
  {2019}{\natexlab{c}})},\ \Eprint {https://arxiv.org/abs/1902.01431}
  {arXiv:1902.01431} \BibitemShut {NoStop}%
\bibitem [{\citenamefont {Gygi}(1993)}]{Gygi1993_PRB_11692}%
  \BibitemOpen
  \bibfield  {author} {\bibinfo {author} {\bibfnamefont {F.}~\bibnamefont
  {Gygi}},\ }\bibfield  {title} {\enquote {\bibinfo {title}
  {Electronic-structure calculations in adaptive coordinates},}\ }\href
  {https://doi.org/10.1103/PhysRevB.48.11692} {\bibfield  {journal} {\bibinfo
  {journal} {Phys. Rev. B}\ }\textbf {\bibinfo {volume} {48}},\ \bibinfo
  {pages} {11692--11700} (\bibinfo {year} {1993})}\BibitemShut {NoStop}%
\bibitem [{\citenamefont {Hamann}(1995)}]{Hamann1995_PRB_7337}%
  \BibitemOpen
  \bibfield  {author} {\bibinfo {author} {\bibfnamefont {D.~R.}\ \bibnamefont
  {Hamann}},\ }\bibfield  {title} {\enquote {\bibinfo {title} {Application of
  adaptive curvilinear coordinates to the electronic structure of solids},}\
  }\href {https://doi.org/10.1103/PhysRevB.51.7337} {\bibfield  {journal}
  {\bibinfo  {journal} {Phys. Rev. B}\ }\textbf {\bibinfo {volume} {51}},\
  \bibinfo {pages} {7337--7340} (\bibinfo {year} {1995})}\BibitemShut {NoStop}%
\bibitem [{\citenamefont {Bates}\ and\ \citenamefont
  {Furche}(2012)}]{Bates2012_JCP_164105}%
  \BibitemOpen
  \bibfield  {author} {\bibinfo {author} {\bibfnamefont {J.~E.}\ \bibnamefont
  {Bates}}\ and\ \bibinfo {author} {\bibfnamefont {F.}~\bibnamefont {Furche}},\
  }\bibfield  {title} {\enquote {\bibinfo {title} {{Harnessing the
  meta-generalized gradient approximation for time-dependent density functional
  theory.}}}\ }\href {https://doi.org/10.1063/1.4759080} {\bibfield  {journal}
  {\bibinfo  {journal} {J. Chem. Phys.}\ }\textbf {\bibinfo {volume} {137}},\
  \bibinfo {pages} {164105} (\bibinfo {year} {2012})}\BibitemShut {NoStop}%
\end{thebibliography}%

\end{document}